\pdfoutput=1 
\documentclass[botnum, fleqn]{unmeethesis}

\usepackage{amsmath,amsfonts,amssymb,amstext}
\usepackage{graphicx}
\usepackage[colorlinks,linkcolor=blue]{hyperref}
\usepackage{multirow}
\usepackage{lmodern}
\usepackage[numbered,autolinebreaks,useliterate]{mcode}
\usepackage{listings}

\newcommand{\ket}[1]{\left|#1\right\rangle}
\newcommand{\bra}[1]{\left\langle#1\right|}
\newcommand{\braket}[2]{\left\langle#1\right|\left.#2\right\rangle}
\newcommand{\ketbra}[2]{\left|#1\right\rangle\!\left\langle#2\right|}

\newcommand{\avg}[1]{\left\langle #1 \right\rangle}

\def\cO{{\cal O}}
\DeclareMathOperator{\Tr}{Tr}

\def\({\left(}
\def\){\right)}

\renewcommand{\vec}[1]{\mathbf{#1}}

\def\su{\mathfrak{su}}

\newcommand{\be}{\begin{equation}}
\newcommand{\ee}{\end{equation}}
\newcommand{\bea}{\begin{eqnarray}}
\newcommand{\eea}{\end{eqnarray}}
\newcommand{\bF}{\begin{figure}}
\newcommand{\eF}{\end{figure}}
\newcommand{\dg}{\dagger}
\newcommand{\bi}{\begin{itemize}}
\newcommand{\ei}{\end{itemize}}
\newcommand{\ud}{\mathrm{d}}
\newcommand{\mbf}[1]{\mathbf{#1}}

\begin{document}

\frontmatter

\title{Continuous Measurement Quantum\\ State Tomography of Atomic Ensembles}

\author{Carlos A. Riofr\'io Almeida}

\degreesubject{Ph.D., Physics}

\degree{Doctor of Philosophy \\ Physics}

\documenttype{Dissertation}

\previousdegrees{F\'isico, Escuela Polit\'ecnica Nacional, Quito, Ecuador, 2003 \\
                 M.S., Physics, University of New Mexico, 2007}

\date{December, \thisyear}

\maketitle

\makecopyright

\begin{dedication}
To my parents Carlos and Cristina,\\
\& to Heather,\\
for their love and support
\end{dedication}

\begin{acknowledgments}
   \vspace{1.1in}
   First and most importantly, I would like to thank my advisor Prof. Ivan Deutsch for having given me the opportunity to work in his group and for being a fundamental part in the research that lead to this dissertation. I would also want to thank my collaborators at the University of Arizona, Prof. Poul Jessen, and his students Aaron Smith, Brian Anderson, and H\'ector Sosa for all their help in getting this project working and for producing an amazing experiment that demonstrated the theory discussed in this work. Part of this dissertation was also in collaboration with Seth Merkel and Steve Flammia, who I also thank for their help, comments and for being always open to discuss with me any doubts and questions I may have had.

During my years as a PhD student in the Group of Quantum Information at the University of New Mexico I had the opportunity to meet and talk to a number of people that in one way or another were of great help in my education. I would like to thank Prof. Carl Caves, Rolando Somma, Prof. Andrew Landahl, Robin Blume-Kohout, Vaibhav Madhok, Alex Tacla, Collin Trail, Brian Mischuck, Josh Combes, Rob Cook, Ben Baragiola, Leigh Norris, Krittika Goyal, Iris Reichenbach, Anil Shaji, and all the CQuIC center.

I also want to thank Ziya Kalay and Steve Tremblay for their friendship through all these years. It would not have been as fun as it was without them. Many thanks also go to Daniel Mirell for helping me so much during my first months in the USA. I want to also thank the Fulbright Program for making possible for me to come to study in the USA and pursue my dreams. Special thanks to my brother Daniel for his support and help in computer related issues. To my sister, Mar\'ia Cristina, and my parents I want to thank for the constant encouragement and words of support. Finally, I want to thank Heather Partner for her love, companionship, and patience during these years.

 \end{acknowledgments}

\maketitleabstract

\begin{abstract}
	Quantum state tomography is a fundamental tool in quantum information processing tasks. It allows us to estimate the state of a quantum system by measuring different observables on many identically prepared copies of the system. Usually, one makes projective measurements of an ``informationally complete'' set of observables and repeats them enough times so that good estimates of their expectation values are obtained. This is, in general, a very time-consuming task that requires a large number of measurements. There are, however, systems in which the data acquisition can be done more efficiently. In fact, an ensemble of quantum systems can be prepared and manipulated by external fields while being continuously probed collectively, producing enough information to estimate its state. This provides a basis for continuous measurement quantum tomography, and is the main topic of this dissertation. This method, based on weak continuous measurement, has the advantage of being fast, accurate, and almost nonperturbative. In this work, we present a extensive discussion and a generalization of the protocol proposed in \cite{silberfarb05}, which was experimentally achieved in \cite{smith06} using cold cesium atoms. In this protocol, an ensemble of identically prepared systems is collectively probed and controlled in a time-dependent manner so as to create an informationally complete continuous measurement record. The  measurement history is then inverted to determine the state at the initial time. To achieve this, we use two different estimation methods: the widely used maximum likelihood and the novel compressed sensing algorithms. The general formalism is applied to the case of reconstruction of the quantum state encoded in the magnetic sub-levels of a large-spin alkali atom, ${}^{133}$Cs. We extend the applicability of the protocol in \cite{silberfarb05} to the more ambitious case of reconstruction of states in the full 16-dimensional electronic-ground subspace ($F=3 \oplus F=4$), controlled by microwaves and radio-frequency magnetic fields. We give detailed derivations of all physical interactions, approximations, numerical methods, and fitting procedures, tailored to the realistic experimental setting. In addition, we numerically study the reconstruction algorithms and determine their applicability and appropriate use. Moreover, in collaboration with the lab of Prof. P. Jessen at the University of Arizona, we present an experimental demonstration of continuous measurement quantum tomography in an ensemble of cold cesium atoms with full control of its 16-dimensional Hilbert space. In this case, we show the exquisite level of control achieved in the lab and the excellent agreement between the theory discussed in this dissertation and the experimental results. This allows us to achieve fidelities $>95\%$ for low complexity quantum states, and $>92\%$ for arbitrary random states, which is a formidable accomplishment for a space of this size. To conclude this work, we study quantum tomography in an abstract system driven by random dynamics and show the conditions for high-fidelity estimation when a single parameter defines the dynamics of the system. This study helps elucidate the reconstruction algorithm and gives rise to interesting questions about the geometry of quantum states.

\clearpage 
\end{abstract}

\tableofcontents
\listoffigures


\mainmatter

\chapter{Introduction}

%

The level of control and manipulation of quantum systems has improved dramatically in recent years. Arbitrary quantum states can be prepared and manipulated in laboratories, which opens new and exciting avenues to applications in quantum information and control science, such as quantum metrology, quantum communication, quantum simulations, and, in general, quantum computing. What these applications have in common is that all need some sort of reliable state initialization, the possibility to implement unitary transformations or quantum gates, and some means to measure the system to extract information. In this dissertation, we will focus in the last assuming the previous are possible and readily available. 

How does one diagnose a quantum state when the measurement itself can cause a collapse of the wave function?  To give an estimate of a quantum state, one requires many identical copies of the system so that one can make many different measurements of observables and then use the statistics of the resulting data to determine the unknown state.  In most cases, for complex quantum systems of even a few qubits, this is a laborious and time-consuming procedure in both data acquisition and data processing. 

The procedure by which the state of a quantum system is estimated from a collection of identically prepared systems is called quantum tomography (QT) \cite{paris04}, and is the main subject of this dissertation. QT is a fundamental and, nowadays, commonly used tool in quantum information science that has been carried out in a variety of systems and in a variety of protocols \cite{paris04, DAriano03}. The essential, standard, procedure of QT is to use the statistics of the measurement results on an ensemble of identical systems to determine an estimate of the prepared state $\rho_0$.  This can be achieved, e.g., through a series of strong projective measurements of a set of Hermitian observables \cite{klose01} or through weak continuous measurement of a time-series of observables \cite{silberfarb05, smith06}. Pioneering work by \cite{vogel89} and \cite{smithey93} demonstrated quantum tomography by homodyne detection of a single mode of the electromagentic field. In quantum optics applications, in general, the Wigner function of the quantum state is reconstructed. In the pioneering work, this was accomplished by a direct tomographic inversion (Radon inversion) of the measured data  as in tomographic reconstruction of a 3D image \cite{lvovsky09}, hence the name of quantum tomography. In more recent applications, however, the name of quantum tomography is still used as a synonym of quantum state estimation even though no standard tomographic inversion takes place. This is the way the term is used throughout this document.

Many examples exist of successful implementation of standard QT in a variety of systems. The state of two entangled qubits encoded in the polarization of two photons generated in a down-conversion experiment was first measured in \cite{james01}. In \cite{haffner05}, for example, QT was used to diagnose the performance in the creation of highly entangled states in up to eight qubit trapped ions. Furthermore, it has been used to measure states encoded in solid state superconductive qubits \cite{neeley10}, in NMR applications \cite{chuang98}, in cold atomic systems \cite{klose01}, and to determine the performance of quantum gates, i.e., quantum process tomography, for example in Rydberg atoms \cite{isenhower10}. In all of these examples a large number of preparation states and long times for post processing were required. 

Recently a new protocol for continuous measurement QT was proposed by Silberfarb \cite{silberfarb05} and has been achieved in \cite{smith06}. The key idea is to make a weak (not projective) measurement on the collective variable on a large ensemble of identically prepared systems by coupling the ensemble to some probe that is then measured. While the measurement takes place, the system is dynamically driven so that new information is continuously mapped onto the probe. At every instant in time, one obtains a good estimate of the expectation value of some observable if sufficient signal-to-noise is available, i.e., one has a sufficiently large ensemble and low noise. If the driving is chosen so that one measures an informationally complete set of observables, then in principle one can invert the signal to estimate the state at initial time.  If the system is ``controllable'', then one can find a time-dependent evolution that generates this informationally complete set.  Essential to this type of procedure is an accurate model of the dynamics so that the only unknown is the initial condition.  This is the major challenge for implementation.  In practice, however, some of the signal can be used to do parameter estimation to nail down some of the unknowns of the dynamics.  Overall the protocol is fast, robust, and accurate when one has access to a large ensemble and has a good model of the dynamics.  It is fast because one can quickly obtain expectation values rather than performing projective measurements on repeatedly prepared systems, and it is robust because one calibrates the system in situ to determine the unknown parameters of the dynamics. Extending the applicability of such continuous measurement protocols is the main subject of this dissertation.

More recently, as the quantum systems implemented have grown in complexity, moving towards quantum many-body systems, the efforts of the community are pointing toward even more efficient implementation of QT procedures. Although in general, arbitrary QT methods are intrinsically exponentially complex for many-body systems, since the dimension of a quantum system grows exponentially with the number of particles, not all states in Hilbert space, but a fraction of them, are relevant for applications in quantum technologies. In general, quantum pure states are preferred as a resource for quantum information tasks. This fact makes it possible to simplify the methods used for QT. The works of \cite{flammia11, dasilva11}, for example, have proposed protocols that attempt to estimate the fidelity of preparation of a quantum pure state. Such protocols achieve high fidelity results while greatly reducing the number of measurements compared to full QT. Moreover, QT protocols have been proposed for many-body systems that can be efficiently represented as matrix product states \cite{cramer10}. Following the same trend, new and more efficient signal processing techniques can be applied to the quantum estimation problem. One such example, which has recently received considerable attention, is the so called compressed sensing algorithm \cite{candes05}, \cite{donoho06}, \cite{candes10a}, \cite{candes10b}, which has been recently applied to QT in \cite{gross10} and \cite{liu11}. This novel idea consists in finding estimates of quantum states assuming they are close to pure, and can be efficiently implemented. In this dissertation, we have extended the compressed sensing protocol to the case of continuous-time QT.  

In this work, we take as our platform for quantum tomography ensembles of ultra-cold atoms whose spin is the quantum system of interest, further developing and extending the applicability of the original work of \cite{silberfarb05} in continuous measurement QT. In a nutshell, we assume we have an ensemble of atomic spins identically initialized being driven by external magnetic fields whose dynamics is very well known. While the system is evolving in a controlled manner, it is weakly coupled to some off-resonant probe laser field whose state is continuously being measured and collected. The state of the transmitted light contains information about the atomic spins. Our job is to try to guess what the initial state of the system was using only whatever information we measured from the system and the fact that the dynamics is known. In some sense, the fact that one has available many spins in the ensemble enhances the overall signal-to-noise ratio. However, we assume the coupling is sufficiently weak so that quantum backaction is negligible over the course of the measurement. In this case, the ensemble is always separable and our reconstruction is of the state of one of the spins in the ensemble.

The success and applicability of this method depends critically on one's ability to precisely simulate the dynamical behavior of a particular system. In particular, fundamental decoherence and dephasing due to inhomogeneous fields have to be taken into account. We have developed an accurate master equation model for this system including the control of the full hyperfine ground spin degrees of freedom and the interaction with the probe field which induces both coherent dynamics (sometimes unwanted) and decoherence on the atoms. 

This dissertation is in large part a research collaboration performed with the experimental group of Prof. Poul S. Jessen, College of Optical Sciences, University of Arizona. Our continuous-time QT protocol requires a close interplay between theory and experiment. At its heart is an accurate dynamical model and the ability for the experimentalists to control the system to an exquisite degree such that the dynamics seen by the atoms in the lab is exactly the one theory demands.  We have achieved this!   In this work, we will present data taken by the team of Aaron Smith, Brian Anderson, and H\'ector Sosa,  which we have analyzed according to our protocol. We used standard maximum likelihood/least squares methods for reconstruction in addition to novel compressed sensing techniques, which are shown to be more robust and reliable in the cases we are interested in this work. We find an average fidelity greater than 92\% for pure states picked at random according to the Haar measure. This is a substantial achievement for QT on such a large, 16-dimensional, Hilbert space of cesium atoms.

Mastery of continuous measurement techniques opens exciting new avenues for quantum information and control diagnostic tools. One can imagine extending the ideas discussed here to much more ambitious settings in which one demands adaptive or real-time quantum tomographic methods that use information more efficiently. Moreover, one can think of applying these ideas to estimating some aspect of the state of a many-body quantum system. The present work is yet another step towards that goal.

The remainder of this dissertation is organized as follows. In Chapter \ref{ch:QT},  we give a detailed review of our weak-continuous-measurement QT protocol for the general reconstruction of a density matrix. We describe the technical details of two methods to achieve high fidelity QT: maximum likelihood/least squares and compressed sensing. In addition, we discuss the considerations needed for the dynamical evolution, in particular, a generalized form of the Heisenberg picture including decoherence. In Chapter \ref{ch:Atoms}, we specialize to the case of hyperfine atomic spin systems, and in particular, the control of the ground-electronic manifold of cesium atoms. The atomic physics of the system is described in detail and controllability is discussed for QT on the full 16-dimensional $F=3\oplus F=4$ electronic ground state subspace, driven by time-dependent radio frequency and microwave magnetic fields. This case is much more ambitious than what was presented in \cite{silberfarb05} and \cite{smith06} where experiments were restricted to controlling one of the two hyperfine ground manifolds of cesium. Since our goal is to be able to implement continuous measurement QT in the laboratory, in Chapter \ref{ch:Simulations}, we discuss the challenges of such an application and detail all that is needed to successfully achieve it. In this sense, we discuss an appropriate operation regime for the experiment and review the parameter estimation steps needed for practical implementation. Moreover, we present simulations of the model of the system and characterize the reconstruction methods being implemented. Next, in Chapter \ref{ch:Results}, we present QT examples using experimental data from Poul Jessen's laboratory which shows successful reconstruction of quantum states in cesium atoms. The examples are chosen to emphasize the extraordinary level of control achieved in the lab and to illustrate how the protocols work. All of the fundamentals of our QT protocol applied to the hyperfine manifold of cesium are discussed in \cite{riofrio11}. Application of the procedure and its implementation in the laboratory are still under investigation and will soon be prepared for publication.

Additionally, in Chapter \ref{ch:RandomUnitaries}, we change gears a little to explore QT in the context of arbitrary random evolution. This work, in collaboration with Seth Merkel and Steven Flammia, involves a theoretical study about QT and information generation through random unitary maps, which appeared in \cite{merkel10}. It opens interesting questions involving quantum chaos and information generation. Finally, Chapter \ref{ch:Conclusion} includes the conclusions and outlook of this work.

\chapter{Quantum State Tomography}\label{ch:QT}


Quantum tomography (QT) is an essential tool in quantum information science \cite{paris04}.  The ability to estimate a quantum state  is required for diagnosing quantum information processors and evaluating the fidelity of a given protocol.  The fundamental information-gain/disturbance tradeoff in a measurement of a quantum system implies that any QT protocol requires multiple, nearly identical copies of the state.   Typically, this procedure is carried out through a series of strong destructive measurements of an informationally complete set of observables acting on repeatedly prepared copies of the system.  For this reason, QT is generally a time consuming and tedious procedure when applied to large dimensional systems \cite{haffner05} and even more so when extended to quantum-process tomography in which a whole collection of quantum states must be analyzed \cite{dariano04}.  

In its simplest form, QT uses the outcomes of many measurements of a series of independent observables to estimate the probability distribution for those outcomes to occur. Since that probability depends on the state of the system, it can be used to estimate such state. This direct method in general produces unphysical states and is not currently used in practice. More reliable methods have been proposed and developed. Maximum likelihood, first proposed in \cite{hradil97}, has become the most widely used method for QT. It has the advantage that it is statistically motivated and can be parametrized to produce estimates that are physical. Other methods that are also statistically motivated include maximum-entropy, in which the estimated state is sought to have the maximum possible von Neuman entropy \cite{buzek04}; Bayesian methods are also used in QT, for example in \cite{blume-kohout10}; and hybrid approaches in which maximum likelihood and maximum entropy are combined to give unbiased estimates are also being proposed \cite{teo11}. For examples of tomography on arbitrary spin-$j$ systems see \cite{klose01, dodonov97, hofmann04}. All of the procedures referred above have the premise that measurements are carried out in different, identically prepared systems where some projective measurements or more general POVM has taken place.

In some platforms, however, one has the ability to probe a large ensemble of identical systems simultaneously.  In this case, one can enhance the collection of statistics, e.g., for estimation of the probability of occupation in an eigenstate under projective measurements.  While such projective measurements on ensembles can be used for QT \cite{klose01}, in principle one can dramatically improve the speed, robustness, and experimental complexity of QT by instead employing weak-continuous measurement.  In a protocol originally developed in \cite{silberfarb05}, based on a maximum likelihood approach, an atomic ensemble undergoes a chosen dynamical evolution to generate an informationally complete measurement record.  An algorithm is then used to invert the measurement history to determine the maximum likelihood of the initial state.  For moderately large ensembles, one can attain a sufficient signal-to-noise ratio to enable extraction of the required information, while simultaneously maintaining the quantum projection noise below the intrinsic noise of the quantum probe.  In this case, quantum backaction is negligible and in principle, one can extract the necessary information for QT in a single run of the experiment on a single ensemble.  

Such a continuous measurement protocol was previously employed to perform QT on the 7-dimensional, $F=3$ atomic hyperfine spin manifold, in an ensemble of cesium atoms \cite{silberfarb05}.  With this tool in hand, the performance of state-to-state quantum maps, which was designed and implemented by optimal control techniques, was diagnosed  experimentally \cite{smith06}.  More recently,  this QT protocol was a central component that enabled the Jessen group to measure the time evolution of the quantum state of the spin undergoing the quantum chaotic dynamics of a nonlinear kicked top \cite{chaudhury09}.  Observing dynamics of a density matrix for any reasonable duration would have been a formidable challenge without an efficient method for QT at each time step.  

In this chapter, we describe in detail the continuous measurement QT methods used in this dissertation. We start by describing the systems in which they are naturally applicable and discuss its technical aspects.

\section{General overview of quantum tomography}
The most general measurement process in quantum mechanics can be described by a POVM (Positive Operator-Valued Measurement) \cite{nielsen00}. A POVM is the set of operators $\{\Pi_i\}$ such that $\Pi_i\ge 0$ (positive operators) and $\sum_i\Pi_i=I$ (resolution of the identity). According to quantum mechanics, measuring POVM element $\Pi_i$ will produce an outcome labeled by $i$ with probability
\be
p_i=\Tr(\rho\Pi_i)
\label{eq:POVMprob}
\ee 
where $\rho$ is the density matrix representing the state of the quantum system. Notice that the conditions mentioned above imply that $p_i\ge 0$ and $\sum_ip_i=1$.

To estimate the probability distribution, one makes $\mathcal{N}$ measurements of the POVM in $\mathcal{N}$ identically prepared copies of the unknown state $\rho$. Thus, the outcome $i$, corresponding to POVM element $\Pi_i$, appears $n_i$ times from which it is clear that $\mathcal{N}=\sum_in_i$. In its simplest form, quantum tomographic methods approximate $p_i$ by $f_i=n_i/\mathcal{N}$, where $f_i$ is the frequency of appearance of outcome $i$ in the ensemble. To determine the unknown density matrix, $\rho$, a simple linear parametrization is used
\be
\rho=\frac{I}{d}+\sum_{\alpha=1}^{d^2-1}r_\alpha E_\alpha
\ee
where $d$ is the dimension of the Hilbert space, $r_\alpha$ are $d^2-1$ real numbers, and $\{E_\alpha\}$ is an orthonormal Hermitian basis of traceless operators. In this way, Eq. (\ref{eq:POVMprob}) can approximately be written 
\be
f_i\approx \frac{\Tr(\Pi_i)}{d}+\sum_{\alpha=1}^{d^2-1}r_\alpha E_\alpha,
\ee
or in matrix form
\be
\mbf{\tilde{f}}\approx \mbf{\tilde{\Pi}}\mbf{r},
\ee
where $\tilde{f}_i=f_i-\Tr(\Pi_i)/d$ and $\tilde{\Pi}_{i\alpha}=\Tr(\Pi_iE_\alpha)$. If the POVM $\{\Pi_i\}$ spans the space of density matrices acting on the Hilbert space, then $\mbf{\tilde{\Pi}}$ is invertible, and the POVM is informationally complete. For the finite dimensional case considered here, the space is the Lie algebra ${\mathfrak u}(d)$, or if we assume the state is normalized and remove the identity from the space, the algebra ${\mathfrak su}(d)$. Thus, the vector $\mbf{r}$ can be found by simply
\be
\mbf{r}=\mbf{\tilde{\Pi}}^{-1}\mbf{\tilde{f}}.
\ee
The estimate quantum state, $\bar{\rho}$, is then given by
\be
\rho=\frac{I}{d}+\sum_{\alpha=1}^{d^2-1}r_\alpha E_\alpha,
\ee
and is by construction Hermitian and normalized, and in the absence of noise, positive semidefinite. In practice, however, there is always noise in the measurements and thus the quantum state found this way will not be positive semidefinite in general. An example of this fact is discussed in \cite{james01} in which a similar method to the one discussed here was used to estimate the state of 2 entangled photons generated in a down-conversion experiment.

In order to avoid estimating non-physical states, several methods have been proposed as indicated above. Here, we review the most popular one (which we use later in the chapter): the maximum likelihood method \cite{hradil97}. First, one needs to define the likelihood function, which is proportional to the probability of getting a particular sequence of outcomes given some quantum state, in other words
\be
P(\{n_i\}|\rho)\propto \mathcal{L}=\prod_i p_i^{n_i}=\prod_i \left[\Tr(\rho\Pi_i)\right]^{n_i},
\ee
where we do not bother to include the unimportant multinomial coefficient of the probability. In practice, the number of measurements made in the system is $\gg 1$ and thus $\mathcal{N}\gg 1$ which allows one to approximate the likelihood function according to the law of large numbers by a multivariate Gaussian distribution of mean $\mathcal{N}f_i=n_i$ and variance $\sigma_i^2=\mathcal{N}f_i(1-f_i)\approx n_i$ for large $\mathcal{N}$. In this approximation, the likelihood function is
\be
\mathcal{L}=\prod_i\exp{\left[\frac{(\mathcal{N}\Tr(\rho\Pi_i)-n_i)^2}{2\sigma_i^2}\right]}=\exp{\left[\sum_i\left(\frac{(\mathcal{N}\Tr(\rho\Pi_i)-n_i)^2}{2\sigma_i^2}\right)\right]}.
\ee
In general, one seeks a parametrization of $\rho$ that guarantees that it is a  positve semidefinite matrix. One such parametrization commonly used is taking $\rho=T^\dagger T$ where $T$ is a lower triangular matrix, i.e., the Cholesky decomposition. The maximum likelihood method consists in finding $T$ that maximizes the value of $\mathcal{L}$ or equivalently minimizes $-\log{(\mathcal{L})}$, which is usually achieved numerically. This method will produce estimates of the state that are physical and compatible with the data $\{n_i\}$.

\section{Continuous measurement quantum state tomography}\label{sec:QTprotocol:basic}
The general setting for our protocol is as follows.  One is given an ensemble of $N$, noninteracting, simultaneously prepared systems in an identical state $\rho_0$ that can be controlled and probed collectively.  We seek to find an estimate of the state of the system by continuously measuring some traceless observable $\cO_0$. We restrict our attention to states in Hilbert spaces of finite dimension $d$ and measure traceless-Hermitian observables in the algebra $\mathfrak{su}(d)$. In practice, we can reach the informationally complete set if the system is ``controllable''. This is true if the available control Hamiltonians are generators of the Lie algebra.  

In an idealized form, the probe performs a QND measurement that couples uniformly to the collective variable across the ensemble and measures $\cO_c=\sum_j^N \cO_0^{(j)}$, where $\cO^{(j)}$ acts on the $j^{th}$ subsystem.  For a sufficiently strong QND measurement, quantum back-action will result in substantial entanglement between the particles.  For example, such a phenomenon has been employed to create spin squeezed states of an ensemble when the fluctuations in projection-valued measurements (``projection noise") can be resolved within fundamental quantum fluctuations in the probe (``shot noise") \cite{kuzmich2000, appel2009, takahashi2009, vuletic2010, mitchell2010}.  We consider the opposite case of a very weak measurement such that back-action noise is negligible compared with the detector noise.   In this case, the procedure can be analyzed as a single atom control problem in which each member of the ensemble evolves under the same dynamics. In more detail, we envision that the collective variable is coupled to a probe observable $P$ according to a QND Hamiltonian, $H_{QND} = \kappa \cO_cP$. Taking the probe to be in the initial fiducial state $\ket{0}$, after the interaction we measure a conjugate variable $X$. According to the theory of completely positive maps, the state of the system conditioned on the measurement is
\be
\rho_X^{(out)}=\frac{A_X\rho^{(in)}A_X^\dagger}{P(X|\rho^{(in)})},
\ee
where $A_X = \bra{X}\exp{(-i\kappa\cO_c P)}\ket{0}$. The probability of measuring outcome $X$ is
\be
P(X|\rho^{(in)})=\Tr(\Pi_X\rho^{(in)})
\ee
where $\Pi_X=A_X^\dagger A_X$ is the POVM element. We will take the probe to be described by the modes of a harmonic oscillator, with $\ket{0}$ being the vacuum state and $X$ and $P$ the canonical phase space variables. In that case, the probe noise is ``shot noise'' (SN), which masks the quantum uncertainty in the state $\rho^{(in)}$ and makes the measurement nonprojective. If the shot noise fluctuations are larger than the uncertainty in the observable measured, then backaction can be neglected, and $\rho_X^{(out)}\approx \rho^{(in)}$.

More precisely, using the position representation of the vacuum we see that
\be
\Pi_X=\frac{1}{\sqrt{\pi}}\exp{\left[-(X-\kappa\cO_c)^2\right]}=\frac{1}{\sqrt{\pi}}\exp{\left[-\kappa^2(\cO_c-X/\kappa)^2\right]}.
\ee

The shot-noise variance that limits the resolution of the collective $\cO_c$ is $\Delta\cO_{c,SN}^2= (2\kappa^2)^{-1}$. Let us write $\cO_c=\avg{\cO_c}+\delta\cO_c$, where $\avg{\cO_c}=\Tr(\cO_c \rho^{(in)})$. The fluctuations $\Delta\cO_{c,PN}^2=\avg{\delta\cO_c^2}$ are known as ``projection noise'' (PN) in the QND measurement because they represent fluctuations seen in the meter due to projections onto different eigenstates of the observable. Backaction is negligible when $\Delta\cO_{c,SN}^2\gg\Delta\cO_{c,PN}^2$. We assume that we start in a separable state of $N$-identical systems, $\rho^{(in)}=\rho_0^{\otimes N}$. Then, the system remains separable and symmetric, so $\Delta\cO_{c,PN}^2=N\Delta\cO_{PN}^2$ , where $\Delta\cO_{PN}^2$ is the single particle variance. The condition that backaction can be neglected is then $\Delta\cO_{PN}^2\ll1/(N\kappa^2)$. We thus assume that we act on an ensemble that is sufficiently large that we have enough signal-to-noise to extract the information we need for QT, but not so large that quantum backaction is an important part of the dynamics. We have such a large ensemble the projection noise fluctuations themselves are Gaussian according to the central limit theorem. In the Gaussian approximation, the measurement outcomes are thus distributed according to the probability distribution
\be
\begin{split}
P(X|\rho^{(in)})=&\frac{1}{\sqrt{\pi}}\exp{\left[-\kappa^2(X-N\avg{\cO}/\kappa)^2\right]}\times\\
			&\Tr\left(\rho_0^{\otimes N}\exp{\left[-\kappa^2(X-N\avg{\cO}/\kappa)\sqrt{N}\delta\cO-N\kappa^2\delta\cO^2\right]}\right).
\end{split}
\ee
In the limit where projection noise fluctuations are negligible, $\Delta\cO_{PN}^2=\avg{\delta\cO^2}\ll1/(N\kappa^2)$, we arrive at the final form of the likelihood function for the measurement given the state. Absorbing constants, and defining the measurement outcome as M, normalized to the number of particles, one can write
\be
P(M|\rho_0)=\frac{1}{\sqrt{2\pi\sigma^2}}\exp{\left[-\frac{(M-\Tr(\cO\rho_0))^2}{2\sigma^2}\right]}
\ee
where $\sigma^2$ are the fluctuations set by shot noise. The time-dependent measurement record then follows under the assumption that a fresh copy of the probe mode comes in and interacts with the ensemble in a given measurement interval. Assuming the measurements are sampled from the Gaussian distribution above, we can model the measurement record through a Weiner process W(t) with zero mean and unit variance,
\begin{equation}
M(t)=\Tr(\cO_0 \rho(t))+\sigma W(t),
\label{eq:measurementconti}
\end{equation}
where we have assumed that the system evolves dynamically.

In order to generate a measurement record $M(t)$ that can be inverted to determine the initial state, one must control the dynamics so as to continuously write new information onto the measured observable.  To do so, the system is manipulated by external fields.  The Hamiltonian of the system, $H(t)=H[\phi_i(t)]$, is a functional of a set of time-dependent control functions, $\phi_i(t)$, which are chosen so that the dynamics produces an informationally complete measurement record $M(t)$.  Since our objective is to estimate the initial state of the system from the measurement record and  our knowledge of the system dynamics, it is more convenient to carry out the procedure in the Heisenberg picture.  Expressed this way, the state is fixed and control is used to generate new observables that we measure.  Note, this is generally different from the standard Heisenberg picture in that we allow for decoherence during the dynamical evolution.  We will return to this issue below.   The measurement record, Eq. (\ref{eq:measurementconti}), is then written $M(t)=\Tr(\cO(t)\rho_0)+\sigma W(t)$. For implementation of the algorithm, a time discretization of the problem is necessary. We sample the measurement record at discrete times so that 
\be
M_i=\Tr(\cO_i\rho_0)+\sigma W_i.
\label{eq:measurement}
\ee 
We have thus reduced the problem of QT to a linear stochastic estimation problem.  The goal is to determine $\rho_0$ given the measurement record $\{M_i\}$ for a well chosen $\{\cO_i\}$ in the presence of noise $\{W_i\}$. We solve this problem using two separate approaches that are common in inverse-problem theory: maximum likelihood and compressed sensing, which we explain in detail in the next sections.

\subsection{Maximum likelihood/least squares method}\label{subsec:ML}
In this section, we discuss a maximum likelihood method applied to continuous measurement quantum state estimation. 

A number of transformations of Eq. (\ref{eq:measurement}) are necessary to increase the numerical stability and reliability of the algorithm. Let $\{E_{\alpha}, I/\sqrt{d}\}$, $\alpha=1,\dots, d^2-1$, be an orthonomal Hermitian basis of matrices, where $I$ is the identity matrix and $\Tr(E_{\alpha})=0$. The unknown initial state, $\rho_0$, can thus be decomposed as
\be
\rho_0=\frac{1}{d}I+\sum_{\alpha=1}^{d^2-1}r_{\alpha}E_{\alpha},
\label{eq:parameterizedrho}
\ee 
where $r_{\alpha}= \Tr(\rho_0 E_{\alpha})$ are real numbers. This parametrization ensures that the density matrix is Hermitian and that its trace is always 1. We can then write Eq. (\ref{eq:measurement}) as
\be
M_i=\sum_{\alpha=1}^{d^2-1}r_{\alpha}\Tr(\cO_iE_{\alpha})+\frac{1}{d}\Tr(\cO_i)+\sigma W_i,
\label{eq:measurement2}
\ee 
or, written in matrix form,
\be
\vec{\tilde{M}}=\mbf{\tilde{\cO}}\mbf{r}+\sigma\vec{W},
\label{eq:measurementVector}
\ee
which in general is an overdetermined set of linear equations with $d^2-1$ unknowns $\mbf{r}=(r_1,\dots,r_{d^2-1})$, where $\tilde{M}_i=M_i-\Tr(\cO_i)/d$, and $\tilde{\cO}_{i\alpha}=\Tr(\cO_iE_{\alpha})$. Note that $\tilde{O}_{i\alpha}$ is not a square matrix. We generally measure many more expectation values in the time history as compared to the dimension $d^2-1$. Also note that the term $\Tr(\cO_i)/d$ would be exactly 0 if the dynamics were to be unitary, given the fact that $\Tr(\cO_0)=0$ by assumption; however, that is not true when more general, non-unitary maps are considered, which is the case in this work.

Eq. (\ref{eq:measurementVector}) explicitly states that the conditional probability of the random variable $\vec{\tilde{M}}$ given the state $\vec{r}$ is the Gaussian distribution
\be
P(\vec{\tilde{M}}|\vec{r})\propto \exp{\left(-\frac{1}{2\sigma^2}(\vec{\tilde{M}}-\mbf{\tilde{\cO}}\vec{r})^T(\vec{\tilde{M}}-\mbf{\tilde{\cO}}\vec{r})\right)}.
\label{eq:measurementdistribution}
\ee
Maximum likelihood methods deal with the maximization of the likelihood function, Eq. (\ref{eq:measurementdistribution}), or equivalently, the minimization of its exponent. To gain insight into this method, we will handle the minimization problem in two different ways. First, we will discuss a two-step approach which will allow us to understand the information content in the measurement record and illustrate the connection between maximum likelihood and the least squares method. Second, we will show a more practical, robust, and stable method, which we call the one-step approach, which is the one we will use in the practical implementation of this protocol.

\subsubsection{A. Two-step approach}

We can use the fact that the argument of the exponent in Eq. (\ref{eq:measurementdistribution}) is a quadratic function of $\vec{r}$ to write the likelihood function
\be
P(\vec{\tilde{M}}|\vec{r})\propto \exp{\left(-\frac{1}{2}(\vec{r}-\vec{r}_{ML})^T\vec{C}^{-1}(\vec{r}-\vec{r}_{ML})\right)},
\label{eq:likelihoodfunction}
\ee
describing a Gaussian function over possible states $\vec{r}$ centered around the most likely state, $\vec{r}_{ML}$, where the unconstrained maximum likelihood solution is given by 
\be
\mbf{r}_{ML}=(\mbf{\tilde{\cO}}^T\mbf{\tilde{\cO}})^{-1}\mbf{\tilde{\cO}}^T\mbf{\tilde{M}},
\label{eq:LSsolution}
\ee 
 with covariance matrix
 \be
 \mbf{C} = \sigma^2(\mbf{\tilde{\cO}}^T\mbf{\tilde{\cO}})^{-1}.  
 \label{eq:Covariance}
 \ee
Note that the covariance matrix is square of dimension $(d^2-1)\times(d^2-1)$. By replacing Eqs. (\ref{eq:LSsolution}) and (\ref{eq:Covariance}) in the exponent of Eq. (\ref{eq:likelihoodfunction}), it is not difficult to see that we recover the exponent of Eq. (\ref{eq:measurementdistribution}) plus a term that does not depend on $\mbf{r}$, which we illustrate below. Let's define $S({\mbf r})=\frac{1}{2}(\vec{r}-\vec{r}_{ML})^T\vec{C}^{-1}(\vec{r}-\vec{r}_{ML})$, which we can write
\be
\begin{split}
S({\mbf r})&=\frac{1}{2\sigma^2}\left[\vec{r}-(\mbf{\tilde{\cO}}^T\mbf{\tilde{\cO}})^{-1}\mbf{\tilde{\cO}}^T\mbf{\tilde{M}}\right]^T\mbf{\tilde{\cO}}^T\mbf{\tilde{\cO}}\left[\vec{r}-(\mbf{\tilde{\cO}}^T\mbf{\tilde{\cO}})^{-1}\mbf{\tilde{\cO}}^T\mbf{\tilde{M}}\right]\\
&=\frac{1}{2\sigma^2}\left[\mbf{\tilde{\cO}}\vec{r}-\mbf{\tilde{\cO}}(\mbf{\tilde{\cO}}^T\mbf{\tilde{\cO}})^{-1}\mbf{\tilde{\cO}}^T\mbf{\tilde{M}}\right]^T\left[\mbf{\tilde{\cO}}\vec{r}-\mbf{\tilde{\cO}}(\mbf{\tilde{\cO}}^T\mbf{\tilde{\cO}})^{-1}\mbf{\tilde{\cO}}^T\mbf{\tilde{M}}\right]\\
&=\frac{1}{2\sigma^2}\left\|\mbf{\tilde{\cO}}\vec{r}-\mbf{\tilde{\cO}}(\mbf{\tilde{\cO}}^T\mbf{\tilde{\cO}})^{-1}\mbf{\tilde{\cO}}^T\mbf{\tilde{M}}\right\|^2\\
&=\frac{1}{2\sigma^2}\left\|\mbf{\tilde{\cO}}\vec{r}-\mbf{\tilde{M}}+({\mbf I}-\mbf{\tilde{\cO}}(\mbf{\tilde{\cO}}^T\mbf{\tilde{\cO}})^{-1}\mbf{\tilde{\cO}}^T)\mbf{\tilde{M}}\right\|^2.
\end{split}
\ee
For simplicity, let ${\mbf K}_1=({\mbf I}-\mbf{\tilde{\cO}}(\mbf{\tilde{\cO}}^T\mbf{\tilde{\cO}})^{-1}\mbf{\tilde{\cO}}^T)\mbf{\tilde{M}}$, which is independent of ${\mbf r}$ and allows us to write
\be
\begin{split}
S({\mbf r})&=\frac{1}{2\sigma^2}\left\|\mbf{\tilde{\cO}}\vec{r}-\mbf{\tilde{M}}+{\mbf K}_1\right\|^2\\
&=\frac{1}{2\sigma^2}\left[ (\mbf{\tilde{\cO}}\vec{r}-\mbf{\tilde{M}})^T(\mbf{\tilde{\cO}}\vec{r}-\mbf{\tilde{M}})+(\mbf{\tilde{\cO}}\vec{r}-\mbf{\tilde{M}})^T{\mbf K}_1+{\mbf K}_1^T(\mbf{\tilde{\cO}}\vec{r}-\mbf{\tilde{M}})+{\mbf K}_1^T{\mbf K}_1\right]\\
&=\frac{1}{2\sigma^2}\left[ (\mbf{\tilde{\cO}}\vec{r}-\mbf{\tilde{M}})^T(\mbf{\tilde{\cO}}\vec{r}-\mbf{\tilde{M}})+(\mbf{\tilde{\cO}}\vec{r})^T{\mbf K}_1+{\mbf K}_1^T(\mbf{\tilde{\cO}}\vec{r})-\mbf{\tilde{M}}^T{\mbf K}_1\right.\\
&-\left.{\mbf K}_1^T\mbf{\tilde{M}}+{\mbf K}_1^T{\mbf K}_1  \right].
\end{split}
\ee
Now, note that $(\mbf{\tilde{\cO}}\vec{r})^T{\mbf K}_1=\left({\mbf K}_1^T(\mbf{\tilde{\cO}}\vec{r})\right)^T=\vec{r}^T\mbf{\tilde{\cO}}^T\mbf{\tilde{M}}-\vec{r}^T(\mbf{\tilde{\cO}}^T\mbf{\tilde{\cO}})(\mbf{\tilde{\cO}}^T\mbf{\tilde{\cO}})^{-1}\mbf{\tilde{\cO}}^T\mbf{\tilde{M}}=0$. Thus, we see that
\be
S({\mbf r})=\frac{1}{2\sigma^2}\left[ (\mbf{\tilde{\cO}}\vec{r}-\mbf{\tilde{M}})^T(\mbf{\tilde{\cO}}\vec{r}-\mbf{\tilde{M}})+{\mbf K} \right]
\ee
where ${\mbf K}={\mbf K}_1^T{\mbf K}_1-\mbf{\tilde{M}}^T{\mbf K}_1-{\mbf K}_1^T\mbf{\tilde{M}}$ is not dependent on ${\mbf r}$.

Since we treat the noise as Gaussian, the maximum likelihood solution corresponds exactly to the least squares solution of the linear system in Eq. (\ref{eq:measurementVector}),  \cite{tarantola05}.  Clearly, the solution found in Eq. (\ref{eq:LSsolution}) minimizes the exponent of the likelihood function. However, this minimization does not, in general, produce physical quantum states and we are forced to further constrain it. In fact, for an informationally incomplete measurement record and/or for finite noise, the unconstrained maximum likelihood solution, Eq. (\ref{eq:LSsolution}), generally produces estimates of the density matrix with negative eigenvalues, i.e., the density matrix thus obtained is not positive semidefinite. To obtain a physical estimate of the state, we, therefore, must impose the constraint that the estimated density matrix be positive semidefinite.  Such a constraint can be enforced through an appropriate parametrization of the unknown initial state, e.g., a Cholesky decomposition $\rho_0=T^\dg T$ where $T$ is a lower diagonal matrix \cite{paris04, klose01}. Although this parametrization has the advantage that the estimated state is Hermitian and positive-semidefinite by definition, it is not compatible with our continuous measurement protocol.  A least squares solution to Eq. (\ref{eq:measurement}) would involve a nonlinear unconstrained optimization for which there is no known efficient solution. We thus turn to constrained numerical optimization to find the ``closest" positive matrix to the unconstrained-maximum-likelihood estimate, i.e., the constrained-maximum-likelihood estimate, $\bar{\rho}$.

The eigenvectors of $\mbf{C}^{-1}$, the inverse of the covariance matrix defined in Eq. (\ref{eq:Covariance}), specify the directions in the operator space $\mathfrak{su}(d)$ that have been measured and its eigenvalues are the squares of the signal-to-noise ratio of those measurements.  The covariance matrix thus allows us to quantify the information extracted from the measurement record.  A sufficient condition for an informationally complete measurement record (though not necessarily unit fidelity due to noise) is one for which $\mbf{C}^{-1}$ is full rank.  This will be true when $\{\cO_i\}$ spans $\mathfrak{su}(d)$, i.e., the measurement must be informationally complete.  In the case of relevance to this work, to achieve an informationally complete measurement record, the quantum system must be {\em controllable} in the sense that we can map any $\cO_0$ to any $\cO_i$ over the Lie algebra. Also, it is essential that the dynamics be sufficiently coherent such that an informationally complete set of observables can be generated before decoherence erases the state. We will discuss these issues in Chapter \ref{ch:Atoms}, but for now, we assume these conditions as given.  

Clearly, the covariance matrix determines a natural cost function metric with which to measure the distance between the unconstrained estimate (which is not physical) and physical estimate we seek to find. The minimization cost function is then defined by
\be \label{E:costnorm}
	\| \mathbf{r}_{\rm ML}-\bar{\mathbf{r}}\|^2 = (\mathbf{r}_{\rm ML}-\bar{\mathbf{r}})^T \mathbf{C}^{-1}(\mathbf{r}_{\rm ML}-\bar{\mathbf{r}}) .
\ee
Technically speaking, this quantity is not a norm but rather a seminorm when we consider informationally incomplete measurements ($\vec{C}$ is not full-rank), meaning that there exist some vectors $\mathbf{v}$ such that $\| \mathbf{v} \| = 0$ but $\mathbf{v} \not=0$ in those cases.  The use of this metric can be justified as follows. The inverse of the covariance matrix, $\mbf{C}^{-1}$, encodes all of the information about the independent directions in operator space that are being measured by our procedure.  A small eigenvalue of  $\mbf{C}^{-1}$ means a low signal-to-noise ratio associated with measurements of the corresponding eigen-operator, and thus that little is known about the trace-projection of the initial state onto that operator direction. The cost function, Eq. (\ref{E:costnorm}), takes into account that different directions in the space $\mathfrak{su}(d)$ are not measured in the same way and weights this in the distance between the initial estimate and the positive state.  In this way, during the numerical optimization, the more uncertain components of $\bar{\rho}$ can be adjusted more freely than the more certain ones, thereby maintaining faithfulness with the measurement record, but ensuring positivity. 

To find the physical estimate we thus solve the following optimization problem:
\be
\begin{split}
&\rm{minimize}~~ \| \mathbf{r}_{\rm ML}-\bar{\mathbf{r}}\|^2\\
&{\rm subject~to}~~ \frac{1}{d}I+\sum_{\alpha=1}^{d^2-1}\bar{r}_{\alpha} E_{\alpha}\ge 0.
\end{split}
\label{eq:convexprogram}
\ee 
While there is generally no analytic solution to this problem, it takes the form of a standard convex program since the matrix $\mbf{C}^{-1}$ is positive semidefinite and both the objective and the constraint are convex functions \cite{boyd08}.  The optimization is a convex program which is efficiently solvable numerically. We implement this in MATLAB using the freely available convex optimization package CVX \cite{cvx11}. Once convex problem Eq. (\ref{eq:convexprogram}) is solved, a set of parameters $\{\bar{r}_\alpha\}$, $\alpha=1,\ldots,d^2-1$, is found and we can calculate the estimate density matrix
\be
\bar{\rho}=\frac{1}{d}I+\sum_{\alpha=1}^{d^2-1}\bar{r}_{\alpha} E_{\alpha},
\label{eq:EstimatedRho}
\ee
which is a Hermitian, unit trace, and positive semidefinite matrix.

This approach gives us knowledge about the covariance matrix and a way of quantifying information in the measurement record; however, it is not practical from the point of view of numerical implementation. Eq. (\ref{eq:LSsolution}) requires the explicit calculation of the inverse of a matrix that may be, in general, very ill-conditioned, which makes taking that inverse numerically unstable. In fact, at early times during the reconstruction procedure, $\mbf{C}^{-1}$ will not be full rank and thus we must use the Moore-Penrose pseudo inverse in Eq. (\ref{eq:LSsolution})  \cite{ben-israel03}. However, even if we do this for very low rank covariance matrices, we still see convergence problems and numerical instablities. A more practical and stable method for solving this problem is discussed below.

\subsubsection{B. One-step approach}
Instead of minimizing the likelihood function, Eq. (\ref{eq:measurementdistribution}), first in an unconstrained way, and then constraining the initial estimate to find the ``closest'' physical state that is compatible with the measured data, it is possible to do it in a single step. In fact, solving the optimization problem
\be
\begin{split}
&{\rm minimize}~~ ~(\vec{\tilde{M}}-\mbf{\tilde{\cO}}\vec{\bar{r}})^T(\vec{\tilde{M}}-\mbf{\tilde{\cO}}\vec{\bar{r}})\\
&{\rm subject~to}~~~ \frac{1}{d}I+\sum_{\alpha=1}^{d^2-1}\bar{r}_{\alpha} E_{\alpha}\ge 0,
\end{split}
\label{eq:OneStepConvex}
\ee
will find an estimate of the density matrix, Eq. (\ref{eq:EstimatedRho}), that is compatible with the measurement record and is Hermitian, unit trace and positive semidefinite. 

The optimization problem shown in Eq. (\ref{eq:OneStepConvex}) is again a convex program that can be easily and  efficiently solved using MATLAB and CVX as mentioned above. Clearly, in this method, we do not have the need to explicitly calculate matrix inverses, which makes it more reliable and robust even when the covariance matrix is not full rank. Moreover, it seems to be the most practical and straightforward way of solving the estimation problem, and it is the preferred way for its application.

\subsection{Compressed sensing method}
In general, estimation problems have no unique solution. Prior information about the estimated parameters or the experimental conditions can be used to help guide the estimator to a desired or better solution. In fact, in the previous section, by using a maximum likelihood method to find a solution for our estimation problem, we assumed Gaussian noise conditions, which is the natural choice in our case. However, what if we know something else about the system? Could we use that information to make better inferences of what the state of the system is? In the remainder of the section, we will discuss the case in which it can be assumed that the quantum state being reconstructed is close to a pure state. This can be justified by the fact that in well controlled experiments it is the case that one tries to produce states that are as pure as possible. Moreover, high purity states are, in general, represented by approximately low rank matrices, for which there are efficient reconstruction methods that we will proceed to briefly review.

Recently, a number of groups have tried to give mathematical foundation to long used heuristic methods to reconstruct vectors using a small number of measurements, see e.g., \cite{candes05}, and \cite{donoho06}.  In those problems, one seeks to find a sparse vector given a small number of ``measured'' entries. Its solution, which involves the minimization of the vector's $l_1$ norm, requires solving a linear/convex optimization program and thus is attractive for practical applications. The techniques developed in those papers constitute the foundation of compressed sensing. Generalization of the sparse vector reconstruction problem to a matrix completion problem in which either a low rank or a sparse matrix is to be reconstructed from a number of measured elements are discussed in \cite{candes10a, candes10b, fazel02, recht10, gross11}. In particular, they have shown that reconstructing a low rank matrix constrained to a convex set amounts to solving the ``trace heuristic'', i.e., minimizing the nuclear norm (also known as the trace norm), which we define below in Eq. (\ref{eq:tracenorm}), of the unknown matrix subject to convex constraints. In contrast to attempting to minimize directly the rank of the unknown matrix, which is an NP hard problem, this procedure is efficiently achievable through standard convex optimization techniques and produces reasonably low rank estimates. Mathematically, in \cite{fazel02} and  \cite{recht10}, they proved that the nuclear norm is a convex envelope of the rank function, meaning that it is the largest convex function that puts a lower bound to the rank function. These results enable application of compressed sensing techniques to reconstruction of low rank matrices, in particular, high purity quantum states.

More recently, \cite{gross10} applied the matrix completion techniques to the problem of QT involving qubits. The key point is that tomography is usual applied to states that are nearly pure and thus low rank, as in the output of a well performing quantum information processor. This is prior information one can use to find better estimates. In this context, they seek to recover a low rank/high purity density matrix of a system composed of $N$ two-level quantum systems. Their scheme, which involves projective measurements of Pauli observables on the qubits to obtain their expectation values, shows that the low rank matrix completion techniques are naturally, efficiently, and successfully applied to QT. Here, following \cite{gross10}, we present a continuous measurement version of a compressed sensing method for quantum state tomography. 

As before, we are interested in developing this method for ensembles of particles that can be probed simultaneously. We begin by finding a parametrization for the unknown density matrix, similar to Eq. (\ref{eq:parameterizedrho}), with the difference that we do not fix the value of its trace
\be
\rho_0=\sum_{\alpha=0}^{d^2-1}r_{\alpha}E_{\alpha},
\label{eq:CSparameterizedrho}
\ee 
where $\{E_{\alpha}\}$, $\alpha=0,\dots, d^2-1$, is an orthonomal basis of Hermitian matrices, where $\Tr(E_{\alpha})=0$, for $\alpha=1,\dots, d^2-1$, and we define explicitly $E_0=I/\sqrt{d}$. Under this consideration, the measurement record, Eq. (\ref{eq:measurement}), can be written
\be
M_i=\sum_{\alpha=0}^{d^2-1}r_{\alpha}\Tr(\cO_iE_{\alpha})+\sigma W_i.
\label{eq:CSmeasurement}
\ee 

We need to define an optimization cost function to solve this overdetermined system of equations. As mention above, in general, compressed sensing methods seek to find a matrix whose nuclear norm (trace norm) is minimum, i.e., the minimization of the ``trace heuristic''. The nuclear norm of a $d\times d$ matrix $A$ is defined by
\be
||A||_*=\Tr(\sqrt{A^\dagger A}).
\label{eq:tracenorm}
\ee
This choice of cost function guarantees that this method will be biased towards low-rank matrices as shown in \cite{fazel02,gross11}. As mentioned above, this cost function is convex and bounds the rank of matrix $A$, giving an approximate solution to the rank minimization problem which is NP hard.

In the case of interest in this work, the matrix $A$ is Hermitian and positive semidefinite, which simplifies the nuclear norm $||A||_*=\Tr(A)$. Let $\lambda_i\ge 0$ be the eigenvalues of $A$ for $i=1,\ldots,d$. In terms of its eigenvalues, the trace norm is simply $||A||_*=\sum_i\lambda_i=\sum_i|\lambda_i|=||\mbf{\lambda} ||_1$, where $||\mbf{\lambda}||_1$ is the $l_1$ norm of the vector of eigenvalues $\boldsymbol{\lambda}$. In compressed sensing, $||A||_*$ is minimized to find $A$. It is known that the use of the $l_1$ norm in this type of minimization problem produces sparse solutions \cite{fazel02}, meaning that the number of non-zero $\lambda_i$ is limited, giving as a result a low rank matrix $A$.

In our case, $||\bar{\rho}||_*=\Tr(\bar{\rho})=\sqrt{d}\bar{r}_0$  . Therefore, we solve the following optimization problem:
\be
\begin{split}
&{\rm minimize}~~~~ \bar{r}_0\\
&{\rm subject~to}~~ \sum_{\alpha=0}^{d^2-1}\bar{r}_{\alpha} E_{\alpha}\ge 0,\\
&{\rm and}~~~~~~~~~~(\vec{M}-\mbf{\tilde{\cO}}\vec{\bar{r}})^T(\vec{M}-\mbf{\tilde{\cO}}\vec{\bar{r}})\le \epsilon,
\end{split}
\label{eq:CSConvex}
\ee
where $\epsilon$ is a quantity that must be calibrated before using this method and depends on the signal-to-noise ratio and any systematic error present in the experiment. Note that the constraints ensure that the estimated density matrix is positive semidefinite and is compatible with the measured data. Again, this type of optimization problem is a convex program that can be easily solved using the same software packages described in the last section. Finally, since we minimize the coefficient $\bar{r}_0$, which is proportional to $\Tr(\bar{\rho})$, the state parametrized by the $\{\bar{r}_\alpha\}$ is generally under-normalized, i.e., $\Tr(\bar{\rho})<1$. Therefore, we must renormalize the estimated state which we can write 

\be
\bar{\rho}=\frac{1}{\bar{r}_0\sqrt{d}}\sum_{\alpha=0}^{d^2-1}\bar{r}_{\alpha}E_{\alpha}.
\ee

Comparing this technique to the maximum likelihood/least squares method discussed in the previous section, we see two interesting advantages. First, in some sense, the fact that compressed sensing is biased towards pure states makes it use the measured information more efficiently when one is reconstructing a state that is close to pure. Second, the other advantage we see is its robustness to errors. The least squares method fits the data, meaning that it also fits any errors present in the data. Fitting the errors will certainly damage the quality of the estimate. In contrast, since compressed sensing finds a high-purity state consistent with the data up to a threshold $\epsilon$, it has a built-in tolerance to errors. In fact, by estimating the appropriate error threshold, one can get better estimates using compressed sensing. A detailed study of their performance is shown in Chapter \ref{ch:Simulations} and its laboratory implementation is discussed in Chapter \ref{ch:Results}. Finally, we want to note that the techniques discussed in this section constitute a continuous measurement generalization of \cite{gross10}. 

\subsection{Control and dynamics: The Heisenberg picture and numerical integration}

An essential component of this protocol is accurate modeling of the dynamical evolution of the observables measured in the continuous signal.  Fundamental to this is decoherence induced while the system is being driven and probed.  Under typical conditions of Markovian evolution, these dynamics are generated by a Lindblad master equation,
\begin{eqnarray}
\frac{d\rho(t)}{dt} & =& \mathcal{L}_t [\rho(t)]  \nonumber \\
&=& -i [H(t), \rho(t)] -\frac{1}{2}\sum_\mu \left( L_\mu^\dg L_\mu \rho(t) + \rho(t) L_\mu^\dg L_\mu  \right) \nonumber \\
&+& \sum_\mu L_\mu  \rho(t) L_\mu^\dg,
\label{eq:mastereq}
\end{eqnarray}
where $H(t)$ is the Hamiltonian of the system, responsible for the control dynamics, and $L_\mu$ are the operators that account for decoherence in the system due to optical pumping.

The formal solution to Eq. (\ref{eq:mastereq}) is a completely positive map on the initial density operator, $\rho(t) = \mathcal{V}_t [\rho(0)]$ with $\mathcal{V}_t$ being the solution to 
\be
\frac{d \mathcal{V}_t }{dt} = \mathcal{L}_t   \mathcal{V}_t \Rightarrow  \mathcal{V}_t = \mathcal{T} \left(\exp \int_0^t \mathcal{L}_s ds \right),
\ee
where $\mathcal{T}$ is the time-ordering operator.  We seek, however, the solution to the {\em Heisenberg} evolution, given formally by the adjoint map $\cO(t)=\mathcal{V}_t^\dag[\cO(0)]$, satisfying 
\be
\frac{d \mathcal{V}_t^\dag}{dt} =   \mathcal{V}_t^\dag \mathcal{L}_t ^\dag.
\label{adjoint}
\ee
Naively, one might assume that the generalization of the master equation for $\rho(t)$, Eq. (\ref{eq:mastereq}), to the Heisenberg evolution for $\cO(t)$ is
\begin{eqnarray}
\frac{d\cO(t)}{dt}   & =& \mathcal{L}_t^\dg [\cO(t)]  \nonumber \\
&=& +i [H(t),\cO(t)] -\frac{1}{2}\sum_\mu \left(  \cO(t) L_\mu^\dg L_\mu + L_\mu^\dg L_\mu \cO(t) \right) \nonumber \\
&+& \sum_\mu L_\mu^\dg \cO(t) L_\mu.
\end{eqnarray}
However, this is {\em not generally true} since $\mathcal{V}_t^\dag \circ \mathcal{L}_t \ne  \mathcal{L}_t \circ \mathcal{V}_t^\dag$.  Note that the correct Heisenberg evolution is
\be
\frac{d \cO(t)}{dt} =   \mathcal{V}_t^\dag \left[ \mathcal{L}_t ^\dag[\cO(0)] \right],
\label{eq:FullHeisenbergEvol}
\ee
and $d \cO(t)/dt \ne  \mathcal{L}_t ^\dag[\cO(t)]$ unless $\mathcal{L}$ is time independent.  The lack of commutativity between the adjoint map and its generator will be the case for the generic time-dependent control Hamiltonians under consideration here. Because of this, the decohering Heisenberg operators do not satisfy a {\em time-local differential equation} \cite{breuer03}.  Since our procedure requires finding a solution of Eq. (\ref{eq:FullHeisenbergEvol}), this severely complicates the efficiency with which we can integrate the dynamics to determine the measurement set $\{\cO_i\}$.  

To deal with this problem in moderately large Hilbert spaces, as we will discuss in Section \ref{sec:waveforms}, we restrict our waveforms so that the control parameters are {\em piecewise constant} over a reasonable duration.  Then, over each interval in which the Hamiltonian is constant we can simply exponentiate the Lindblad generator of the superoperator map.  In addition, the operators, $\mathcal{O}_i$, are ``vectorized" to a large column of dimension $d^2-1$ and the superoperator, $\mathcal{V}_{t_i}$, is a large $(d^2-1)\times(d^2-1)$ matrix expanded in the basis $E_{\alpha}$.  Using curved bra-(row) ket-(column) notation for the supervectors and superoperators, our integration then takes the form
\be
\left(\cO_i \right| = \left(\cO_0 \right| \mathcal{V}_{t_i},
\label{eq:supopevolutionO}
\ee
where 
\be
\mathcal{V}_{t_{i+1}} = e^{\mathcal{L}_{t_i}\delta t}\, \mathcal{V}_{t_i}.
\label{eq:supopevolutionV}
\ee
For non-piecewise-constant controls, this corresponds to an Euler integration of the completely positive map.  Such an approximation will be very inefficient and numerically unstable for large dimensional systems.  For this reason a piecewise constant control is best suited to our protocol.

\section{Further technical considerations}

Beyond decoherence, an essential ingredient for accurate modeling of the dynamics is parameter estimation. The ability to reach high fidelities for the estimated states relies on the assumption that we know exactly how the system is evolving at the time that the data is taken so that we know exactly which operators $\{\cO_i\}$ are being measured.  This means that all of the parameters in the Hamiltonian must be precisely known before the quantum state reconstruction is even possible.  In practice, many such parameters can be precalibrated.  However, other parameters, such as background fields, may be unknown, and some parameters may be inhomogeneous across the ensemble in ways that are not known a priori.  Our protocol is robust because, unlike other optimal control tasks, such as state-to-state mapping, the exact parameters of the experiment need not be fixed.  As long as we can determine, a posteriori, the operating conditions via parameter estimation, and the signal is informationally complete, we can extract the quantum state with high fidelity.  

An additional parameter we must fix is the initial observable being measured.  Though abstractly we have called this $\cO_0$, in practice the true observable may delicately depend on special alignment of the apparatus.  We will see how we can use parameter estimation as well to fix this observable and the overall calibration of the signal in physical units when compared with the dimensionless units treated here.  

Finally, one technical detail that we have not discussed so far is the signal-to-noise ratio (SNR). In this particular case, we define the SNR
\be
SNR=\frac{{\rm Signal~Power}}{{\rm Noise~Power}}=\frac{(1/T)\int_0^TM(t)^2dt}{\sigma^2},
\label{eq:SNR}
\ee
where the numerator and denominator are the squares of the RMS value of the signal $M(t)$ and the noise signal, $\sigma W(t)$, respectively.

Even under ideal conditions, the performance of the QT is fundamentally limited by the quantum noise of the probe (shot noise).  For Gaussian white noise, it is essential to limit the bandwidth in which we analyze the measurement record and the dynamics must be chosen so that the relevant information about the state is contained in a limited frequency band. In addition, $1/f$ noise in the detector dictates that the relevant signal be sufficiently far from DC.  To maximize our SNR for a given experimental setup, we pass the digitized measurement signal through a digital bandpass filter that matches the signal and noise power spectral distributions at hand.  Choosing the low- and high-frequency cutoffs of this filter well within the passband of the physical detector response function ensures that the filtering of the measured signal is accurately modelled. We discuss this in detail in Section \ref{sec:Filter}.  In numerical modelling of the experiment, the simulated signal is passed through an identical digital filter.  This effectively removes any uncertainty associated with the physical detector response.

In the next section, we will apply our QT protocols to the reconstruction of states encoded in atomic hyperfine spins.  After defining the system, we can simulate a measurement record in the presence of noise, decoherence and errors for a given initial state $\rho_0$ and use it to run it through the algorithm to find the estimate $\bar{\rho}$.  In order to quantify the performance of our method, we calculate the fidelity \cite{jozsa94}
\be
\mathcal{F}(\bar{\rho},\rho_0)=\left[\rm{Tr}\left(\sqrt{\sqrt{\bar{\rho}}\rho_0\sqrt{\bar{\rho}}}\right)\right]^2.
\label{eq:fidelity}
\ee
Good performance is judged by high fidelity averaged across a collection of randomly sampled states.  
Notice that we define the fidelity of reconstruction as the square of what e.g. \cite{nielsen00} or \cite{zyczkowski2006} have defined. In fact, there is no uniformity in the literature about this definition, see e.g., \cite{jozsa94}. The reason we do it this way is that when one considers pure states, this definition gives the overlap squared of the two states, i.e., it gives a proper probability. 

In later chapters, we will use the concept of purity of a quantum state, $\mathcal{P}(\rho)$, to determine the range of applications of the methods described in this chapter. For completeness, we give its definition here \cite{nielsen00}
\be
\mathcal{P}(\rho)=\Tr{(\rho^2)}\le 1,
\label{eq:purity}
\ee
with equality reached if $\rho$ is a pure state.

\chapter{Hyperfine Spin Systems: Alkali Atoms}\label{ch:Atoms}


	In this chapter, we review the appropriate atomic physics of alkali atoms necessary to implement QT in atomic ensembles. Our platform is the hyperfine manifold of magnetic sublevels associated with the ground-electronic state of laser-cooled alkali-metal atoms, providing a Hilbert space of dimension $d=(2S+1)(2I+1)$ where $S=1/2$ is the single valence electron spin and $I$ is the nuclear spin. In particular, we work with $^{133}$Cs, whose nuclear spin is $I=7/2$, yielding hyperfine coupled spins of magnitude $F_\pm=3,~4$ and a total Hilbert space of dimension $d=16$. A detailed discussion of the relevant interactions and controls is given including all required tools to accurately model the system for experimental application. Furthermore, following \cite{deutsch10}, we discuss the decoherence model relevant in this work.

\section{Hyperfine Hamiltonian and interaction with static magnetic fields}\label{sec:staticfield}
For the applications of interest in this work, it will be necessary to consider the interaction of alkali atoms with static magnetic fields. We define the free Hamiltonian, $H_0$, as the the hyperfine interaction Hamiltonian plus the interaction Hamiltonian with external magnetic fields. Written in its most general form
\be
H_0 = A\,\mbf{I}\cdot\mbf{S} -\boldsymbol{\mu}\cdot\mbf{B_0}=A\,\mbf{I}\cdot\mbf{S} +\mu_B(g_e\mbf{S}+g_{L}\mbf{L}+g_{I}\mbf{I})\cdot\mbf{B_0},
\label{eq:GeneralBreit-RabiHamiltonian}
\ee
where $A$ is the hyperfine coupling constant, $\mbf{I}$, $\mbf{L}$, $\mbf{S}$ are the nuclear, orbital, and electronic angular momentum operators, respectively, $\mu_B$ is the Bohr magneton, $g_I=g_N\mu_N/\mu_B$, $g_L$, $g_e$ are the nuclear, orbital and electronic $g$-factors respectively, and $\mbf{B_0}$ is a static magnetic field which we call a bias field.

We now consider the case in which we are interested only in the ground state of alkali atoms, $\mbf{L}=0$, and assume that the bias field is in the $z$-direction, i.e., $\mbf{B_0}=B_0\mbf{z}$. With these assumptions, Eq. (\ref{eq:GeneralBreit-RabiHamiltonian}) simplifies to the well known Breit-Rabi Hamiltonian \cite{steck10}
\be
H_0 = A\,\mbf{I}\cdot\mbf{S} +\mu_B B_0 (g_eS_z+g_{I}I_z).
\label{eq:Breit-RabiHamiltonian}
\ee
This Hamiltonian is exactly diagonalizable in the same basis in which the $z$ component of the total angular momentum, $\mbf{F}=\mbf{S}+\mbf{I}$, is diagonal. Calling such basis $\{\ket{F,m_F}\}$ for $F=I\pm 1/2$ and $m_F=-F,\ldots,F$, the eigenvalues of Eq. (\ref{eq:Breit-RabiHamiltonian}) are given by the well known Breit-Rabi formula (see for example \cite{steck10})
\be
E_{m_F} = -\frac{\omega_{HF}}{2(2I+1)}+g_I\mu_B B_0 m_F \pm \frac{\omega_{HF}}{2}\left( 1+\frac{4m_F}{2I+1}x+x^2 \right)^{1/2}
\label{eq:Breit-RabiFormula}
\ee
where $\omega_{HF}=A(I+1/2)$ is the hyperfine splitting and $x\equiv(g_e-g_I)\mu_BB_0/\omega_{HF}$.

In this work, we will need to consider Eq. (\ref{eq:Breit-RabiFormula}) up to second order Zeeman corrections. This is because the condition $\mu_BB_0\ll\omega_{HF}$, necessary for the linear Zeeman regime, is not fulfilled by the bias field needed for the technical reasons that will become clear in Section \ref{sec:RWA}. Finding the second order Zeeman correction is simply achieved by Taylor expanding Eq. (\ref{eq:Breit-RabiFormula}) in terms of $x$ and keeping terms up to $x^2$
\be
E_{m_F} \approx -\frac{\omega_{HF}}{2(2I+1)}+g_I\mu_B B_0 m_F \pm \frac{\omega_{HF}}{2}\left( 1+\frac{2m_F}{2I+1}x+\frac{1}{2}\left(1-\frac{4m_F^2}{(2I+1)^2} \right)x^2 \right).
\label{eq:Breit-RabiSecondOrder}
\ee

Eq. (\ref{eq:Breit-RabiSecondOrder}) can be written in operator form by using the basis $\{\ket{F,m_F}\}$. By neglecting its first term, which is an overall constant energy shift, and thus irrelevant to the dynamics, we can write
\be
H_0=\frac{\omega_{HF}}{2}\left(1+\frac{x^2}{2}\right)(P_+-P_-)+\Omega_0(F_z^{(+)}-g_r F_z^{(-)})-\alpha((F_z^{(+)})^2-(F_z^{(-)})^2),
\label{eq:H0}
\ee
where $P_F$ is the projection operator onto the $F=I\pm 1/2$ manifolds, $F_z^{(\pm)}=\sum_{m_F}m_F\ket{F=I\pm 1/2,m_F}\bra{F=I\pm 1/2,m_F}$ is the projection of the $z$ component of the total angular momentum $\mbf{F}$ onto the manifolds with $F=I\pm 1/2$, $\Omega_0=g_{+}\mu_BB_0$ is the Larmor frequency produced by the bias field $B_0$, $g_r=|g_-/g_+|$ is the ratio of g-factors between the $F=I\pm 1/2$, and the Land\'e g-factors are
\be
g_+=\frac{g_e}{2I+1}+\frac{2Ig_I}{2I+1},
\label{eq:gplus}
\ee
and 
\be
g_-=-\frac{g_e}{2I+1}+\frac{2(I+1)g_I}{2I+1}.
\label{eq:gminus}
\ee
The last term in Eq. (\ref{eq:H0}) is the quadratic Zeeman shift in which we have defined $\alpha=x^2 \omega_{HF}/(2I+1)^2$. 

\section{Cs atom: atomic structure and relevant parameters}\label{sec:Cs}
\begin{figure}[t]
\begin{center}
\includegraphics[width=15.1cm,clip]{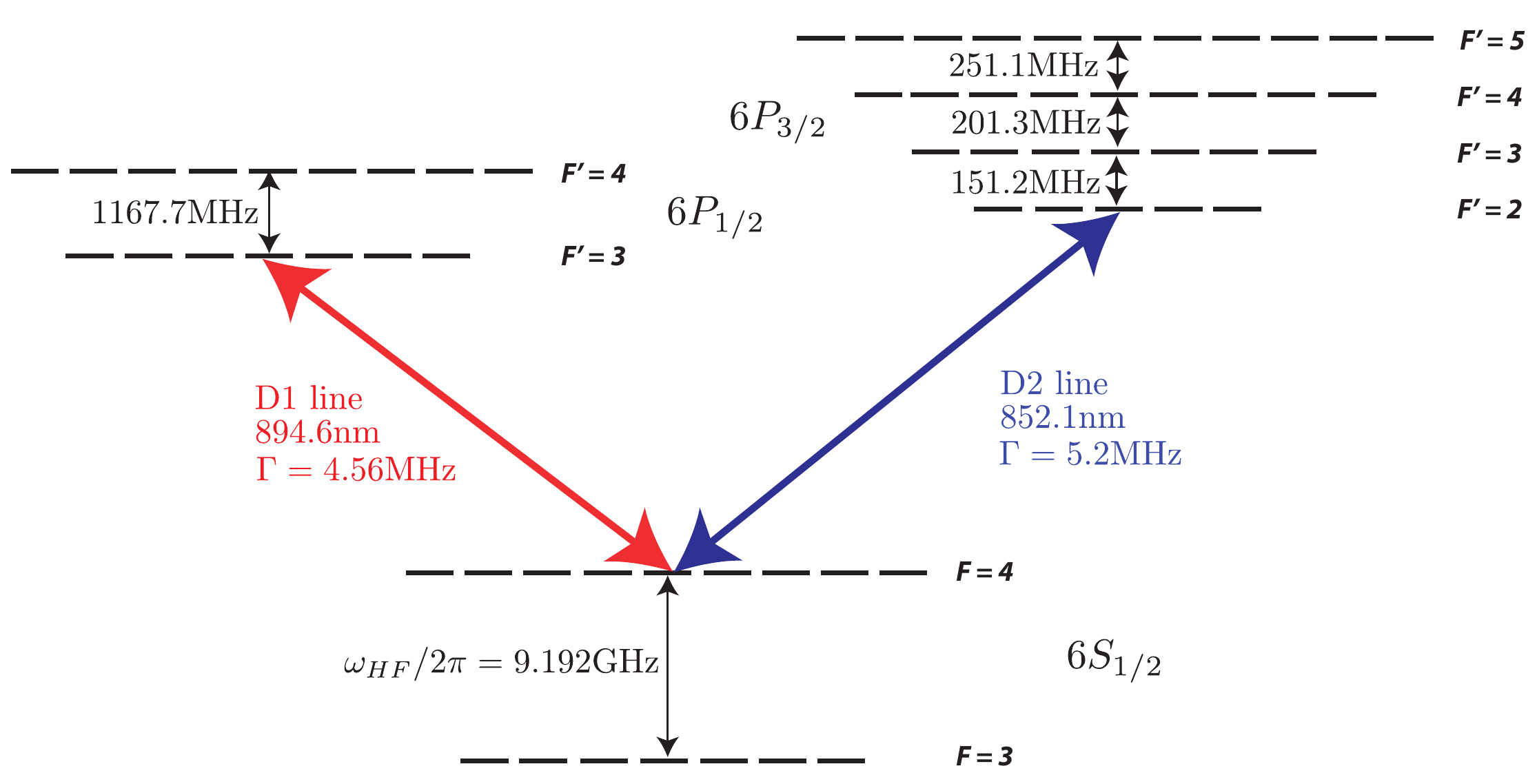}
\caption{Relevant electronic hyperfine structure of $^{133}$Cs (not to scale). The D1, $6P_{1/2}$, and D2, $6P_{3/2}$, transitions are shown. The ground state, $6S_{1/2}$ has two hyperfine manifolds with total angular momentum $F_+=4$ and $F_-=3$, which we control in our scheme.}
\label{F:CesiumLevels}
\end{center}
\end{figure}

In this section, we specialize the general discussion of alkali atoms to the cesium atom $^{133}$Cs, which is the atomic spin we will use for the remainder of this dissertation. This particular isotope of Cesium has a nuclear spin $I=7/2$, which in turn produces two electronic ground-hyperfine manifolds with angular momentum $F_+=4$ and $F_-=3$. Thus, the ground state provides us with a 16-dimensional hyperfine ($F_-=3~\oplus ~F_+=4$) Hilbert space. The relevant transitions and energy levels of $^{133}$Cs are shown in Fig. \ref{F:CesiumLevels}. The control scheme we will describe in next sections considers only the ground manifold, $6S_{1/2}$, and the D1 transition. We give the relevant parameters for that case here. The hyperfine splitting for cesium is $\omega_{HF}/2\pi \approx~9.19$GHz, and the linewidth for the D1 transition is $\Gamma/2\pi\approx~4.561$MHz. 

In addition to the parameters shown in Fig. \ref{F:CesiumLevels}, we must specify the nuclear and electronic g-factors for cesium, which are $g_I\approx -0.0004$ and $g_e\approx 2.0023$, respectively and thus, from Eqs. (\ref{eq:gplus}) and (\ref{eq:gminus}), we have $g_-\approx 0.2499$, $g_+\approx -0.2507$ and their ratio $g_r=|g_-/g_+|\approx 1.0032$. Although the relative g-factor is not much different from unity, accounting for the small difference is of fundamental importance in the applications we will discuss in subsequent chapters. 

Finally, the last quantity we need to define is the saturation intensity for unit oscillator strength for cesium whose value is $I_{sat}=0.8352~mW/cm^2$ for the D1 transition. For more detailed information on cesium transitions and atomic parameters, see for example \cite{steck10}.

\section{Hyperfine spin dynamics}

In this section, we develop the tools necessary for reconstructing the entire 16-dimensional hyperfine ground state manifold of cesium. The spins can be controlled through a combination of magnetic interactions and off-resonant optical coupling from a laser field.  The fundamental Hamiltonian is
\be
H(t) = H_0 -\boldsymbol{\mu}\cdot\mbf{B}(t) -\frac{1}{4}E_i^*  \alpha_{ij} E_j,
\label{eq:controlH}
\ee
where $H_0$ is given in Eq. (\ref{eq:H0}),  $\boldsymbol{\mu}$ is the atomic magnetic moment operator, and $\alpha_{ij}$ is the atomic dynamic polarizability operator, both depending on the atomic spin degrees of freedom.   Here and throughout, we take the laser field complex amplitude $\mbf{E}$ to be fixed, and control is accomplished through time-variation of the magnetic field, $\mbf{B}(t)$. The last term in Eq. (\ref{eq:controlH}) is the light-shift Hamiltonian, $H_{LS}$. Under typical operating conditions, where the hyperfine coupling dominates over all other forces, the total spin angular momentum, $F=I\pm1/2 =3,4$ and its projection along a quantization axis, $m_F$, are approximate good quantum numbers and define the basis of states in the Hilbert space we seek to control.  

A central component of our QT protocol is quantum controllability.  A finite dimensional system with a generic Hamiltonian of the form $H(t) =  \sum_j^k \lambda_j(t) H_j$, with external fields determined by $\lambda_j(t)$, is said to be controllable if $\{H_j\}$ generates the Lie algebra of the relevant group of unitary matrices on the space \cite{schirmer02}.  As we generally do not measure the trace of the density operator, we restrict our attention to the Lie algebra $\mathfrak{su}(d)$.  For situations in which we seek control in a single irreducible manifold $F$, the relevant algebra is $\mathfrak{su}(2F+1)$, $F=3,4$; on the the entire hyperfine manifold the algebra is $\mathfrak{su}(16)$.  For $F>1/2$, this requires Hamiltonians that are not linear in all of the components of $\mbf{F}$, as explored in \cite{silberfarb05, smith06} or a type of control that goes beyond simple rotations, as the one discussed below in Section \ref{sec:Control}.

Full controllability, and thus continuous measurement quantum state tomography, is achieved by external microwave ($\mu$w) and radio-frequency (RF) modulated magnetic fields that are used to drive the atoms, as we detail in Section \ref{sec:Control}. Since the light-shift interaction is not necessary for controllability, in principle, the control can be achieved in a decoherence free way \cite{merkel08}. However, in practice, since a laser is used to measure the system, some decoherence will be present that we must include in our model. In the next section, we describe the master equation formalism to take into account the effects of decoherence in the system.
\subsection{Master equation dynamics}\label{sec:MasterEquation}
The light-shift interaction induces dynamics on the atomic spin in addition to decoherence, depending on the probe's polarization $\boldsymbol{\epsilon}$. The combination of coherent evolution and decoherence due to photon scattering can be modeled by a master equation of the form \cite{deutsch10}
\be
\begin{split}
\frac{d\rho(t)}{dt}&=-i\left(H_{\rm eff}(t)\rho(t)-\rho(t)H_{\rm eff}^{\dg}(t)\right)+\Gamma\sum_{q}\left(\sum_{F,F_1}W_q^{F F_1}\rho^{F_1 F_1}(t) W_q^{F F_1\dg}\right.\\
                                &+\left.\sum_{F_1\ne F_2}W_q^{F_2 F_2}\rho^{F_2 F_1}(t) W_q^{F_1F_1\dg}\right).
\end{split}
\label{eq:csmaster}
\ee
In this equation, projections of operators onto subspaces with a given $F$ are denoted $A^{F_1 F_2} = P_{F_1} A P_{F_2} = \sum_{m_{1} m_{2}} \ket{F_1,m_{1}}\bra{F_1,m_{1}}A \ket{F_2,m_{2}}\bra{F_2,m_{2}}$.  The total effective Hamiltonian is given by $H_{\rm eff}(t) = H_{0}+H_B(t)+H^{LS}_{\rm eff}$, where $H_{0}$ is the hyperfine and static field interactions, as given in Eq. (\ref{eq:H0}), $H_B$ is the control magnetic field interaction, and the effective (non-Hermitian) Hamiltonian accounting for light-shift and optical pumping is
\be
H^{LS}_{\rm eff}=\frac{\Omega^2}{4}\sum_{FF'}\frac{(\boldsymbol{\epsilon}^*\cdot\vec{D}_{FF'})(\vec{D}_{F'F}^{\dg}\cdot\boldsymbol{\epsilon})}{\Delta_{F'F}+i\Gamma/2}.
\label{eq:lshamiltonian}
\ee
Here, $\Omega$ is the laser Rabi frequency for a unit oscillator strength, which is related to its intensity, $I_{probe}$, by
\be
\Omega=\Gamma\sqrt{\frac{I_{probe}}{2I_{sat}}},
\label{eq:RabiFeqIntensity}
\ee
where $I_{sat}$ is the saturation intensity for unit oscillator strength, and $\Gamma$ is the linewidth of the transition.

The strength of the transitions for  $\boldsymbol{\epsilon}$-polarized light are accounted for by the dimensionless dipole raising operator,
\be
\vec{e}_q \cdot \vec{D}_{F'F}^{\dg}=\sum_m \mathcal{K}_{JF}^{J'F'}\braket{F'm+q}{Fm;1q}\ketbra{F'm+q}{Fm}
\ee
where the coefficient $\mathcal{K}_{JF}^{J'F'}$ is given in terms of a Wigner 6j symbol
\be
\mathcal{K}_{JF}^{J'F'}=(-1)^{F'+I+J'+1}\sqrt{(2J'+1)(2F+1)}\left\{\begin{array}{ccc}F' & I & J'\\ J & 1 & F\end{array}\right\}.
\ee
The Lindblad jump operators are given by
\be
W_q^{F_b F_a}=\sum_{F'}\frac{\Omega/2}{\Delta_{F'F_a}+i\Gamma/2}(\vec{e}_q^*\cdot\vec{D}_{F_b F'})(\vec{D}_{F'F_a}^{\dg}\cdot\boldsymbol{\epsilon}),
\label{eq:jump}
\ee
describing absorption of a photon with polarization $\boldsymbol{\epsilon}$, emission of a photon with polarization $q$, and optical pumping between hyperfine manifolds $F_a$ and $F_b$.  Transfer of population between sublevels by optical pumping occurs at a rate $\gamma_{F_a m_a \rightarrow F_b m_b} = \sum_q |\bra{F_b m_b} W^{F_b F_a}_q \ket{F_a m_a}|^2$.  The final term in the master equation, Eq. (\ref{eq:csmaster}), proportional to $\rho^{F_2 F_1}$, represents transfer of coherences that may exist between hyperfine manifolds, but are preserved in spontaneous emission when the detuning of the light is sufficiently large.

\subsection{Control Hamiltonian}\label{sec:Control}
An important question that has not yet been addressed is the way in which we drive the dynamics to generate the measurement record.  As discussed in Chapter \ref{ch:QT}, a sufficient condition is that the dynamics generate an informationally complete set of observables $\{\cO_i\}$, meaning that they span the Lie algebra $\mathfrak{su}(d)$.  The quantum dynamics must thus be ``controllable" in the Lie algebraic sense \cite{schirmer02}.

\begin{figure}[t]
\begin{center}
\includegraphics[width=15.1cm,clip]{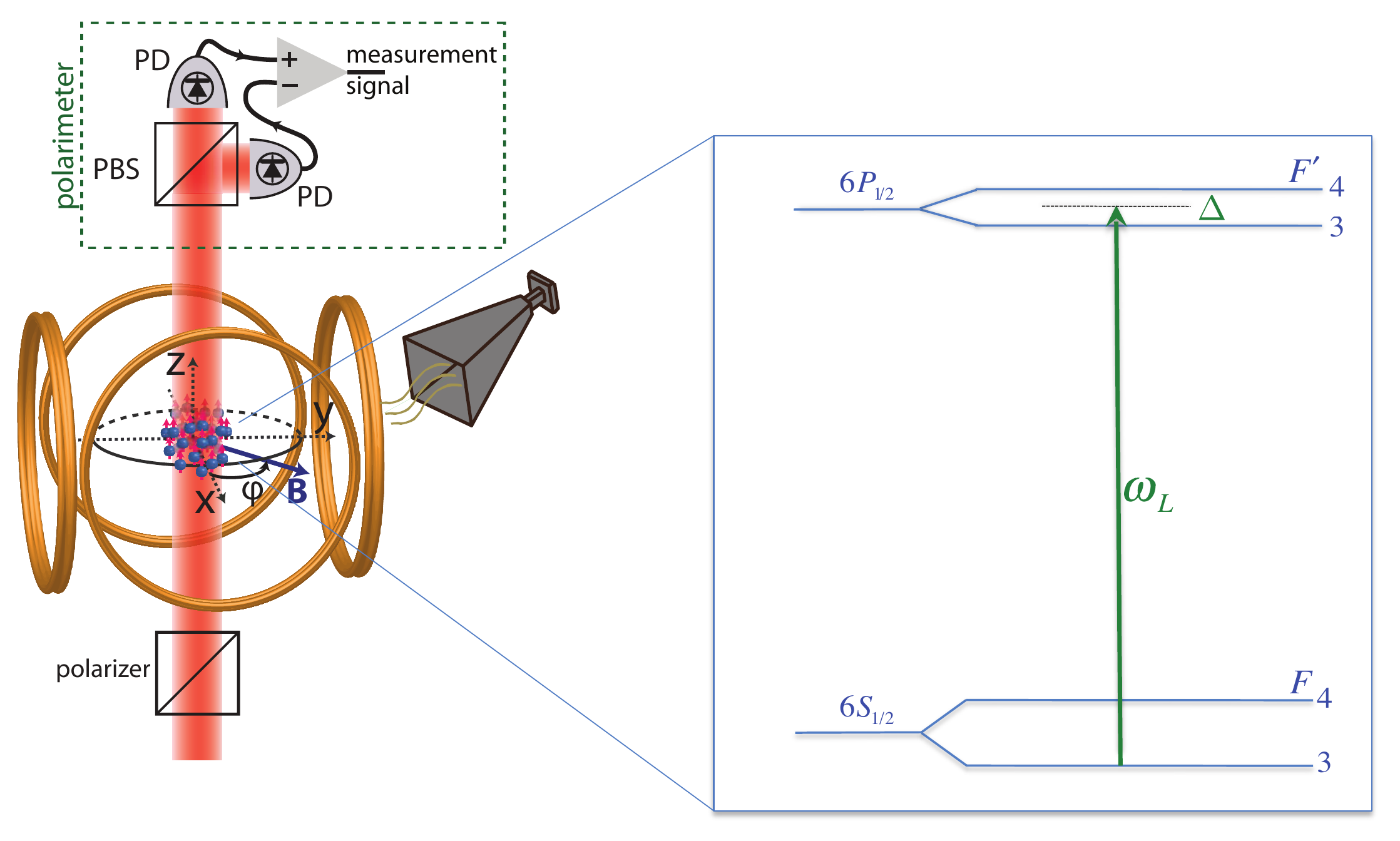}
\caption{Schematic of our system geometry.  A cold gas of atoms is collected from a magneto-optic-trap/optical molasses, and optically pumped to form a nearly pure ensemble of identical spins.  The spins are controlled through a combination of light-shift interaction, magnetic fields produced by pairs of Helmholtz coils, and microwave fields.  A measurement of the spins is performed by polarization analysis of the transmitted probe.  A sketch of the atomic level structure for the D1 transition in $^{133}$Cs is shown inset (not to scale). This picture was taken from \cite{deutsch10} and modified to fit our current experimental scheme.}
\label{F:setup}
\end{center}
\end{figure}

\begin{figure}[t]
\begin{center}
\includegraphics[width=8.7cm,clip]{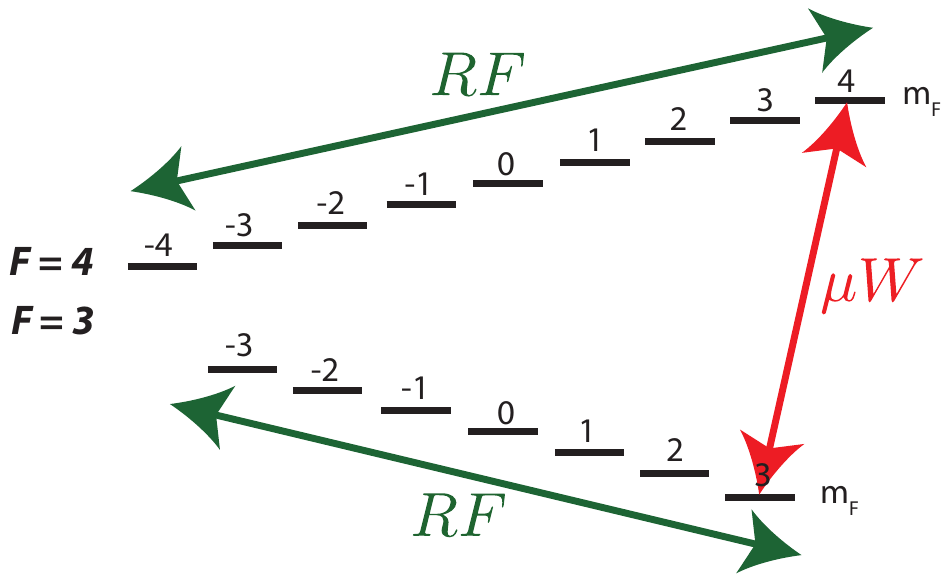}
\caption{Ground electronic-hyperfine manifolds of cesium. Each level corresponds to an element of the basis $\{\ket{F,m_F}\}$ of the 16-dimensional Hilbert space. Energy degeneracy has been broken by the bias field, $B_0$, interaction. The arrows show the levels that are connected by the $RF$ and $\mu w$ fields. The $RF$ fields cause the atoms to Larmor precess independently for $F_+=4$ and $F_-=3$. The $\mu W$ field connects resonantly the stretched states $\ket{3,3}$ and $\ket{4,4}$ making the system fully controllable.}
\label{F:CSGround}
\end{center}
\end{figure}

To achieve controllability, we make use of the type of control developed in \cite{merkel08}, which employs microwave ($\mu$w) and radio-frequency (RF) modulated external magnetic fields to drive the atoms. The RF-fields drive rotations, i.e., Larmor precession on the $F_- = 3$ and $F_+ = 4$ manifolds independently, whereas the microwave fields drive a resonant transition between two Zeeman levels in $F_- = 3$ and $F_+ = 4$, thus making the system fully controllable. A schematic of the system, including atomic level structure, control, and measurement components, is shown in Fig. \ref{F:setup}. For illustration, Fig. \ref{F:CSGround} shows in detail the ground manifolds of cesium with arrows connecting the levels that are driven by the RF (green) and $\mu$w (red) fields.

The fundamental Hamiltonian, Eq. (\ref{eq:controlH}), can be written in this case $H(t)=H_0+H_{RF}+H_{\mu w}+H_{LS}$. The free Hamiltonian, $H_0$, as defined in Section \ref{sec:staticfield}, includes the hyperfine interaction and the Zeeman shift produced by the bias magnetic field interaction, which is necessary to define the quantization axis and to resolve microwave-induced transitions of a pair of magnetic sublevels. 

Following the control scheme studied in \cite{merkel08}, we can write the control magnetic field $\mbf{B}(t)=\mbf{B}_{RF}(t)+\mbf{B}_{\mu w}(t)$ where $\mbf{B}_{RF}(t)$ and $\mbf{B}_{\mu w}(t)$ are the RF- and $\mu$w- modulated magnetic fields. We now assume that the RF magnetic field has two orthogonal components along the $x$ and $y$ directions and has the form
\be
\mbf{B}_{RF}(t)=B_{x}(t)\cos{(\omega_{RF}t-\phi_x(t))}\mbf{x}+B_{y}(t)\cos{(\omega_{RF}t-\phi_y(t))}\mbf{y}
\ee
where $\phi_x(t)$, and $\phi_y(t)$ are the control phases of the magnetic field in the $x$ and $y$ directions respectively and $\omega_{RF}$ is the frequency at which the RF fields are modulated.

In normal operation, the magnitude of the RF fields is much smaller than the hyperfine splitting, $\mu_BB_{x}(t),~\mu_BB_{y}(t)\ll\omega_{HF}$; therefore, the RF part of the control Hamiltonian can be written in the standard first order Zeeman approximation
\be
\begin{split}
H_{RF}&=\Omega_x(t)\cos{(\omega_{RF}t-\phi_x(t))}(F_x^{(+)}-g_r F_x^{(-)})\\
             &+\Omega_y(t)\cos{(\omega_{RF}t-\phi_y(t))}(F_y^{(+)}-g_r F_y^{(-)}),
\end{split}
\label{eq:Hrf}
\ee
where the Larmor frequency is defined as $\Omega_{i}(t)=g_+\mu_{B}B_{i}(t)$, for $i=x,y$. This Hamiltonian allows for independent ${\sf SU}(2)$ rotations within the $F_-=3$ and $F_+=4$ manifolds.

For control via application of microwave radiation, we consider a purely $\sigma_+$-polarized field.  While in practice the polarization of microwaves is not well controlled at the position of the atoms, this is not critical, since ultimately we will drive a selected two-level transition through its unique resonance frequency. We account for the effect of the relevant off-resonant ac-Stark shifts caused by microwaves and choose one polarization to analyze, for simplicity.  Under this assumption, the microwave control Hamiltonian is
\be
H_{\mu w}=\Omega_{\mu w}(t)\cos{(\omega_{\mu w}t-\phi_{\mu w}(t))}\sum_{m=-3}^{3}\braket{4,m+1}{3,m;1,1}\sigma_x^{(m)},
\label{eq:Huw}
\ee
where the bare microwave Rabi frequency is $\Omega_{\mu w}(t)=\sqrt{\frac{7}{8}}\mu_{B}B_{\mu w}(t)$, $\phi_{\mu w}(t)$ is its control phase, $\braket{4,m+1}{3,m;1,1}$ is the Clebsch-Gordan coefficient associated to the transition $|3,m\rangle\rightarrow|4,m+1\rangle$, and  $\sigma_x^{(m)}=\ket{4,m+1}\bra{3,m}+\ket{3,m}\bra{4,m+1}$. This Hamiltonian couples Zeeman levels in the two different manifolds, taking into account the resonant driving as well as the off-resonant level shifts. 

 It is convenient to use the effective, nonHermitian, light-shift Hamiltonian, Eq. (\ref{eq:lshamiltonian}), instead of $H_{LS}$ to account for decoherence. For linear polarization of the laser probe along $x$, the effective light-shift Hamitonian Eq. (\ref{eq:lshamiltonian}) can be expressed in irreducible tensor components as \cite{deutsch10}
\begin{equation}
H_{{\rm eff}}^{LS}=\gamma_{sc} \sum_{F}\left[\left(\beta^{(0)}_F - \beta^{(2)}_F \frac{F(F+1)}{3} \right) I_F + \beta^{(2)}_F F_x^2 \right]
\label{eq:effhamiltonianls}
\end{equation}
where $I_F$ is the identity operator on the hyperfine manifold $F$. The term proportional to $I_F$ is the ``scalar light shift" and the term proportional to $F_x^2$ is the ``tensor light shift". Here
\begin{equation}
\beta^{(K)}_F =\frac{2 \Delta_c^2}{\Gamma^2} \sum_{F'}C^{(K)}_{F'F} \frac{\Gamma/2}{ \Delta_{F'F} +i \Gamma/2}
\label{eq:betas}
\end{equation}
are complex coupling coefficients depending on the rank-$K$ atomic polarizability, and where  
\begin{subequations}
\begin{align}
C^{(0)}_{F'F} &=(-1)^{3F-F'+1}\frac{1}{\sqrt{3}}\frac{2F'+1}{\sqrt{2F+1}}\left\{\begin{array}{ccc}F & 1 & F'\\ 1 & F & 0\end{array}\right\}\left|\mathcal{K}_{JF}^{J'F'}\right|^2\\
C^{(1)}_{F'F} &=(-1)^{3F-F'}\sqrt{\frac{3}{3}}\frac{2F'+1}{\sqrt{F(F+1)(2F+1)}}\left\{\begin{array}{ccc}F & 1 & F'\\ 1 & F & 1\end{array}\right\}\left|\mathcal{K}_{JF}^{J'F'}\right|^2\\
\begin{split}
C^{(2)}_{F'F} &=(-1)^{3F-F'}\frac{\sqrt{30}(2F'+1)}{\sqrt{F(F+1)(2F+1)(2F-1)(2F+3)}}\\
                      &\times\left\{\begin{array}{ccc}F & 1 & F'\\ 1 & F & 2\end{array}\right\}\left|\mathcal{K}_{JF}^{J'F'}\right|^2
\end{split}                      
\end{align}
\label{eq:tensorcoeff}
\end{subequations}
are the irreducible tensor coefficients evaluated in $J=1/2$ and $J'=1/2$ corresponding to the transition between the ground state and some excited state in the D1 line transition ($S_{J=1/2}$ to $P_{J'=1/2}$), respectively. The real part of Eq. (\ref{eq:effhamiltonianls}) leads to the light shift and the imaginary part causes decoherence via photon scattering.  For emphasis, we have explicitly factored out the characteristic photon scattering rate $\gamma_{sc} = (\Omega^2  \Gamma)/(4 \Delta_c^2)$, which sets the time scale for dynamics on the atom-photon interaction. For the remainder of this dissertation, the characteristic detuning, $\Delta_c$ is defined for the transition $(6S_{1/2})F=3$ to $(6P_{1/2})F'=3$.
\subsection{Rotating wave approximation}\label{sec:RWA}
With the control Hamiltonian in hand, we employ the rotating wave approximation (RWA) to eliminate its explicit time dependence. 
In general, for applicable parameters, we must go beyond the usual linear Zeeman effect, as shown in Eq. ({\ref{eq:H0}}), and first order rotating wave approximation, which substantially complicates the Hamiltonian beyond what was presented in \cite{merkel08}.  To begin, we transform to a frame that is rotating at the frequency of the control fields, according to the unitary transformation $U(t)=U_{RF}U_{\mu w}$, where 
\begin{subequations}
\begin{align}
U_{RF}&=\exp{[-i\omega_{RF}t(F_z^{(4)}-F_z^{(3)})]}\\
U_{\mu w}&=\exp{[-i\frac{\theta t}{2}(P_4-P_3)]}
\end{align}
\label{eq:rotatingframeU}
\end{subequations}
with $\theta=\omega_{\mu w}-(m_++m_-)\omega_{RF}$, where $m_+$ and $m_-$ label two Zeeman levels corresponding to the $F_+=4$ and $F_-=3$ manifolds respectively. It then follows from  Eq. (\ref{eq:H0}), 
\be
\begin{split}
H'_{0}&=U^{\dagger}(t)H_0U(t)-iU^{\dagger}\frac{dU}{dt}\\
           &= \left(\frac{3\Omega_0}{2}(1-g_r)+\frac{25}{2}\alpha+\frac{1}{2}(7\Delta_{RF}-\Delta_{\mu w})\right )(P_+-P_-)\\
           &-\Delta_{RF} F_z^{(+)} +\left(  \Delta_{RF} +\Omega_0(1-g_r) \right) F_z^{(-)}-\alpha ((F_z^{(+)})^2-(F_z^{(-)})^2),
\end{split}
\label{eq:AppH0R}
\ee
where we have chosen $m_+=4$, $m_-=3$, $\Delta_{RF}=\omega_{RF}-\Omega_0$, $\Delta_{\mu w}=\omega_{\mu w}-\omega_0$, with $\omega_0=\omega_{HF}+(4+3g_r)\Omega_0+7\alpha$ being the on-resonant transition for the two-level system formed by the stretched states $|3,3\rangle$ and $|4,4\rangle$. Although our goal is to be as close to resonance as possible ($\Delta_{RF}=\Delta_{\mu w}=0$), in practice we must also account for nonzero detunings that might result, e.g., from gradients in $\Omega_0$ across the ensemble.

Going to the rotating frame, the RF Hamiltonian in Eq. (\ref{eq:Hrf}) transforms to $H'_{RF}(t)=U^{\dagger}(t)H_{RF}(t)U(t)$, yielding
\be
\begin{split}
H'_{RF}(t)&=\frac{\Omega_x(t)}{2}(\cos{(2\omega_{RF}t-\phi_x(t))}+\cos{(\phi_x(t)}))(F_x^{(4)}-g_r F_x^{(3)})\\
	        &-\frac{\Omega_x(t)}{2}(\sin{(2\omega_{RF}t-\phi_x(t))}+\sin{(\phi_x(t)}))(F_y^{(4)}+g_r F_y^{(3)})\\
                  &+\frac{\Omega_y(t)}{2}(\cos{(2\omega_{RF}t-\phi_y(t))}+\cos{(\phi_y(t)}))(F_y^{(4)}-g_r F_y^{(3)})\\
                  &+\frac{\Omega_y(t)}{2}(\sin{(2\omega_{RF}t-\phi_y(t))}+\sin{(\phi_y(t)}))(F_x^{(4)}+g_r F_x^{(3)})
\end{split}
\label{eq:HrfRot}
\ee

In the same manner, we can write the microwave Hamiltonian, Eq. (\ref{eq:Huw}), in the rotating frame $H'_{\mu w}(t)=U^{\dagger}(t)H_{\mu w}(t)U(t)$
\be
\begin{split}
H'_{\mu w}(t)&=\frac{\Omega_{\mu w}(t)}{2}(\cos{(2\omega_{\mu w}t-\phi_{\mu w}(t))}+\cos{(\phi_{\mu w}(t)}))\sigma_x\\
                        &+\frac{\Omega_{\mu w}(t)}{2}(\sin{(2\omega_{\mu w}t-\phi_{\mu w}(t))}+\sin{(\phi_{\mu w}(t)}))\sigma_y\\
                        &+\frac{\Omega_{\mu w}(t)}{2}\sum_{m\neq 3}\braket{3,m;1,1}{4,m+1}(\cos{(2\omega_{\mu w}t+2(m-3)\omega_{RF}t-\phi_{\mu w}(t))}\\
                        &+\cos{(2(m-3)\omega_{RF}t+\phi_{\mu w}(t)}))\sigma_x^{(m)}\\
                  &+\frac{\Omega_{\mu w}(t)}{2}\sum_{m\neq 3}\braket{3,m;1,1}{4,m+1}(\sin{(2\omega_{\mu w}t+2(m-3)\omega_{RF}t-\phi_{\mu w}(t))}\\
                  &+\sin{(2(m-3)\omega_{RF}t+\phi_{\mu w}(t)}))\sigma_y^{(m)}
\end{split}
\label{eq:HuwRot}
\ee
where $\sigma_y^{(m)}=-i\ket{3,m}\bra{4,m+1}+i\ket{4,m+1}\bra{3,m}$. Note that we have explicitly separated the resonant terms from the off-resonant ones. The off-resonant interaction produces an AC-Zeeman shift of the magnetic levels that must be accounted for in the regime we consider in our simulations. We have defined for the resonant transition $\sigma_x=\sigma_x^{(3)}$ and $\sigma_x=\sigma_y^{(3)}$.

Finally, the effective light-shift Hamiltonian, given by Eq. (\ref{eq:lshamiltonian}) written in its irreducible tensor representation (as in Eq. (\ref{eq:effhamiltonianls})), can be expressed in the rotating frame as $H'^{LS}_{{\rm eff}}(t)=U^{\dagger}(t)H^{LS}_{{\rm eff}}U(t)$, yielding 
\be
\begin{split}
H'^{LS}_{{\rm eff}}=&\gamma_{sc}\sum_F\left[\left(\beta_F^{(0)}-\beta_F^{(2)}\frac{F(F+1)}{3}\right)I_F\right.\\ 
                              &+\left.\beta_F^{(2)} (F_x^{(F)}\cos{(\omega_{RF}t)}+F_y^{(F)}\sin{(\omega_{RF}t)})^2\right],
\end{split}
\label{eq:LShamiltonianRot}
\ee
where $\beta_{F}^{(K)}$ is given in Eqs. (\ref{eq:betas}).

Given the Hamiltonian in the rotating frame, we proceed to apply the RWA. Typically this is a straightforward task, equivalent to dropping the rapidly oscillating counter-rotating terms. This is true for the case of the $\mu$w and effective light-shift Hamiltonians in the frame rotating at $\omega_{\mu w}$, however, for the case of the RF and $\mu$w Hamiltonians in the frame rotating at $\omega_{RF}$, since the RF-Larmor and $\mu$w-Rabi frequencies of the control magnetic fields, $\Omega_x$, $\Omega_y$, and $\Omega_{\mu w}$, may not be much larger that the driving frequency, $\omega_{RF}$, a second order correction of the RWA is needed to keep the model as general as possible. For that reason, we leave the discussion of the RWA for the RF and $\mu$w Hamiltonians to the next section.

As stated above, the RWA, in the frame rotating at $\omega_{\mu w}$, for the microwave Hamiltonian is straightforward since in general $\Omega_{\mu w} \ll \omega_{\mu w}$. However, in order to keep the correct off-resonant terms that lead to microwave-induced AC Stark shifts, we will need to do a second order RWA correction in the frame oscillating at $\omega_{RF}$, which we describe in the next section. Here, we only take the average of Eq. (\ref{eq:HuwRot}) with respect to the terms oscillating at $\omega_{\mu w}$ while leaving the terms oscillating at $\omega_{RF}$ alone. The resulting microwave Hamiltonian is

\be
\begin{split}
H'_{\mu w}(t)&=\frac{\Omega_{\mu w}(t)}{2}(\cos{(\phi_{\mu w}(t)})\sigma_x+\sin{(\phi_{\mu w}(t)}))\sigma_y\\
                        &+\frac{\Omega_{\mu w}(t)}{2}\sum_{m\neq 3}\braket{3,m;1,1}{4,m+1}\cos{(2(m-3)\omega_{RF}t+\phi_{\mu w}(t)})\sigma_x^{(m)}\\
                  &+\frac{\Omega_{\mu w}(t)}{2}\sum_{m\neq 3}\braket{3,m;1,1}{4,m+1}\sin{(2(m-3)\omega_{RF}t+\phi_{\mu w}(t)})\sigma_y^{(m)}.
\end{split}
\label{eq:HuwRotOnlyuW}
\ee

The RWA for the effective light-shift Hamiltonian, Eq. (\ref{eq:LShamiltonianRot}), is an excellent approximation in this case since generally we will have parameters such that the Zeeman splitting is much larger than the rate of coherent coupling induced by the light shift, $\Omega_0 \gg \text{Re}(\beta^{(2)}) \gamma_{sc}$.  Thus, averaging over the rapidly varying terms, we obtain the effective light-shift Hamiltonian 
\be
H'^{LS}_{{\rm eff}}=\gamma_{sc}\sum_F \left[\left( \beta_F^{(0)} + \beta_F^{(2)} \frac{F(F+1)}{6}  \right) I_F - \frac{\beta_F^{(2)}}{2} F_z^2\right].
\label{eq:LShamiltonianRWA}
\ee
Note, under the RWA, the light-shift does not drive coherences between magnetic sublevels defined by the quantization axis.  Such coherent couplings are no longer resonant in the presence of a strong bias field.  Also note that in considering dynamics over the full hyperfine manifold, we must retain the real part of the scalar light shift, since generally $\beta_3^{(0)} \ne \beta_4^{(0)}$.  The scalar contribution to the light shift thus drives coherences between the $F_-=3$ and $F_+=4$ manifolds.  

\subsection{RWA corrections}

The case of the RF and $\mu$w Hamiltonians is much more complex as mentioned above.  In order to maintain rapid control, the RF-Larmor rotation frequencies must be sufficiently large.  However, if the bias field, or equivalently $\omega_{RF}$, is not sufficiently large, the condition $\Omega_x$, $\Omega_y$, $\Omega_{\mu w}$ $\ll \omega_{RF}$ will not be fulfilled.  In this case, we must consider higher order corrections to the RF and $\mu$w Hamiltonians in the RWA. To do this, we follow \cite{fox87} and use the method of averages for ordinary differential equations, which we briefly review below, to provide the required correction to the RWA. 

Given a set of first order differential equations of the form 
\be
\frac{d{\bf x}}{dt}=\epsilon~ {\bf f}({\bf x},t,\epsilon),
\ee
where ${\bf x}$ represents the state of the system, ${\bf f}({\bf x},t,\epsilon)$ is a periodic function with period $T$, and $\epsilon$ is a small parameter. We seek an approximate solution of the equivalent averaged system ${\bf y}$ under the transformation ${\bf x}={\bf y}+\epsilon~{\boldsymbol \omega}({\bf y},t,\epsilon)$, where ${\boldsymbol \omega}({\bf y},t,\epsilon)$ is also periodic with period $T$ and is given by 
\be
{\boldsymbol \omega}({\bf y},t')=\int_0^{t'}({\bf f}({\bf y},t'')-{\bf\bar{f}}({\bf y}))dt'',
\ee
where 
\be
{\bf \bar{f}}({\bf y})=\frac{1}{T}\int_0^T{\bf f}({\bf y},t,\epsilon)dt.
\ee
The averaging theorem says that the equations of motion of the equivalent system are 
\be
\frac{d{\bf y}}{dt}=\epsilon~ {\bf \bar{f}}({\bf y})+\epsilon^2~{\bf f}_1({\bf y},t,\epsilon)+O(\epsilon^3), 
\ee
where 
\be
{\bf f}_1({\bf y},t,\epsilon)=\nabla {\bf f}({\bf y},t,\epsilon){\boldsymbol \omega}({\bf y},t,\epsilon)-\nabla {\boldsymbol \omega}({\bf y},t,\epsilon){\bf f}({\bf y},t,\epsilon).
\ee

In the next 2 sections we apply this method to the $\mu$w and RF Hamiltonians separately.

\subsubsection{Microwave Hamiltonian}

In this case, it is convenient to define the small parameter $\epsilon=\Omega_{\mu w}(t)/\omega_{RF}$.  Turning the RF Hamiltonian off ($\Omega_{x}=\Omega_{y}=0$), the Hamiltonian restricted to the states connected by the microwave field can be written
\be
\begin{split}
\tilde{H}'_{\mu w}(t)&=\sum_{m\neq 3} \frac{\Delta_{\mu w}^{(m)}}{2}\sigma_z^{(m)}+\frac{\Omega_{\mu w}(t)}{2}\left[\cos{\phi_{\mu w}(t)}\sigma_x+\sin{\phi_{\mu w}(t)}\sigma_y\right]\\
                        	         &+\frac{\Omega_{\mu w}(t)}{2}\sum_{m\neq 3}\braket{3,m;1,1}{4,m+1}\cos{(2(m-3)\omega_{RF}t+\phi_{\mu w}(t)})\sigma_x^{(m)}\\
                  			&+\frac{\Omega_{\mu w}(t)}{2}\sum_{m\neq 3}\braket{3,m;1,1}{4,m+1}\sin{(2(m-3)\omega_{RF}t+\phi_{\mu w}(t)})\sigma_y^{(m)}
\end{split}
\label{eq:HuWrestricted}
\ee
where $\sigma_z^{(m)}=\ket{3,m}\bra{3,m}-\ket{4,m+1}\bra{4,m+1}$ and where we define
\be
\Delta_{\mu w}^{(m)}=\frac{\Delta_{\mu w}}{2}-\frac{\Omega_0(1-g_r)(3-m)}{2}-(12-m^2-m)\alpha-\Delta_{RF}(3-m).
\ee
Notice that the microwave field couples a series of independent 2-level systems labeled by $m$. Thus, choosing a generic subsystem would suffice for our analysis. In that context, further projection of Eq. (\ref{eq:HuWrestricted}) into the subspace defined by $\{ \ket{4,m+1}, \ket{3,m} \}$, for $m\neq 3$ is
\be
\tilde{H}'^{(m)}_{\mu w}(t)= \frac{\Delta_{\mu w}^{(m)}}{2}\sigma_z^{(m)}+\omega_{RF}\frac{\chi^{(m)}(t)}{2}\sigma_x^{(m)}+\omega_{RF}\frac{\upsilon^{(m)}(t)}{2}\sigma_y^{(m)}
\label{eq:HuWMorerestricted}
\ee

where 
\be
\chi^{(m)}(t)=\frac{\Omega_{\mu w}(t)}{\omega_{RF}}\braket{3,m;1,1}{4,m+1}\cos{(2(m-3)\omega_{RF}t+\phi_{\mu w}(t)}),
\ee
and
\be
\upsilon^{(m)}(t)=\frac{\Omega_{\mu w}(t)}{\omega_{RF}}\braket{3,m;1,1}{4,m+1}\sin{(2(m-3)\omega_{RF}t+\phi_{\mu w}(t)}).
\ee

The Heisenberg equations of motion for the components of the pseudo spin angular momentum, $d\sigma^{(m)}_k/dt=i[\tilde{H}'_{\mu w}(t),\sigma^{(m)}_k]$, can then be easily written
\begin{subequations}
\begin{align}
\frac{d\sigma_x^{(m)}}{dt'}&=\epsilon\left(\upsilon^{(m)}(t')\sigma_z^{(m)}-\frac{\Delta_{\mu w}^{(m)}}{\Omega_{\mu w}(t)}\sigma_y^{(m)}\right),\\
\frac{d\sigma_y^{(m)}}{dt'}&=\epsilon\left(-\chi^{(m)}(t')\sigma_z^{(m)}+\frac{\Delta_{\mu w}^{(m)}}{\Omega_{\mu w}(t)}\sigma_x^{(m)}\right),\\
\frac{d\sigma_z^{(m)}}{dt'}&=\epsilon\left(\chi^{(m)}(t')\sigma_y^{(m)}-\upsilon^{(m)}(t')\sigma_x^{(m)}\right),
\end{align}
\end{subequations}
where we have scaled the time so that $t'=\omega_{RF}t$. This system of differential equations is in the form needed to apply the averaging theorem when we note that
\be
{\bf x}\rightarrow\left[\begin{array}{c} \sigma_x^{(m)}\\\sigma_y^{(m)}\\\sigma_z^{(m)}\end{array}\right],~ {\bf f}({\bf x},t')\rightarrow\left[\begin{array}{c} \upsilon^{(m)}(t')\sigma_z^{(m)}-\tilde{\Delta}^{(m)}\sigma_y^{(m)}\\-\chi^{(m)}(t')\sigma_z^{(m)}+\tilde{\Delta}^{(m)}\sigma_x^{(m)}\\ \chi^{(m)}(t')\sigma_y^{(m)}-\upsilon^{(m)}(t')\sigma_x^{(m)}   \end{array}\right],
\ee
where, for convenience, we have defined $\tilde{\Delta}^{(m)}=\Delta_{\mu w}^{(m)}/\Omega_{\mu w}(t)$.
Transforming the original system to the averaged equivalent one, we have
\be
{\bf y}\rightarrow\left[\begin{array}{c} \bar{\sigma}_x^{(m)}\\ \bar{\sigma}_y^{(m)}\\ \bar{\sigma}_z^{(m)}\end{array}\right],~ {\bf \bar{f}}({\bf y})\rightarrow\left[\begin{array}{c}-\tilde{\Delta}^{(m)}\bar{\sigma}_y^{(m)}\\\tilde{\Delta}^{(m)}\bar{\sigma}_x^{(m)}\\ 0   \end{array}\right].
\ee
The only terms that participate in the averaging process are the fast oscillating ones, while the slow varying terms are treated as constant.

We now proceed to calculate the function ${\boldsymbol \omega}({\bf y},t')$ by again integrating only over the fast varying terms
\be
{\boldsymbol \omega}({\bf y},t')\rightarrow\left[\begin{array}{c} \Upsilon(t')\bar{\sigma}_z^{(m)}\\-{\rm X}(t')\bar{\sigma}_z^{(m)}\\ {\rm X}(t')\bar{\sigma}_y^{(m)}-\Upsilon(t')\bar{\sigma}_x^{(m)}   \end{array}\right]
\ee
where 
\be
{\rm X}(t)=\frac{\braket{3,m;1,1}{4,m+1}}{2(m-3)}(\sin{(2(m-3)\omega_{RF}t+\phi_{\mu w}(t))}-\sin{(\phi_{\mu w}(t))}),
\ee
and
\be
\Upsilon(t)=\frac{\braket{3,m;1,1}{4,m+1}}{2(m-3)}(-\cos{(2(m-3)\omega_{RF}t+\phi_{\mu w}(t))}+\cos{(\phi_{\mu w}(t))}).
\ee
We can thus write
\be
\overline{{\bf f}_1({\bf y},t')}\rightarrow \Lambda\left[\begin{array}{c} -\bar{\sigma}_y^{(m)}\\ \bar{\sigma}_x^{(m)}\\ 0   \end{array}\right]+\Xi\left[\begin{array}{c} -\sin{(\phi_{\mu w}(t))}\bar{\sigma}_z^{(m)}\\ \cos{(\phi_{\mu w}(t))}\bar{\sigma}_z^{(m)}\\ \sin{(\phi_{\mu w}(t))}\bar{\sigma}_x^{(m)}-\cos{(\phi_{\mu w}(t))}\bar{\sigma}_y^{(m)}   \end{array}\right]
\ee
where
\be
\Lambda = \frac{|\braket{3,m;1,1}{4,m+1}|^2}{4(m-3)},~{\rm and}~~\Xi = \frac{\braket{3,m;1,1}{4,m+1}\tilde{\Delta}^{(m)}}{2(m-3)}.
\ee

Putting all together, the Heisenberg equations for the pseudo spin components, up to second order correction of the RWA, are
\be
\begin{split}
\frac{d}{dt'}\left[\begin{array}{c} \bar{\sigma}_x^{(m)}\\ \bar{\sigma}_y^{(m)}\\ \bar{\sigma}_z^{(m)}\end{array}\right]&=\epsilon\left[\begin{array}{c} -\Delta_{RF}\bar{\sigma}_y^{(m)}\\\Delta_{RF}\bar{\sigma}_x^{(m)}\\ 0  \end{array}\right]\\
&+\epsilon^2\left\{\Lambda\left[\begin{array}{c} -\bar{\sigma}_y^{(m)}\\ \bar{\sigma}_x^{(m)}\\ 0   \end{array}\right]+\Xi\left[\begin{array}{c} -\sin{(\phi_{\mu w}(t))}\bar{\sigma}_z^{(m)}\\ \cos{(\phi_{\mu w}(t))}\bar{\sigma}_z^{(m)}\\ \sin{(\phi_{\mu w}(t))}\bar{\sigma}_x^{(m)}-\cos{(\phi_{\mu w}(t))}\bar{\sigma}_y^{(m)}   \end{array}\right]\right\}.
\end{split}
\label{eq:RWAsecondorderuW}
\ee
Equivalently, using Eq. (\ref{eq:RWAsecondorderuW}), we can write the Hamiltonian, Eq. (\ref{eq:HuWMorerestricted}), up to second order correction
\be
\begin{split}
\tilde{H}'^{(m)}_{\mu w}(t) &= \frac{\Delta_{\mu w}^{(m)}}{2}\sigma_z^{(m)}-\frac{\Omega_{\mu w}^2(t)}{\omega_{RF}}\frac{|\braket{3,m;1,1}{4,m+1}|^2}{8(m-3)}\sigma_z^{(m)}\\
&-\frac{\Omega_{\mu w}(t)}{\omega_{RF}}\frac{\braket{3,m;1,1}{4,m+1}\Delta_{\mu w}^{(m)}}{4(m-3)}(\cos{(\phi_{\mu w}(t))}\sigma_x^{(m)}+\sin{(\phi_{\mu w}(t))}\sigma_y^{(m)})
\end{split}
\ee

Putting all the second order correction terms together, the complete $\mu$w control Hamiltonian for the full ground manifold, in the RWA, corrected up to second order, i.e., keeping the correct off-resonant terms that lead to microwave-induced AC Stark shifts, can be written
\be
\begin{split}
H'_{\mu w}(t)&\approx\frac{\Omega_{\mu w}(t)}{2}\left[\cos{\phi_{\mu w}(t)}\sigma_x+\sin{\phi_{\mu w}(t)}\sigma_y\right]\\
		        &+\frac{\Omega_{\mu w}^2(t)}{8\Omega_{0}}\sum_{m\neq 3}\frac{|\braket{3,m;1,1}{4,m+1}|^2}{3-m}\sigma_z^{(m)}\\
		        &-\frac{\Omega_{\mu w}(t)}{\omega_{RF}}\frac{\braket{3,m;1,1}{4,m+1}\Delta_{\mu w}^{(m)}}{4(m-3)}(\cos{(\phi_{\mu w}(t))}\sigma_x^{(m)}+\sin{(\phi_{\mu w}(t))}\sigma_y^{(m)}).
\end{split}
\label{eq:Hmicrowave}
\ee
This differs from the Hamiltonian given in \cite{merkel08} in the fact that we maintain terms of order $\Omega_{\mu w}^2/\omega_{RF}$, which lead to Bloch-Seigert-like shifts and extra corrections due to counter-rotating terms. 

\subsubsection{RF Hamiltonian}

Noting that when only the RF part of the Hamiltonian is present in the problem, the complete dynamics of the system can be described in the $ {\sf SU}(2)$ group, and thus, all the dynamics of the system can be described by the Heisenberg equations of motion of $F_x$, $F_y$ and $F_z$. We carry out this calculation for the $F_+=4$ and $F_-=3$ manifolds separately since there is no coupling between them in the absence of the microwaves. Moreover, we assume a small enough bias field $B_0$ so that we can neglect the quadratic Zeeman shift introduced in Eq. (\ref{eq:H0});  for a very large bias the standard RWA is sufficient. For illustration, we discuss in detail the second order correction to the $F_+=4$ manifold RF Hamiltonian.

In this case, it is convenient to define the small parameter $\epsilon=\epsilon_0/\omega_{RF}$ where $\epsilon_0=\sqrt{\Omega_x^2(t)+\Omega_y^2(t)}$ to allow the RF Larmor frequencies to be different.  Turning the microwave Hamiltonian off ($\Omega_{\mu w}=0$) and neglecting the second order Zeeman shift, the Hamiltonian restricted to the $F=4$ manifold can be written
\be
\begin{split}
H^{(4)}(t)&=\left(\frac{3\Omega_0}{2}(1-g_r)+\frac{1}{2}(7\Delta_{RF}-\Delta_{\mu w})\right)I^{(4)}\\
	        &+\epsilon_0\frac{\chi(t)}{2}F_x^{(4)}+\epsilon_0\frac{\upsilon(t)}{2}F_y^{(4)}-\Delta_{RF}F_z^{(4)}
\end{split}
\label{eq:H4}
\ee
where 
\be
\begin{split}
\chi(t)&=\frac{\Omega_x(t)}{\epsilon_0}(\cos{(2\omega_{RF}t-\phi_x(t))}+\cos{(\phi_x(t))})\\
	  &+\frac{\Omega_y(t)}{\epsilon_0}(\sin{(2\omega_{RF}t-\phi_y(t))}+\sin{(\phi_y(t))}),
\end{split}
\ee
and
\be
\begin{split}
\upsilon(t)=&-\frac{\Omega_x(t)}{\epsilon_0}(\sin{(2\omega_{RF}t-\phi_x(t))}+\sin{(\phi_x(t))})\\
	  &+\frac{\Omega_y(t)}{\epsilon_0}(\cos{(2\omega_{RF}t-\phi_y(t))}+\cos{(\phi_y(t))}).
\end{split}
\ee

The Heisenberg equations of motion for the components of the total angular momentum can then be easily written
\begin{subequations}
\begin{align}
\frac{dF_x^{(4)}}{dt'}&=\epsilon\left(\frac{\upsilon(t')}{2}F_z^{(4)}+\frac{\Delta_{RF}}{\epsilon_0}F_y^{(4)}\right),\\
\frac{dF_y^{(4)}}{dt'}&=-\epsilon\left(\frac{\chi(t')}{2}F_z^{(4)}+\frac{\Delta_{RF}}{\epsilon_0}F_x^{(4)}\right),\\
\frac{dF_z^{(4)}}{dt'}&=\epsilon\left(\frac{\chi(t')}{2}F_y^{(4)}-\frac{\upsilon(t')}{2}F_x^{(4)}\right),
\end{align}
\end{subequations}
where we have scaled the time so that $t'=\omega_{RF}t$. This system of differential equations is in the form needed to apply the averaging theorem when we note that
\be
{\bf x}\rightarrow\left[\begin{array}{c} F_x^{(4)}\\F_y^{(4)}\\F_z^{(4)}\end{array}\right],~ {\bf f}({\bf x},t')\rightarrow\frac{1}{2}\left[\begin{array}{c} \upsilon(t')F_z^{(4)}+2\tilde{\Delta}F_y^{(4)}\\-\chi(t')F_z^{(4)}-2\tilde{\Delta}F_x^{(4)}\\ \chi(t')F_y^{(4)}-\upsilon(t')F_x^{(4)}   \end{array}\right],
\ee
where, for convenience, we have defined $\tilde{\Delta}=\Delta_{RF}/\epsilon_0$.
Transforming the original system to the averaged equivalent one, we have
\be
{\bf y}\rightarrow\left[\begin{array}{c} \bar{F}_x^{(4)}\\ \bar{F}_y^{(4)}\\ \bar{F}_z^{(4)}\end{array}\right],~ {\bf \bar{f}}({\bf y})\rightarrow\frac{1}{2}\left[\begin{array}{c} \bar{\upsilon}\bar{F}_z^{(4)}+2\tilde{\Delta}\bar{F}_y^{(4)}\\-\bar{\chi}\bar{F}_z^{(4)}-2\tilde{\Delta}\bar{F}_x^{(4)}\\ \bar{\chi}\bar{F}_y^{(4)}-\bar{\upsilon}\bar{F}_x^{(4)}   \end{array}\right]
\ee
where
\be
\bar{\chi}=\frac{\Omega_x(t)}{\epsilon_0}\cos{(\phi_x(t))}+\frac{\Omega_y(t)}{\epsilon_0}\sin{(\phi_y(t))},
\ee
and
\be
\bar{\upsilon}=-\frac{\Omega_x(t)}{\epsilon_0}\sin{(\phi_x(t))}+\frac{\Omega_y(t)}{\epsilon_0}\cos{(\phi_y(t))}.
\ee
The only terms that participate in the averaging process are the fast oscillating ones, while the slow varying terms are treated as constant.

We now proceed to calculate the function ${\boldsymbol \omega}({\bf y},t')$ by again integrating only over the fast varying terms
\be
{\boldsymbol \omega}({\bf y},t')\rightarrow\frac{1}{4}\left[\begin{array}{c} \Upsilon(t')\bar{F}_z^{(4)}\\-{\rm X}(t')\bar{F}_z^{(4)}\\ {\rm X}(t')\bar{F}_y^{(4)}-\Upsilon(t')\bar{F}_x^{(4)}   \end{array}\right]
\ee
where 
\be
\begin{split}
{\rm X}(t)&=\frac{\Omega_x(t)}{\epsilon_0}(\sin{(2\omega_{RF}t-\phi_x(t))}+\sin{(\phi_x(t))})\\
	  &+\frac{\Omega_y(t)}{\epsilon_0}(-\cos{(2\omega_{RF}t-\phi_y(t))}+\cos{(\phi_y(t))}),
\end{split}
\ee
and
\be
\begin{split}
\Upsilon(t)&=\frac{\Omega_x(t)}{\epsilon_0}(\cos{(2\omega_{RF}t-\phi_x(t))}-\cos{(\phi_x(t))})\\
	  &+\frac{\Omega_y(t)}{\epsilon_0}(\sin{(2\omega_{RF}t-\phi_y(t))}+\sin{(\phi_y(t))}).
\end{split}
\ee
We can thus write
\be
\overline{{\bf f}_1({\bf y},t')}\rightarrow\frac{\bar{\Lambda}}{8}\left[\begin{array}{c} -\bar{F}_y^{(4)}\\ \bar{F}_x^{(4)}\\ 0   \end{array}\right]+\frac{\tilde{\Delta}}{4}\left[\begin{array}{c} -\bar{\rm X}\bar{F}_z^{(4)}\\ -\bar{\Upsilon}\bar{F}_z^{(4)}\\ \bar{\rm X}\bar{F}_x^{(4)}+\bar{\Upsilon}\bar{F}_y^{(4)}   \end{array}\right]
\ee
where
\be
\begin{split}
\bar{\Lambda}&=\frac{1}{2}-\frac{\Omega^2_x(t)}{\epsilon^2_0}\cos{(2\phi_x(t))}-\frac{\Omega^2_y(t)}{\epsilon^2_0}\cos{(2\phi_y(t))}\\
	  &+\frac{\Omega_x(t)\Omega_y(t)}{\epsilon^2_0}\sin{(\phi_x(t)-\phi_y(t))},
\end{split}
\ee
\be
\bar{\rm X}=\frac{\Omega_x(t)}{\epsilon_0}\sin{(\phi_x(t))}+\frac{\Omega_y(t)}{\epsilon_0}\cos{(\phi_y(t))},
\ee
and
\be
\bar{\Upsilon}=-\frac{\Omega_x(t)}{\epsilon_0}\cos{(\phi_x(t))}+\frac{\Omega_y(t)}{\epsilon_0}\sin{(\phi_y(t))}.
\ee

Putting all together, the Heisenberg equations for the components of the total angular momentum, up to second order correction of the RWA, are
\be
\begin{split}
\frac{d}{dt'}\left[\begin{array}{c} \bar{F}_x^{(4)}\\ \bar{F}_y^{(4)}\\ \bar{F}_z^{(4)}\end{array}\right]&=\frac{\epsilon}{2}\left[\begin{array}{c} \bar{\upsilon}\bar{F}_z^{(4)}+2\Delta_{RF}\bar{F}_y^{(4)}\\-\bar{\chi}\bar{F}_z^{(4)}-2\Delta_{RF}\bar{F}_x^{(4)}\\ \bar{\chi}\bar{F}_y^{(4)}-\bar{\upsilon}\bar{F}_x^{(4)}   \end{array}\right]\\
&+\epsilon^2\left\{\frac{\bar{\Lambda}}{8}\left[\begin{array}{c} -\bar{F}_y^{(4)}\\ \bar{F}_x^{(4)}\\ 0   \end{array}\right]+\frac{\tilde{\Delta}}{4}\left[\begin{array}{c} -\bar{\rm X}\bar{F}_z^{(4)}\\ -\bar{\Upsilon}\bar{F}_z^{(4)}\\ \bar{\rm X}\bar{F}_x^{(4)}+\bar{\Upsilon}\bar{F}_y^{(4)}   \end{array}\right]\right\}.
\end{split}
\label{eq:RWAsecondorder}
\ee
Equivalently, using Eq. (\ref{eq:RWAsecondorder}), we can write the Hamiltonian, Eq. (\ref{eq:H4}), up to second order correction
\be
\begin{split}
H^{(4)}(t)&\approx\left(\frac{3\Omega_0}{2}(1-g_r)+\frac{1}{2}(7\Delta_{RF}-\Delta_{\mu w})\right)I^{(4)}\\
	       &+\left(\frac{\Omega_x(t)}{2}\left(\cos{(\phi_x(t))}+\frac{\Delta_{RF}}{2\omega_{RF}}\sin{(\phi_x(t))}\right)\right.\\
	       &+\left.\frac{\Omega_y(t)}{2}\left(\sin{(\phi_y(t))}+\frac{\Delta_{RF}}{2\omega_{RF}}\cos{(\phi_y(t))}\right)\right)F_x^{(4)}\\
	       &+\left(-\frac{\Omega_x(t)}{2}\left(\sin{(\phi_x(t))}+\frac{\Delta_{RF}}{2\omega_{RF}}\cos{(\phi_x(t))}\right)\right.\\
	       &+\left.\frac{\Omega_y(t)}{2}\left(\cos{(\phi_y(t))}+\frac{\Delta_{RF}}{2\omega_{RF}}\sin{(\phi_y(t))}\right)\right)F_y^{(4)}\\
	       &+\frac{1}{16\omega_{RF}}\bigg(\Omega_x^2(t)\Big(1-2\cos{(2\phi_x(t))}\Big)+\Omega_y^2(t)\Big(1-2\cos{(2\phi_y(t))}\Big)\\
	       &+ 2\Omega_x(t)\Omega_y(t)\sin{(\phi_x(t)-\phi_y(t))}\bigg)F_z^{(4)}.
\end{split}
\ee
Using a similar procedure to the one detailed above, a second order correction for the Hamiltonian acting on the $F_-=3$ manifold can also be obtained. Putting all the second order correction terms together, the RF control Hamiltonian for the full ground manifold, in the RWA, corrected up to second order can be written
\be
\begin{split}
H'_{RF}(t)&=\frac{\Omega_x(t)}{2}\left[\cos{(\phi_x(t))}\left(F_x^{(4)}-g_r\left(1-\frac{\Omega_0(1-g_r)}{2\omega_{RF}}\right)F_x^{(3)}\right)\right.\\
	       &\left.-\sin{(\phi_x(t))}\left(F_y^{(4)}+g_r\left(1+\frac{\Omega_0(1-g_r)}{2\omega_{RF}}\right)F_y^{(3)}\right)\right]\\ 
	       &+\frac{\Omega_x(t)}{2}\left[\frac{\Delta_{RF}}{2\omega_{RF}}\left(\sin{(\phi_x(t))}F_x^{(4)}-g_r\cos{(\phi_x(t))}F_x^{(3)}\right)\right.\\
	       &\left.-\frac{\Delta_{RF}}{2\omega_{RF}}\left(\cos{(\phi_x(t))}F_y^{(4)}+g_r\sin{(\phi_x(t))}F_y^{(3)}\right)\right]\\
	       &+\frac{\Omega_y(t)}{2}\left[\cos{(\phi_y(t))}\left(F_y^{(4)}-g_r\left(1-\frac{\Omega_0(1-g_r)}{2\omega_{RF}}\right)F_y^{(3)}\right)\right.\\
	       &\left.+\sin{(\phi_y(t))}\left(F_x^{(4)}+g_r\left(1+\frac{\Omega_0(1-g_r)}{2\omega_{RF}}\right)F_x^{(3)}\right)\right]\\ 
	       &+\frac{\Omega_y(t)}{2}\left[\frac{\Delta_{RF}}{2\omega_{RF}}\left(\cos{(\phi_y(t))}F_x^{(4)}+g_r\sin{(\phi_y(t))}F_x^{(3)}\right)\right.\\
	       &\left.+\frac{\Delta_{RF}}{2\omega_{RF}}\left(\sin{(\phi_y(t))}F_y^{(4)}+g_r\cos{(\phi_y(t))}F_y^{(3)}\right)\right]\\
	       &+\frac{1}{16\omega_{RF}}\bigg(\Omega_x^2(t)\Big(1-2\cos{(2\phi_x(t))}\Big)+\Omega_y^2(t)\Big(1-2\cos{(2\phi_y(t))}\Big)\\
	       &+2\Omega_x(t)\Omega_y(t)\sin{(\phi_x(t)-\phi_y(t))}\bigg)F_z^{(4)}\\
	       &-\frac{g_r^2}{16\omega_{RF}}\bigg(\Omega_x^2(t)\Big(1-2\cos{(2\phi_x(t))}\Big)+\Omega_y^2(t)\Big(1-2\cos{(2\phi_y(t))}\Big)\\
	       &-2\Omega_x(t)\Omega_y(t)\sin{(\phi_x(t)-\phi_y(t))}\bigg)F_z^{(3)}.
\end{split}
\label{eq:Hrfcomplete}
\ee
This differs from the Hamiltonian given in \cite{merkel08} in two ways.  We account for the relative magnitudes of the $g$-factors in the upper and lower manifolds due to the small nuclear magneton, which implies $g_r \neq 1$.  Additionally, we maintain terms of order $\Omega_{i}^2/\omega_{RF}$, $i=x,y$, which lead to Bloch-Seigert-like shifts and extra corrections due to counter-rotating terms. 

 
 \subsection{A note about the RWA and decoherence terms}
 Special care must be taken when considering the full master equation. All operators, including the Lindblad jump operators, must be written in the rotating frame and the RWA should be applied accordingly. This is essential in order to account for spontaneous emission processes that become distinguishable once the energy degeneracy is broken by the shift produced by the bias field. The RWA in the master equation is achieved by explicitly calculating  the transformation $U^{\dagger}(t)W_q^{F_bF_a}U(t)$, and averaging the superoperator map over the rapid oscillations. Due to the difficulty of doing these transformations analytically, we do them numerically in our simulation code.

\section{Measurement observables}
Continuous measurement of the system is carried out through polarization spectroscopy of a probe laser beam that passes through the ensemble while it is being controlled.  The atoms induce a polarization-dependent index of refraction in a manner depending on their spin state according to the light-shift interaction \cite{deutsch10}.  In the limit of negligible backaction, the effect of the interaction is a rotation of the probe's Stokes vector $\vec{\mathcal{S}}$ on the Poincar\'{e} sphere according to the rotation operator $U_R = \exp \left( -i \chi_0 \avg{\mbf{\mathcal{O}}} \cdot \vec{\mathcal{S}} \right)$, where $\chi_0= OD_0 (\Gamma/2\Delta_c)$ is the characteristic rotation angle depending on the resonant optical density, $OD_0$, and a characteristic detuning from resonance, $\Delta_c$.  This interaction has the form of the QND Hamiltonian discussed in Section \ref{sec:QTprotocol:basic}. Taking the $z$-axis along the direction of propagation of the probe, the components of the vector of atomic observables that generate the rotations about the three axes of the Poincar\'{e} sphere are,
\begin{subequations}
\begin{align}
\vec{\mathcal{O}}\cdot\vec{e}_1 &= \sum_{F,F'} C^{(2)}_{F'F} \frac{\Delta_c}{\Delta_{F'F}}\left( \frac{F_x^2 - F_y^2}{2} \right) \\
\vec{\mathcal{O}}\cdot\vec{e}_2 &= \sum_{F,F'} C^{(2)}_{F'F} \frac{\Delta_c}{\Delta_{F'F}}\left( \frac{F_x F_y + F_y F_x}{2} \right) \\
\vec{\mathcal{O}}\cdot\vec{e}_3 &= \sum_{F,F'} C^{(1)}_{F'F} \frac{\Delta_c}{\Delta_{F'F}} F_z 
\label{eq:FaradayObservable}
\end{align}
\end{subequations}
where $C^{(K)}_{F'F}$ are coupling constants that depend on the irreducible rank-$K$ tensor polarizability for the given probe detuning $\Delta_{F'F}$ from the ground $(nS_{1/2})F$  to the excited $(nP_J')F'$ manifold \cite{deutsch10}, as given in Eqs. (\ref{eq:tensorcoeff}).  For weak interactions under consideration here, $\chi_0 \ll 1$, this rotation corresponds to a small local displacement.  Measurement of the Stokes vector component along the direction $\hat{n}$ then correlates with a measurement of the atomic operator $\hat{n}\cdot \mbf{\mathcal{O}}$. This has the same structure of the general POVM discussed in Section \ref{sec:QTprotocol:basic}. Thus, preparing the probe initially linearly polarized along the $\mbf{e}_1$ of the Poincar\'{e} sphere, and analyzing along the direction, $\mbf{n} = \cos{\theta}\mbf{e}_2+\sin{\theta} \mbf{e}_3$, the general measurement record will be of the form
\be
M(t)= a \avg{F_z}_t + b \avg{F_x F_y + F_y F_x}_t+ \sigma W(t),
\label{eq:fullrecord}
\ee
where $a$ and $b$ are constants that depend on the vector and tensor contributions to the polarizability for the given detuning, as well as the polarization analysis direction, $\vec{n}$. The first term arises from the Faraday effect whereas the second arises from the birefringence in the index of refraction.

Since all our analysis is carried out in the rotating frame defined by Eqs. (\ref{eq:rotatingframeU}), we see that the relevant measurement operators written in this frame are
\be
\begin{split}
U(t)^{\dagger}\vec{\mathcal{O}}\cdot\vec{e}_2U(t) &= (2\cos{(\omega_{RF}t)}^2-1)\sum_{F,F'} C^{(2)}_{F'F} \frac{\Delta_c}{\Delta_{F'F}}\left( \frac{F_x F_y + F_y F_x}{2} \right) \\										 &+(\sin{(2\omega_{RF}t)})\sum_{F,F'} C^{(2)}_{F'F} \frac{\Delta_c}{\Delta_{F'F}}(-1)^F\left( \frac{F_y^2 -  F_x^2}{2} \right), 
\end{split}
\label{eq:birefringenceRot}
\ee
and
\be
U(t)^{\dagger}\vec{\mathcal{O}}\cdot\vec{e}_3U(t) = \sum_{F,F'} C^{(1)}_{F'F} \frac{\Delta_c}{\Delta_{F'F}} F_z. 
\label{eq:FaradayRot}
\ee

Applying the RWA in the previous equations, we see how the birefringence, Eq. (\ref{eq:birefringenceRot}) averages to zero, while the Faraday rotation, Eq. (\ref{eq:FaradayRot}) remains the same. With these considerations, we can write the most general form of the measurement record as
 \be
M(t)= a \avg{F_z}_t+ \sigma W(t).
\label{eq:fullrecordRot}
\ee
 
With this general framework in hand, we have the tools necessary for our QT protocol: control and continuous measurement.  We  apply this formalism in the results shown in Chapters \ref{ch:Simulations} and \ref{ch:Results} in which we achieve quantum state reconstruction in ensembles of cesium atoms.

\chapter{Simulations and Experimental Considerations}\label{ch:Simulations}


	In this chapter, we discuss all the technical details of the protocol that are necessary to achieve high-fidelity continuous measurement quantum state tomography in practical applications. First we define an appropriate operation regime for the experiment by choosing adequate control parameters, including the laser field detuning and intensity. Then, we illustrate the performance of QT in simulation using a particular application, and finally, we perform some basic numerical benchmarking of what to expect from our QT methods.

\section{Control parameters}\label{sec:waveforms}
The Hamiltonian that governs the dynamics is a functional of a set of control waveforms such as externally applied fields parametrized by frequencies, amplitudes, and phases, as described in Section \ref{sec:Control}.  Our task is then to choose these waveforms to generate an  informationally complete set of observables $\{\cO_i\}$ in the desired time.   In practice, we fix the duration of the measurement record as determined by the characteristic time scales for evolution, dictated by both the Hamiltonian evolution for the given power in the controls, and by decoherence.  We choose the total time $T$ to be such that we can attain a good approximation to any unitary evolution matrix in ${\sf SU}(d)$.  The total time is then coarse-grained into slices of duration $\delta t$, consistent with the slew rates and bandwidth constraints of the waveform drivers in the laboratory.  We thus reduce the problem to specification of a discrete set of waveform values compatible with experimental constraints.  The translation of the discretely sampled parameters to the continuous-time waveform depends on the characteristics of the physical drivers and the challenges of numerical integration, as mentioned in Chapter \ref{ch:QT}.

With the specific Hamiltonian in hand, we must choose the control parameters.  There is no unique solution; any choice that yields an informationally complete set $\{\cO_i\}$ in the given time series will suffice.  In principle, one would like to optimize the information gain over time $T$.  This amounts to optimizing the entropy associated with the eigenvalues of the covariance matrix.  We have found empirically that the landscape for performing such an optimization is not favorable, and this approach becomes intractable, even for moderately sized Hilbert spaces ($d>9$).  Instead, our numerical studies show that one can achieve the required high-fidelity measurement record by choosing the control parameters {\em randomly} over a designated interval.  We will demonstrate this below for the specific example of control and measurement of atomic hyperfine spins.  A more rigorous justification of this approach is still under consideration.  We have seen a connection between evolution via random unitary dynamics and the generation of an informationally complete measurement record \cite{merkel10}, which is the subject of Chapter \ref{ch:RandomUnitaries}.   This may give us clues to optimally designing the control waveforms.

\begin{figure}[t]
\begin{center}
\includegraphics[width=8.7cm,clip]{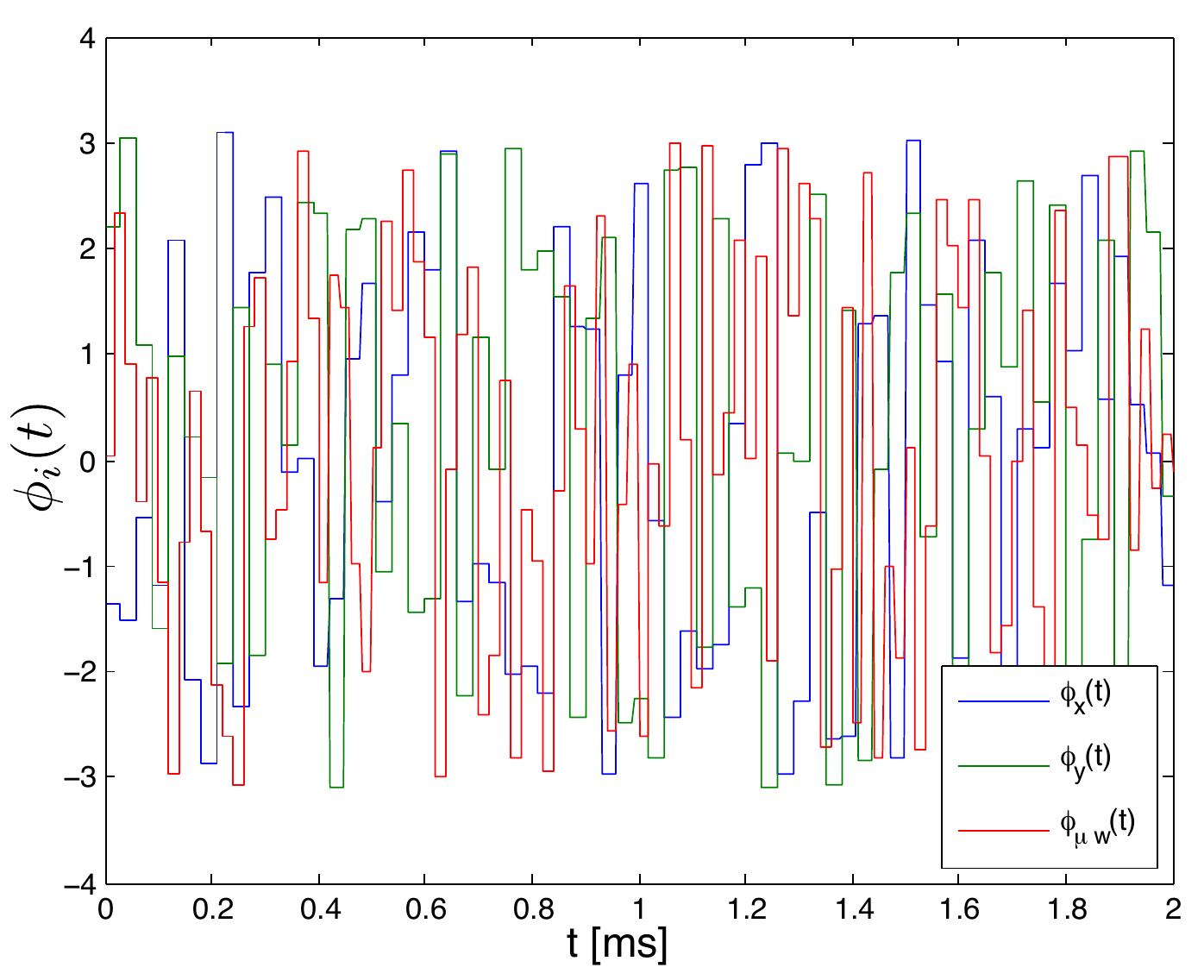}
\caption{Randomly sampled control waveforms that determine the phases of the applied RF and microwave magnetic fields as given in Eqs. (\ref{eq:Hmicrowave}) and (\ref{eq:Hrfcomplete}). These waveforms produce high fidelity estimates with high numerical stability.}
\label{F:RFuwcontrols}
\end{center}
\end{figure}

As discussed in Section \ref{sec:Control} and shown in \cite{merkel08}, full controllability of the system can be achieved by keeping the RF-Larmor and $\mu$w-Rabi frequencies, $\Omega_x$, $\Omega_y$, and $\Omega_{\mu w}$, constant in time, while varying the control phases $\phi_x(t)$, $\phi_y(t)$, and $\phi_{\mu w}(t)$. Due to the size of the Hilbert space, finding a set of control waveforms is a very challenging task.  Optimizing the entropy of the Gaussian probability distribution, as mentioned in \cite{silberfarb05}, is generally an intractable problem. Instead, we choose the control waveforms as piecewise random functions. Intuitively, and partially inspired by the results in \cite{merkel10}, this choice is justified by the fact that for tomography to work with high fidelity, we need to measure all, or almost all, independent directions in operator space. Moreover, we need to do it in a way that does not depend on the particular state being reconstructed and is unbiased, which we believe we partially achieve by using this choice. Furthermore, we have found in our numerics that this is sufficient to generate an informationally complete measurement record. For the RF ($\mu$w) waveforms, the phase is chosen uniformly between $-\pi$ and $\pi$ and kept constant over intervals of 30$\mu$s (20$\mu$s). After a total time of 2 ms, we ensure that we have sufficient information in the measurement for QT. Fig. \ref{F:RFuwcontrols} shows an example of the control phases used in this work. In general, almost all the control waveforms designed in this random way will produce informationally complete measurement records for most initial states. However, numerical stability of the QT algorithms may become an issue for certain waveforms. We thus choose a set of waveforms that produce the most stable results by repeating the design procedure several times.  However, an optimal, faster, and robust control waveform would be preferred, which is still an open problem. 

To complete the control Hamiltonian, we choose the following parameters, representing the control magnetic fields: $\Omega_{x}/2\pi=\Omega_{y}/2\pi=9$ kHz, $\Omega_{\mu w}/2\pi=27.5$ kHz, $\Omega_0/2\pi=1.0$ MHz. These fields are chosen according to the capability of Poul Jessen's laboratory to produce them. In fact, we try to make them as high as is technically possible since they define the controllability time scales of the problem. In general, faster control is preferred since we need to generate the informationally complete measurement record before decoherence erases all the information about the initial state of the system.

\section{Probe parameters}

To complete our protocol, we must choose the detuning $\Delta_c$ and the intensity of the laser probe. In previous work \cite{silberfarb05}, the detuning and intensity were chosen to maximize the nonlinear light shift relative to photon scattering, which was essential in their controllability scheme.  In the current context, we have much more flexibility, since full control of the Hilbert space can be achieved without the light shift.  There are, however, many technical considerations that inform the choice of detuning and probe intensity.  Firstly, the measurement strength is proportional to $\gamma_{sc}$, which is proportional to the probe intensity and inversely proportional to the detuning squared, so we can never make the measurement record free of decoherence by detuning further off resonance while still getting enough signal \cite{smith03}. In fact, at very large detunings, in order to maintain a reasonably large $\gamma_{sc}$, we would need a large probe intensity for which shot-noise-limited detection is difficult.  For these reasons, the light-shift-driven dynamics must be included in the analysis, as detailed in Section \ref{sec:MasterEquation}.  An additional technical issue is the effect of inhomogeneity in the light intensity across the ensemble.  Indeed, the difficulty in estimating the distribution of intensities caused substantial complexity in the reconstruction algorithm \cite{silberfarb05}, and ultimately limited the fidelity of the protocol.  Mitigating this effect would greatly improve the performance.  

We start by choosing a typical laser intensity that can be achieved in the laboratory, so that we have enough signal-to-noise ratio but are not limited by decoherence. We choose a nominal probe intensity commonly achievable of $I_{probe}=0.98~mW/cm^2$. With this choice, the laser Rabi frequency, given by Eq. (\ref{eq:RabiFeqIntensity}), is $\Omega/2\pi\approx 3.5$ MHz. This choice, along with the choice of detuning described below, will ensure an appropriate regime of operation of the experiment in terms of signal-to-noise ratio and low enough decoherence. Moreover, this ensures that we can extract complete information from the system before decoherence erases it.

\begin{figure}[t]
\begin{center}
\includegraphics[width=8.7cm,clip]{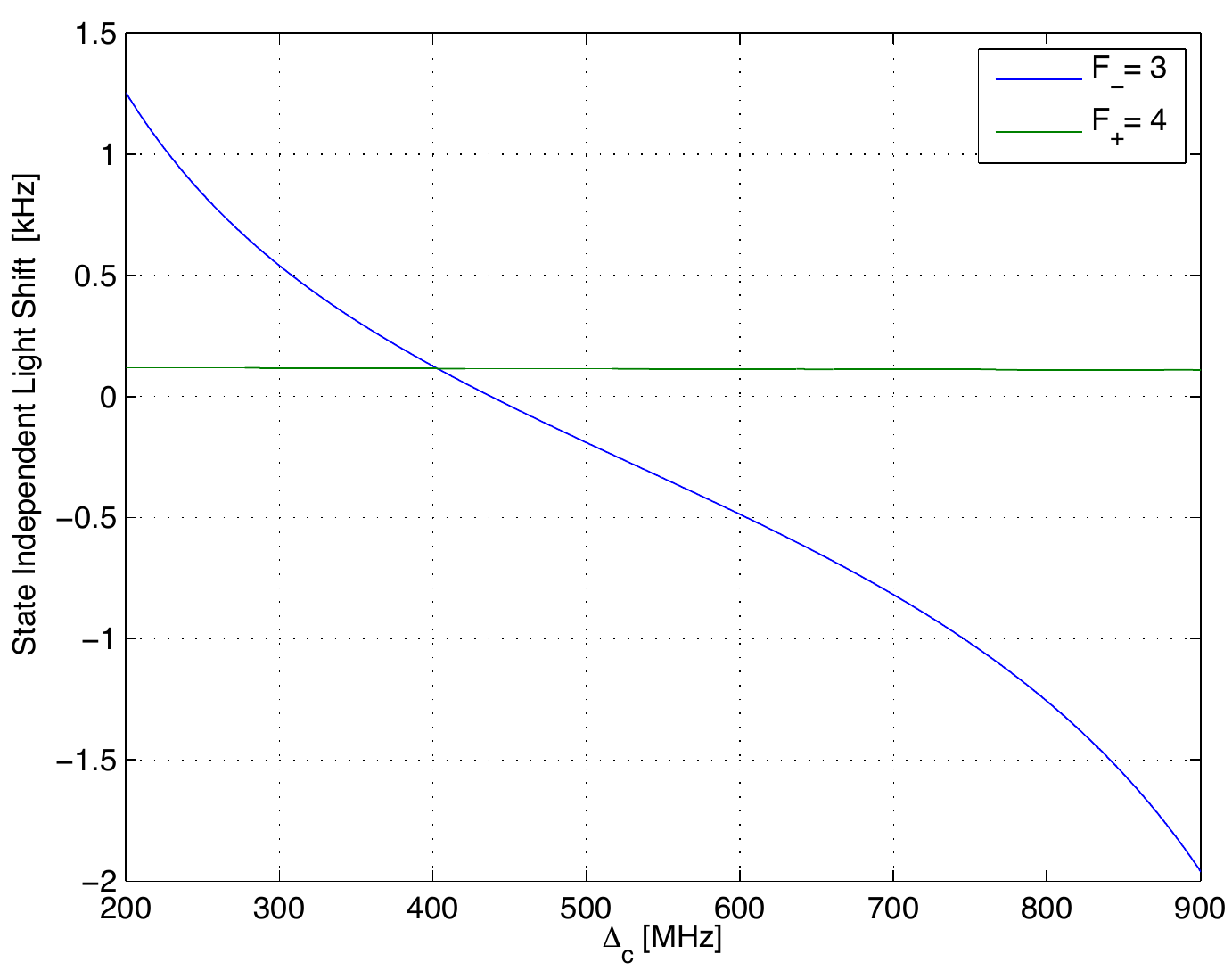}
\caption{State independent light shift, $\gamma_{sc}{\rm Re}\left( \beta_F^{(0)} + \beta_F^{(2)} \frac{F(F+1)}{6}\right)$, for $F_-=3$ and $F_+=4$ as a function of the characteristic detuning $\Delta_c$. For the detunings depicted, only the $F_-=3$ light shift can be cancelled, which occurs at the ``magic detuning'' $\Delta_c=437.8$MHz.}
\label{F:LightShiftVsDetuning}
\end{center}
\end{figure}

Finally, we choose the detuning in such a way that the system is more robust to inhomogeneities in light intensity. We will choose a relatively small detuning. This ensures that the light shift dynamics are not dominating over the photon scattering that gives rise to the signal. From Eq. (\ref{eq:LShamiltonianRWA}), we see that the scalar, $\gamma_{sc}\text{Re}(\beta^{(0)})$ and tensor, $\gamma_{sc}\text{Re}(\beta^{(2)})$, light shift components are responsible for the introduction of inhomogeneity into the problem. While there is no choice of detuning that makes both terms exactly zero, the state independent light shift, i.e., the real part of the term proportional to the identity operator in Eq. (\ref{eq:LShamiltonianRWA}), $\gamma_{sc}{\rm Re}\left( \beta_F^{(0)} + \beta_F^{(2)} \frac{F(F+1)}{6}\right)$, will cause the largest problems, and our goal is to cancel it for $F_-=3$. Fig. \ref{F:LightShiftVsDetuning} shows the appropriate region of the state independent light shift for both $F_+=4$ and $F_-=3$ as a function of the characteristic detuning $\Delta_c$. Putting all of these considerations together, we choose a relatively small detuning where the measurement strength can still be large at low intensity. We choose $\Delta_c/2\pi\approx 437.8$ MHz, defined for the $(6S_{1/2})F = 3$ to $(6P_{1/2})F' = 3$ transition, in between the two excited states in the D1 line, which is the ``magic detuning'' that nulls the light shift for the $F_-=3$ manifold. For such a detuning, only one $F$ manifold is effectively coupled to the light, and the other is so far from resonance that its coupling is very small. In fact, from Fig. \ref{F:LightShiftVsDetuning}, we clearly see that the light shift at this detuning for $F_+=4$ is small, and thus will cause a negligible negative impact in the measurement signal. 

With these choices of detuning and light intensity, the characteristic photon scattering rate is then $\gamma_{sc}/2\pi\approx 73$ Hz. The residual light shift, due to the tensor term in Eq. (\ref{eq:LShamiltonianRWA}), is $\gamma_{sc}\text{Re}(\beta^{(2)}_3/2)/2\pi\approx 117$ Hz, which together with microwave and RF fields, drives the spin dynamics during the course of the measurement.

\begin{figure}[t]
\begin{center}
\includegraphics[width=8.7cm,clip]{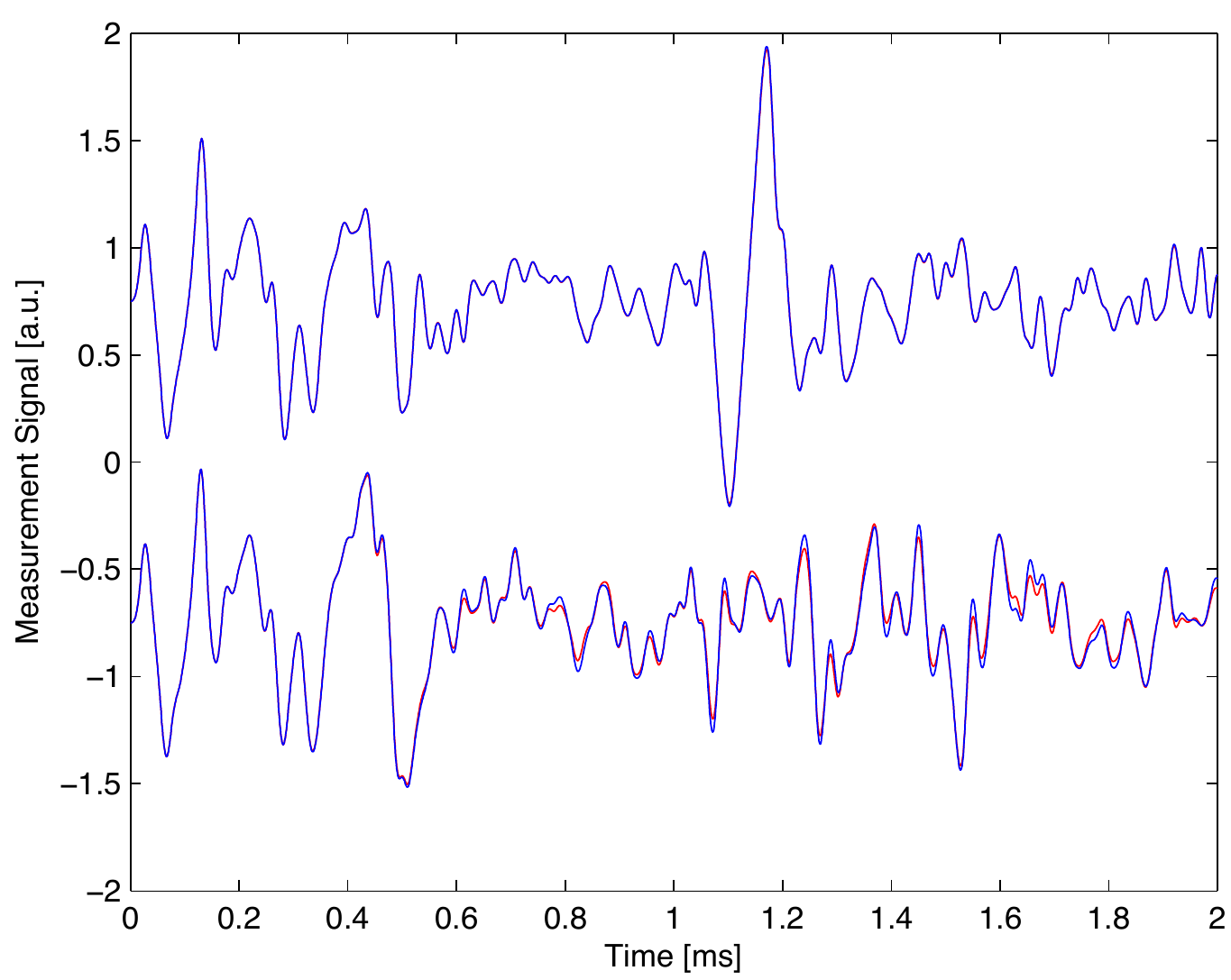}
\caption{(Color online) Simulated measurement signal for a random pure state chosen from the Haar measure for different choices of detuning of the laser probe. (Top) $\Delta_{c}/2\pi=437.8$MHz, the ``magic wavelength'' at which the state independent light shift is set to zero for $F=3$. The red line is an averaged signal over a Gaussian distribution of intensity with a typical spread of $15.5\%$ of the mean. The blue line is the signal that a system would produce if it evolves under the nominal value of intensity. (Bottom) $\Delta_{c}/2\pi=700$MHz, arbitrary detuning that does not cancel the light shift. The red line is an averaged signal over a Gaussian distribution of intensity with the same standard deviation. The blue line is the signal that a system would produce if it evolves under the nominal value of intensity.  Clearly, the signal is more robust at the ``magic wavelength".}
\label{F:signalcomparison}
\end{center}
\end{figure}

As a first check, we confirm that our choice of detuning and intensity makes our system more robust to light-shift inhomogeneity, by adding Gaussian fluctuations in the intensity across the ensemble, and then averaging the result for simulated data.  For example, if we choose an arbitrary detuning, $\Delta_c/2\pi=700$ MHz, for which the state independent light shift is not zero for both $F$-manifolds, we see that  the averaged signal leads to fidelities $\le 0.80$, whereas a similar simulation with the optimized detuning, $\Delta_c/2\pi=437.8$ MHz produces fidelities $\ge 0.90$.  Fig. \ref{F:signalcomparison} shows qualitatively a comparison between the simulated measurement records, averaged over a Gaussian distribution of intensities for these two detunings  and the simulated signals with a fixed, nominal value of intensity.  It is clear that the optimized detuning produces much better results, making both averaged and nominal signals look very similar. This will simplify our procedure for estimating the intensity distribution seen by the atomic ensemble. A fit to a Gaussian distribution will be sufficient to capture the effects of the inhomogeneous light shift. 

\section{Filter parameters}\label{sec:Filter}
\begin{figure}[t]
\begin{center}
\includegraphics[width=15.1cm,clip]{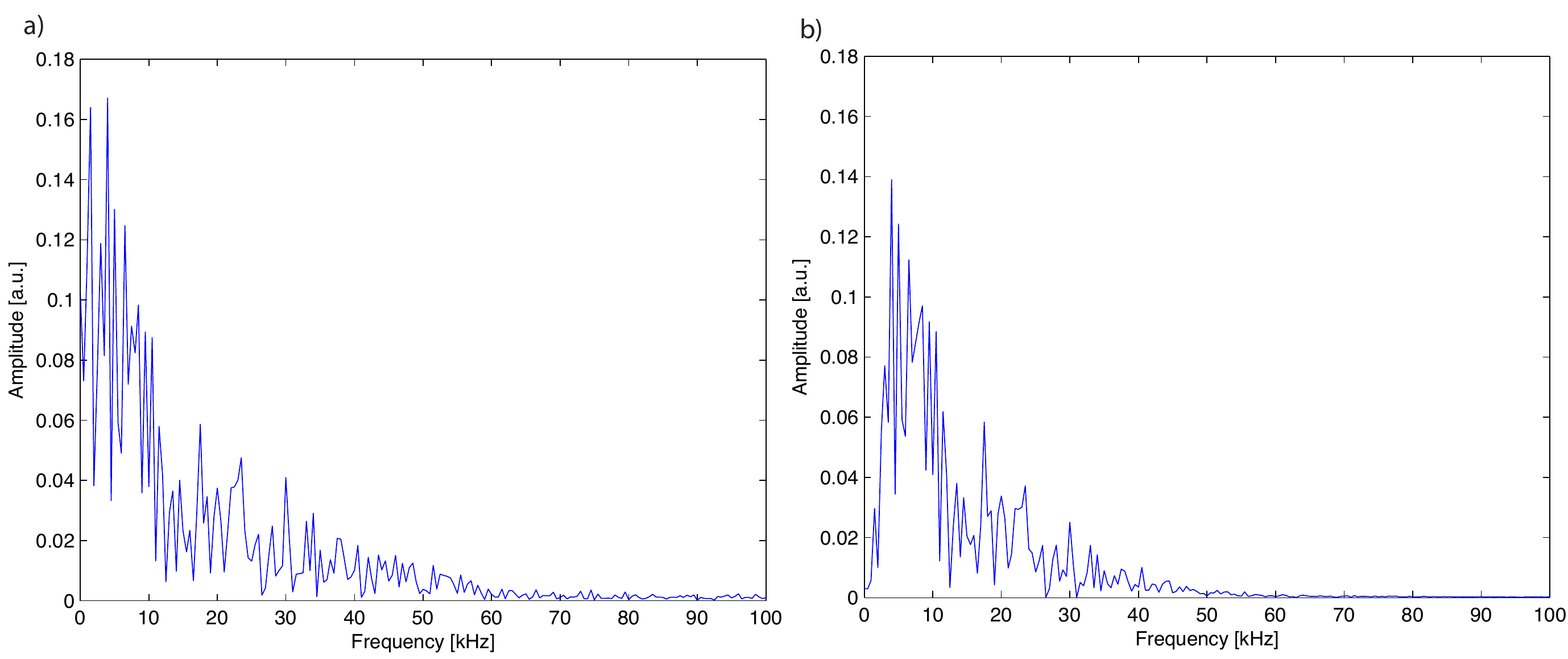}
\caption{Spectrum of the measurement record simulated for a random pure state. a) Unfiltered, b) filtered measurement record. The effect of the filter is to eliminate the noise at low and high frequencies so the signal-to-noise ratio is increased.}
\label{F:PowerSpectrum}
\end{center}
\end{figure}

As mentioned in Chapter \ref{ch:QT}, the measurement signal has fundamental noise due to the quantum fluctuations of the state of the probe (shot noise) in addition to technical noise inherent to the electronic controllers ($1/f$ noise). The presence of noise in the measurement affects the performance of our QT protocols in the sense that it decreases their ability to distinguish states that produce measurement records that are close enough, i.e., which are within the noise levels. In other words, noise erases information about the state of the system. In the laboratory, all possible efforts are made to maintain the signal-to-noise ratio adequately high; however, to improve reconstruction fidelities of the protocols, we must filter the measured signal and the Heisenberg picture operators $\cO_i$. Due to the way we carry out the control, we find that the information content of the measurement signals lies in a narrow frequency band. This fact allows us to easily choose a bandpass filter for the appropriate frequency band. Clearly, eliminating the noise on frequencies outside of information-carrying narrow band will increase the signal-to-noise ratio. Fig. \ref{F:PowerSpectrum} shows a comparison between a typical power spectrum of an unfiltered signal (\ref{F:PowerSpectrum}\textcolor{blue}{a}) and that of a filtered one (\ref{F:PowerSpectrum}\textcolor{blue}{b}) using a fourth order bandpass digital Bessel filter with cutoff frequencies of $2$kHz to $40$kHz. The signal was produced by a Haar-random pure state and the signal-to-noise ratio was chosen to be $46.5$ to better illustrate the effect of the filter. We use this bandpass digital filter to achieve better signal-to-noise ratios and apply it to both the theoretical studies and the actual data processing.

\section{Parameter estimation}\label{sec:ParameterEstimation}

The success and applicability of the continuous measurement QT protocols described in this dissertation depend strongly in our ability to model very closely what the actual quantum system is doing at any given time. In other words, practical application of the protocol requires accurate knowledge of the various parameters that characterize the experiment.  In fact, since we assume complete knowledge of the dynamics of the system, we need to very precisely know all the parameters that are part of the Hamiltonian as well as the decoherence terms. In practical situations, it is impossible to precisely know the probe intensity and the magnitude of the control and bias magnetic fields that the atoms see a priori. Moreover, those field parameters may fluctuate or drift by non-negligible amounts in time scales of minutes or hours, which makes it a necessity to carry out parameter estimation runs of the experiment before trying to use our quantum tomography techniques.

As mentioned above, a number of calibration errors are possible in this system. Our simulations suggest that we will need to fit the light intensity and its inhomogeneity, the RF and $\mu$w magnetic field amplitudes, $\Omega_{x}$, $\Omega_y$, and $\Omega_{\mu w}$, and, most importantly, the bias field, $\Omega_0$ and its inhomogeneity. Moreover, we must fit the overall constant $a$, in Eq. (\ref{eq:fullrecordRot}), which depends on both the power of the probe beam and the number of atoms in the ensemble, in addition to the particular units conversion factor required to transform the model's angular momentum units to the current or voltage measured in the laboratory. Least squares techniques, which we describe in detail in the rest of this section, should suffice to achieve good accuracy parameter estimation and high-fidelity reconstructions. 

Due to the size of the Hilbert space, it is neither practical nor tractable to determine all the unknown parameters in a single experiment. For this reason, we break the parameter estimation problem into different experiments that are simple, well understood and easy to analyze. Thus,  the parameters are estimated through a series of independent calibration runs prior to using the QT protocol in an actual experiment.

To perform parameter estimation, in general, we employ a least-squares fit between the measured signal and simulated measurement record for a well known initial state. In particular, to fit the probe intensity, and the RF Larmor frequencies, we use the cost function 
\be
\mathcal{C} =\sum_i (a M_i-\bar{M}_i(\Omega_k, \gamma_{sc},\sigma_I))^2. 
\label{eq:LScostFunctionFit1}
\ee
Here $M_i$ is the time-sampled data point from the calibration run, $\Omega_k$ represents any of the two RF magnetic fields, $a$ is the units conversion factor, and 
\be
\begin{split}
\bar{M}_i(\Omega_k,\gamma_{sc}) &= \int \avg{\cO_i (\Omega_k, \xi\gamma_{sc})} f(\sigma_I,\xi) d\xi\\
						        &\approx \frac{\Delta\xi}{2}\sum_n(\avg{\cO_i \left(\Omega_k, \xi_{n+1}\gamma_{sc})} f(\sigma_I,\xi_{n})+\avg{\cO_i (\Omega_k, \xi_{n}\gamma_{sc})} f(\sigma_I,\xi_{n}) \right)
\end{split}
\label{eq:FitParams1}						      
\ee
is the simulated measurement time-series.  The unknown parameters are the Larmor frequency, $\Omega_k$, photon scattering rate, $\gamma_{sc}$, the intensity inhomogeneity spread $\sigma_I$, and the units factor. We account for inhomogeneity in the laser intensity through the distribution function $f(\sigma_I,\xi)$, where $\xi$  is the ratio between the nominal scattering rate and the local scattering rate at the position of the atom.  The parameter $\gamma_{sc}$ then represents the scattering rate at $\xi=1$ where $f(\xi)$ is peaked. Note, we take $f(\xi)$ to be a Gaussian distribution with mean 1 and standard deviation $\sigma_I$.  The overall scale, $a$, determines the conversion between the simulated dimensionless signal and the laboratory measurement record.

Note that for simplicity, we calculate the integral in Eq. (\ref{eq:FitParams1}) using a linear interpolation method with $n=1,\ldots,12$. With all these considerations, we minimize Eq. (\ref{eq:LScostFunctionFit1}) in terms of $\Omega_k$, $a$, $\gamma_{sc}$, and $\sigma_I$. Such a minimization is a nonlinear problem that requires integrating the master equation, Eq. (\ref{eq:csmaster}), many times, which can be a time consuming task. However, since we do this for a simple dynamics, i.e., we assume a time-independent Hamiltonian for a simple Larmor precession experiment, we can solve this problem efficiently. Moreover, using parallel computing techniques, this problem can be solved in a time scale $\le 10$min in a 12-core MacPro running at 2.66 GHz using Matlab.

We use a similar procedure to the one described above to fit part of the remaining unknown parameters: $\Omega_{\mu w}$ for which we assume a Gaussian inhomogeneity distribution $f(\sigma_{\mu w},\xi)$. These parameters are fitted by using a simple Rabi flopping experiment with only the bias and $\mu$w fields turned on. Additional parameter estimation is performed for the bias field power $\Omega_0$, its Gaussian inhomogeneity $f(\sigma_{0},\xi)$, and the microwave detuning $\Delta_{\mu w}$. For these parameters, a random state is prepared and driven by the reconstruction control waveforms while its measurement record is acquired. Then, its fidelity of reconstruction is optimized as a function of the unknown parameters until a maximum value is obtained. Note that all of the subsequent calibrations runs are done in such a way that the previously calibrated parameters, i.e., $\gamma_{sc}$, $a$, etc, are assumed known and constant. This is done under the assumption that the parameters will not change in a time scale of minutes, which is reasonable for the types of experiments in which we are interested. Examples of the performance of these procedures are shown in Chapter \ref{ch:Results} using experimental data.

For application to the compressed sensing techniques discussed in Chapter \ref{ch:QT}, we need to estimate one final parameter, the error threshold parameter $\epsilon$ defined in Eq. (\ref{eq:CSConvex}). The way we do this in practice is, as before, by preparing a well known initial state and letting the system evolve under the complicated control phases defined in Section \ref{sec:waveforms}, while measuring the Faraday rotation of the probe. We then compare the experimental data thus obtained with a simulation produced by our model, after all the inhomogeneities and parameters are calibrated, and compute
\be
\epsilon \approx \sum_i (a M_i-\bar{M}_i)^2
\ee
with $\bar{M}_i$ being the simulated measurement signal as described in Eq. (\ref{eq:FitParams1}). This procedure estimates the parameter $\epsilon$ that is used for all subsequent compressed sensing reconstructions regardless of the state being measured.

\section{Simulations}\label{sec:CsRFUW}

We now have all the ingredients to proceed with our simulations, which will enable us to show the applicability of quantum tomography in ensembles of cesium atoms for the chosen regime of operation. We remember that the system evolves according to the master equation, Eq. (\ref{eq:csmaster}), with the effective Hamiltonian expressed in the RWA and the control Hamiltonian described in Chapter \ref{ch:Atoms}. For the simulations shown in this section, we have chosen our initial observable to be the Faraday operator $\cO_0= \vec{\mathcal{O}}\cdot\vec{e}_3$ as given in Eq. (\ref{eq:FaradayObservable}). Furthermore, we have added Gaussian white noise to the signal so that the signal-to-noise ratio is finite with a fixed, arbitrary, but reasonable noise standard deviation of $\sigma=0.03$.  

First, we illustrate the general behavior of the QT methods discussed in this dissertation in a qualitative way that shows how the methods converge and to visualize the method's performance, hence the abundant number of figures shown in this section. Second, to study the performance of our protocol in a more quantitative way, we performed numerical simulations of the expected measurement signal for random pure states sampled from the Haar measure, see for example \cite{zyczkowski2006} or \cite{mezzadri07}, in addition to random mixed states sampled from the Hilbert-Schmidt measure \cite{zyczkowski11}.  
 
We have run several simulations to test the performance and efficiency of this protocol. We numerically generated a measurement record for different initial states and different noise realizations according to Eq. (\ref{eq:measurement}). Then a Bessel bandpass filter from 2 to 40 kHz was applied to the simulated measurement record in order to limit the noise in frequency components that are not present in the measurement. The same filter is applied to the Heisenberg picture observable $\cO(t)$ to account for all dynamical effects the signal undergoes. Once this is done, Eqs. (\ref{eq:OneStepConvex}) and (\ref{eq:CSConvex}) are used to find the physical density matrix that best represents the measured data. In order to quantify the performance of our method, we calculate the fidelity between the initial and estimated state,  Eq. (\ref{eq:fidelity}).

\subsection{An illustrative example}
	
\begin{figure}[t]
\begin{center}
\includegraphics[width=15.1cm,clip]{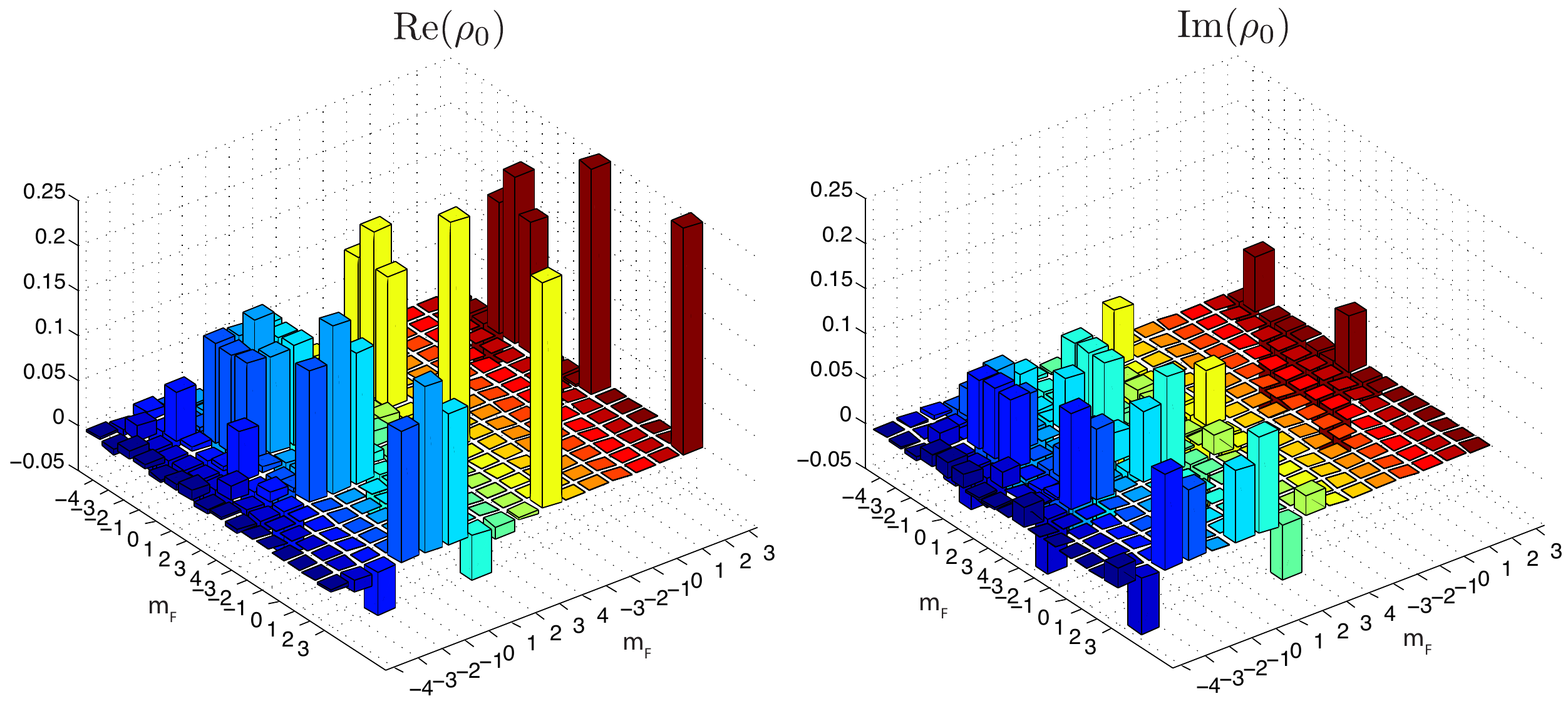}
\caption{Real and imaginary parts of the elements of $\rho_0$, the initial state used in the simulations. This is a non-trivial state used for illustration, $\ket{\psi}=\left( \ket{\psi^{(+)}_{sq}}+\ket{\psi^{(-)}_{cat}} \right) / \sqrt{2}$, consisting of an equal superposition of a spin squeezed state in the $F_+=4$ manifold, $\ket{\psi^{(+)}_{sq}}=\exp\left\{-i0.5F_z^2\right\} \ket{F=4,m_x=4}$ and a ``cat state" in the $F_-=3$ manifold, $\ket{\psi^{(-)}_{cat}}=\left(\ket{F=3,m_z=3}+\ket{F=3,m_z=-3}\right)/\sqrt{2}$.}
\label{F:InitialState}
\end{center}
\end{figure}

\begin{figure}[t]
\begin{center}
\includegraphics[width=8.7cm,clip]{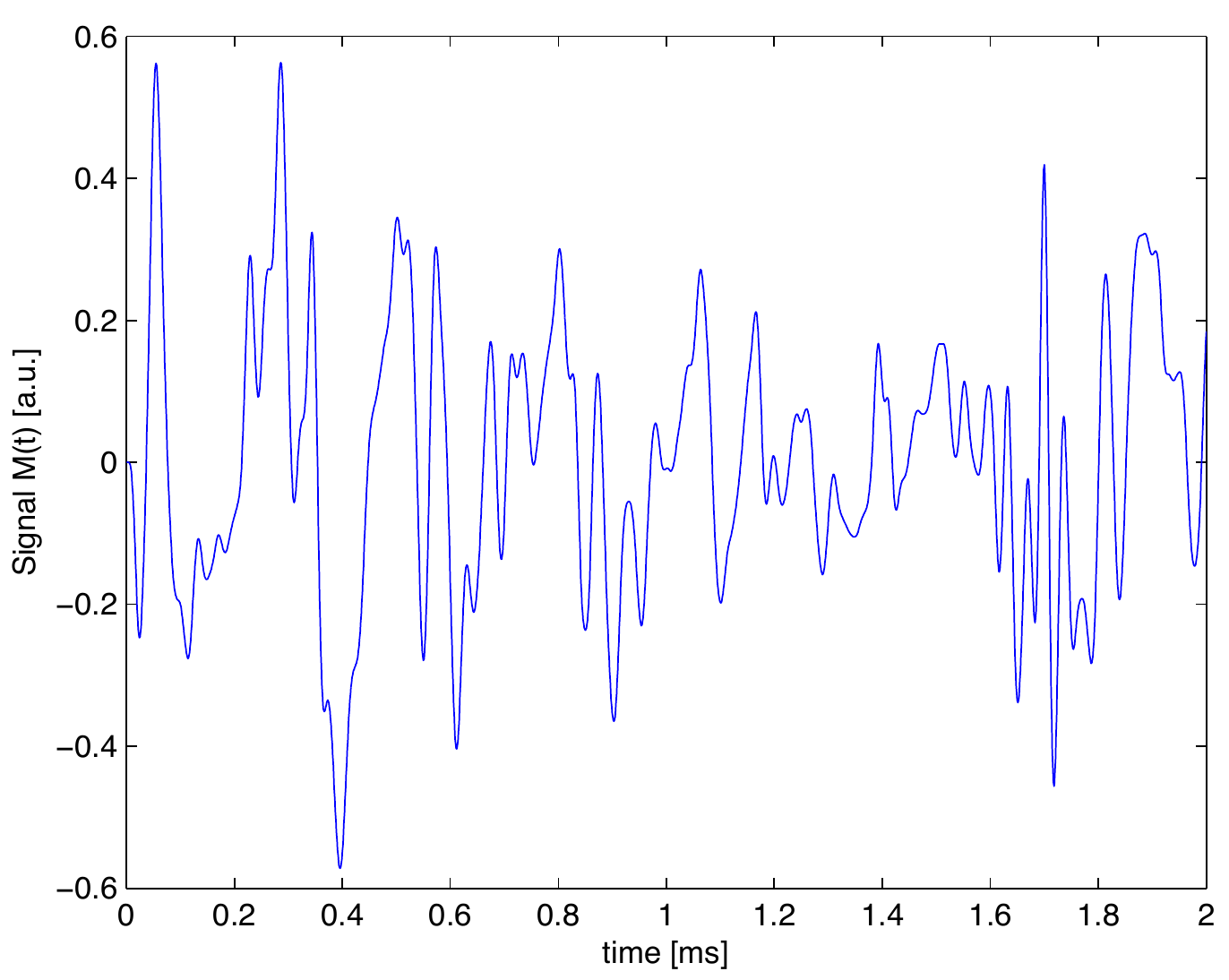}
\caption{Simulated Faraday rotation measurement record for the initial state depicted in Fig. \ref{F:InitialState}. The signal-to-noise ratio for this simulation was $\sim 91$ and the bandpass filter from $2$ to $40$ kHz. We use this measurement record to illustrate the performance of our state tomographic methods.}
\label{F:MeasurementRecord}
\end{center}
\begin{center}
\includegraphics[width=8.7cm,clip]{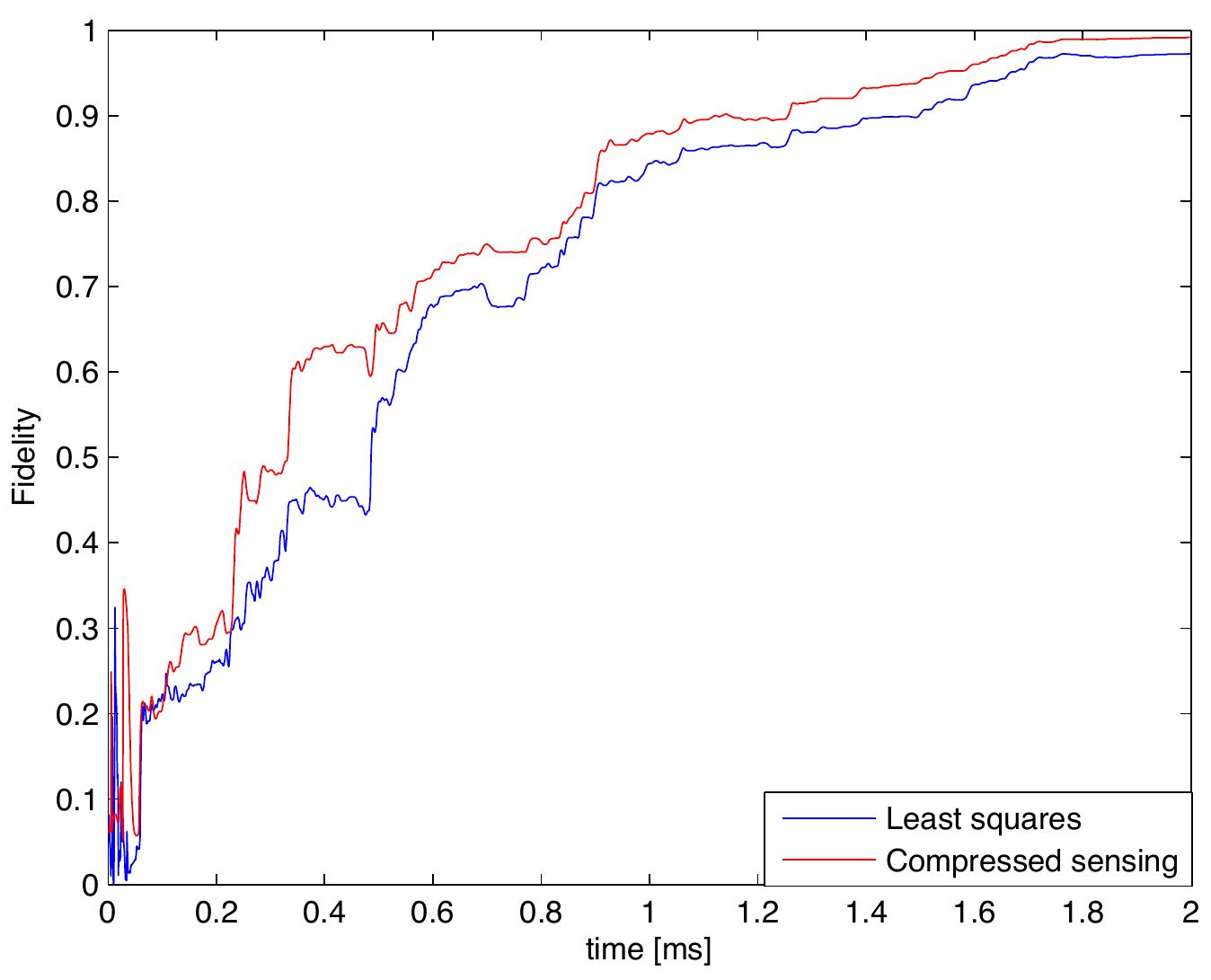}
\caption{Fidelity of reconstruction for least squares and compressed sensing calculated with the simulated data shown in Fig. \ref{F:MeasurementRecord} as a function of time. At earlier times, while there is not enough information to reconstruct the state, both methods give estimates with poor fidelities. At later times, however, as more information is gathered, higher fidelities are achieved, compressed sensing being the method that achieves the highest in the case of pure states.}
\label{F:ReconFidelity}
\end{center}
\end{figure}

As an example to illustrate how the whole procedure works and performs, we simulate the reconstruction of the nontrivial state, $\ket{\psi}=\left( \ket{\psi^{(+)}_{sq}}+\ket{\psi^{(-)}_{cat}} \right) / \sqrt{2}$, shown in Fig. \ref{F:InitialState}, consisting of an equal superposition of a spin squeezed state in the $F_+=4$ manifold, $\ket{\psi^{(+)}_{sq}}=\exp\left\{-i0.5F_z^2\right\}\ket{F=4,m_x=4}$ and a ``cat state"  in the $F_-=3$ manifold $\ket{\psi^{(-)}_{cat}}=\left(\ket{F=3,m_z=3}+\ket{F=3,m_z=-3}\right)/\sqrt{2}$. The bar plots represent the real and imaginary parts of the estimated density matrix elements, which is a common way to visualize matrix elements. This particular state is chosen to visually emphasize the performance of the protocols and to show our ability to reconstruct nontrivial quantum states in the 16-dimensional ground manifold of cesium. The simulated measurement record with added noise and bandpass filtered as described above is shown in Fig. \ref{F:MeasurementRecord}. This signal is used as the ``finger print'' of the quantum state we want to reconstruct. The signal is processed, using no knowledge about the initial state of the system, by our two reconstruction methods described in previous chapters: least squares and compressed sensing. We show in Fig. \ref{F:ReconFidelity} the fidelity of reconstruction as a function of time for both methods. At earlier times, while there is not enough information to reconstruct the state, both methods give estimates with poor fidelities. At later times, however, as more information is gathered, higher fidelities are achieved, compressed sensing being the method that achieves the highest, $0.9915$ compared to $0.9727$ that least squares produces. This is because the input state is a pure state, which compressed sensing is optimized for. The reason that neither method achieves unit fidelity is the finite signal-to-noise ratio. For artificially higher SNR, we see that both methods achieve full fidelity; of course, that case is of no practical interest. Moreover, if we let the simulation run for times larger than the 2ms shown in the picture, higher fidelities are achieved, and reach unity asymptotically, which is expected by this model. However, such long measurement periods will eventually collapse the state into a random eigenvalue of the observable operator, see for example \cite{wiseman10}, which our model does not consider.  

\newpage

%

\begin{figure}[h!]
\begin{center}
\includegraphics[width=11.0cm,clip]{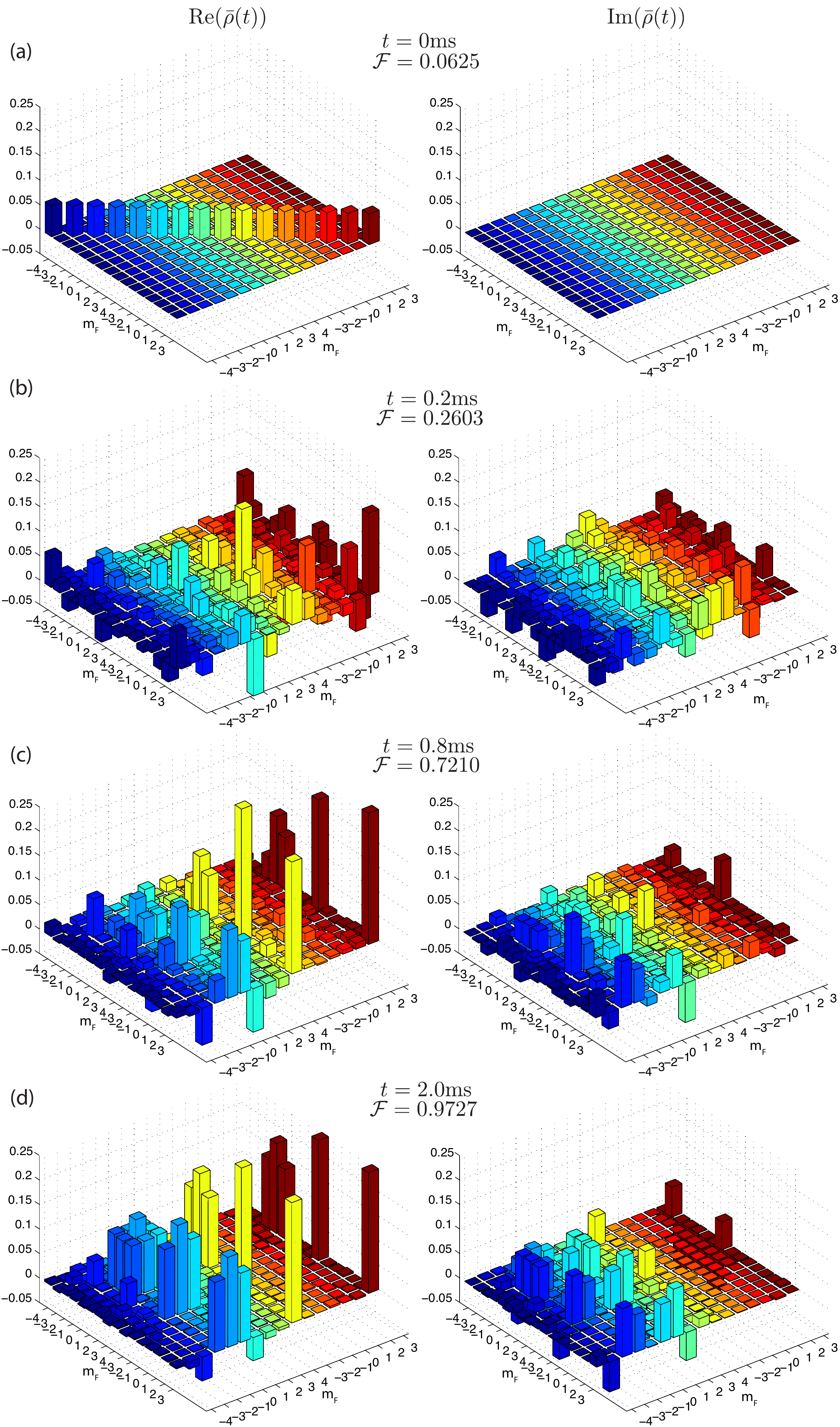}
\caption{Real and imaginary parts of the estimated initial state $\bar{\rho}$ using the least squares method. (a) When no information has been collected, the maximally mixed state is guessed by the algorithm. (b) and (c) More information is acquired as time passes and higher fidelities are obtained. (d) A high fidelity estimate is obtained after 2.0ms of simulation.}
\label{F:Frames}
\end{center}
\end{figure}

\begin{figure}[h!]
\begin{center}
\includegraphics[width=11.0cm,clip]{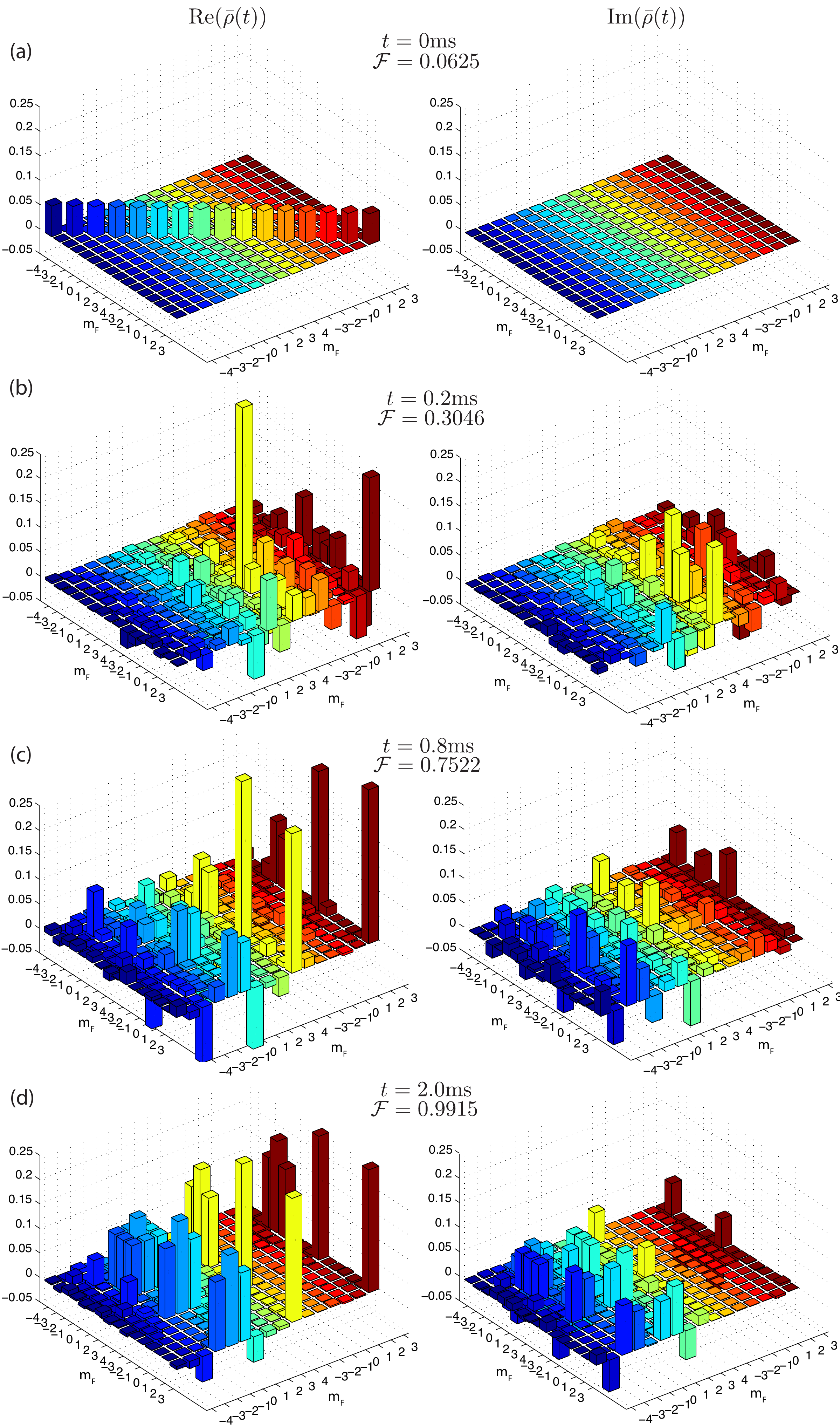}
\caption{Real and imaginary parts of the estimated initial state $\bar{\rho}$ using the compressed sensing method. (a) When no information has been collected, the maximally mixed state is guessed by the algorithm. (b) and (c) More information is acquired as time passes and higher fidelities are obtained. (d) A high fidelity estimate is obtained after 2.0ms of simulation.}
\label{F:Frames2}
\end{center}
\end{figure}

For visualization purposes, in Figs. \ref{F:Frames} and \ref{F:Frames2}, we show how our QT procedures converge as a function of time to an estimate of the initial state with high fidelity. Estimates of the initial quantum state, $\bar{\rho}(t)$, are shown for $t=0,~0.2,~0.8,~{\rm and}~2$ ms with the intention to display the convergence path that QT takes for this particular example. Within the few first microseconds of the simulation there is little information, and both protocols return the maximally mixed state as the estimation of $\rho_0$, which is shown in part (a) of both plots. However, as time passes, more information about the informationally complete set of observables is acquired and the protocols make better guesses of the initial state.  For a SNR of $\sim 91$ simulated here, within 2 ms, a fidelity $>0.97$ is achieved for least squares and a fidelity $>0.99$ is achieved for compressed sensing. As we will see later in this chapter, the trend of compressed sensing performing better for pure states will be confirmed as a general feature.

\clearpage
\subsection{General performance}

The plots from the last section show the performance of the reconstruction protocols for a particular state, which was useful as a descriptive and pedagogical example. In this section, however, we study the general performance of both methods for random pure and mixed states. We generate random pure states using the Haar measure, whereas we sample mixed states using the Hilbert-Schmidt measure. As before, we use these states to generate a valid measurement record with added Gaussian noise of standard deviation $\sigma=0.03$ and process the signals to try to learn the initial state of the system. 

This numerical study shows the average performance of our methods for arbitrary random states. In general, states found this way will have support in all the 16 ground state sub labels $\ket{F,m_F}$. Moreover, the SNR for each of these states is variable since we have kept the size of the noise $\sigma$ constant, and the RMS value of the measurement signal strongly depends on the initial quantum state.

\begin{figure}[t]
\begin{center}
\includegraphics[width=15.1cm,clip]{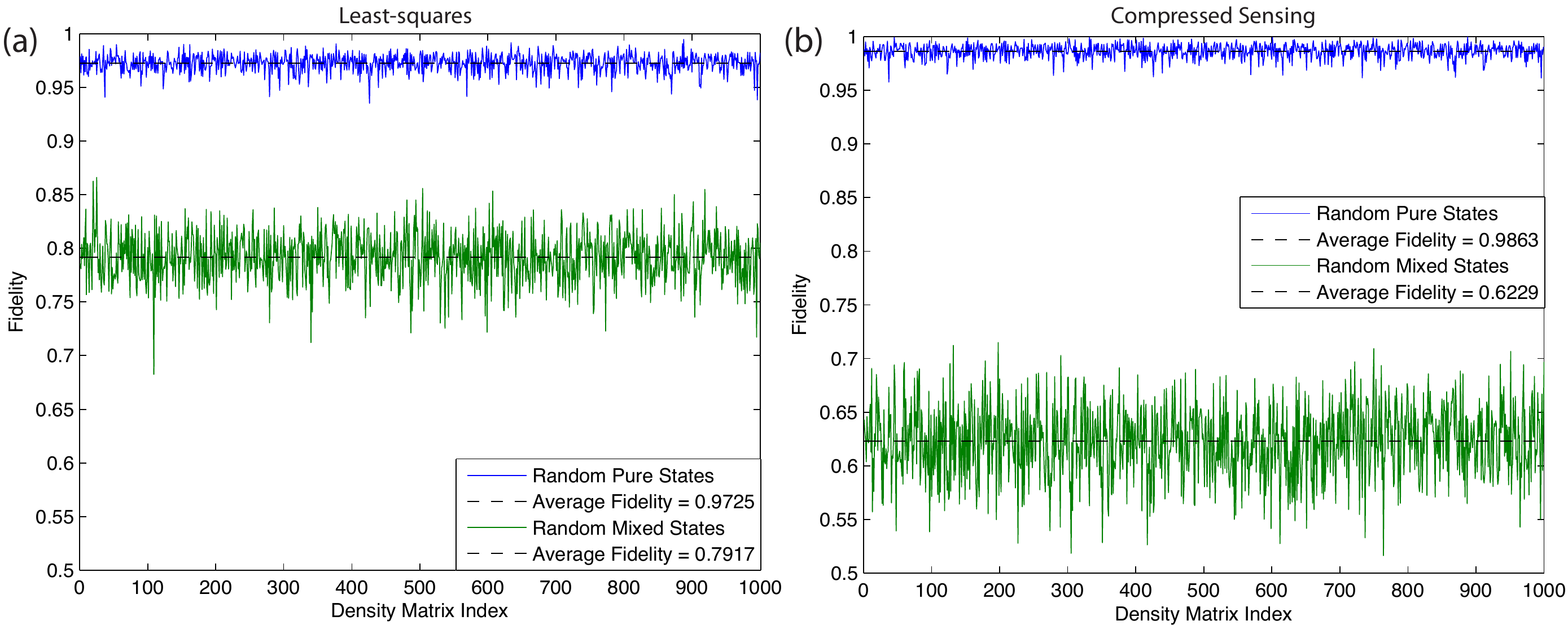}
\caption{(Color online) Fidelity of reconstruction for random states using (a) least squares and (b) compressed sensing. The blue line represent the fidelities for random pure states sampled from the Haar measure whereas the green line is for random mixed states sampled from the Hilbert-Schmidt measure. For pure states, compressed sensing shows a superior performance than least squares. However, for mixed states, although both methods performances are far from good, least squares seems to be a more advantageous method. }
\label{F:FidelityRandomStates}
\end{center}
\end{figure}

Fig. \ref{F:FidelityRandomStates} shows the fidelity of reconstruction of 1000 random pure states sampled from the Haar measure and a 1000 random mixed states sampled from the Hilbert-Schmidt measure, processed by our least squares (left) and compressed sensing (right) methods. We see that we achieve average fidelities, for pure states, 0.9725 (standard deviation 0.0083) and 0.9863 (standard deviation 0.0066) for least squares and compressed sensing, respectively. For mixed states, however, we obtain fidelities of 0.7917 (standard deviation 0.0233) and 0.6229 (standard deviation 0.0342) for least squares and compressed sensing, respectively. From these numerical experiments, we can conclude that, as expected, compressed sensing methods do perform better than least squares in general when the states being reconstructed are pure. In fact, the error (measured by $1-\mathcal{F}$) for compressed sensing is about 50$\%$ smaller than that of least squares.  For mixed states, however, the situation is quite different: least squares performance, although far from optimal, is vastly superior to compressed sensing. As we will see in the next section, this is because compressed sensing tries to find the purest possible state compatible with the data, while least squares just finds the state that best fits the data.

%
%

\subsection{Rank vs purity: understanding the behavior of compressed sensing}\label{sec:RankVsPurity}

The assumption made by compressed sensing methods is that the state being reconstructed is low rank or has high purity. In this section, we proceed to study the difference between the least squares and compressed sensing methods described in Chapter \ref{ch:QT} when we vary either the rank or the purity of the initial states, which we achieve by a series of numerical experiments.

First, we generate random initial density matrices, in a Hilbert space of dimension $d=16$, with fixed rank and random purity. Second, we generate random states with fixed purity and random rank. In both cases, we use the random states to generate a simulated measurement record as in Eq. (\ref{eq:measurement}), with a variable signal-to-noise ratio corresponding to a fixed-variance Gaussian noise, $\sigma=0.03$, and use that measurement record to find what the initial state of the system was. We used both least squares and compressed sensing methods to analyze the simulated data and find different estimates of the initial state. 

\subsubsection{A. Fixed rank random states}
For our first numerical experiment we fix the rank of the density matrix, i.e., we fix the number of non zero eigenvalues, and choose its eigenvectors randomly from the Haar measure. Algorithmically, the process of finding such states is described in Appendix \ref{App:FixedRankRandom}.

In Fig. \ref{F:FixedRankyHistograms}, we show the performance of this procedure for a typical run in which we obtained an ensemble of 1000 density matrices. The histograms show, for different ranks, $r=2,\ldots,6$, how the purities are distributed within the ensemble. Moreover, the mean purity, $\bar{\mathcal{P}}$, defined in Eq. (\ref{eq:purity}), computed for these states is: for $r=2$, $\bar{\mathcal{P}}= 0.6204$; for $r=3$, $\bar{\mathcal{P}} = 0.4320$; for $r=4$, $\bar{\mathcal{P}}= 0.3337$; for $r=5$, $\bar{\mathcal{P}} = 0.2637$; and for $r=6$, $\bar{\mathcal{P}}= 0.2187$. Clearly, this method of obtaining such random states is not uniform over purity and is biased towards low purity states. However, it will be useful to illustrate key differences between the least squares and compressed sensing methods. 

\begin{figure}[t]
\begin{center}
\includegraphics[width=15cm,clip]{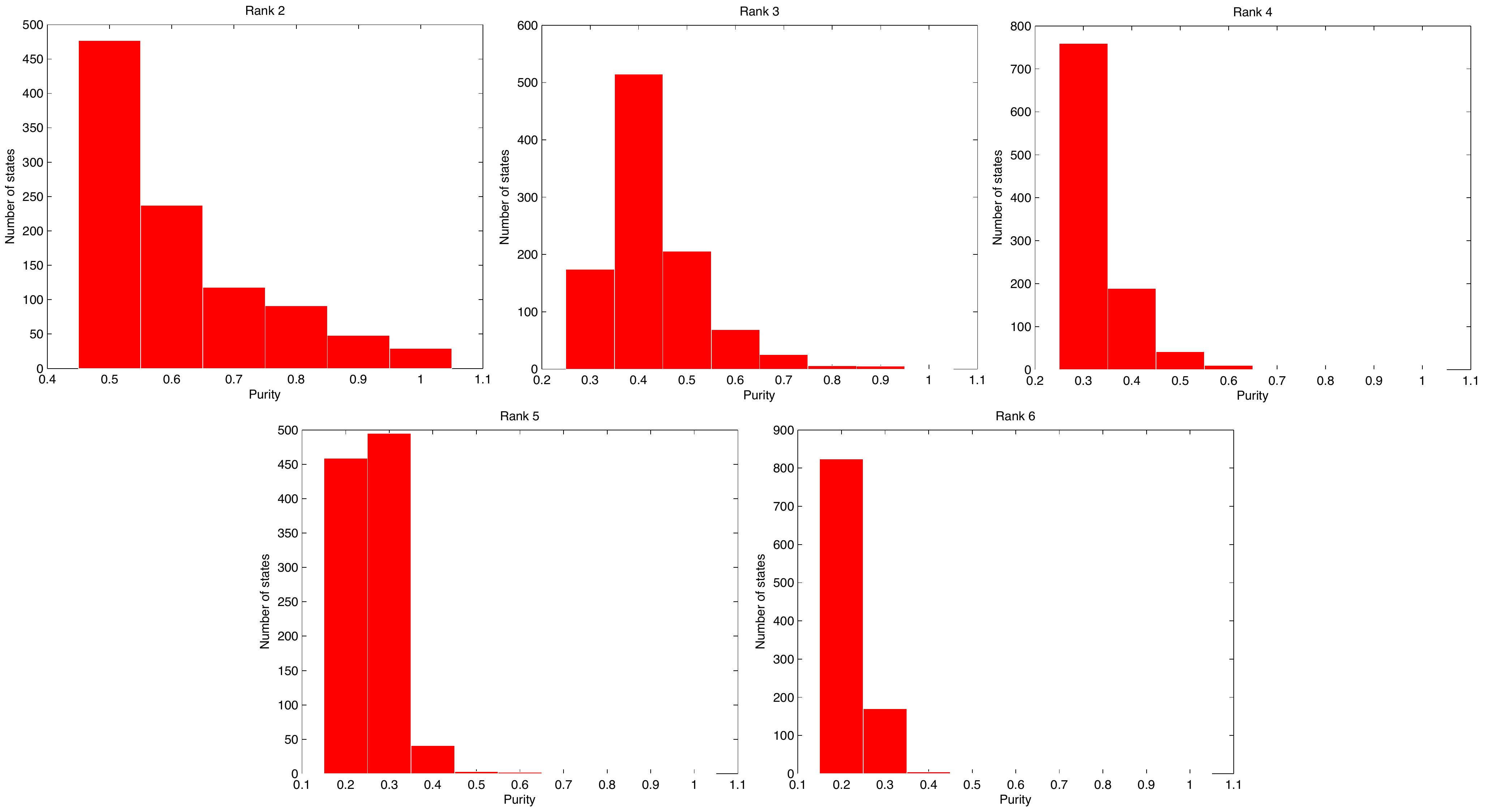}
\caption{Random mixed states with fixed rank. The histograms show the way the purity of the random states is distributed for each fixed rank. Clearly, the method outlined in the text used to generate such random states seems to find them concentrated into low purity states. Although not uniformly sampled over purity, these states will help us illustrate important differences between the least squares and compressed sensing reconstruction methods.}
\label{F:FixedRankyHistograms}
\end{center}
\end{figure}

\begin{figure}[t]
\begin{center}
\includegraphics[width=15.1cm,clip]{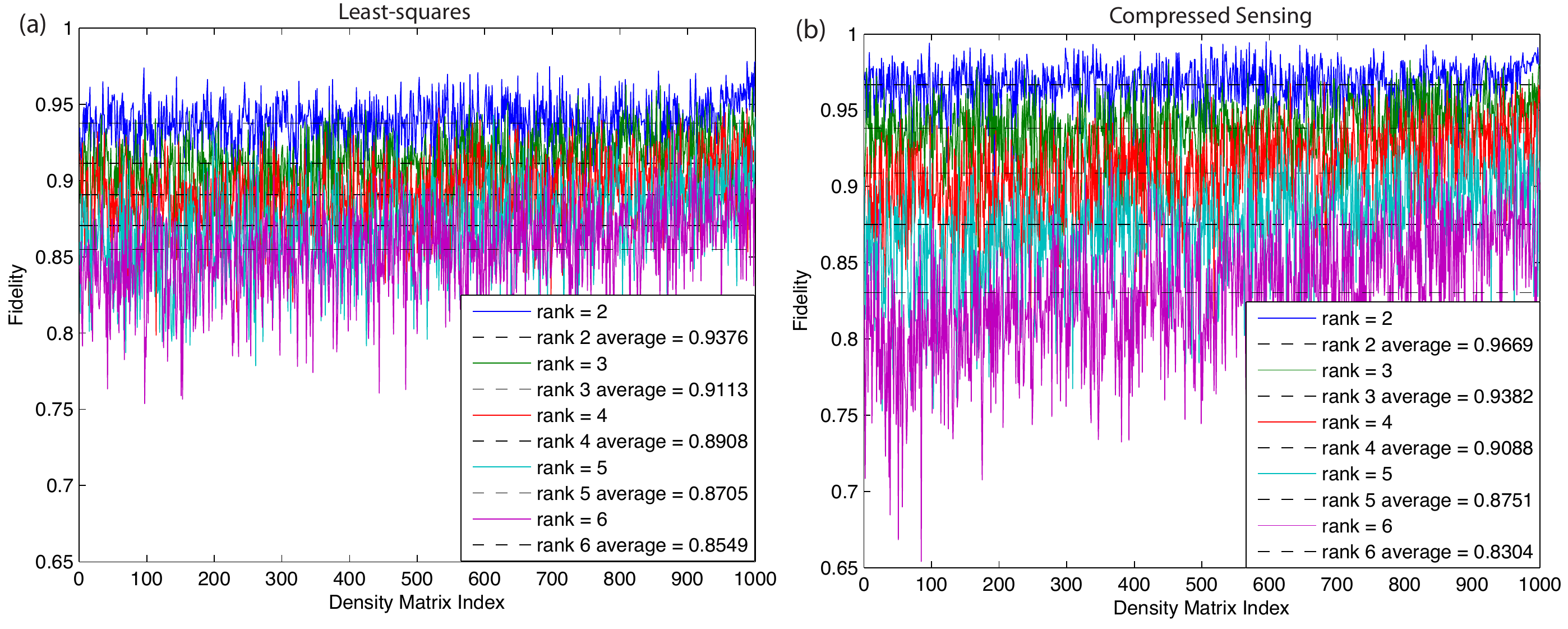}
\caption{Fidelity of reconstruction for random mixed states with fixed rank for (a) least squares and (b) compressed sensing. The quantum states are ordered from low purity to high purity to improve visibility. There is a strong dependence between the purity/rank of the state and the performance of compressed sensing, whereas that correlation is less visible for the case of least squares. Average fidelity for CS seems to be higher than for LS until the rank of the states is sufficiently large, in which case the tendency is reversed.}
\label{F:FidelityRandomPurity}
\end{center}
\end{figure}

Fig. \ref{F:FidelityRandomPurity} shows the fidelity of reconstruction, Eq. (\ref{eq:fidelity}), for the random states with fixed rank described above, processed by both least squares and compressed sensing, as a function of an arbitrary number that indexes a particular density matrix. The random density matrices were ordered from low purity to high purity from left to right to improve the readability of the plots. For the case of the least squares method, Fig. \ref{F:FidelityRandomPurity}\textcolor{blue}{a}, we see, in general, lower average fidelities than for the case of compressed sensing method, Fig. \ref{F:FidelityRandomPurity}\textcolor{blue}{b}, for low rank states. However, that tendency changes for states of higher rank, which clearly indicates that compressed sensing methods are more effective at reconstructing low rank/high purity states, as intuitively we expected. Moreover, we see a strong correlation between the performance of compressed sensing and the purity of the initial state. This is illustrated by the fact that the fidelity of reconstruction seems to increase with purity for the states of higher rank. This correlation is less strong for the least squares method, though clearly there is still some correlation between the fidelity and rank for this method as well.

\begin{figure}[t]
\begin{center}
\includegraphics[width=15.1cm,clip]{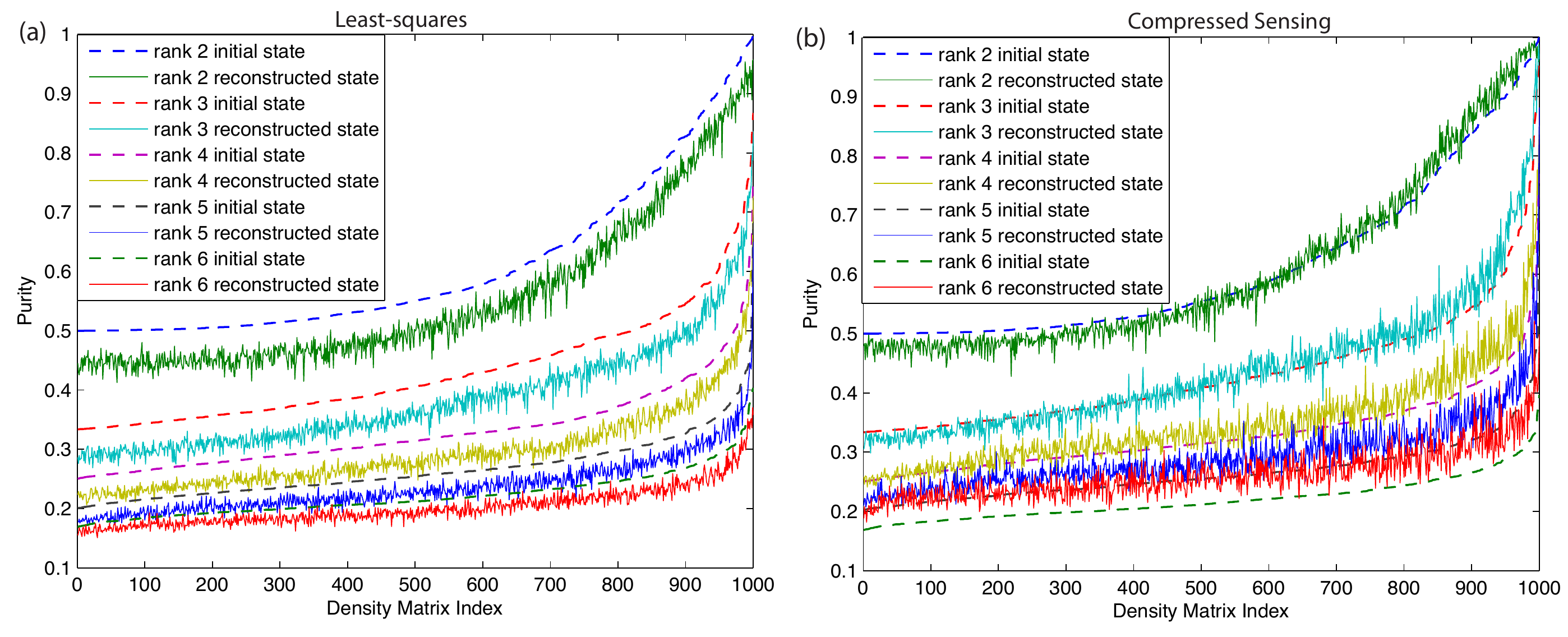}
\caption{Purity of the reconstructed quantum states for random states of fixed rank processed by (a) least squares, (b) compressed sensing. The dotted lines represent the purity of the initial quantum state for the 5 different ranks considered in this numerical experiment. The solid lines are the purities of the reconstructed states. The states were ordered from low to high purity to improve visualization. The least squares method for reconstruction seems to almost always estimate states that are more mixed, i.e., lower purity, than the actual input states, whereas compressed sensing does a good job estimating the actual purity of the states of rank $\le 4$. It, however, has the tendency to estimate purer than the input states for ranks $\ge 5$. }
\label{F:PurityReconAndInitial}
\end{center}
\end{figure}

Fig. \ref{F:PurityReconAndInitial} shows the purity of the reconstructed quantum state compared to the purity of the actual initial state. Again, to improve the readability of the plots, the $x$ axis index was ordered from low to high purity density matrices. In this plot, we can see how our two reconstruction methods perform in determining the actual purity of an unknown quantum state. Clearly, the least squares method, Fig. \ref{F:PurityReconAndInitial}\textcolor{blue}{a}, generally estimates states that are more mixed than the actual input state regardless of the rank of the states. This occurs due to the fact that this method finds a physical state that fits the noisy measured signal, regardless of its purity. By doing so, it also fits the noise, which in turn is fitted by some mixed state. We can see that the compressed sensing method, Fig. \ref{F:PurityReconAndInitial}\textcolor{blue}{b}, for low rank states, estimates their purity in a better way, however, it becomes clear that for higher rank/lower purity states, this method always estimates states that are more pure than the actual initial states. These observations are in agreement with the assumed prior information that the states are low rank or almost pure.

\begin{figure}[t]
\begin{center}
\includegraphics[width=15.1cm,clip]{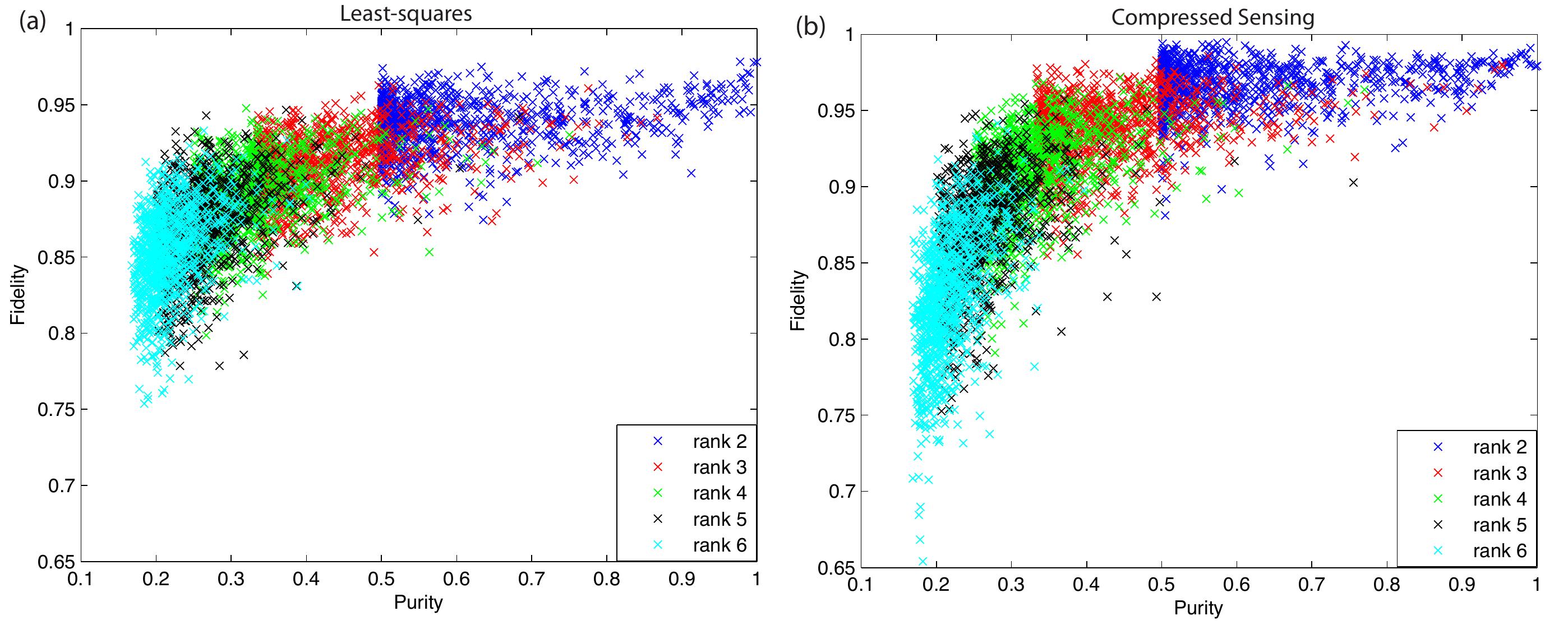}
\caption{Scatter plots of fidelity of reconstruction as a function of purity of the initial states for (a) least squares, and (b) compressed sensing. We see in this type of plot the concentration and spread of fidelity for the same purity and different ranks. We also see how compressed sensing performs better than least squares for states with low rank/purity, as expected.}
\label{F:CSScatterFidelityVsPurity}
\end{center}
\end{figure}

Finally, in Fig. \ref{F:CSScatterFidelityVsPurity}, we show the same data discussed above in scatter plots in order to better explain the differences between methods. In these plots, we have the purity of the initial density matrices on the $x$ axis and the fidelity of reconstruction on the $y$ axis. We see how, even for states with the same purity, both methods perform in different ways giving different reconstruction fidelities. We attribute this to different noise realizations and the fact that different states, even if they have the same purity, can produce measurement signals that are different in amplitude. This makes their signal-to-noise ratios different which affects the quality of the reconstruction. Another aspect we notice is that compressed sensing, generally, performs better than least squares, in terms of fidelity of reconstruction, for states of low rank and high purity.

In summary, from this particular numerical experiment, we can conclude that in general compressed sensing does better in cases of mixed states with low rank and high purity than least squares. However, when we, a priori, know that we need to distinguish states that may be high-rank or very mixed, we should use the least squares method, which does not assume anything about the purity or the rank of the unknown states.

\subsubsection{B. Fixed purity random states}

The question of how the two methods perform for a given purity remains unanswered. We need some kind of method that generates random density matrices for a given purity. Although not optimal, we illustrate the method employed here in Appendix \ref{App:FixedPurityRandom}.

\begin{figure}[t]
\begin{center}
\includegraphics[width=15cm,clip]{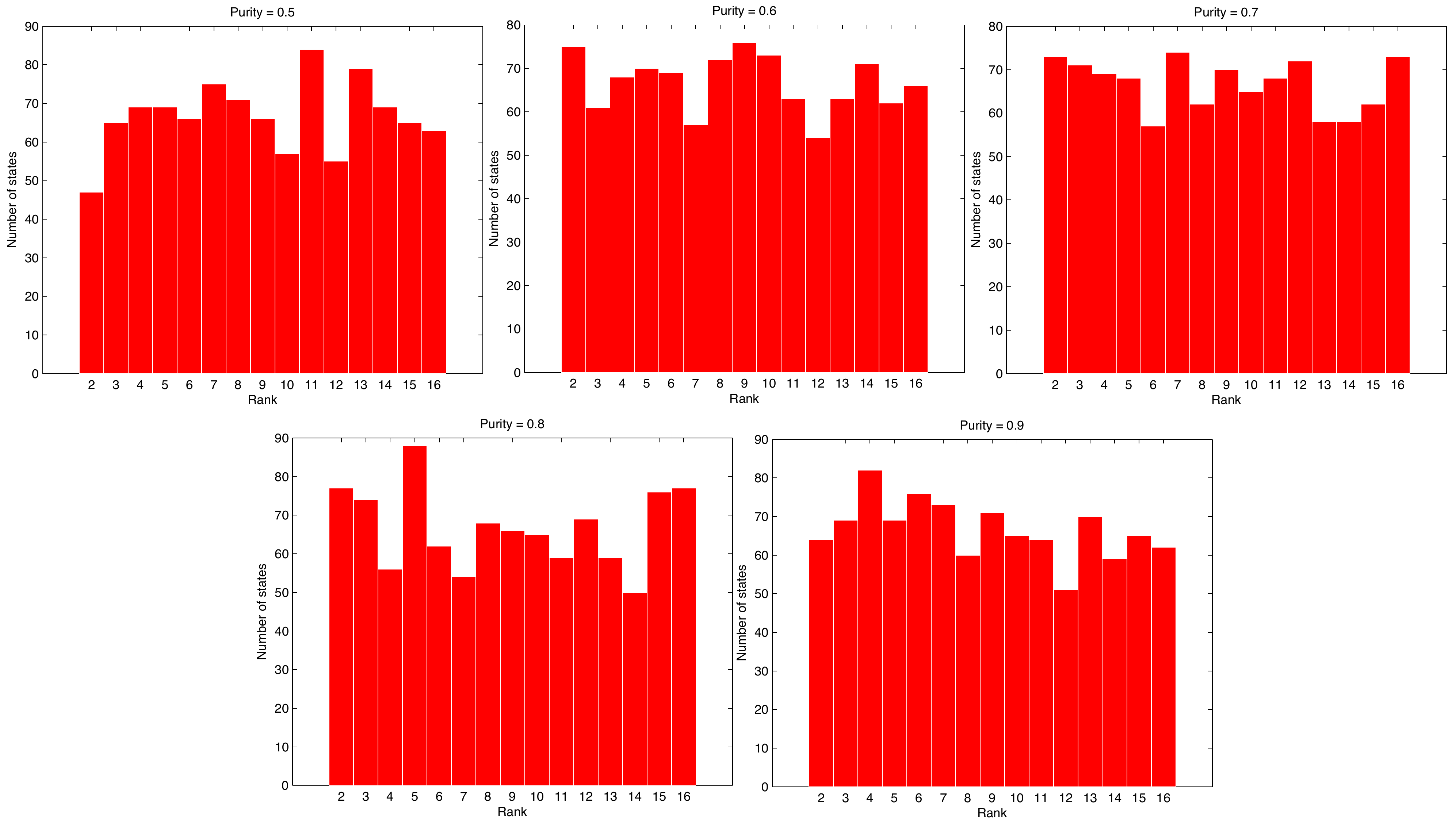}
\caption{Random mixed states with fixed purity. The histograms show the way the rank of the random states is distributed for each fixed purity. The method outlined in the text used to generate such random states seems to find them fairly uniformly distributed over rank. These states are used to study the performance of both QT methods and will help us illustrate important differences between them.}
\label{F:FixedPurityHistograms}
\end{center}
\end{figure}

We generate 1000 random density matrices according to the method mentioned above for 5 different purities $\mathcal{P}=0.5,~0.6,~0.7,~0.8,~{\rm and},~0.9$. In Fig. \ref{F:FixedPurityHistograms}, we show the histograms of the number of density matrices distributed over their rank for a typical run of the procedure. The method discussed in Appendix \ref{App:FixedPurityRandom} seems to generate fairly uniform samples of states, which we will use to simulate measurement records and do quantum tomography.

\begin{figure}[t]
\begin{center}
\includegraphics[width=15.1cm,clip]{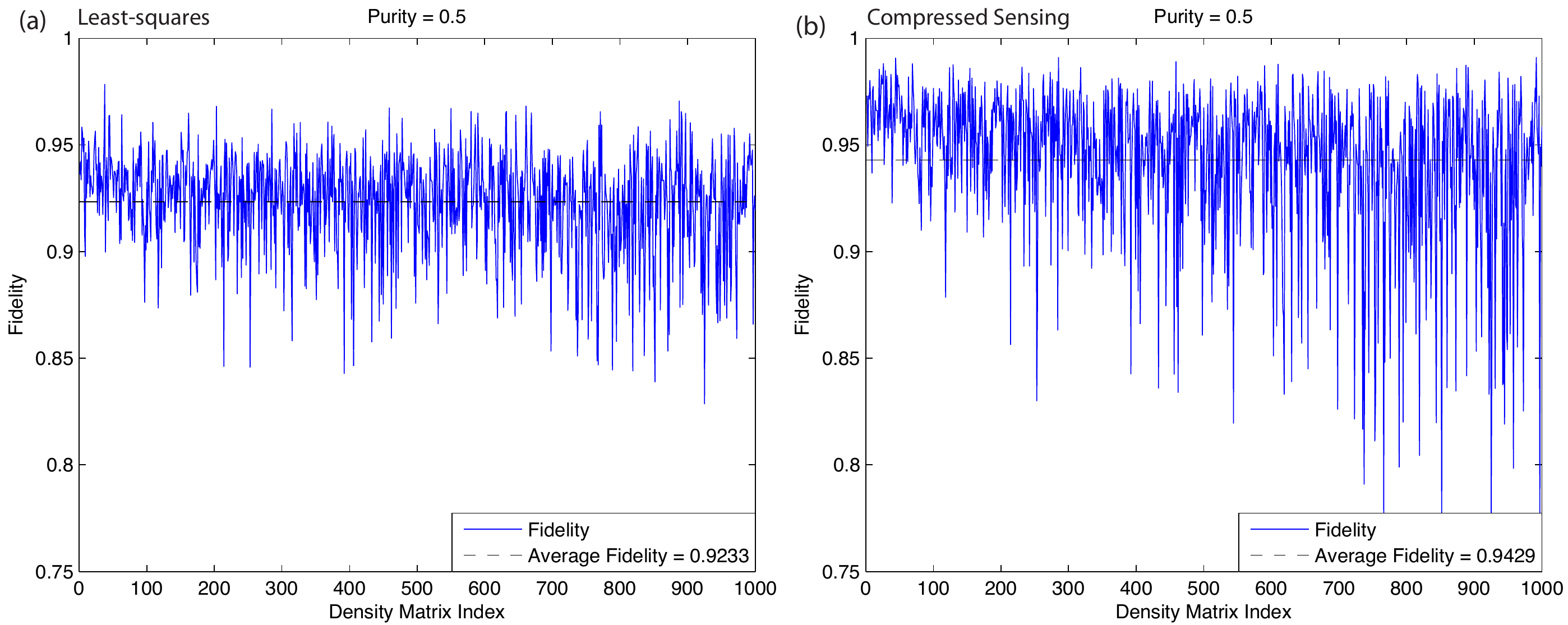}
\caption{Fidelity of reconstruction for random states of purity 0.5 calculated with (a) least squares, and (b) compressed sensing. The density matrices are ordered from low to high rank in each plot to improve visualization. Although the average fidelity is higher for compressed sensing, we clearly see that the spread of fidelities increases with the rank of the states. This fact is not seen for the least squares data shown on the left. }
\label{F:Purity0p5Fidelity}
\end{center}
\begin{center}
\includegraphics[width=15.1cm,clip]{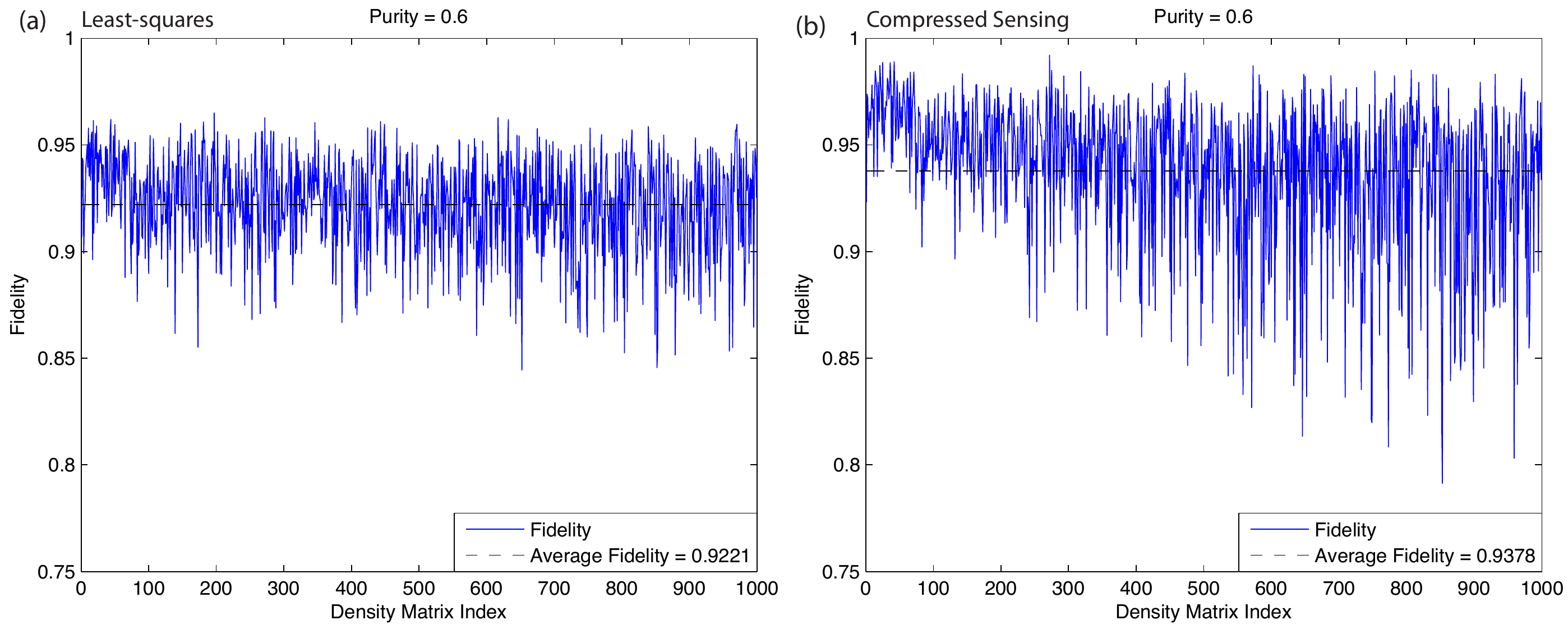}
\caption{Fidelity of reconstruction for random states of purity 0.6 calculated with (a) least squares, and (b) compressed sensing. The density matrices are ordered from low to high rank in each plot to improve visualization. Although the average fidelity is higher for compressed sensing, we clearly see that the spread of fidelities increases with the rank of the states. This fact is not seen for the least squares data shown on the left.}
\label{F:Purity0p6Fidelity}
\end{center}
\end{figure}

\begin{figure}[t]
\begin{center}
\includegraphics[width=15.1cm,clip]{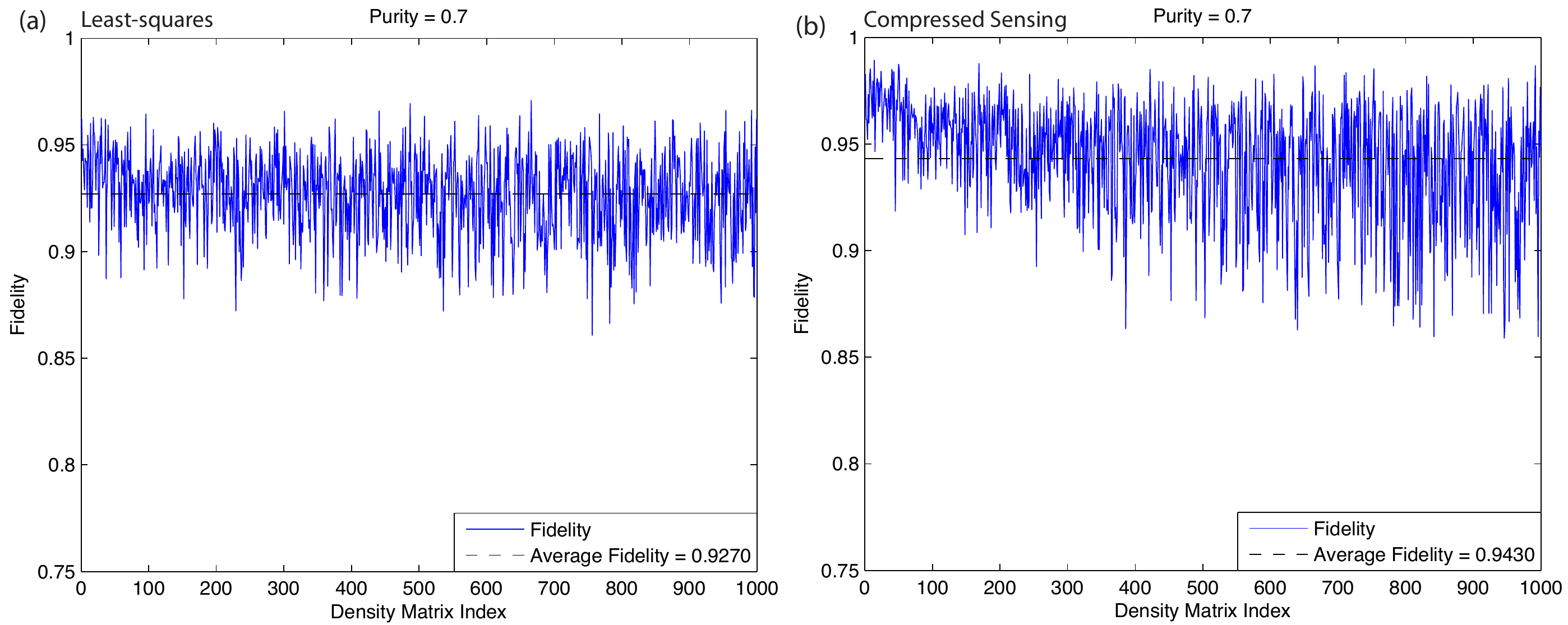}
\caption{Fidelity of reconstruction for random states of purity 0.7 calculated with (a) least squares, and (b) compressed sensing. The density matrices are ordered from low to high rank in each plot to improve visualization. The average fidelity is higher for compressed sensing and we clearly see that the spread of fidelities increases with the rank of the states as we saw in Figs. \ref{F:Purity0p5Fidelity} and \ref{F:Purity0p6Fidelity}. However, the spread in fidelity is greatly decreased due to the fact that the states are more pure. The least squares estimates (a) seem not to be affected by the rank of the matrices.}
\label{F:Purity0p7Fidelity}
\end{center}
\begin{center}
\includegraphics[width=15.1cm,clip]{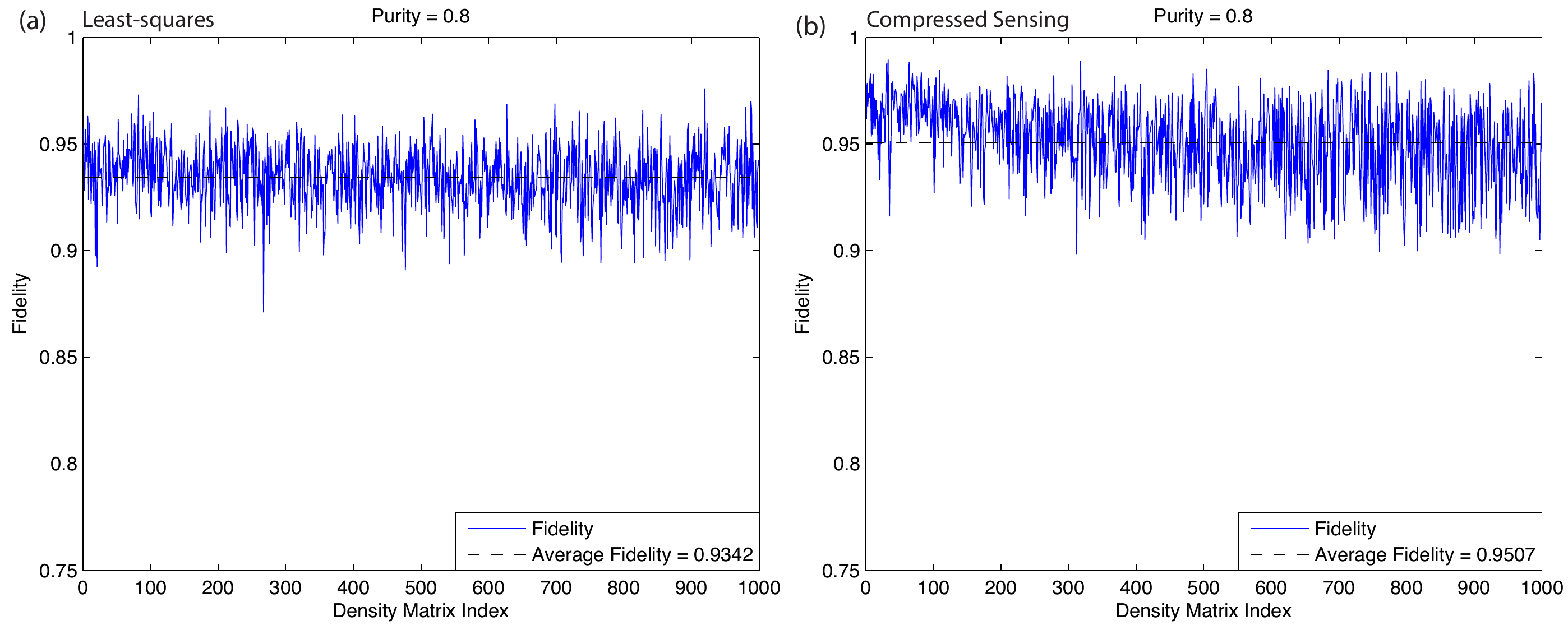}
\caption{Fidelity of reconstruction for random states of purity 0.8 calculated with (a) least squares, and (b) compressed sensing. The density matrices are ordered from low to high rank in each plot to improve visualization. The average fidelity is higher for compressed sensing and has increased for both methods with respect to the previous plots. Again, we see that the spread of fidelities increases with the rank of the states as we saw before. However, the spread in fidelity is greatly decreased due to the fact that the states are more pure. The least squares estimates (a) seem not to be affected by the rank of the matrices.}
\label{F:Purity0p8Fidelity}
\end{center}
\end{figure}

\begin{figure}[t]
\begin{center}
\includegraphics[width=15.1cm,clip]{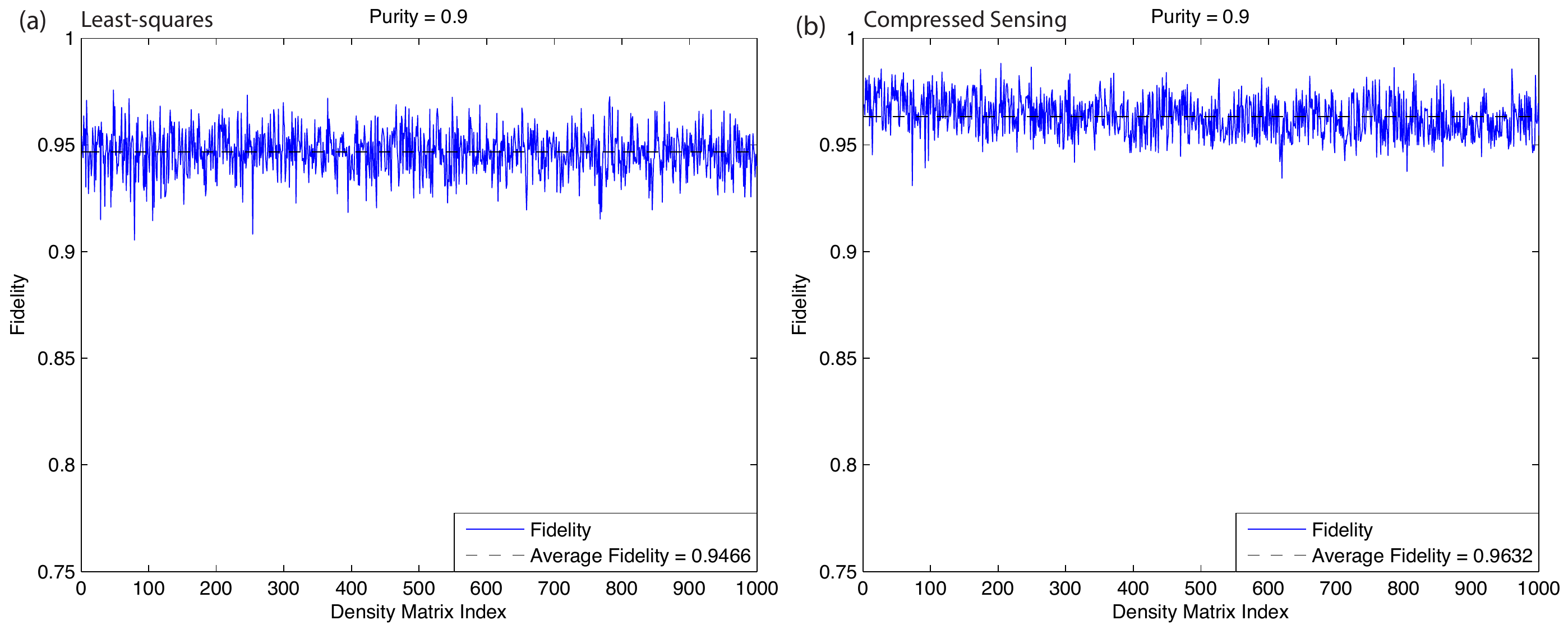}
\caption{Fidelity of reconstruction for random states of purity 0.9 calculated with (a) least squares, and (b) compressed sensing. The density matrices are ordered from low to high rank in each plot to improve visualization. The average fidelity is higher for compressed sensing and has increased for both methods with respect to the previous plots. The increase in spread of fidelities with the rank of the states is less apparent than before meaning that for high enough purities the actual rank of the density matrix is irrelevant for CS estimates. As shown in previous plots, the least squares estimates (a) seem not to be affected by the rank of the matrices.}
\label{F:Purity0p9Fidelity}
\end{center}
\end{figure}

To complete the numerical experiment, we calculate the fidelity of reconstruction for the random states with fixed purity and compare our least squares and compressed sensing methods. We show the results of these experiments in Figs. \ref{F:Purity0p5Fidelity} to \ref{F:Purity0p9Fidelity}, in which the fidelities of reconstruction are given for a simulation time of 2 ms. To make the plots more readable, the density matrices are ordered from low to high rank. The first thing we notice is that for low purity states, $\mathcal{P}\le 0.7$, there is a clear correlation between the rank of the state and the fidelity for the case when the compressed sensing method was used. This is seen by realizing that the spread of fidelities increases with the rank of the density matrices. This observation confirms the intuition that for low purity states, the rank of the density matrix matters when compressed sensing is used. Another interesting feature we observe is that the least squares method seems not to depend on what the rank of the matrices is, although its average performance is better for purer states. We attribute this to two causes. First, mixed states in general produce signals with smaller amplitude (or equivalently, RMS value) and thus lower signal-to-noise ratio, which affects the quality of reconstruction. Second, purer states are more highly constrained than mixed states and thus are easier to reconstruct. As purity increases, however, we see that the rank of the initial states seems to be of less importance in terms of fidelity of reconstruction, and when the states are pure enough, $\mathcal{P}=0.9$, as shown in Fig. \ref{F:Purity0p9Fidelity}, it seems not to matter at all. These simulations confirm the fact that the compressed sensing method outperforms the standard least squares method for high purity states regardless of the rank, in general, and should be the preferred method to reconstruct quantum states that are known to have these properties.

In summary, with these two numerical experiments we have studied the behavior of compressed sensing and least squares applied to QT. While least squares does not assume anything about the purity of the state being reconstructed, we see that its estimates are generally more mixed than the actual state. We think this happens because of the noise added to the signal. Since least squares fits the measurement record directly, it actually finds the best quantum state that is compatible with the measured data. Due to the fact that the signal is noisy, this method not only fits the data but also the noise, giving as a result estimates that are mixed. However, this method implicitly works better for purer states because of the geometry of positive matrices. Purer states require less information to be reconstructed and positivity constrains the data even further to be compatible with a small set of physical states that are close to pure. On the other hand, compressed sensing methods do not fit the data; they find the purest/lowest rank state that can reproduce the data within a finite error. In this sense, they are most robust to noise and systematic errors; however, they assume that the state is pure or close to pure, which can be disadvantageous for  applications that require determining mixed states. 

In the context of the type of experiments we discussed in this work, in which we are interested in determining if a particular state preparation task was carried out successfully, and great care is taken in making the states as pure as possible, we believe that the compressed sensing protocol is the appropriate choice as a tool for continuous measurement QT, as we will see in the next chapter.

\chapter{Experimental Results}\label{ch:Results}
	In this chapter, we present and discuss the experimental application of our continuous measurement QT protocols. All plots show experimental data taken by the team of Aaron Smith, Brian Anderson, and H\'ector Sosa in the lab of Prof. P. Jessen, and was analyzed using the algorithms and model discussed in previous chapters. In general, an ensemble of cesium atoms is prepared in a particular target state using the optimal control techniques detailed in \cite{merkel08}. The system is immediately driven by the control waveforms described in Chapter \ref{ch:Simulations} while the polarization of the detuned laser field is continuously  measured. That signal is then input into our Matlab code that inverts it and gives an estimate of the target state based solely on the data and the known dynamics. The only part of the analysis that uses knowledge about the target state is the computation of the fidelity of reconstruction Eq. (\ref{eq:fidelity}), which we use as a measure of the quality of the tomographic procedure.

We start by describing the parameter estimation/calibration of the experiment which measures the relevant control parameters needed as input to the reconstruction algorithms. Next, we show the reconstruction of four hand-picked target states created in the lab to illustrate the performance of QT. Finally, we give a more quantitative discussion of the experimental performance of our algorithms where we show their average behavior over 49 Haar-random pure states.

\section{Calibration runs}
\begin{figure}[tbp]
\begin{center}
\includegraphics[width=15.1cm,clip]{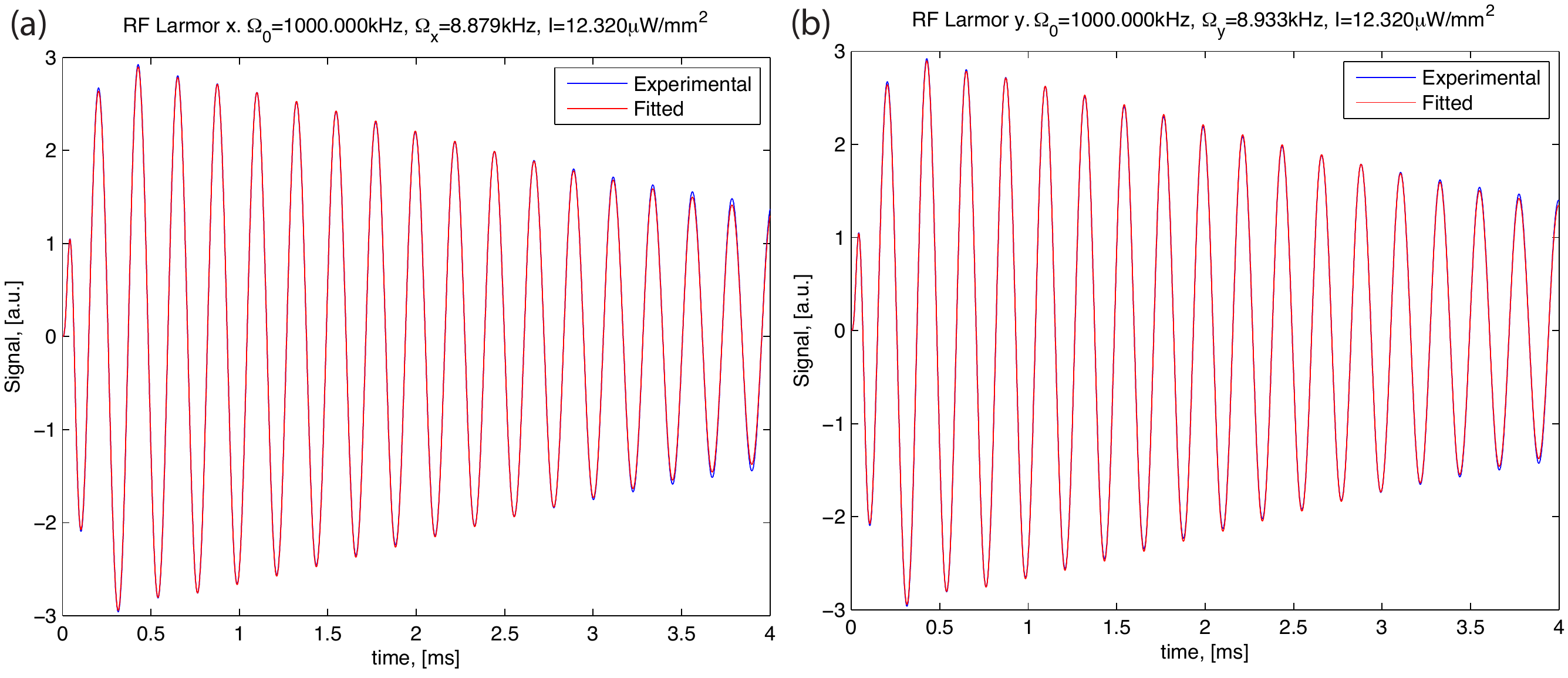}
\caption{Comparison between the experimental Faraday rotation signal (blue) and the signal fitted by our model (red) for the case of Larmor precession in the presence of the nonlinear light-shift and intensity inhomogeneity.  This signals are used to calibrate the magnitude of the RF field (a) $\Omega_{x}$, (b) $\Omega_{y}$ as well as the intensity of the light and its inhomogeneous distribution seen by the atoms.}
\label{F:RFLarmor}
\end{center}
\end{figure}

In order to estimate the relevant control parameters of the experiment, for any given data set, we use as our starting point the state $\ket{F=3,m=3}$, achieved with high fidelity through optical pumping to the state $\ket{F=4,m=4}$ followed by a microwave $\pi$-pulse. Then, a series of simple experiments are carried out whose measurement records are fitted by our model. First, Figure \ref{F:RFLarmor} shows two continuous Faraday measurement signals (Eq. (\ref{eq:fullrecordRot})) of the Larmor precession of this state with a single constant-amplitude RF magnetic field along (a) $x$ ($\Omega_y=0$) and (b) $y$ ($\Omega_x=0$) where in both cases there are no mirowaves, i.e. $\Omega_{\mu w}=0$. Decoherence via photon scattering as well as the inhomogeneous intensity of the laser probe results in the decay of the overall amplitude. The signal also shows inhomogeneous broadening due to variations in the probe intensity across the ensemble. In addition, another component of the signal modulation is due to nonlinear spin dynamics that we model together with photon scattering, which makes it possible for us to determine the laser intensity inhomogeneity. Finally, technical issues such as finite response time affect the time origin of the measurement record.  All of these features must be accurately included in our model of the Heisenberg evolution in order to obtain high-fidelity QT. 

Using the techniques described in Section \ref{sec:ParameterEstimation}, we determine the control parameters. For the signals shown in Fig. \ref{F:RFLarmor}, we obtained Larmor frequencies of $\Omega_{x}/2\pi=8.879$ kHz and $\Omega_{y}/2\pi=8.933$ kHz, which are slightly different from the nominal value of $9$ kHz. In addition, we find that the mean intensity of the laser probe is $12.320~\mu {\rm W/mm^2}$ with about $19\%$ inhomogeneous spread $\sigma_I$. Moreover, we fitted the overall units factor $a=2.13$, which we use for all the signals processed in the rest of the chapter. In these fits we have kept the bias field magnitude constant and equal to $\Omega_0/2\pi=1$ MHz due to the fact that the Larmor precession signals are not sensitive to small changes in the RF detuning induced by small variations in the bias field magnitude.

\begin{figure}[tbp]
\begin{center}
\includegraphics[width=8.7cm,clip]{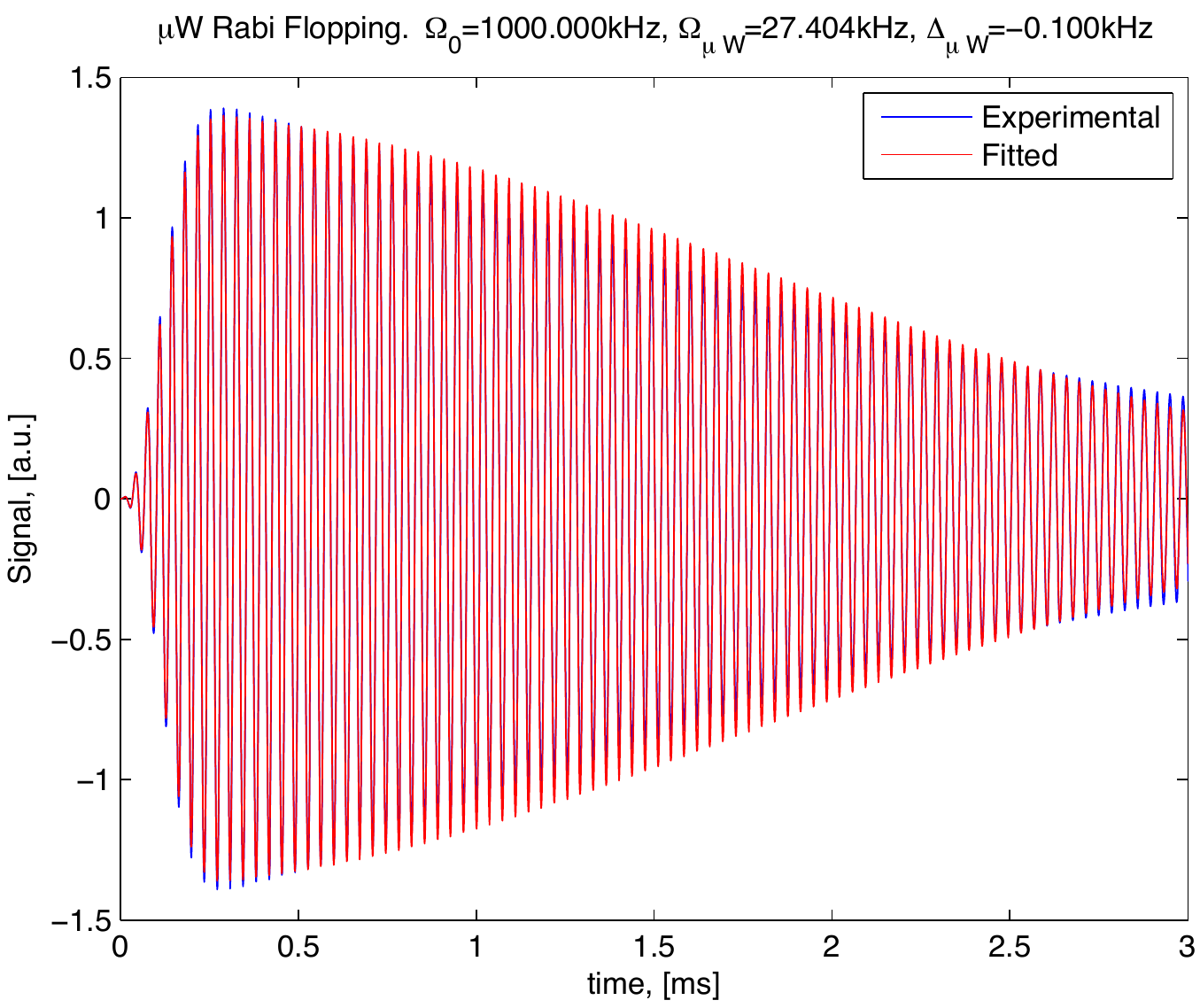}
\caption{Comparison between the experimental Faraday rotation signal (blue) and the signal fitted by our model (red) for the case of Rabi flopping for the pseudo spin system $\ket{3,3}$, $\ket{4,4}$ in the presence of the nonlinear light-shift and inhomogeneity in the power of the microwave field $\Omega_{\mu w}$.  This signal is used to calibrate the nominal value of $\Omega_{\mu w}$ and its inhomogeneity, the microwave detuning and the inhomogeneity in the bias field magnitude as seen by the atoms.}
\label{F:RabiuW}
\end{center}
\end{figure}

Additionally, an extra calibration run must be carried out to determine the magnitude of the microwave power (i,e, the microwave Rabi frequency) and its inhomogeneity as well as the microwave detuning $\Delta_{\mu w}$, defined in Eq. (\ref{eq:AppH0R}), and the inhomogeneity in the bias field magnitude. As before, the system is initially prepared in the state $\ket{F=3,m_F=3}$, and it is evolved through a simple Rabi flopping experiment for the 2-level spin system $\{\ket{F=3,m_F=3},\ket{F=4,m_F=4}\}$. Thus, for this experiment we choose $\Omega_x=\Omega_{y}=0$ and $\phi_x(t)=\phi_y(t)=\phi_{\mu w}(t)=0$ in the model. Fig. \ref{F:RabiuW} shows a comparison between the experimental data and the fitted signal for the Rabi flopping experiment described above. The signal shows the typical behavior of Rabi oscillations and decay due to photon scattering and inhomogeneity of the Rabi frequency and the bias field power. By fitting the data to our model, we calibrate the mean Rabi frequency $\Omega_{\mu w}/2\pi=27.405$ kHz with an inhomogeneous spread of $0.32\%$. In addition, as described in Section \ref{sec:ParameterEstimation}, for a random pure state, we find that $\Delta_{\mu w}/2\pi\approx-100$ Hz and the inhomogeneity spread in the bias field amplitude is about $80$ Hz.

The last parameter we need to estimate is the error threshold for compressed sensing, $\epsilon$, as defined in Eq. \ref{eq:CSConvex}, using the technique discussed in Section \ref{sec:ParameterEstimation}. For a random pure state, we determine the optimal error threshold $\epsilon=1.2$ for the final time $T$ of the experiment and use it throughout all states we want to reconstruct for a particular data set. It is important to note that there is not a single error threshold, but a continuous function $\epsilon(t)$, at different times of the measurement record. We linearly interpolate $\epsilon(t)$ between $\epsilon(0)=0$ and $\epsilon(T)=\epsilon$ in order to give the correct expected functional form.  

With all these parameters in hand, we can proceed to use our model for quantum state estimation of arbitrary states. In the next section, we show the performance of QT for 4 different states which we picked in order to illustrate different aspects of the protocols.


\section{Experimental examples}
In this section, we show the performance of our reconstruction algorithms for several hand-picked states. The quantum states shown in the following examples are meant to illustrate the quality and performance in practical applications of the ideas described in this dissertation, i.e., our quantum tomographic procedures applied to reconstruct states encoded in the 16-dimensional hyperfine Hilbert space of an ensemble of cesium atoms.

Each measurement signal was produced as follows. In the laboratory, after optical pumping purifies a particular fiducial state, the atomic ensemble is prepared in a desired target state, $\ket{\Psi_0}$. The state preparation stage is done by a set of carefully optimized control phases of the RF and $\mu$w magnetic fields and takes about $210~\mu s$ for the Zeeman sub-levels, and $300~\mu s$ for more general random pure states, when robust control is used \cite{smith11}. The probe laser is turned off during this procedure since its presence is not needed for controllability. Then, the laser probe is turned on and the system is evolved according the control phases described in Section \ref{sec:waveforms} while the polarization of the probe is measured. Furthermore, although the procedure for QT can, in principle, find high-fidelity estimates in a single shot of the experiment, the data is averaged 25 times for every measured signal to improve the signal-to-noise ratio. Moreover, the signals are filtered with the bandpass filter described in Section \ref{sec:Filter}. We analyze the data with both least squares and compressed sensing.

\subsection{Reconstruction of an eigenvector of $F_z$ in the $F_+=4$ manifold}
\begin{figure}[b]
\begin{center}
\includegraphics[width=15.1cm,clip]{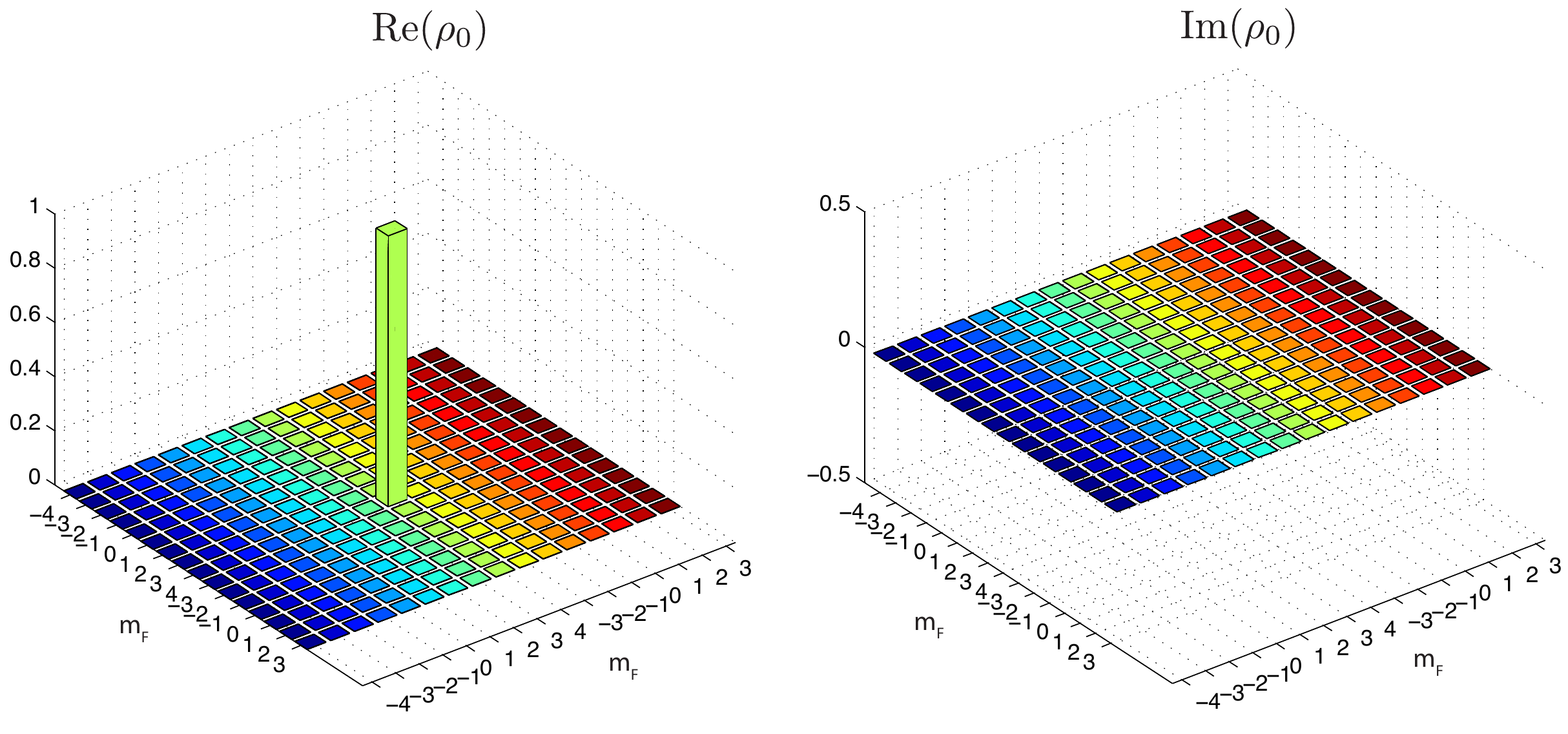}
\caption{Initial target state $\ket{\Psi_0}=\ket{4,4}$.}
\label{F:Initial44}
\end{center}
\end{figure}

\begin{figure}[H!]
\begin{center}
\includegraphics[width=8.7cm,clip]{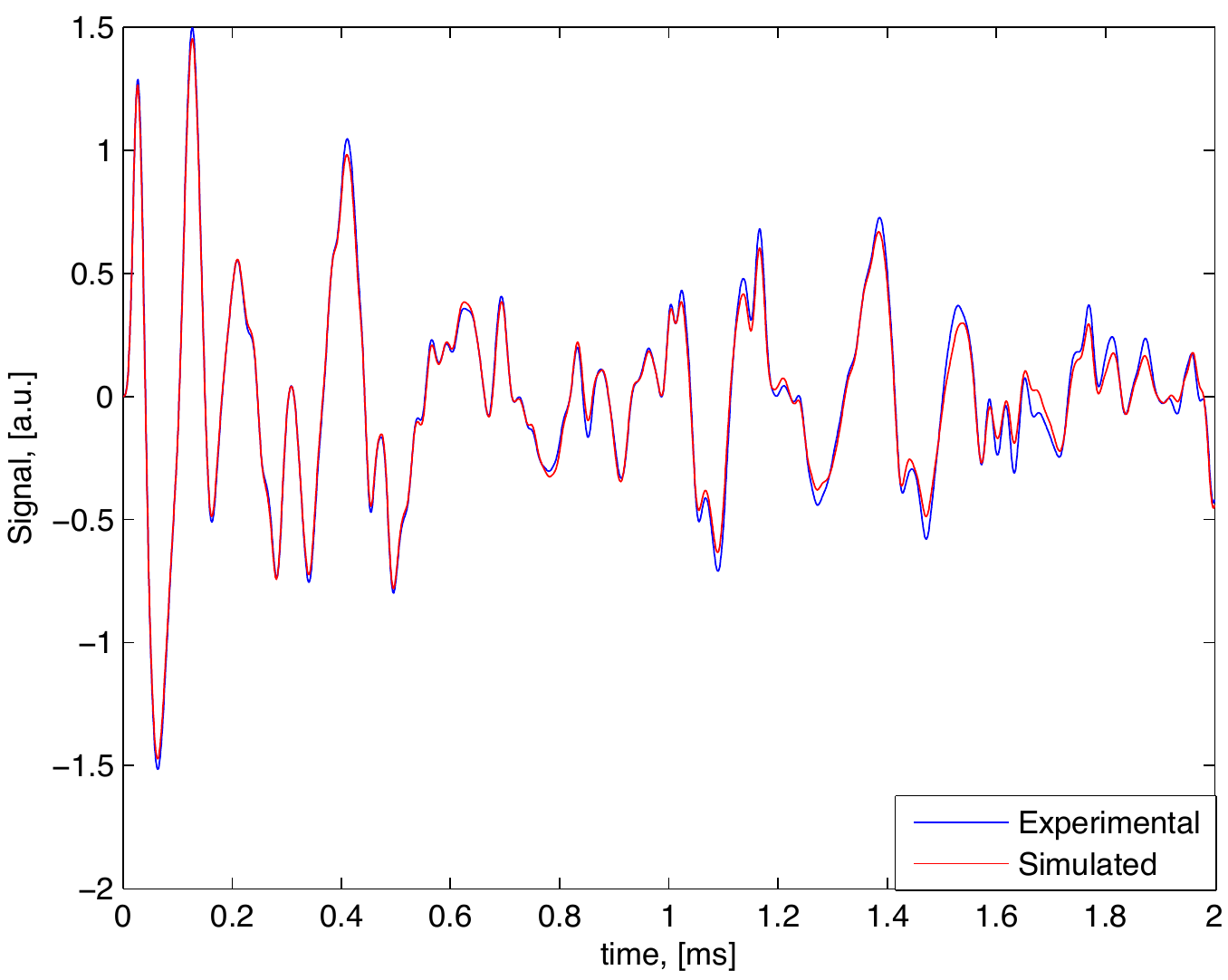}
\caption{Comparison between experimental measured signal (blue) and a simulated measurement record (red) produced by the initial state depicted in Fig. \ref{F:Initial44}. There is an excellent agreement between theory and experiment, which allows for successful quantum state tomography.}
\label{F:Signal44}
\end{center}
\begin{center}
\includegraphics[width=8.7cm,clip]{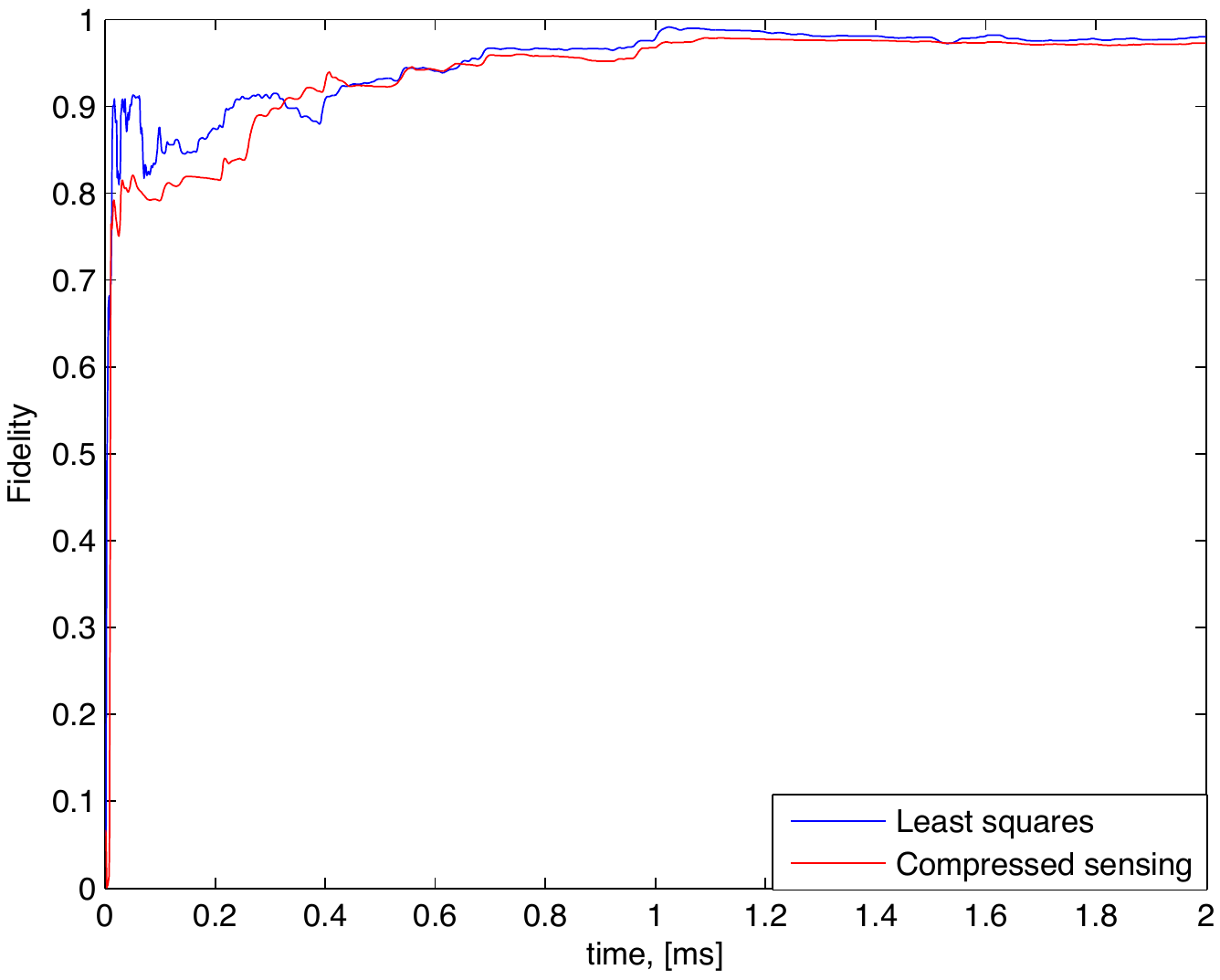}
\caption{Fidelity of reconstruction as a function of time for least squares (blue) and compressed sensing (red). Surprisingly, compressed sensing did not achieve a higher fidelity than least squares in this case.}
\label{F:Fidelity44}
\end{center}
\end{figure}

For our first example, we choose to prepare an eigenvector of $F_z$ as the initial state of the system. In particular, we choose $\ket{\Psi_0}=\ket{4,4}$ since it can be simply obtained through optical pumping with high fidelity. A bar plot of this state is shown in Fig. \ref{F:Initial44}. Fig. \ref{F:Signal44} shows a comparison between the experimental and simulated measurement record. Although the simulation of the measurement record produced by the target state is not necessary for tomography, we decided to show it to emphasize the extraordinary agreement between our model and the experiment. 

The data shown in Fig. \ref{F:Signal44} is then processed by our least squares and compressed sensing methods whose reconstruction fidelities as a function of time are shown in Fig. \ref{F:Fidelity44}. In this case, both methods achieve very high fidelities being least squares the one that obtains the highest, surprisingly. This is probably due to the fact that the error threshold $\epsilon$ used by compressed sensing is incorrect for this particular state. After 2 ms, we see that the compressed sensing fidelity of reconstruction is 0.9724, compared with 0.9799 for least squares. 

This particular state is relatively simple to reconstruct, which is seen in Fig. \ref{F:Fidelity44}. The fidelity jumps to a high fidelity estimate within the first microseconds of the experiment. This is due to the fact that there is only one state whose expectation value is the maximum projection along the $z$ direction (since we are measuring $F_z$). In theory, only one measurement would be necessary to identify this state. 

\begin{figure}[t]
\begin{center}
\includegraphics[width=15.1cm,clip]{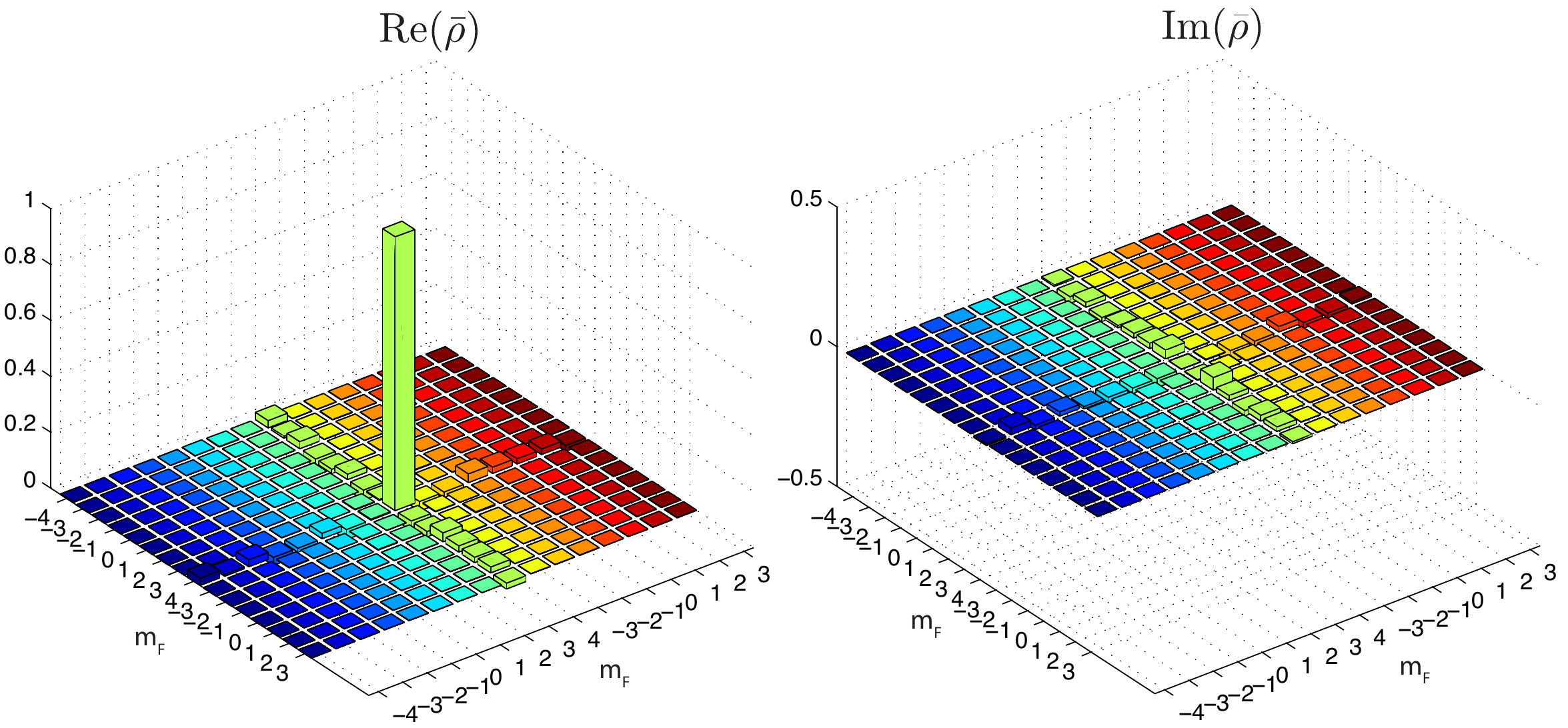}
\caption{Reconstructed quantum state using the least squares method after 2 ms. The fidelity of reconstruction is $\mathcal{F}(\rho_0,\bar{\rho})=0.9799$, and its purity $\mathcal{P}(\bar{\rho})=0.9893$, which is more mixed than expected given the quality of the state mapping used.}
\label{F:LSRho44}
\end{center}
\begin{center}
\includegraphics[width=15.1cm,clip]{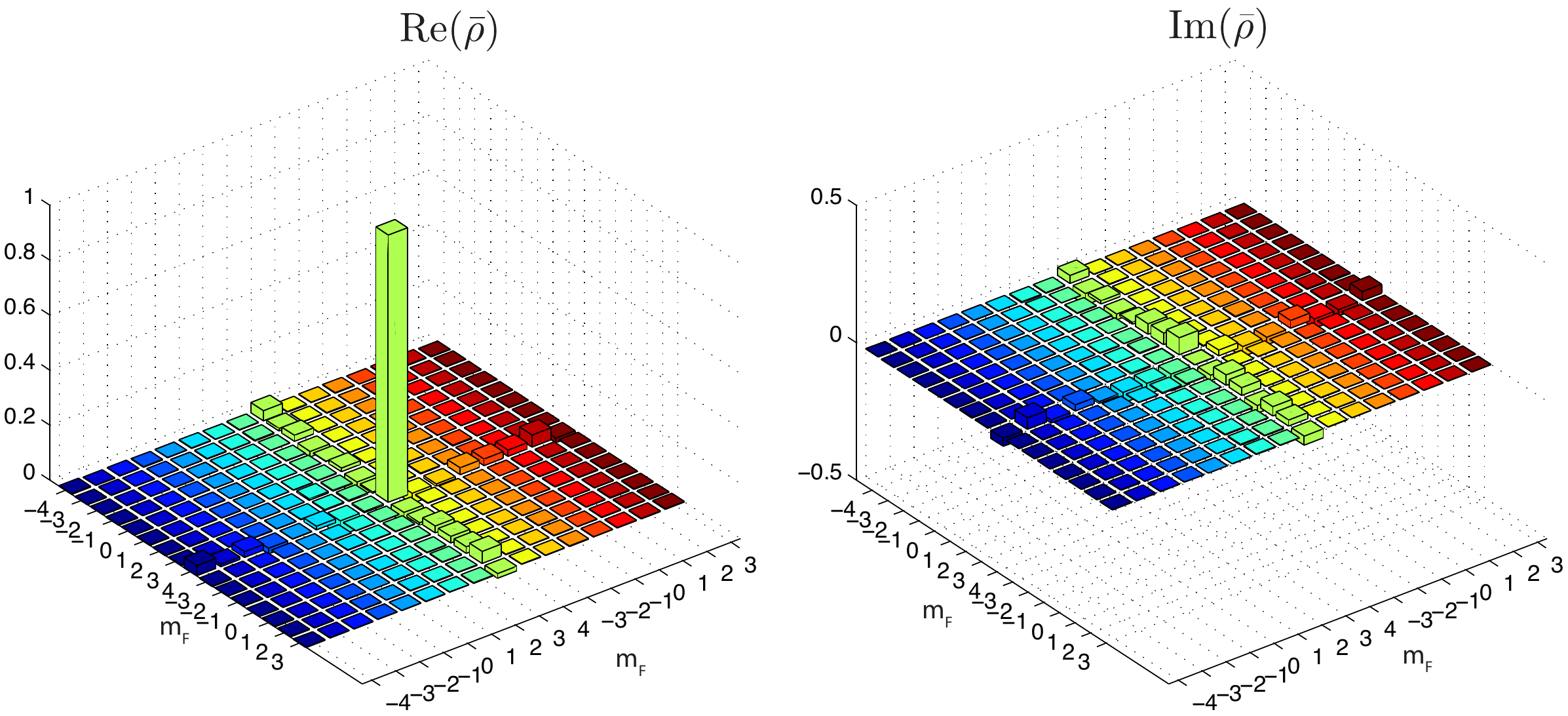}
\caption{Reconstructed quantum state using the compressed sensing method after 2 ms. The fidelity of reconstruction is $\mathcal{F}(\rho_0,\bar{\rho})=0.9724$, and its purity $\mathcal{P}(\bar{\rho})=0.9999$, which shows how compressed sensing tends to find the purest state compatible with the measurement record.}
\label{F:CSRho44}
\end{center}
\end{figure}

Figs. \ref{F:LSRho44} and \ref{F:CSRho44} show bar plots of the reconstructed density matrices for least squares and compressed sensing, respectively, after 2 ms of the experiment. As we saw in our simulations, least squares tends to find states that are more mixed than what is expected. In this case, the reconstructed purity $\mathcal{P}(\bar{\rho})=0.9893$ for least squares and $\mathcal{P}(\bar{\rho})=0.9999$ for compressed sensing, which has found a state with high purity as expected.

%

\clearpage
\subsection{Reconstruction of a \emph{cat state} in the $F_-=3$ manifold}
\begin{figure}[b]
\begin{center}
\includegraphics[width=15.1cm,clip]{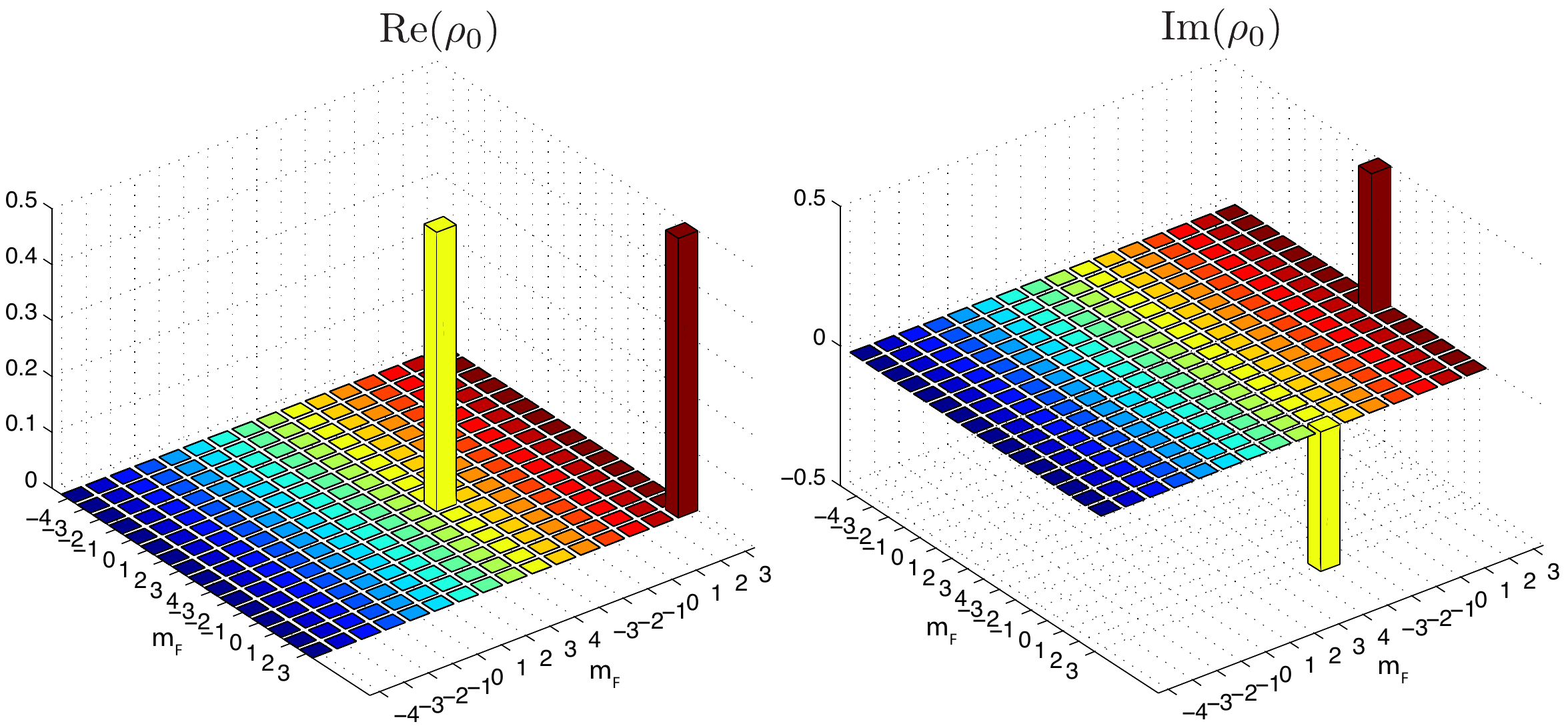}
\caption{Initial target state $\ket{\Psi_0}=\frac{1}{\sqrt{2}}(\ket{3,3}+i\ket{3,-3})$.}
\label{F:InitialCat3i}
\end{center}
\end{figure}

In our second example, we want to explore a particular superposition state. We are going to restrict ourselves to the $F_-=3$ manifold and choose to prepare the experiment in the target state $\ket{\Psi_0}=\frac{1}{\sqrt{2}}(\ket{3,3}+i\ket{3,-3})$, which has a relative phase and whose bar plot is shown in Fig. \ref{F:InitialCat3i}. In the quantum optics jargon, states of this form are referred as \emph{cat states}.  

\begin{figure}[t]
\begin{center}
\includegraphics[width=8.7cm,clip]{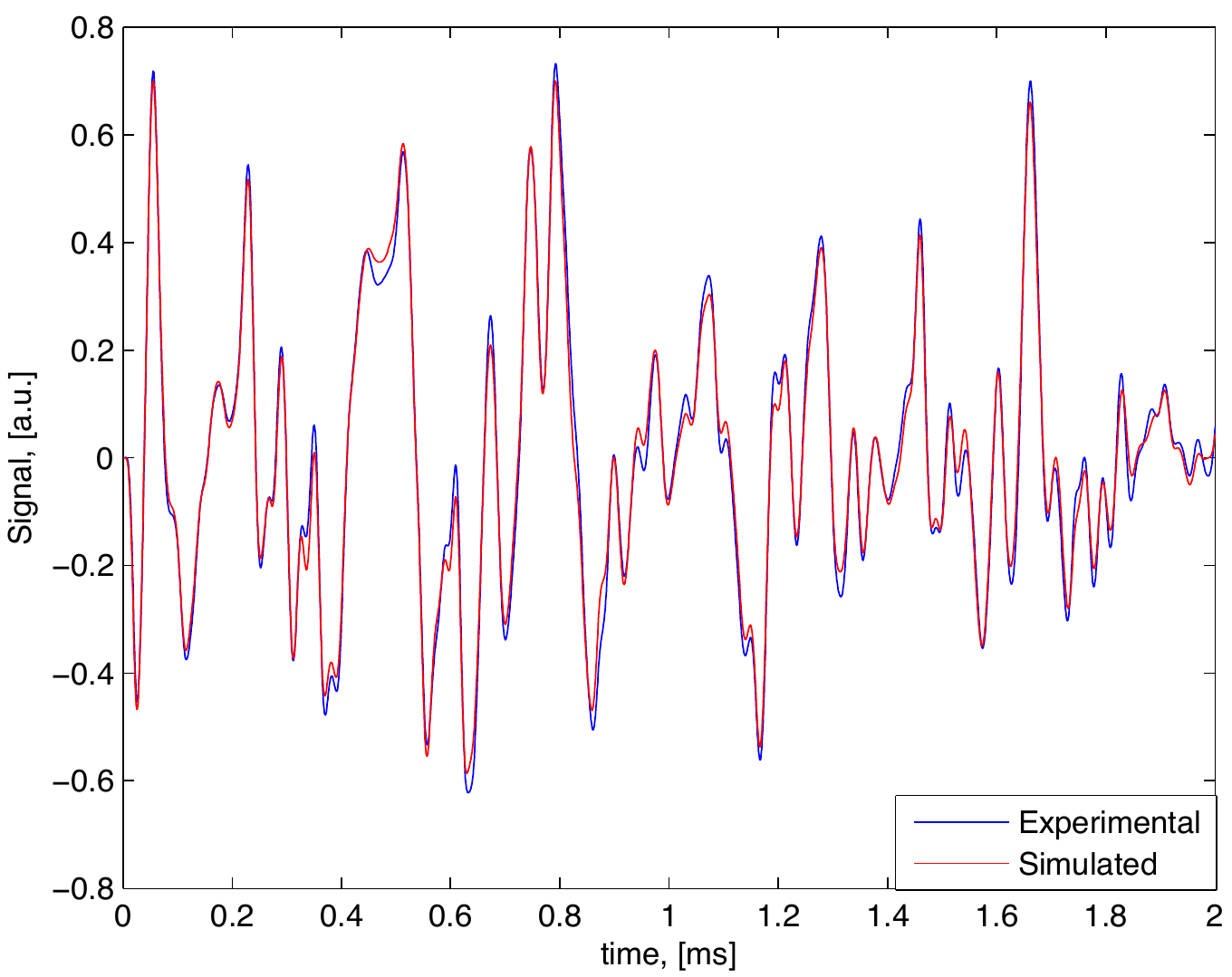}
\caption{Comparison between experimental measured signal (blue) and a simulated measurement record (red) produced by the initial state depicted in Fig. \ref{F:InitialCat3i}. There is an excellent agreement between theory and experiment, which allows for successful quantum state tomography.}
\label{F:SignalCat3i}
\end{center}
\begin{center}
\includegraphics[width=8.7cm,clip]{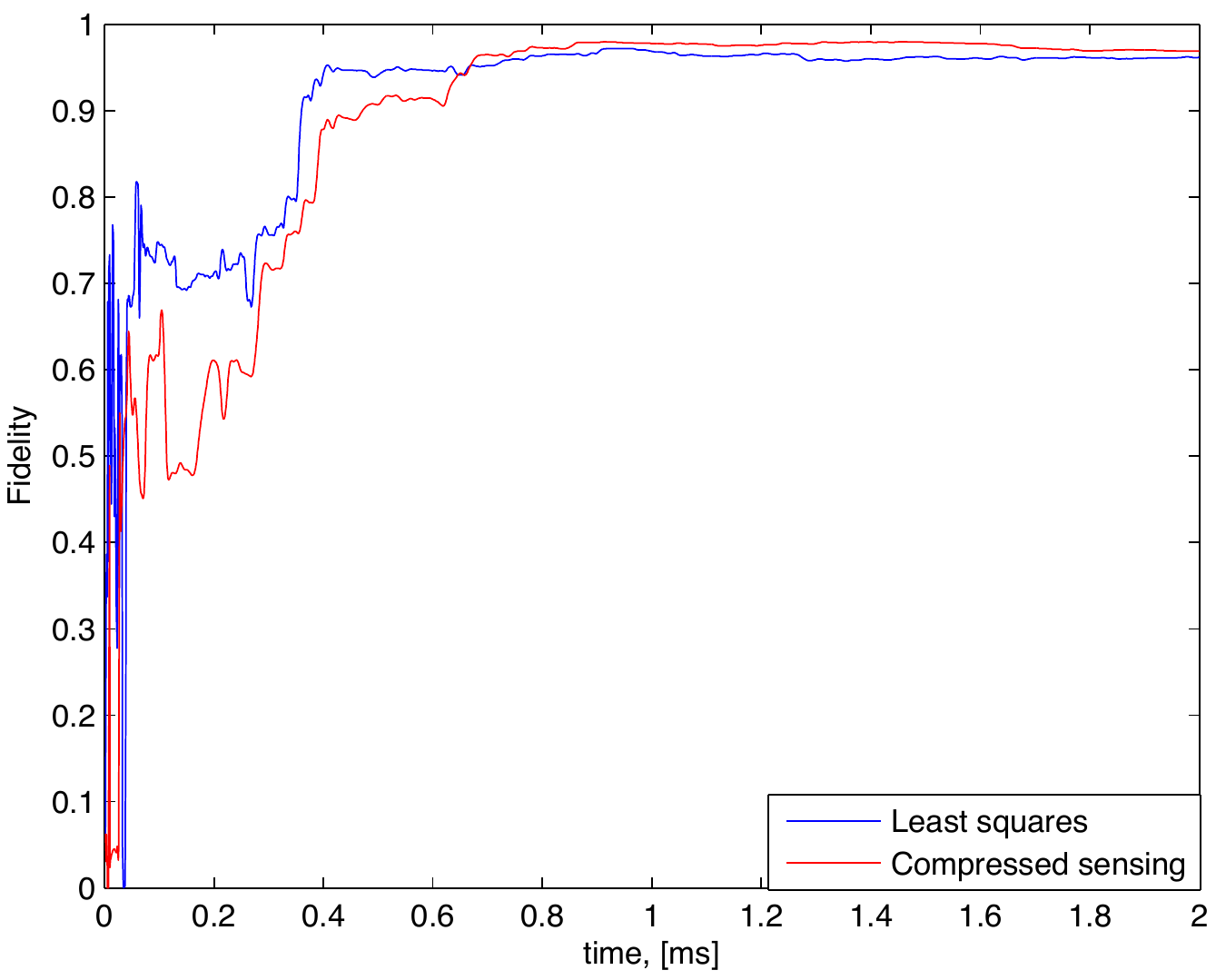}
\caption{Fidelity of reconstruction as a function of time for least squares (blue) and compressed sensing (red). Compressed sensing achieves higher fidelities than least squares.}
\label{F:FidelityCat3i}
\end{center}
\end{figure}

Fig. \ref{F:SignalCat3i} shows a comparison between the experimental and simulated measurement record. Again, we show the simulated signal for the target state to emphasize the agreement between our model and the data. As before, the data is processed by our least squares and compressed sensing methods whose reconstruction fidelities as a function of time are shown in Fig. \ref{F:FidelityCat3i}. The reconstruction fidelity for compressed sensing is slightly higher than for least squares as seen in our simulations. In fact, after 2 ms, we see that the fidelity of reconstruction is 0.9686, compared with 0.9615 for least squares. The overall lower fidelities observed so far in this chapter, in comparison to our simulations, are expected in practical situations in which we are probably limited by systematic and miscalibration errors and not shot noise as assumed in the model.

Judging from the rate of increase of the fidelity, Fig. (\ref{F:FidelityCat3i}), we see that this state is more difficult to reconstruct than the one shown in the previous section. We see, however, that our methods are still able to find excellent estimates after 1 ms. The difference in performance between least squares and compressed sensing is not so impressive in this example. We believe that this is due to the fact that the dynamics is modeled very well in this case and in the limit of low errors, we would expect both methods to perform roughly the same.

\begin{figure}[t]
\begin{center}
\includegraphics[width=15.1cm,clip]{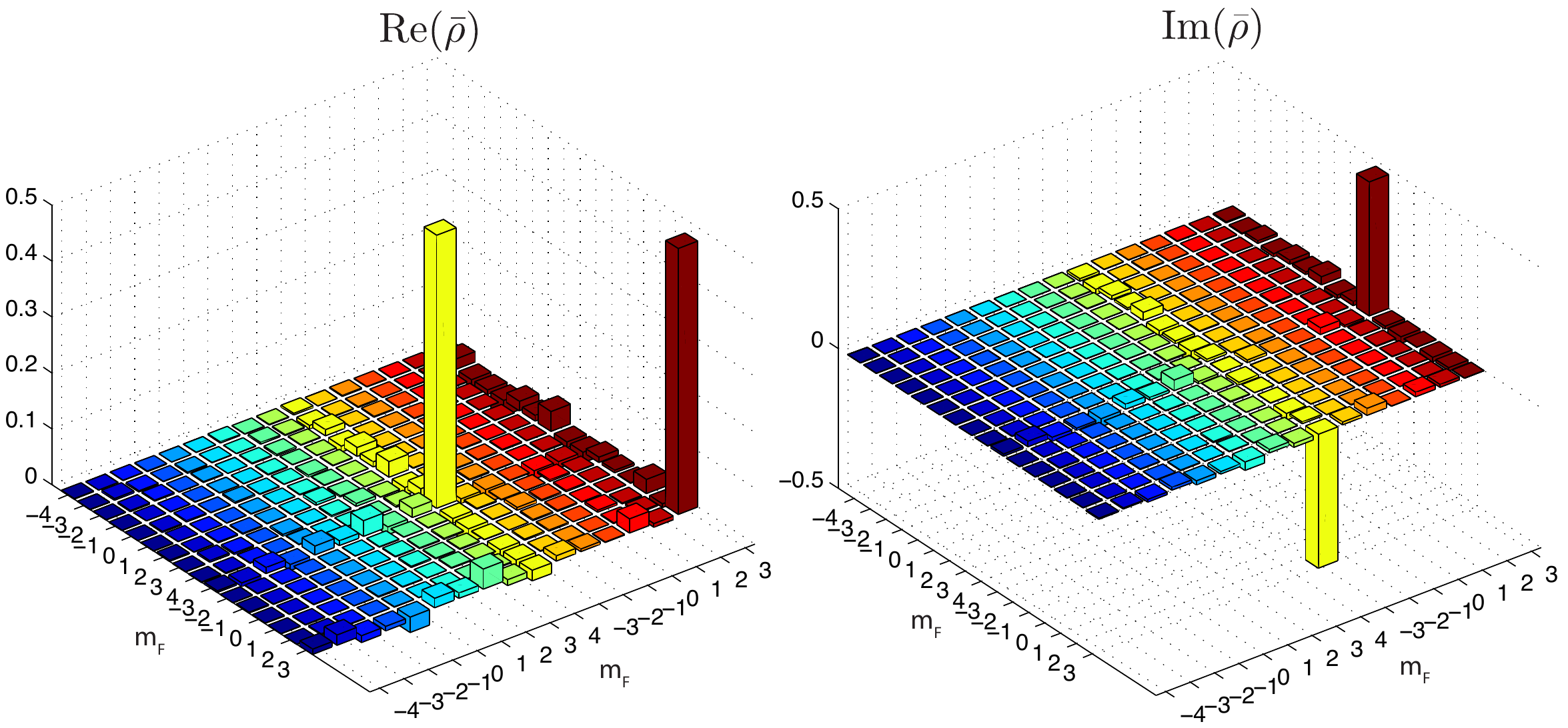}
\caption{Reconstructed quantum state using the least squares method after 2 ms. The fidelity of reconstruction is $\mathcal{F}(\rho_0,\bar{\rho})=0.9615$, and its purity $\mathcal{P}(\bar{\rho})=0.9533$, which is more mixed than expected given the quality of the state mapping used.}
\label{F:LSRhoCat3i}
\end{center}
\end{figure}

\begin{figure}[t]
\begin{center}
\includegraphics[width=15.1cm,clip]{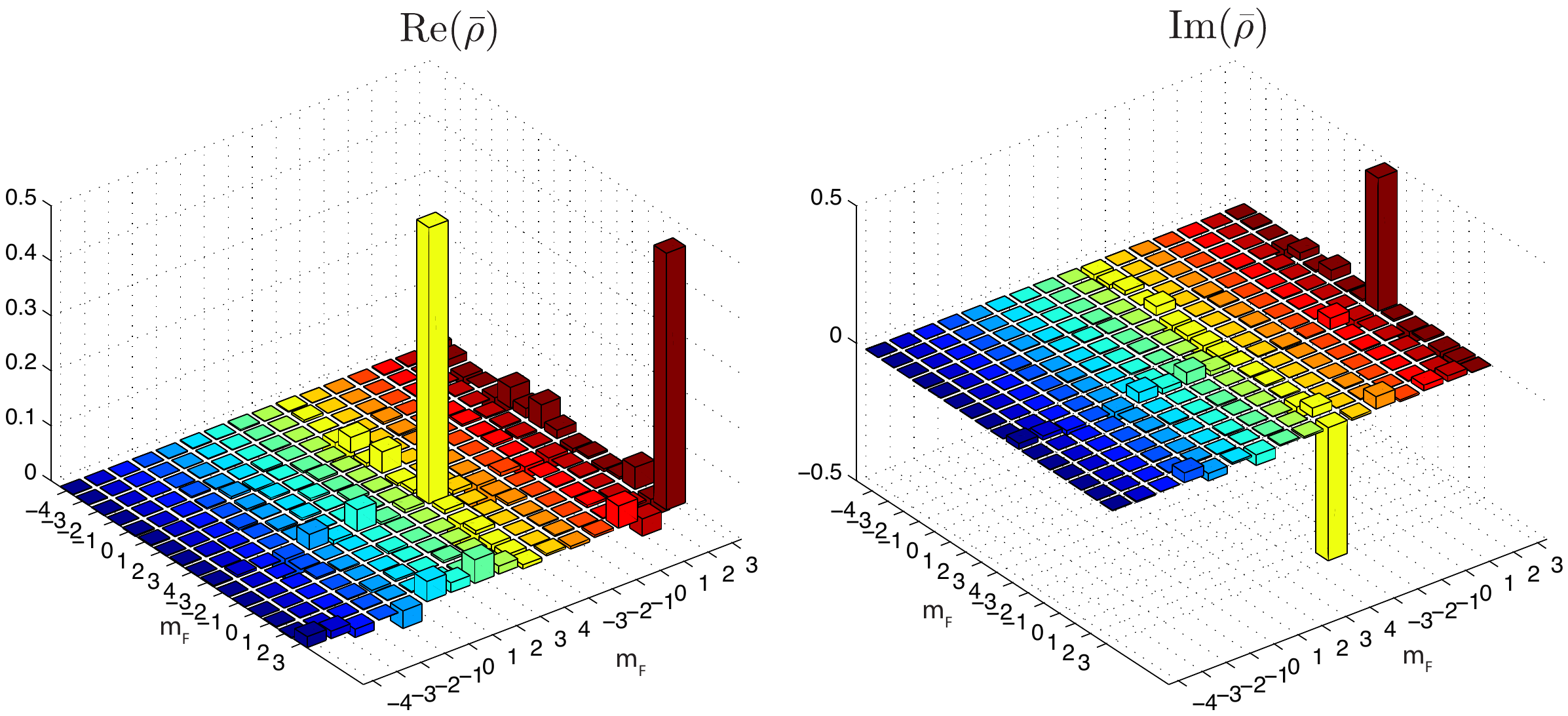}
\caption{Reconstructed quantum state using the compressed sensing method after 2 ms. The fidelity of reconstruction is $\mathcal{F}(\rho_0,\bar{\rho})=0.9686$, and its purity $\mathcal{P}(\bar{\rho})=0.9999$, which shows how CS tends to find the purest state compatible with the measurement record.}
\label{F:CSRhoCat3i}
\end{center}
\end{figure}

Figs. \ref{F:LSRhoCat3i} and \ref{F:CSRhoCat3i} show bar plots of the reconstructed density matrices for least squares and compressed sensing, respectively, after 2 ms of the experiment. As we saw in our simulations, least squares tends to find states that are more mixed than what is expected. In this case, the reconstructed purity $\mathcal{P}(\bar{\rho})=0.9533$ for least squares and $\mathcal{P}(\bar{\rho})=0.9999$ for compressed sensing.

%

\clearpage
\subsection{Reconstruction of a superposition state in $F_-=3$ and $F_+=4$ manifolds}\label{sec:ExampleSup1}
\begin{figure}[b]
\begin{center}
\includegraphics[width=15.1cm,clip]{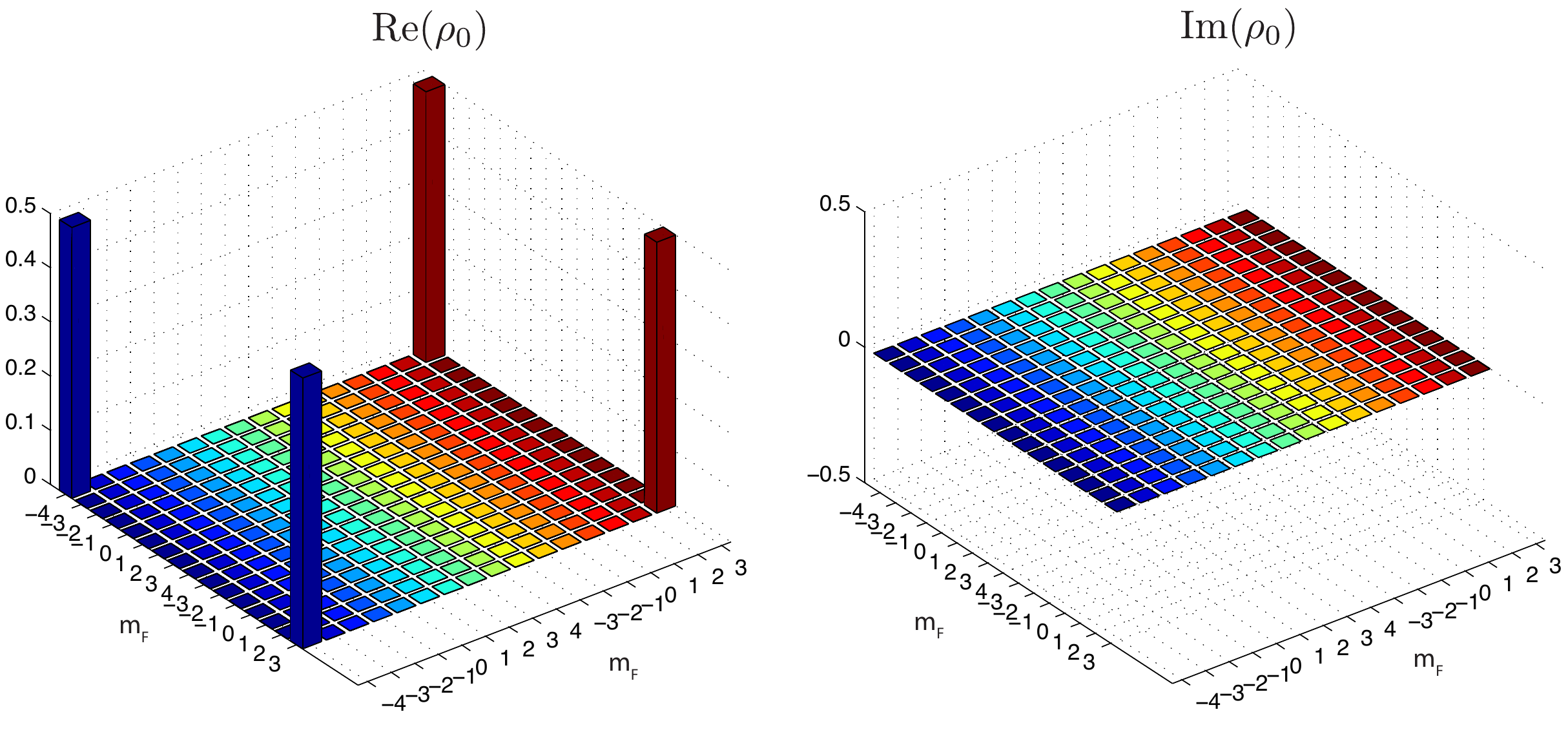}
\caption{Initial target state $\ket{\Psi_0}=\frac{1}{\sqrt{2}}(\ket{4,-4}+\ket{3,3})$. }
\label{F:Initialsup1}
\end{center}
\end{figure}

We have restricted the target states to single manifolds so far. For our next example, we decide to explore more challenging situations in which the state has support in both manifolds $F_-=3$ and $F_+=4$. The inhomogeneous light-shift makes reconstructing such states more complicated. Since the light-shift Eq. (\ref{eq:effhamiltonianls}) is different for different manifolds, its inhomogeneity will lead to a spread of different local Hamiltonians acting on the different atoms. This fact will result in the decrease of the coherences between the two manifolds, averaged across the ensemble, if not properly modeled. We will see this effect in this section by artificially excluding the inhomogeneity in the light-shift from the model and compare the performance of our methods with the case in which we model it properly. For this purpose, we choose our target state to be $\ket{\Psi_0}=\frac{1}{\sqrt{2}}(\ket{4,-4}+\ket{3,3})$, whose bar plot is shown in Fig. \ref{F:Initialsup1}. This state has support in both manifolds and thus is sensitive to inhomogeneity in the light-shift.

\subsubsection{A. Complete model: bias and light-shift inhomogeneities included}
\begin{figure}[t]
\begin{center}
\includegraphics[width=8.7cm,clip]{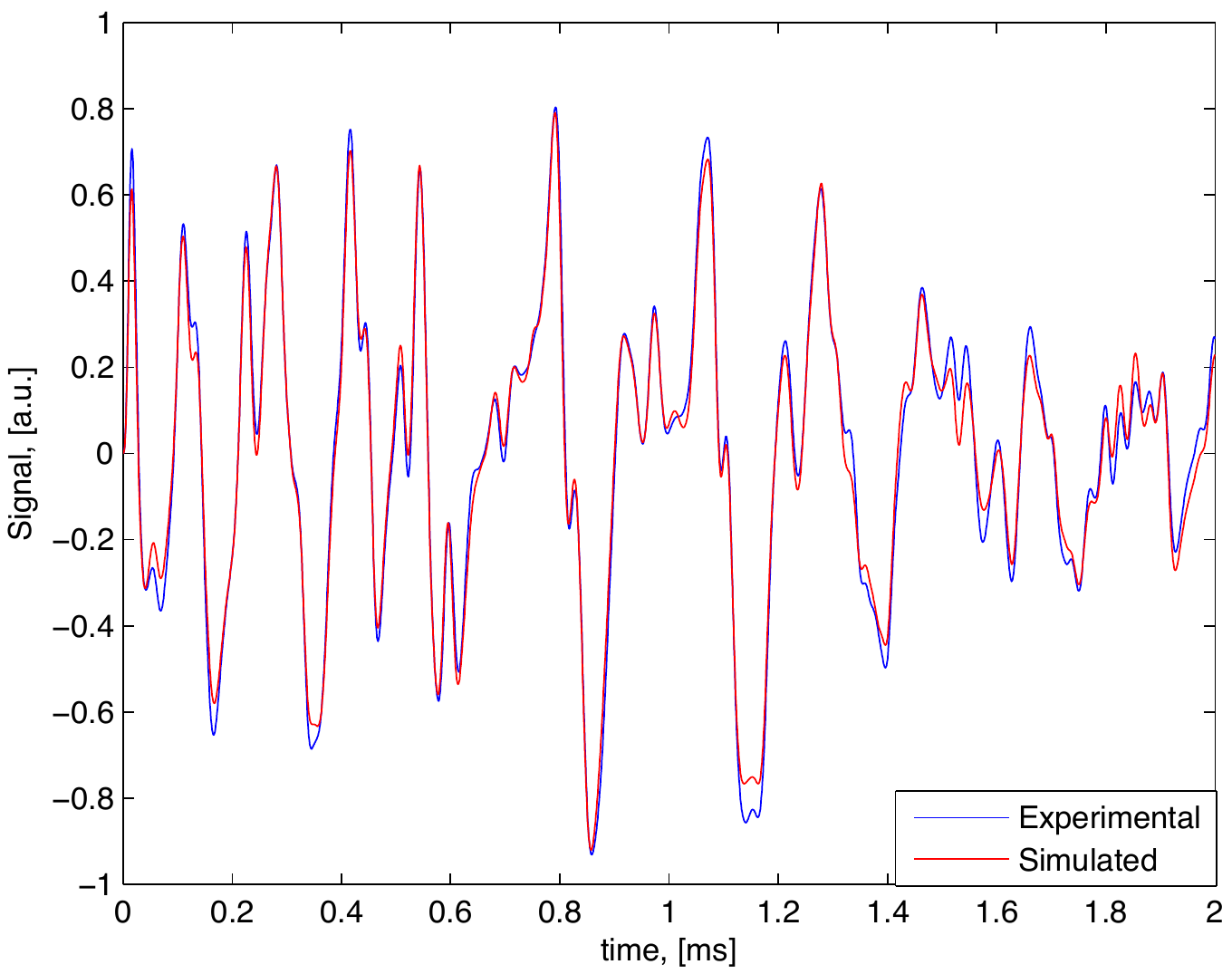}
\caption{Comparison between experimental measured signal (blue) and a simulated measurement record (red) produced by the initial state depicted in Fig. \ref{F:Initialsup1}. There is an excellent agreement between theory and experiment, which allows for successful quantum state tomography.}
\label{F:Signalsup1}
\end{center}
\begin{center}
\includegraphics[width=8.7cm,clip]{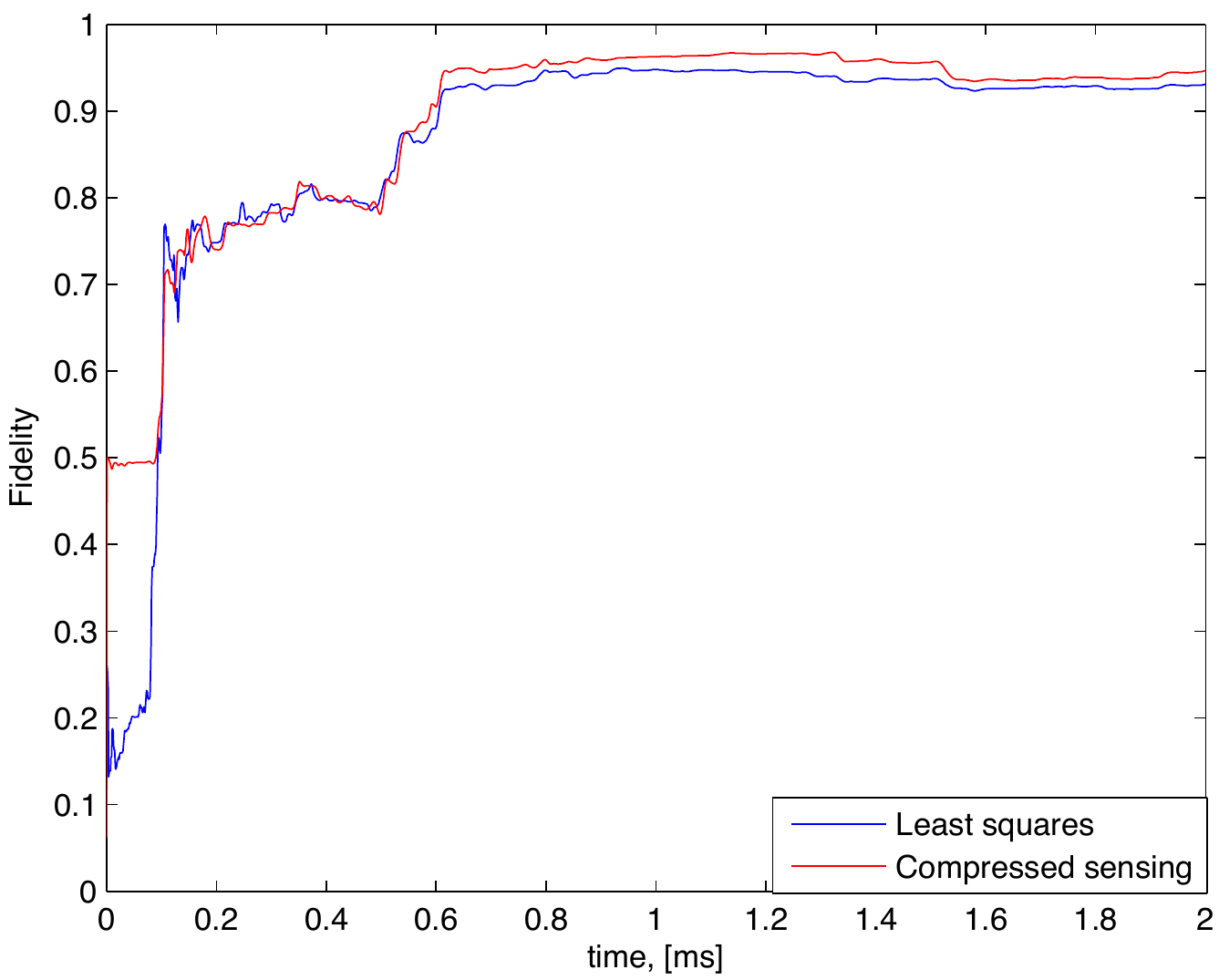}
\caption{Fidelity of reconstruction as a function of time for least squares (blue) and compressed sensing (red). Compressed sensing achieves higher fidelities than least squares.}
\label{F:Fidelitysup1}
\end{center}
\end{figure}

With all inhomogeneous parameters in place, we proceed to analyze the data. Fig. \ref{F:Signalsup1} shows a comparison between the experimental and simulated measurement record for this target state, in which we see an excellent agreement between theory and experiment. As in all previous examples, the data is then processed by our least squares and compressed sensing methods whose reconstruction performances as functions of time are shown in Fig. \ref{F:Fidelitysup1}.

%

\begin{figure}[t]
\begin{center}
\includegraphics[width=15.1cm,clip]{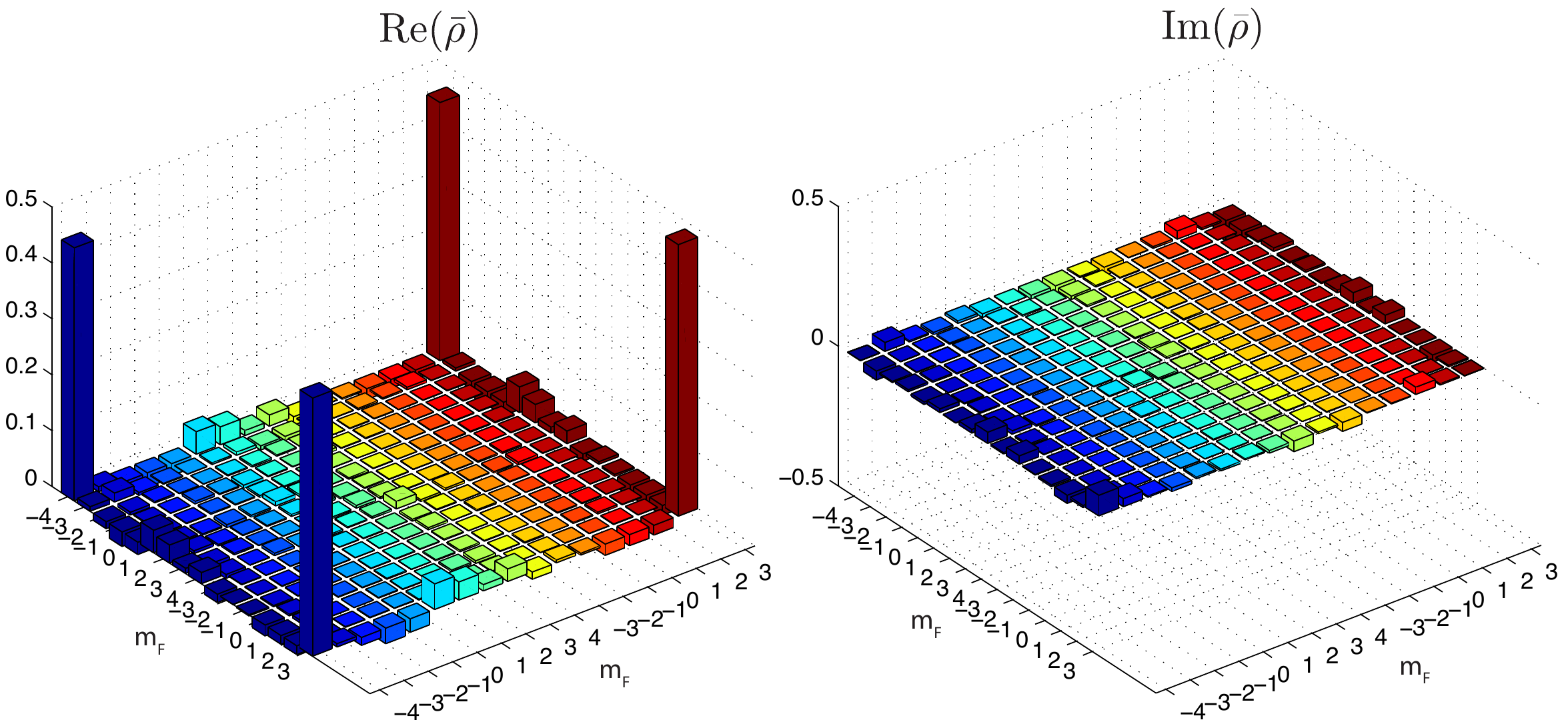}
\caption{Reconstructed quantum state using the least squares method after 2 ms. The fidelity of reconstruction is $\mathcal{F}(\rho_0,\bar{\rho})=0.9308$, and its purity $\mathcal{P}(\bar{\rho})=0.9229$, which is more mixed than expected given the quality of the state mapping used.}
\label{F:LSRhosup1}
\end{center}
\end{figure}

\begin{figure}[t]
\begin{center}
\includegraphics[width=15.1cm,clip]{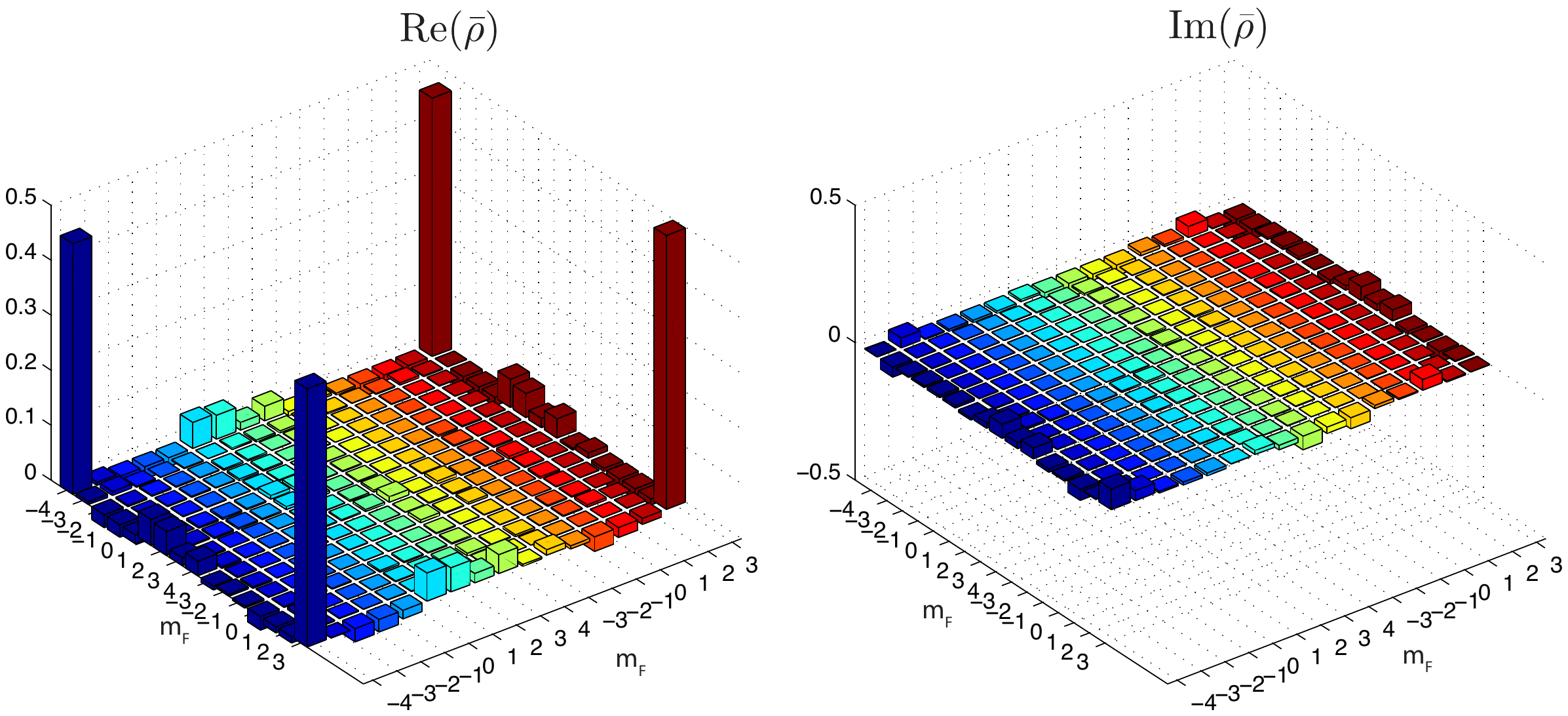}
\caption{Reconstructed quantum state using the compressed sensing method after 2 ms. The fidelity of reconstruction is $\mathcal{F}(\rho_0,\bar{\rho})=0.9467$, and its purity $\mathcal{P}(\bar{\rho})=0.9746$, which shows how compressed sensing tends to find the purest state compatible with the measurement record.}
\label{F:CSRhosup1}
\end{center}
\end{figure}

The reconstruction fidelity for compressed sensing is slightly higher than that of least squares. After 2 ms, we see a compressed sensing fidelity of reconstruction of 0.9467, compared with 0.9308 for least squares. Figs. \ref{F:LSRhosup1} and \ref{F:CSRhosup1} show bar plots of the reconstructed density matrices for least squares and compressed sensing, respectively. As we saw in our simulations, least squares tends to find states that are more mixed than what is expected. In this case, the reconstructed purity $\mathcal{P}(\bar{\rho})=0.9229$ for least squares and $\mathcal{P}(\bar{\rho})=0.9746$ for compressed sensing, which has found a state with higher purity as expected.

\clearpage
\subsubsection{B. Incomplete model: only inhomogeneous bias field included}
To illustrate the robustness of compressed sensing methods to certain types of errors, we decide not to include in the model the effect of the light-shift inhomogeneity and carry out the reconstruction process. We, however, include the effects of inhomogeneity in the bias field power, since our system is extremely sensitive to it.

\begin{figure}[t]
\begin{center}
\includegraphics[width=8.7cm,clip]{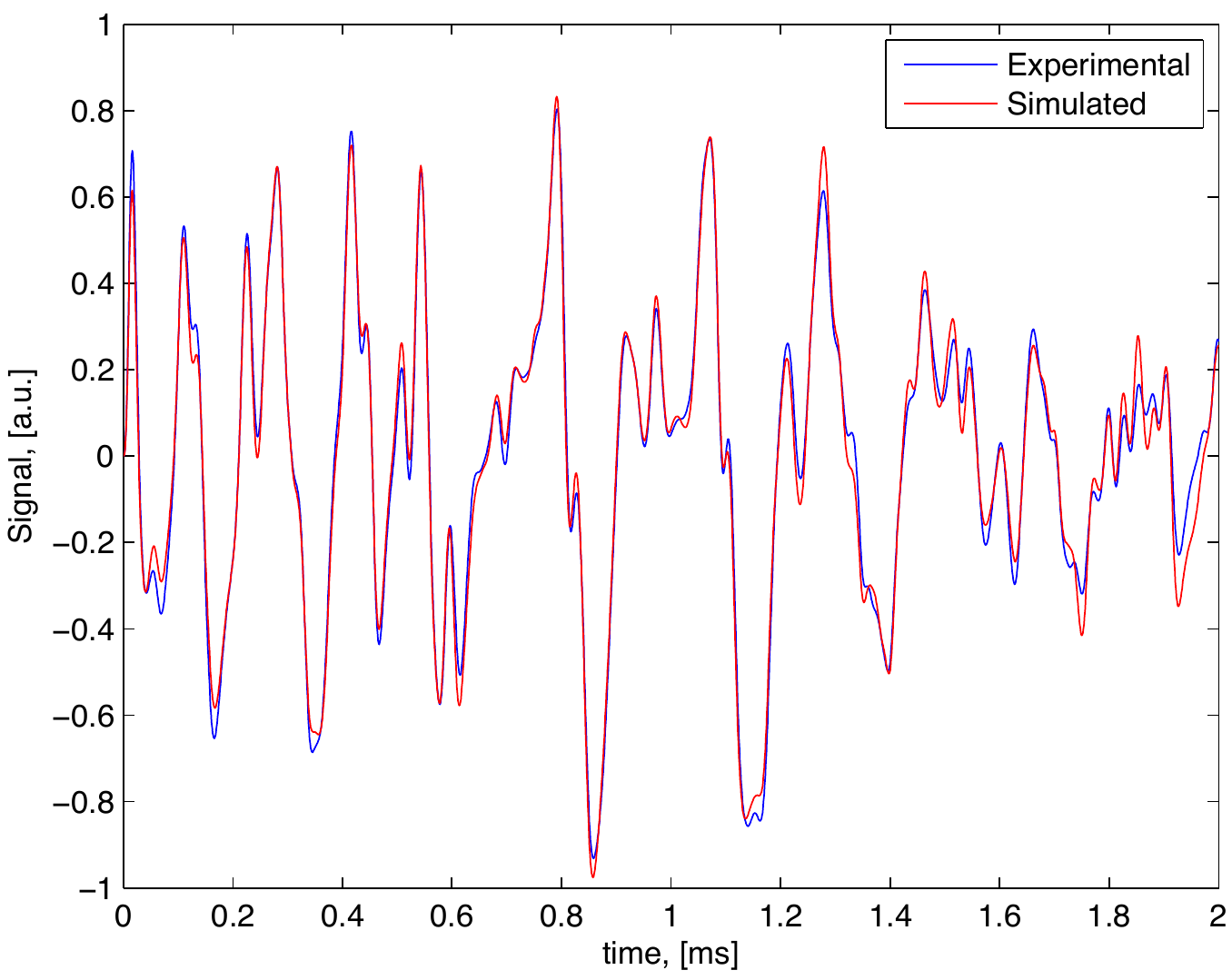}
\caption{Comparison between experimental measured signal (blue) and a simulated measurement record (red) produced by the initial state depicted in Fig. \ref{F:Initialsup1} without including in the model the effect of the light-shift inhomogeneity.}
\label{F:Signalsup1BiasOnly}
\end{center}
\begin{center}
\includegraphics[width=8.7cm,clip]{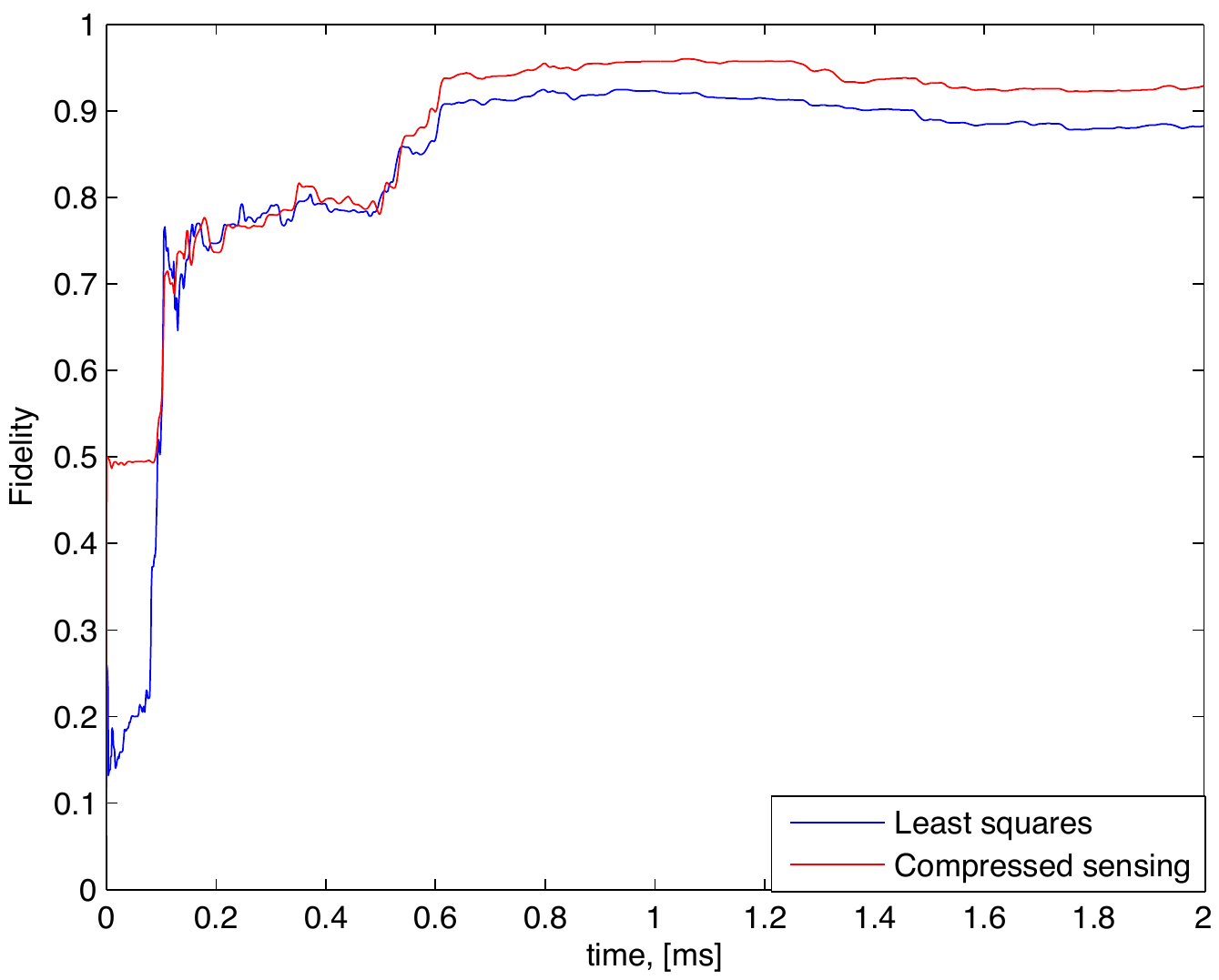}
\caption{Fidelity of reconstruction as a function of time for least squares (blue) and compressed sensing (red). In this example, compressed sensing achieves much higher fidelities than least squares after 2 ms indicating robustness to improper modeling.}
\label{F:Fidelitysup1BiasOnly}
\end{center}
\end{figure}

Fig. \ref{F:Signalsup1BiasOnly} shows a comparison between the experimental measurement record and the simulated one for this target state. Although not strikingly different from the full model prediction, Fig. \ref{F:Signalsup1}, the two simulated signals differ by about $5\%$ with respect to the actual experimental signal, i.e., their least squares residues are $5\%$ different, being the full model simulated signal the closest to the data. Given this difference, it is not surprising that the overall performance of the QT decreases. In fact, as seen in Fig. \ref{F:Fidelitysup1BiasOnly}, which shows the fidelity of reconstruction as a function of time for least squares and compressed sensing, we obtain a compressed sensing fidelity of 0.9287 and a lower least squares fidelity of 0.8825. Clearly in this example, compressed sensing appears to be more robust to this particular error in the model, which is an important feature when some uncertainties are difficult to determine or model.

\begin{figure}[t]
\begin{center}
\includegraphics[width=15.1cm,clip]{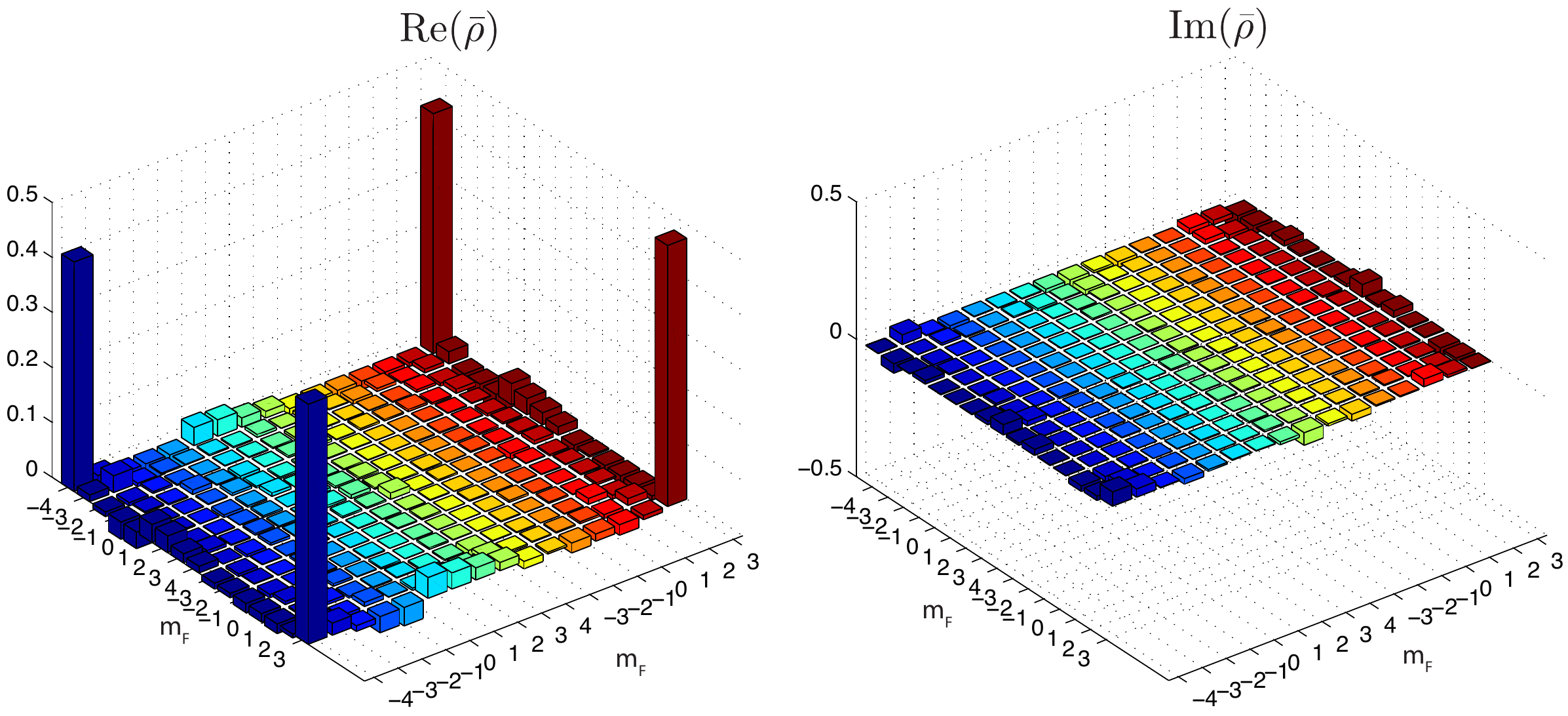}
\caption{Reconstructed quantum state using the least squares method after 2 ms. The fidelity of reconstruction is $\mathcal{F}(\rho_0,\bar{\rho})=0.8825$, and its purity $\mathcal{P}(\bar{\rho})=0.8342$.}
\label{F:LSRhosup1BiasOnly}
\end{center}
\end{figure}

\begin{figure}[t]
\begin{center}
\includegraphics[width=15.1cm,clip]{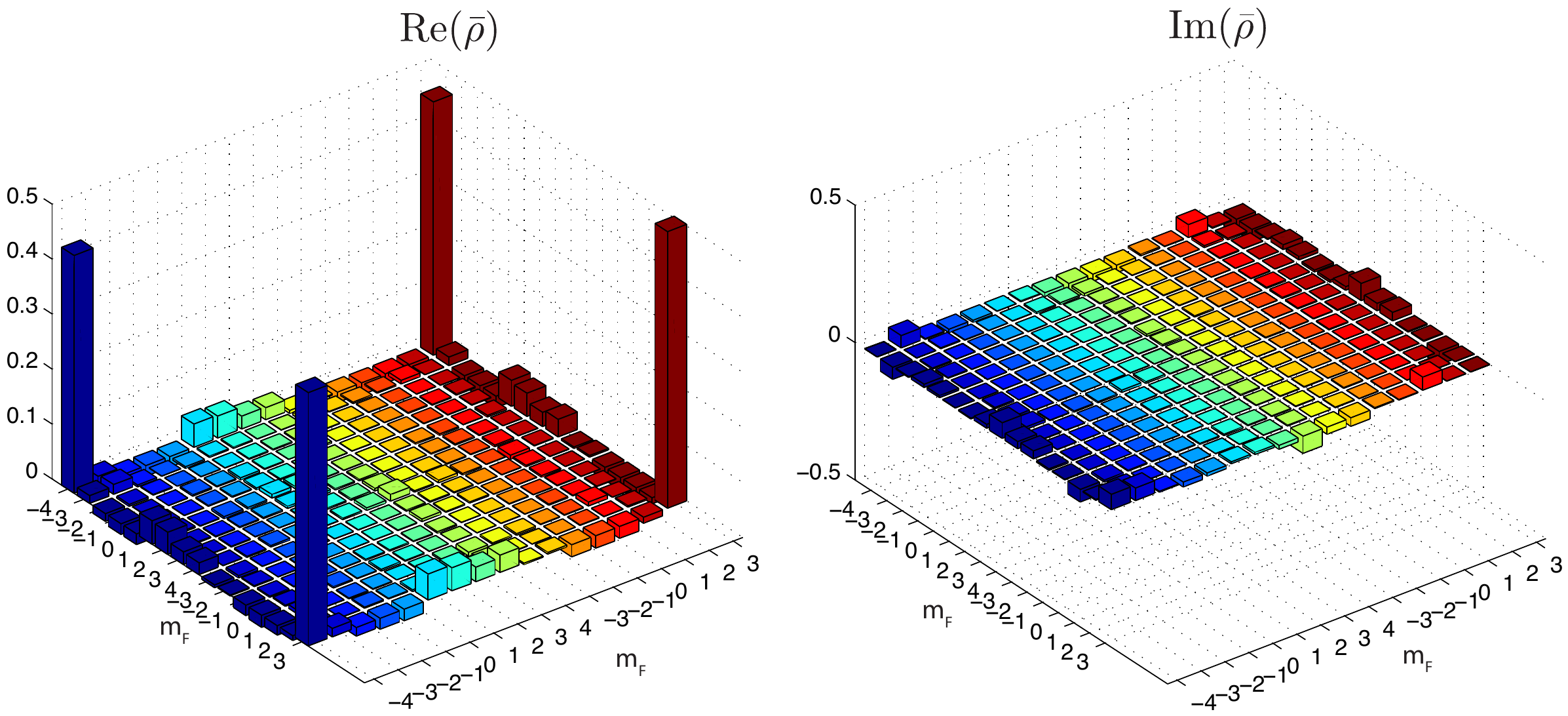}
\caption{Reconstructed quantum state using the compressed sensing method after 2 ms. The fidelity of reconstruction is $\mathcal{F}(\rho_0,\bar{\rho})=0.9287$, and its purity $\mathcal{P}(\bar{\rho})=0.9444$.}
\label{F:CSRhosup1BiasOnly}
\end{center}
\end{figure}

Finally, for completeness, we show the bar plots of the reconstructed quantum state using least squares Fig. \ref{F:LSRhosup1BiasOnly} and compressed sensing \ref{F:CSRhosup1BiasOnly}.

\clearpage
\subsection{Reconstruction of a random pure state in the $F_-=3$ and $F_+=4$ manifolds}
\begin{figure}[b]
\begin{center}
\includegraphics[width=15.1cm,clip]{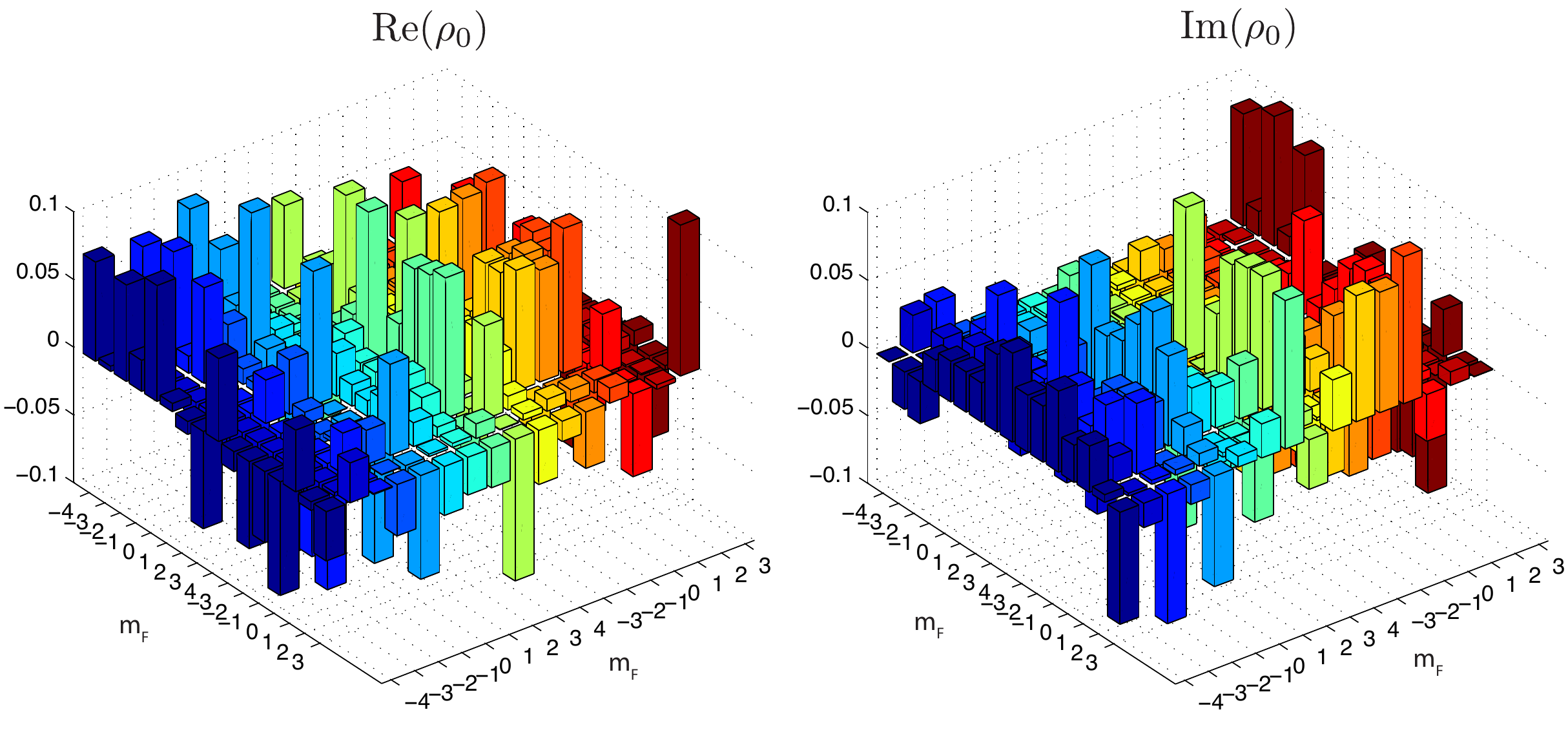}
\caption{The initial target state is a complicated random pure state sampled from the Haar measure. It has support in all the 16 Zeeman sub-levels  of the system.}
\label{F:InitialRand23}
\end{center}
\end{figure}

For our last example, we choose an arbitrary pure state sampled from the Haar measure, $\ket{\Psi_0}$. This state has support in all the 16 Zeeman sub-levels as we see in the bar plot shown in Fig. \ref{F:InitialRand23}.

\begin{figure}[t]
\begin{center}
\includegraphics[width=8.7cm,clip]{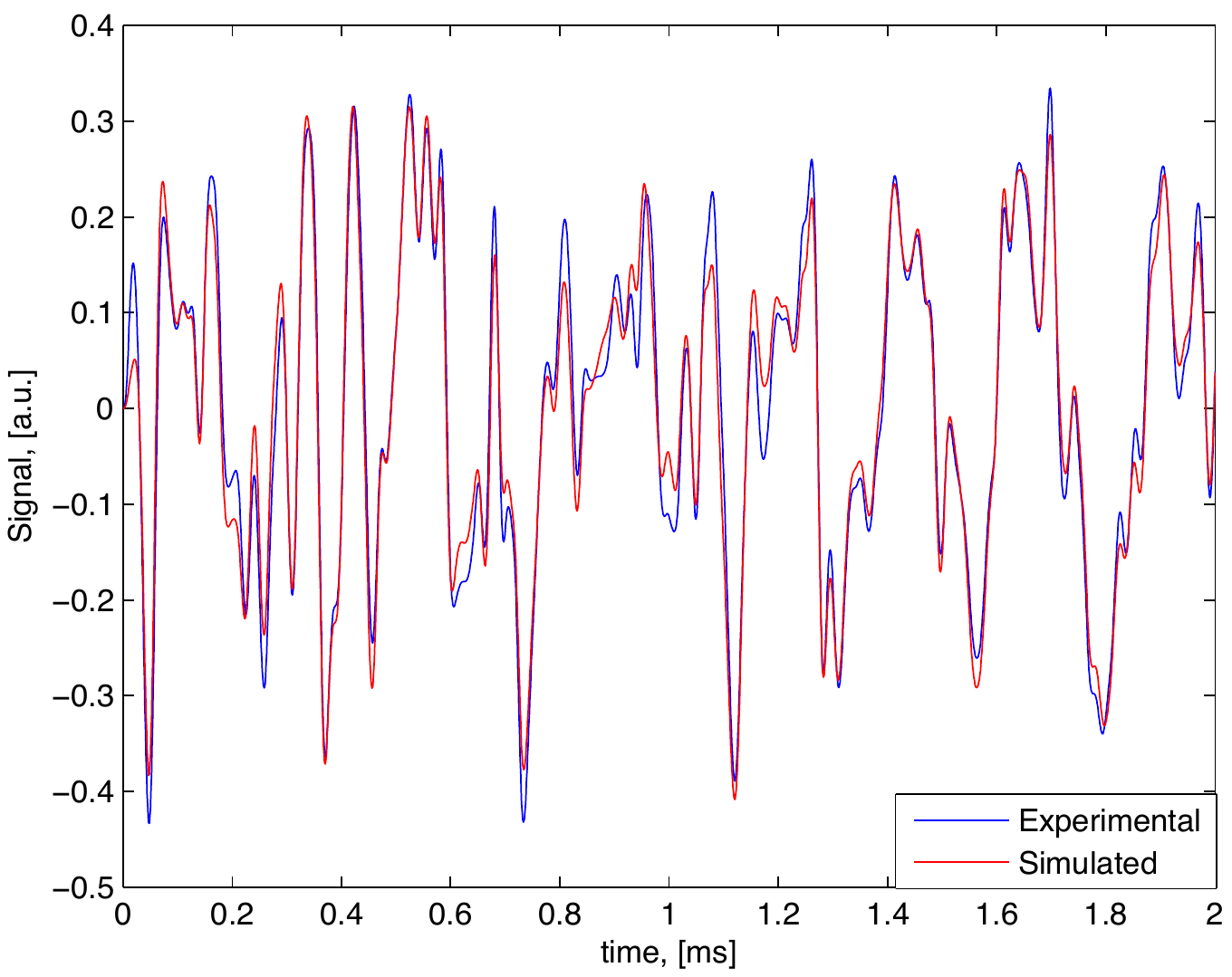}
\caption{Comparison between experimental measured signal (blue) and a simulated measurement record (red) produced by the initial state depicted in Fig. \ref{F:InitialRand23}. Although not as impressive as previous examples, there is still a very good agreement between theory and experiment, which allows for successful quantum state tomography.}
\label{F:SignalRand23}
\end{center}
\begin{center}
\includegraphics[width=8.7cm,clip]{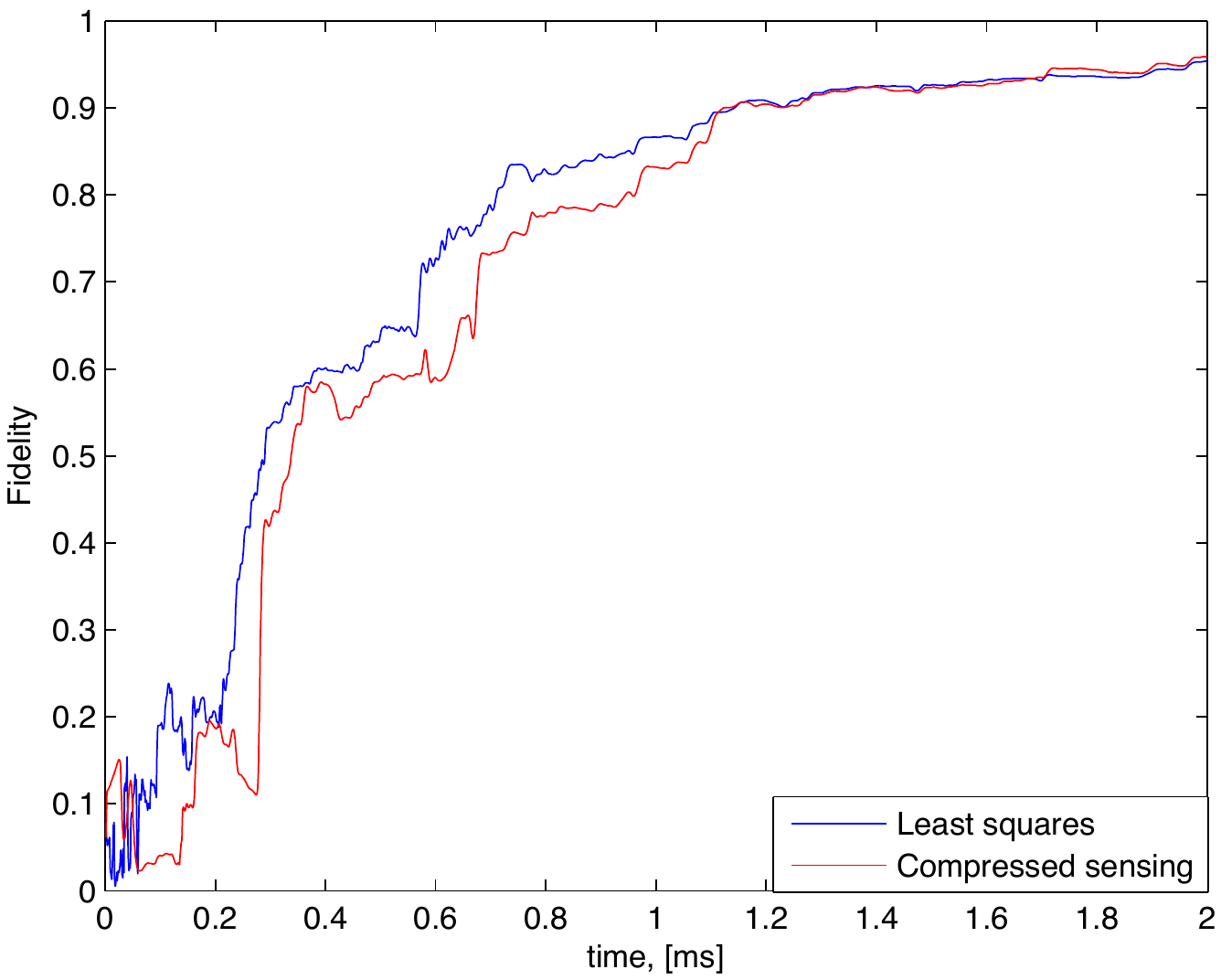}
\caption{Fidelity of reconstruction as a function of time for least squares (blue) and compressed sensing (red). In this example, compressed sensing and least squares achieve roughly the same fidelity after 2 ms.}
\label{F:FidelityRand23}
\end{center}
\end{figure}

Fig. \ref{F:SignalRand23} shows a comparison between the experimental and simulated measurement record. In general, for more complicated states like this the experimental data differs a little more from the theoretical prediction, however, we still see a very good agreement. Moreover, for states like this, the signal-to-noise ratio is generally worse than that of more regular states because the population in each sub-level is smaller and thus produces less signal, and about half of the population is in $F_+=4$ where the signal is a factor of $\sim$17 smaller. As before, we process the data with our least squares and compressed sensing protocols and show their fidelity of reconstruction as a function of time in Fig. \ref{F:FidelityRand23}. 

In this case, the levels of performance of least squares and compressed sensing are very comparable. In fact, after 2 ms, we see that the compressed sensing fidelity of reconstruction is 0.9593, compared to 0.9538 for least squares. As we see in the rate of increase of the fidelity, this state is much harder to reconstruct than previous examples. In fact, more information is needed to recover all elements of the density matrix since the state has support in all standard basis directions.

%

\begin{figure}[t]
\begin{center}
\includegraphics[width=15.1cm,clip]{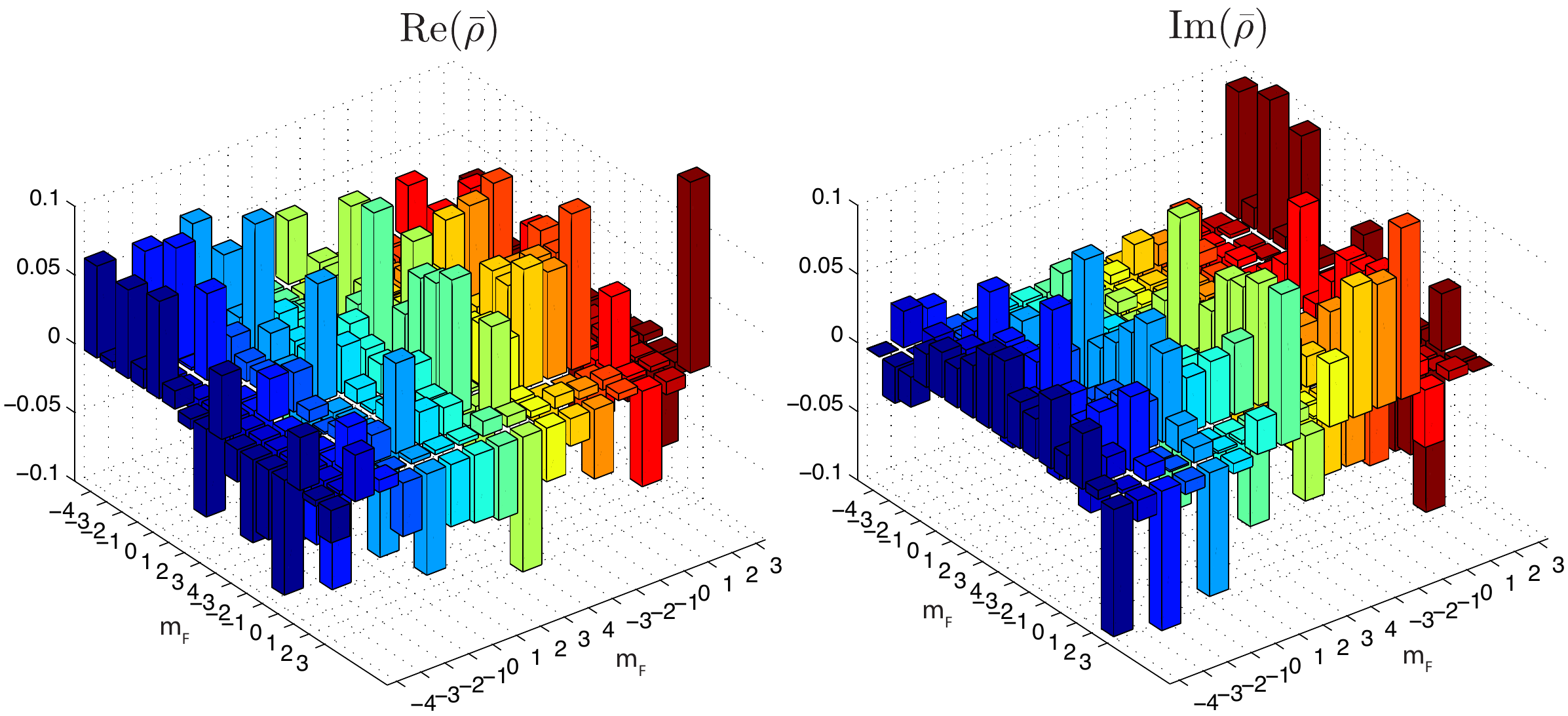}
\caption{Reconstructed quantum state using the least squares method after 2 ms. The fidelity of reconstruction is $\mathcal{F}(\rho_0,\bar{\rho})=0.9538$, and its purity $\mathcal{P}(\bar{\rho})=0.9345$, which is more mixed than expected given the quality of the state mapping used.}
\label{F:LSRhoRand23}
\end{center}
\end{figure}

\begin{figure}[t]
\begin{center}
\includegraphics[width=15.1cm,clip]{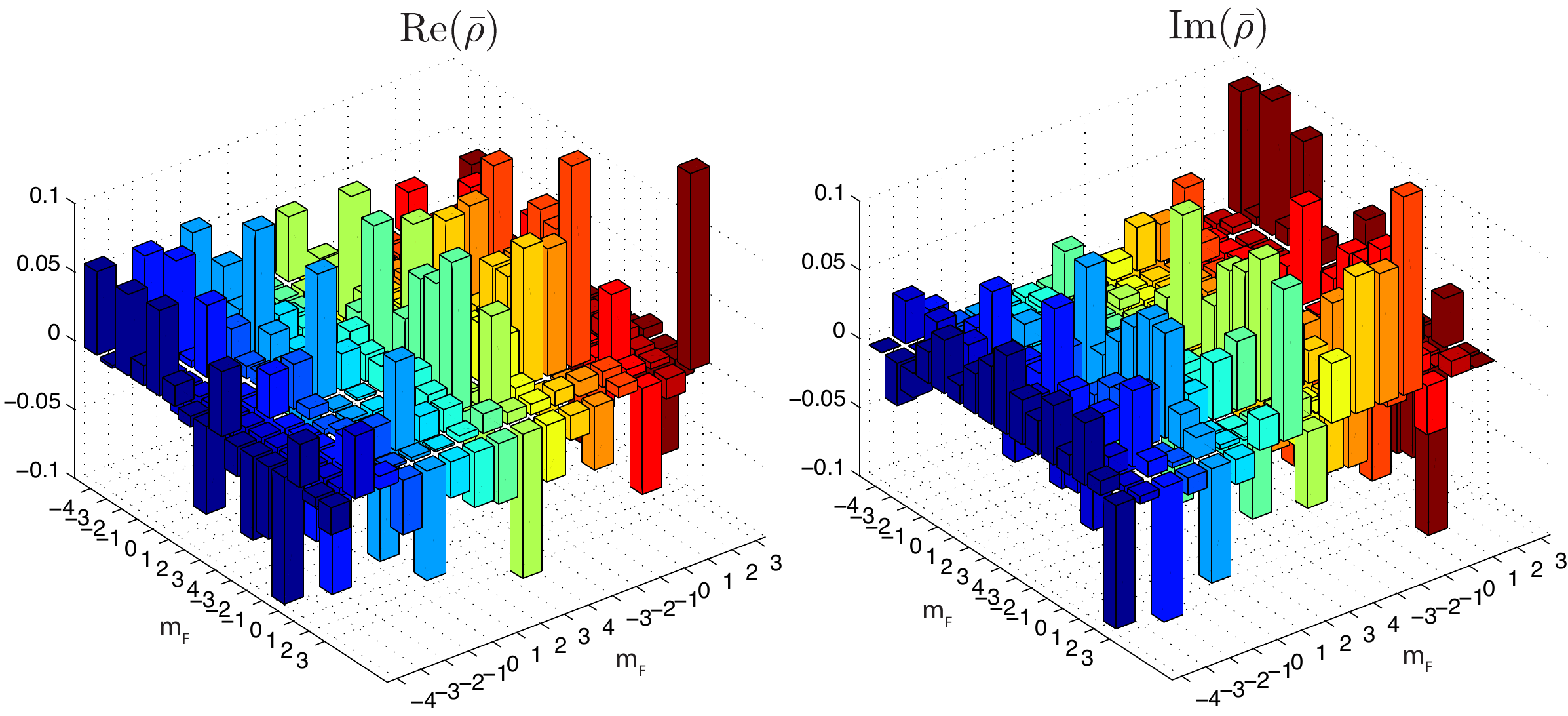}
\caption{Reconstructed quantum state using the compressed sensing method after 2 ms. The fidelity of reconstruction is $\mathcal{F}(\rho_0,\bar{\rho})=0.9593$, and its purity $\mathcal{P}(\bar{\rho})=0.9660$, which shows that compressed sensing tends to find the purest state compatible with the measurement record.}
\label{F:CSRhoRand23}
\end{center}
\end{figure}

Figs. \ref{F:LSRhoRand23} and \ref{F:CSRhoRand23} show bar plots of the reconstructed density matrices for least squares and compressed sensing, respectively. As we saw in our simulations, least squares tends to find states that are more mixed than what is expected. In this case, the reconstructed purity $\mathcal{P}(\bar{\rho})=0.9345$ for least squares and $\mathcal{P}(\bar{\rho})=0.9660$ for compressed sensing, which has found a state with higher purity.


\clearpage
\section{Average performance}
After demonstrating continuous measure QT for a handful of states of diverse complexity, we test our methods with an ensemble of arbitrary quantum states. In this context, we generated 49 random pure states sampled from the Haar measure, similar to the one depicted in Fig. \ref{F:InitialRand23}, which were prepared in the laboratory. The measurement records obtained from each of these Haar-random states were processed by our algorithms and the average fidelity of reconstruction was calculated for two cases: including only bias field inhomogeneity and including both bias and light-shift inhomogeneities. We do this in order to emphasize the robustness properties of the compressed sensing method.

\begin{figure}[t]
\begin{center}
\includegraphics[width=8.7cm,clip]{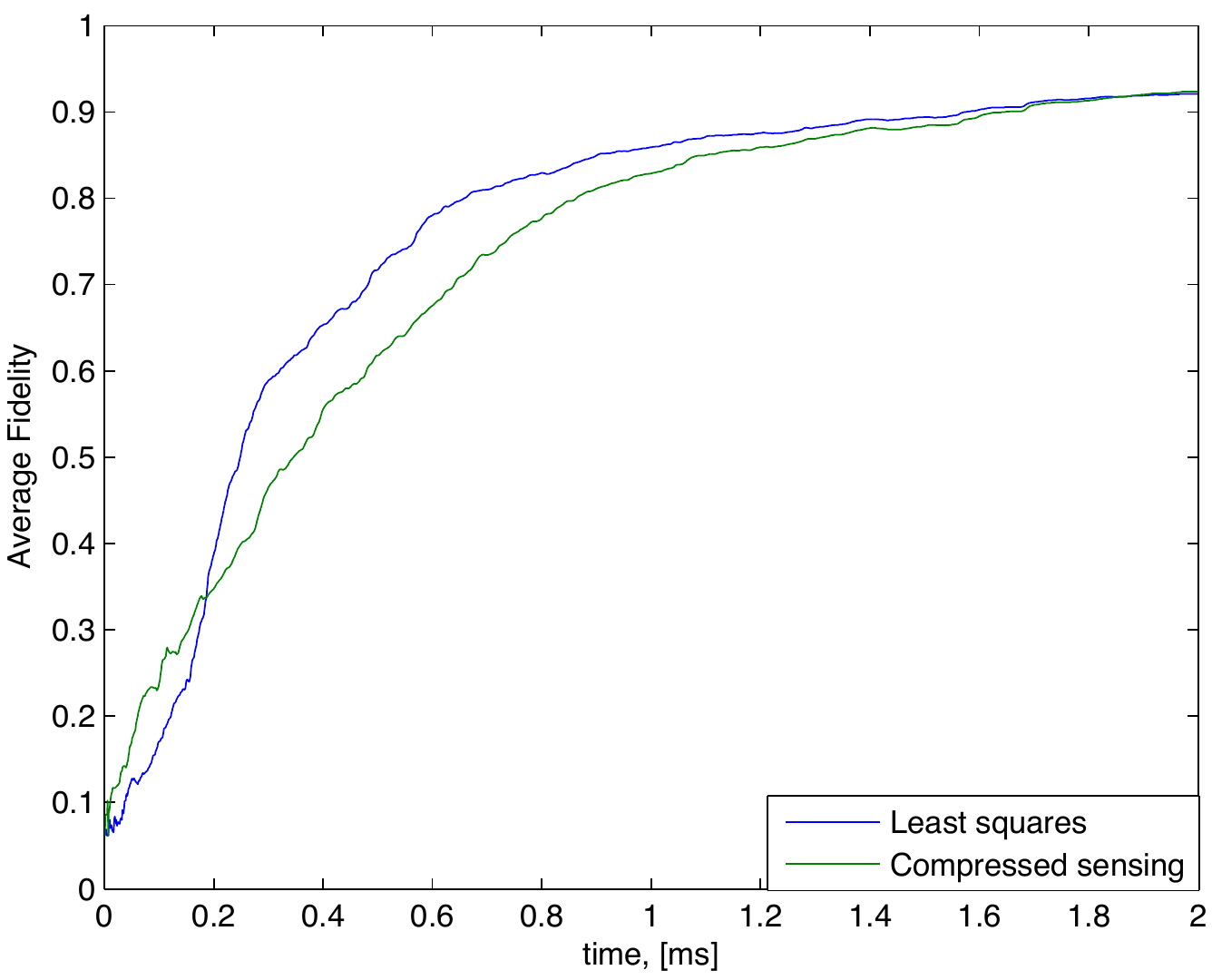}
\caption{Average fidelity as a function of time for 49 random pure states sampled from the Haar measure and processed using least squares (blue), and compressed sensing (green). In the analysis, we include in the model the effects of both bias field and light-shift inhomogeneities. On average, this data set does not show an important difference between the two methods at time t=2 ms.}
\label{F:AVGFidelityHaarRandomComplete}
\end{center}
\end{figure}

Fig. \ref{F:AVGFidelityHaarRandomComplete} shows the fidelity of reconstruction as a function of time using the complete model, i.e., considering both bias field and light-shift inhomogeneities. In this case, we do not see a significant improvement in the performance of compressed sensing. In fact, the average fidelity for compressed sensing is 0.9238 while the least squares gives a fidelity of $0.9211$; basically the same. We also calculated the average reconstruction purity of the 49 random states used in this study. We find that $\mathcal{P}(\bar{\rho})=0.9333$, for compressed sensing, and $\mathcal{P}(\bar{\rho})=0.8925$ for least squares, showing that compressed sensing finds purer estimates.  

\begin{figure}[b]
\begin{center}
\includegraphics[width=8.7cm,clip]{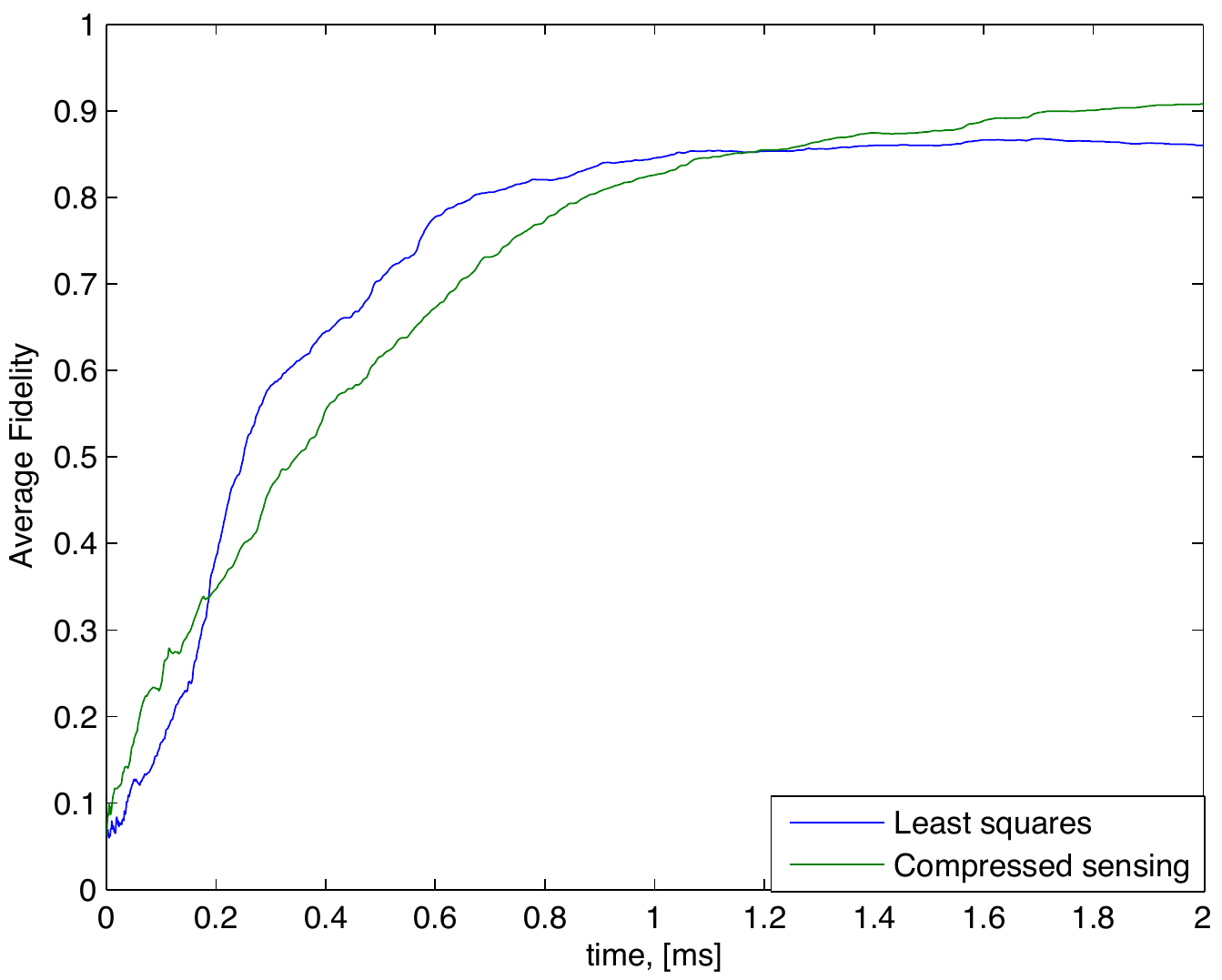}
\caption{Average fidelity as a function of time for 49 random pure states sampled from the Haar measure and processed using least squares (blue), and compressed sensing (green). In the analysis, we included in the model only the effects of bias field inhomogeneity and neglected the inhomogeneous light-shift. On average, this data shows the robustness of compressed sensing to this type of error, giving an improvement in fidelity of about $5\%$ over least squares at t=2 ms.}
\label{F:AVGFidelityHaarRandomIncomplete}
\end{center}
\end{figure}

Fig. \ref{F:AVGFidelityHaarRandomIncomplete} shows the average fidelity of reconstruction for the 49 random states now considering a partial model in which we neglected the inhomogeneity in the light-shift in the same way we did in the example of Section \ref{sec:ExampleSup1}. We clearly see the robustness of compressed sensing in this case to this specific type of error. After 2 ms, compressed sensing achieves a fidelity of 0.9077 while least squares obtains an average fidelity of 0.8601. This represents an improvement in fidelity of about $5\%$ or a reduction in the infidelity error of about $64\%$. Moreover, we see a relatively small reduction in fidelity in compressed sensing of $1.6\%$ while a much greater reduction is seen for least squares of $6.1\%$. 

These results confirm to a degree the robustness of compressed sensing when improper modeling of the system occurs. In fact, it seems from the data that it would be the preferred choice of QT method when it is known that the quantum system is close to a pure state. If there is no prior information about the purity of states, however, it seems that least squares methods would give fair and accurate estimates when the system is properly modeled.  

\chapter{Random Unitary Evolution for Quantum State Tomography}\label{ch:RandomUnitaries}
	
The successful use of random control phases for our continuous measurement quantum tomography schemes opens the question for different QT protocols for which similar choices are more natural, i.e., systems that have chaotic behavior such as the quantum kicked top \cite{haake}. In this chapter, we take a more abstract path to the problem of quantum tomographic reconstruction. We try to answer the question of how good fidelities can be achieved for systems driven through random/chaotic evolution for a simple one-paramater QT protocol. We abstract ourselves form the physical system described in previous chapters and are concerned solely with finite dimension Hilbert spaces under unitary evolution. 

High fidelity QT typically requires an ``informationally complete" measurement record.  One can obtain informational completeness by measuring the expectation values of a set of Hermitian operators that span an operator basis for $\rho_0$, or some more general operator ``frame" \cite{DAriano01,roy07}.  Restricting our attention to a Hilbert space of finite dimension $d$, and fixing the normalization of $\rho_0$, the set of Hermitian operators must form a basis for the Lie algebra $\su(d)$.  Laboratory realization of such a record is intimately tied to {\em controllability}, i.e., the ability to reconfigure the apparatus in such a way as to generate arbitrary unitary maps.  In the continuous measurement context, when the system is controllable it is possible to choose control fields for the system such that, when viewed in the Heisenberg picture, the observables evolve over the span of the algebra.  While the necessity of information completeness is rigorous if one requires high fidelity for the reconstruction of all arbitrary states, in a number of situations this condition can be substantially relaxed.  Examples include schemes that are designed to achieve high performance of the reconstruction on average \cite{Aaronson2006}, or over only some restricted set of the state space \cite{gross10,Shabani2009b}.  For these protocols, the performance of a restricted set of measurements is often nearly as good as for an informationally complete set of measurements, and yet require dramatically fewer measurement resources. 

In this chapter, we study another example of informationally incomplete measurements that nonetheless can be used in a high-fidelity QT --- measurement of a time-series of operators generated by a single-parameter random evolution.  As in previous chapters, we consider weak continuous measurement of an observable, $\cO$, through a meter that couples to an ensemble of $N$ identical systems.  The members of the ensemble undergo identical, separable time evolution in a well chosen manner.  Assuming the subsytems remain in a product state, we can write our measurement record quite generally as defined in Eq. (\ref{eq:measurementconti})
\be
M(t) = \avg{ \cO}(t) + \sigma W(t),
\ee 
where,  as before, $\sigma W(t)$ describes the deviation from the mean value arising from noise in the detection system. As in previous chapters, the QT problem is to retrodict the initial state of the system $\rho_0$ from the signal $M(t)$. For unitary evolution we have, 
\be
\avg{\cO}(t)= \Tr\left(U^{\dagger}(t) \cO U(t) \rho_0\right),
\ee
which contrasts with the complicated Linblad evolution described in Chapters \ref{ch:QT} and \ref{ch:Atoms}.  We can simplify further the analysis of this problem by considering a discrete set of measurements at intervals $\delta t$, $\left\{ \cO_n \equiv  U^{\dagger}(n \delta t) \cO U(n\delta t)  \right\}$. In this context, the ultimate fidelity of the QT will be limited by the finite signal-to-noise ratio.  While the choice of unitary evolution necessary to determine an arbitrary $\rho_0$ from $M_n$ is not unique, a necessary and sufficient condition is that the set $\left\{ \cO_n \right\}$  be informationally complete.  A good strategy is to choose the dynamics such that for each $n$, $U(n\delta t)$ is a random matrix, chosen from an appropriate Haar measure.  In that case our measurement record is not only provably informationally complete but is also unbiased over time.  Suppose, however, we choose $U(n\delta t)=\left(U_0\right)^n$, where $U_0$ is a {\em fixed} unitary matrix.   In this case, the observable series $\cO_n$ traces out a single orbit in operator space; we call this a one-parameter measurement record.  As we will show, the record is not informationally complete, but nevertheless can lead to high fidelity QT for all states but a set of small measure if $U_0$ is a random unitary, especially for large dimensional spaces.  These results elucidate the connection between random evolution and information gain at the quantum level.

In the next section, we show that the measurement operators generated from a single parameter trajectory cannot span the entirety of the operator algebra, $\su(d)$, but that the operators that lie outside of the subspace of the measurement record are a vanishingly small fraction in the limit $d \rightarrow \infty$.  Next, we study the performance of QT using the weak continuous measurement protocols detailed in Chapter \ref{ch:QT} for these incomplete measurement records.  We show that even at small $d$, when one includes the physical constraint of {\em positivity} of the density matrix, QT performs surprising well for almost all quantum states, well beyond that expected if one had solely considered the vector space geometry of the Lie algebra.  Finally, we connect these abstract results to physical realizations using the unitary Floquet maps of the quantum kicked top \cite{haake} whose associated classical dynamics is chaotic.  As quantum chaos is associated with pseudorandom matrix statistics, this protocol provides intriguing new signatures of quantum chaos in QT.

\section{One-parameter measurement records}\label{sec:span}

In this section, we study whether or not information completeness for QT is achievable from a one-parameter measurement record.  The one-parameter orbit in operator space is defined by the time-series $\cO_n = (U_0^{\dag})^n \cO (U_0^{\phantom \dag})^n $, where $\cO$  is a Hermitian operator and  $U_0$ is a fixed unitary matrix.  We will restrict the observable $\cO$ to have zero trace since the component proportional to the identity gives no useful information in QT.  We thus ask, is it possible to reconstruct a generic quantum state $\rho_0$ if one can measure the expectation values of all of the observables in the time series?  To answer this, we consider $\mathcal{A}\equiv\text{span}\left\{ \cO_n \right \}$, and determine the size  the orthocomplement subspace with respect to the trace inner product,  $\mathcal{A}_\perp$; operators in this set are not measured in the time-series. Such missing information renders the measurement incomplete, and thus incompatible with perfect QT, no matter what signal-to-noise ratio is available in the laboratory.

To find the dimension of $\mathcal{A}$, consider the subspace of operators that are preserved under conjugation by $U_0$, $\mathcal{G} \equiv \left\{ g \in \su(d) \left|\, U_0^{\phantom \dag} g U_0^{\dagger} = g \right\}\right.$, i.e., the space of operators that commute with $U_0$. Let $\mathcal{B} =\left\{ g \in \mathcal{G} \left|\, \Tr(g \cO) = 0\right\}\right.$.  It thus follows that $\mathcal{B} \subseteq \mathcal{A}_\perp$ since  $\forall g \in \mathcal{B}$  
\be
\Tr(\cO_n g) = \Tr \left( (U_0^{\dagger})^n \cO(U_0^{\phantom \dag})^n g \right) = \Tr(\cO g)  =0.
\ee
As the two spaces are orthogonal, $\dim\mathcal{A} + \dim \mathcal{B} \le \dim (\su(d)) =d^2-1$.  Now,  if $U_0$ has nondegenerate eigenvalues, $\mathcal{G}$ will be isomorphic to the the largest commuting subalgebra of $\su(d)$ (the Cartan subalgebra),  but for degenerate $U_0$, $\mathcal{G}$ will contain additional elements.  Since the Cartan subalgebra has dimension $d-1$, $\dim \mathcal{G} \geq d-1$.  By definition, $\mathcal{B}$ is obtained from $\mathcal{G}$ by projecting out one direction in operator space, and thus  $\dim \mathcal{B} =\dim \mathcal{G}-1 \ge  d-2$.  It follows that 
\be
\dim \mathcal{A} \le \dim (\su(d))-\dim \mathcal{B} \le d^2-d +1.  
\label{eq:bound}
\ee

This is the first principal result -- a one-parameter measurement record is not informationally complete since $\dim \mathcal{A}_\perp>0$ (when $d > 2$).  However, it remains to be seen how much the missing information impacts the fidelity of QT.  An immediate question is to determine the conditions on $U_0$ and $\cO$ required to saturate bound in Eq.~(\ref{eq:bound}).    Since $U_0$ is a unitary matrix it is always diagonalizable as
\be
 U_0 = \sum_{j=1}^d e^{-i \phi_j} \ket{j}\!\bra{j},
\ee 
and in this basis
$\cO_n$ has the representation 
\be
 \cO_n = \sum_{j,k=1}^d e^{-i n (\phi_j -\phi_k)} \bra{k} \cO \ket{j} \ket{k}\!\bra{j}. 
\ee
The diagonal component has no $n$-dependence, so it is useful to rewrite $\cO_n$ as 
\be 
\cO_n =\sum_{j=1}^d \bra{j} \cO \ket{j} \ket{j}\! \bra{j}   + \sum_{j \neq k }^d e^{-i n(\phi_j - \phi_k)} \bra{k} \cO \ket{j} \ket{k}\!
\bra{j}.\label{eq:On} 
\ee
To show that $\mathcal{A}$ is spanned by $d^2-d+1$ linearly independent matrices we must have that
\be
\sum_{n=0}^{d^2-d} a_n \cO_n = 0 \qquad \textrm{iff} \qquad a_n = 0 ~~ \forall n.
\ee
We can write this condition out explicitly using Eq. (\ref{eq:On}) as
\be
\left( \sum_{n=0}^{d^2-d} a_n \right)  \sum_{j=1}^d \bra{j} \cO \ket{j} \ket{j}\!\bra{j}   +\sum_{j \neq k}^d \left( \sum_{n=0}^{d^2-d} a_n e^{-i n (\phi_j - \phi_k)} \right) \bra{k} \cO \ket{j} \ket{k}\!\bra{j} = 0.
\ee
The system of equations is underconstrained if either $\bra{j} \cO \ket{j}  =0$ for all $j$ or $\bra{j} \cO \ket{k}  =0$ for any $j \neq k$.  Assuming this is not the case, the condition for linear dependence is given by a set of linear equations on $a_n$ of the form
\be\label{eq:systemeq}
\underbrace{\left(\begin{array}{ccccc}
1 & x_0^{\phantom 2} & x_0^2 &\cdots & x_0^{d^2-d}\\
1 & x_1^{\phantom 2} & x_1^2 &\cdots & x_1^{d^2-d}\\
\vdots & \vdots &\vdots &\ddots& \vdots\\
1 & x_{d^2-d}^{\phantom 2} & x_{d^2-d}^2 &\cdots & x_{d^2-d}^{d^2-d}
\end{array}\right) }_V
\left( \begin{array}{c}
a_0\\
a_1\\
\hdots \\
a_{d^2-d}
\end{array} \right) =  0.
\ee
Here we have written $x_0 =1$ and $x_m = e^{-i (\phi_j -\phi_k)}$, for some indexing of the pairs $(j,k)$ to $1\leq m \leq d^2-d$.  
The condition for linear independence is simply $\det V \neq 0$.  Expressed as above, one can see that $V$ is an instance of a Vandermonde matrix, whose determinant is easy to evaluate through the formula \cite{hornandjohnson} 
\be
\det V = \prod_{0 \leq j < k \leq d^2 - d} (x_k - x_j).
\ee 
For our system of equations to become linearly dependent before saturating the previous bound, we would need that $e^{-i (\phi_j - \phi_k)} =
e^{-i (\phi_{j'} - \phi_{k'})}$ for some distinct pair of the couples $(j,k)$ and $(j',k')$, or $e^{-i (\phi_j - \phi_k)} = 1$ for some $(j,k)$. 

In summary, in order for the dimension of the span of a one-parmeter measurement record to saturate the bound of $\dim \mathcal{A} = d^2 -d +1$,  the eigenphases, $\phi_j$, and the eigenvectors, $\ket{j}$, of $U$ must satisfy the following constraints:
\bea
\label{eq:conditions}
	1.&& \exists \ j \ \textrm{s.t.}, \quad \bra{j} \cO \ket{j} \neq 0 \nonumber\\
	2.&& \forall \ j \neq k, \quad \bra{k} \cO \ket{j} \neq 0 \nonumber\\
	3.&& \forall \ j \neq j', \quad \phi_j - \phi_{k} \neq \phi_{j'} - \phi_{k'} ~ (\textrm{mod} ~2\pi)
\eea
Note that the third condition enforces that both the eigenphases, as well as their pairwise differences, must be distinct.  There is an interesting interpretation of these conditions from the perspective of universal control.  While $U_0$ defines a one-parameter trajectory, the set $\{ U_0, e^{i \mathcal{O}} \}$ defines a universal set of unitary matrices that can generate arbitrary maps, if the conditions in Eq.~(\ref{eq:conditions}) are satisfied \cite{altafini2002}.  

If $U_0$ is a unitary matrix chosen randomly from the Haar measure on ${\sf SU}(d)$, the saturation conditions will almost surely be satisfied, independent of $\cO$.  Therefore, a generic unitary evolution will almost always generate a measurement record that spans the full $d^2-d+1$ operators.  In fact, the typical members of many types of pseudorandom ensembles of unitary matrices satisfy these
constraints, e.g. unitaries drawn from $t$-designs or approximate $t$-designs~\cite{Ambainis2007, Gross2007, Dankert2009, Roy2009}, random quantum circuits~\cite{Harrow2009, Brown2009}, as well as unitary evolutions
that possess globally chaotic dynamics in the classical limit \cite{ haake, Emerson03, Scott03}.  These types of pseudorandom evolutions are more
readily available in practical situations, providing possible avenues to test these results in laboratory implementations.

The results of this section show that a one-parameter evolution generates a measurement record that misses a subspace of dimension $d-2$ out of the full $\su(d)$ algebra whose dimension is $d^2-1$.  For very large Hilbert space dimensions, the implication is that all but a vanishing fraction of the information regarding measurements on the quantum system is contained in this type of record. It is not clear that the fidelity of reconstruction, Eq. (\ref{eq:fidelity}), will be directly related to the fraction of operator space spanned by our set of observables.    Surprisingly, the situation is in fact more favorable than this na\"{\i}ve assumption.  In the next section we will see that merely requiring the reconstructed density matrix be positive provides a powerful constraint, allowing us to use a one-parameter measurement record induced by the orbit of a single pseudorandom unitary matrix to perform very high fidelity reconstructions even for small dimensional Hilbert spaces, for all but a very small subset of states.

\section{Density matrix reconstruction from an incomplete measurement}\label{S:Reconstruct}
In this section, we consider the type of reconstruction protocols described in Chapter \ref{ch:QT} in which one has access to an ensemble of  $N$ identically prepared systems all initialized to the same state, $\rho_0$.  The system is weakly measured yielding the record given in Eq.~(\ref{eq:measurement}).  For sufficiently weak coupling, the deviation of the measurement result from the quantum expectation value is dominated by the noise on the detector (e.g., shot noise of a laser probe) rather than the quantum fluctuations of measurement outcomes intrinsic to the state (known as projection noise).  In this case, there is negligible backaction on the quantum state during the course of the measurement, and the ensemble remains factorized.  As before, we treat the detector noise as Gaussian white noise.  

We examine a stroboscopic time-series of the measurement record, where the observables evolve according to the one-parameter trajectory discussed in the previous section.  At discrete times $t=n \delta t$, the measurement record is
\be
M_n= \Tr (\cO_n \rho_0) + \sigma W_n,
\label{eq:record}
\ee
where $\cO_n$ and $\rho_0$ are Heisenberg operators. We use this type of measurement record together with the tomographic methods described in Chapter \ref{ch:QT} to study the reconstruction fidelities that a one-parameter measurement record achieves.

We are now prepared to quantitatively analyze the performance of our QT protocols in the case of the one-parameter measurement record arising from an incomplete set of observables that satisfy Eq.~(\ref{eq:conditions}).  As our system, we consider a atomic system with total spin $F$  described by a Hilbert space of dimension $d=2F+1$.  We will fix $\cO = F_z$ and select a random unitary matrix $U_0$ from the Haar measure on ${\sf SU}(d)$.   Such a random matrix will almost always satisfy the constraints of Eq.~(\ref{eq:conditions}), except for a set of measure zero.  As our goal is to determine how the information missing in a subspace of observables impacts the QT fidelity, we will simplify the analysis by assuming that noise on the measurement  is vanishingly small.  We study the performance of different classes of states, randomly chosen by an appropriate measure.  For each set of states, we will look at the average fidelity between the initial and reconstructed states, $\langle \mathcal{F} \rangle = \int \ud \rho_0 \mathcal{F}(\bar{\rho}, \rho_0)$, where $\ud \rho_0$ is a measure on the space of density operators.

\begin{figure}[t]
\begin{center}
\includegraphics[width=8.7cm,clip]{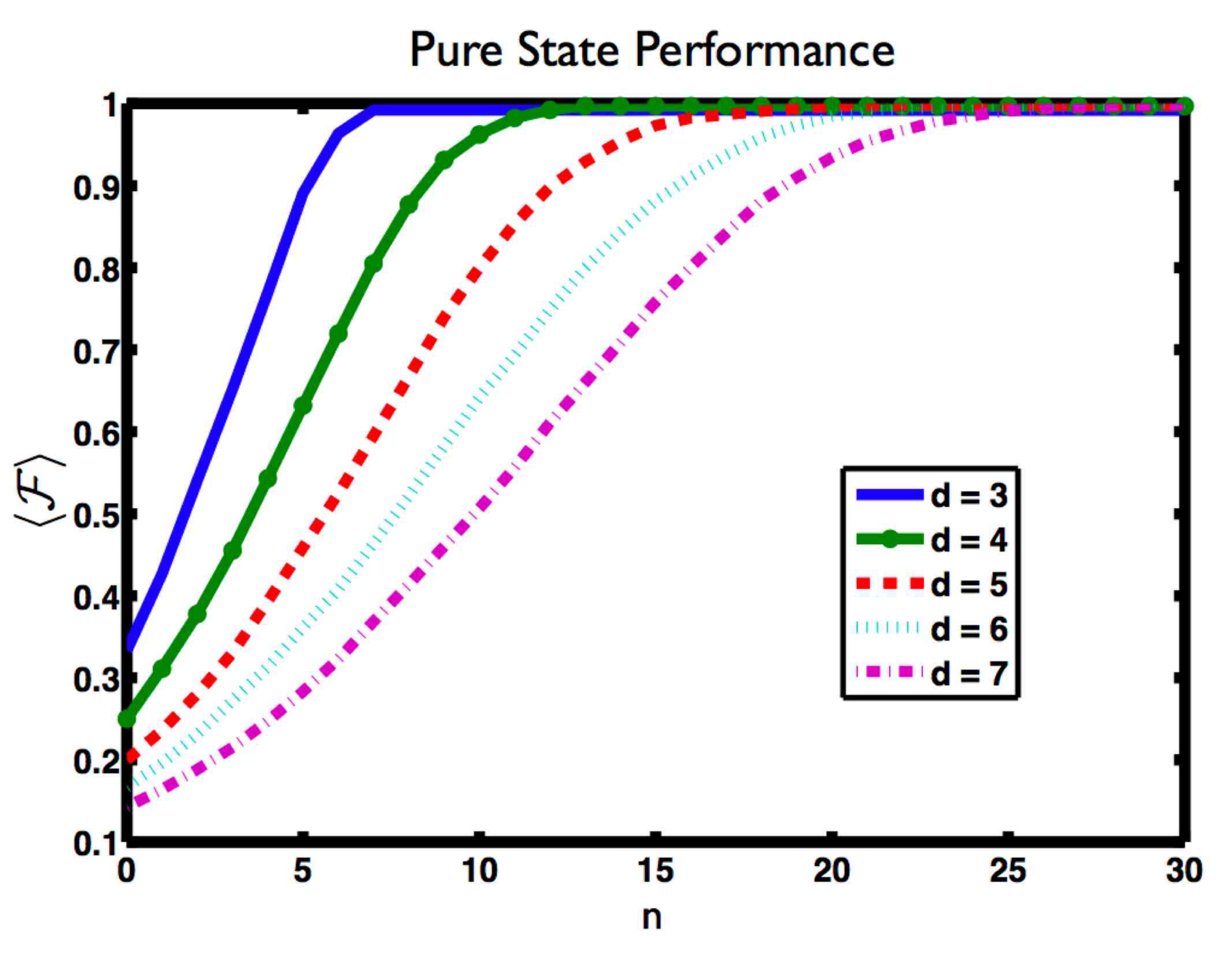}
\caption{Numerical simulations of the QSR protocol for pure states as a function of $n^{th}$ expectation value measured in the time-series, and for different dimensions of the Hilbert space, $d$. Each data point represents the average reconstruction fidelity of 100 pure states drawn from the Fubini-Study measure, additionally averaged over measurement records derived from ten different Haar-random unitary propagators. }
\label{F:pure}
\end{center}
\end{figure}

The simplest case to analyze is when we have prior information that $\rho_0$ is a pure state, $\ket{\psi_0}$.  Figure~\ref{F:pure} shows the average fidelity as one sequentially measures the expectation value of the $n^{\textrm{th}}$ observable in the series, for different dimensions of the Hilbert space $d$.  Averages are taken for 10 choices of random unitary matrices, each of which is averaged over 100 random pure states distributed on the Fubini-Study measure \cite{zyczkowski2006}.  Two striking features are seen in these plots:  (i) unit fidelity is achieved for any $d$ even though the record was said to be informationally incomplete; (ii) the protocol reconstructs the state well before we measure all $d^2-d+1$ independent observables. The inclusion of positivity dramatically improves the reconstruction fidelity for pure states.  In fact, a one-parameter measurement record generated by a random $U_0$ can be used to reconstruct almost all pure states perfectly in the absence of noise.  

The performance of the QT protocol can be understood given the prior information we have assumed.  A pure state is specified by $2d-2$ real parameters, whereas we measure $d^2-d+1$ expectation values.  Thus, it should come as no surprise that the measurement record contains enough information to reconstruct the state.   In fact, in this case one can use positivity to explicitly recover the missing information exactly from the measurement record, without resorting to the numerical convex program discussed in Chapter \ref{ch:QT}.  In general, the missing information is associated with matrices that commute with $U_0$.  Thus, when expressed in the eigenbasis of $U_0$, only the diagonal matrix elements of the density operator might not be estimated.  A necessary (but not generally sufficient) condition for a matrix $\rho_0$ to be positive semidefinite is that its matrix elements must satisfy the following set of inequalities: $\rho_{ii} \rho_{jj} - |\rho_{ij}|^2 \geq 0 $, i.e.~all of the $2 \times 2$ matrix minors must be positive semidefinite \cite{hornandjohnson}.  If additionally the state is pure,  these inequalities become equalities.  Therefore, if any of the off-diagonal matrix elements are nonzero, we can completely determine all of the diagonal elements via the equations  $\rho_{ii} = \left(|\rho_{ij}| |\rho_{ik}|\right)/|\rho_{jk}|$.  A special case is if all of the off-diagonal matrix elements of $\rho_0$ are zero. Then the state must be one of the eigenvectors of $U_0$, but such states lie in a set of measure zero.  Measurements that are informationally complete solely for pure states are called PSI-compete~\cite{Flammia2005, Finkelstein2004}. 

While one can easily explain the high-fidelity performance of the QT protocol in the case of a pure state, for mixed states, this is far from clear, and the power of the positivity constraint comes fully to the fore.  For mixed states, and $d>2$, the average fidelity of our reconstruction will never reach unity because some density matrices cannot be reconstructed from the information in our incomplete measurement record.  For example, some convex combinations of eigenstates of $U_0$ are indistinguishable from the maximally mixed state, even though the fidelity between the two can be very small.  Nonetheless, as we see below, the one-parameter measurement record generated by a single random unitary still performs very well on average, even for generic mixed states.

\begin{figure}[t]
\begin{center}
\includegraphics[width=8.7cm,clip]{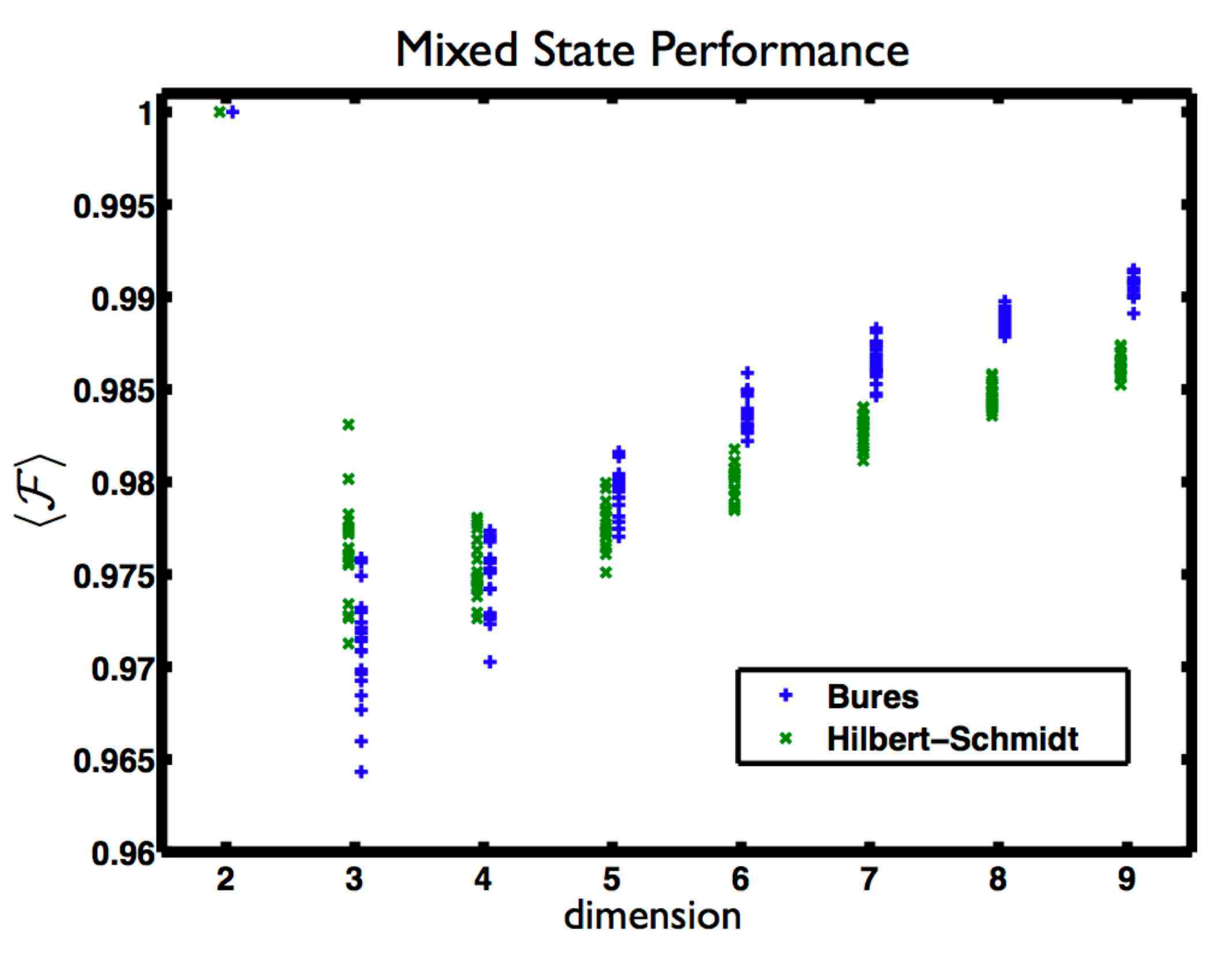}
\caption{Fidelity of the QSR protocol for mixed states as a function of the dimension of Hilbert space $d$.  Each point represents the average reconstruction fidelity from a measurement record of length $10(d^2-d+1)$ (a long-time limit) generated from a different random unitary propagator from the Haar measure.   We average over 200 density matrices drawn from the Bures (blue crosses) or Hilbert-Schmidt (green x's) measures.  For each dimension we show the average fidelities from twenty of these measurement procedures.  }
\label{F:Bures}
\end{center}
\end{figure}

Figure~\ref{F:Bures} shows the average fidelity for two choices of measures on density matrices, the Bures measure and the Hilbert-Schmidt measure \cite{zyczkowski2006, braunstein94}, with states sampled according to the construction provided by Osipov, Sommers and \.{Z}yczkowski \cite{Osipov09}.  For both distributions we look at a long-time limit of the time-series, here $10 (d^2-d+1)$ measurement steps, and plot the average fidelity as a function of the dimension of the Hilbert space rather than $n$.  In the limiting case of negligible noise on the measurement, we have already extracted all the possible information about the state after $d^2-d+1$ measurements.  In practice, increasing the measurement record serves to smear out the information over the measured observables, leading to a more uniform distribution for the non-zero eigenvalues of the covariance matrix $\mbf{C}$, Eq. (\ref{eq:Covariance}), which is numerically favorable. As seen in these two plots, on average, the one-parameter measurement records perform surprisingly well.  In all cases the mean fidelity is greater than $0.96$ with a minimum around $d=3$ or $d=4$.  After this dip, the minimum of the fidelity looks to be monotonically increasing with the size of the Hilbert space.  Additionally, the particular instantiation of the random unitary map appears to make very little difference (less than 0.01 fidelity), with the residual difference decreasing as the dimension increases. 

The condition of positivity is a powerful constraint that describes correlations between observables that can lie along orthogonal directions in operator space.  For example, in the case of a 2-level quantum system, if $\avg{\sigma_z} = 1$, positivity implies $\avg{\sigma_x} = \avg{\sigma_y}=0$, fully specifying the state from a single expectation value.  In the context of noisy measurements, the positivity constraint allows us to perform high fidelity QT in the face of uncertainty by enforcing consistency conditions on our measurement outcomes.  When we consider incomplete measurement records as above, positivity can place bounds on the means of observables which otherwise would be completely undetermined.  This can greatly increase the fidelity of QT.  Intuitively, while many vectors $\bar{\mathbf{r}}$ might minimize, for example, Eq.~(\ref{eq:OneStepConvex}), only very few of these are also compatible with positivity.  This is the reason we observe high fidelities even in the case of random mixed states in the context of a one-parameter measurement record generated by a single random matrix: the requirement of positivity provides substantial additional information leading to very high fidelity of QT, well beyond what one would na\"{\i}vely predict.

In the next section, we consider a physical application of the one-parameter measurement record protocol using the quantum kicked top system, which exhibits chaotic dynamics in its classical realization. We will see that in that type of system, the ideas discussed in this section naturally apply.

\section{Example: quantum kicked-top}

In Sec.~\ref{sec:span}, we discussed that the conditions given in Eq.~(\ref{eq:conditions}) can be satisfied by a pseudorandom unitary matrix, instead of a true random matrix sampled from the Haar measure.  One such class of matrices are the Floquet maps associated with ``quantum chaos", i.e., periodic maps whose classical dynamical description shows a globally chaotic phase space.  An example is the quantum kicked-top  \cite{haake}, a system that recently has been realized in a cold atomic ensemble \cite{chaudhury09}.  In this section, we explore how our QT protocol performs in this context, providing a possible route to laboratory studies, and novel signatures of chaos in quantum information.

The standard quantum kicked top (QKT) dynamics, considered here, consists of a constant quadratic twisting of a spin (``top"), punctuated by a periodic train of delta-kicks of the spin around an orthogonal axis.  The Floquet operator for this perodic map is typically written as the product of noncommuting unitary matrices,
\be
U_{\rm QKT} = e^{- i \phi F_z^2 / F} e^{-i \theta F_x}.\label{eq:floquet}
\ee
The parameters $\theta$ and $\phi$ represent the angles of linear and nonlinear rotation respectively.  The dynamics exhibit a classically chaotic phase space for an appropriate choice of these parameters  \cite{haake}.  The connection between chaos in this system and random matrices has been well studied, particularly, the relationship between the level statistics of the Floquet eigenvalues, chaos, and symmetry.   Floquet maps associated with global chaos are random matrices that divide into different classes.  If the Floquet operator is time-reversal invariant, the level statistics are that of the circular orthogonal ensemble (COE); without additional symmetry they are members of the circular unitary ensemble (CUE). The latter group is ${\sf U}(d)$ or ${\sf SU}(d)$ depending on the context.  The measurement records generated from matrices chosen from either the COE or CUE will satisfy the eigenvalue conditions described in Eq.~(\ref{eq:conditions}) almost surely.  The QKT is known to have a time-reversal symmetry and classically chaotic dynamics. It still does not have COE statistics in the full $d=2F=F+1$ Hilbert space, however, due to an additional symmetry. The QKT map is invariant under a $\pi$-rotation about the $x$-axis, leading to a parity symmetry.    This system therefore has a doubly degenerate eigenspectrum, breaking the conditions in Eq.~(\ref{eq:conditions}).  While such Floquet operators generate a measurement record that has much less information relative to an arbitrary state, we can perform high-fidelity QT for states restricted to a subspace defined by the additional symmetry, here, the states that have even parity under reflection around the $x$-axis.  To do this we require that our initial operator $\cO$ is also symmetric under reflection, e.g., $\cO = F_x$.

\begin{figure}[t]
\begin{center}
\includegraphics[width=8.7cm,clip]{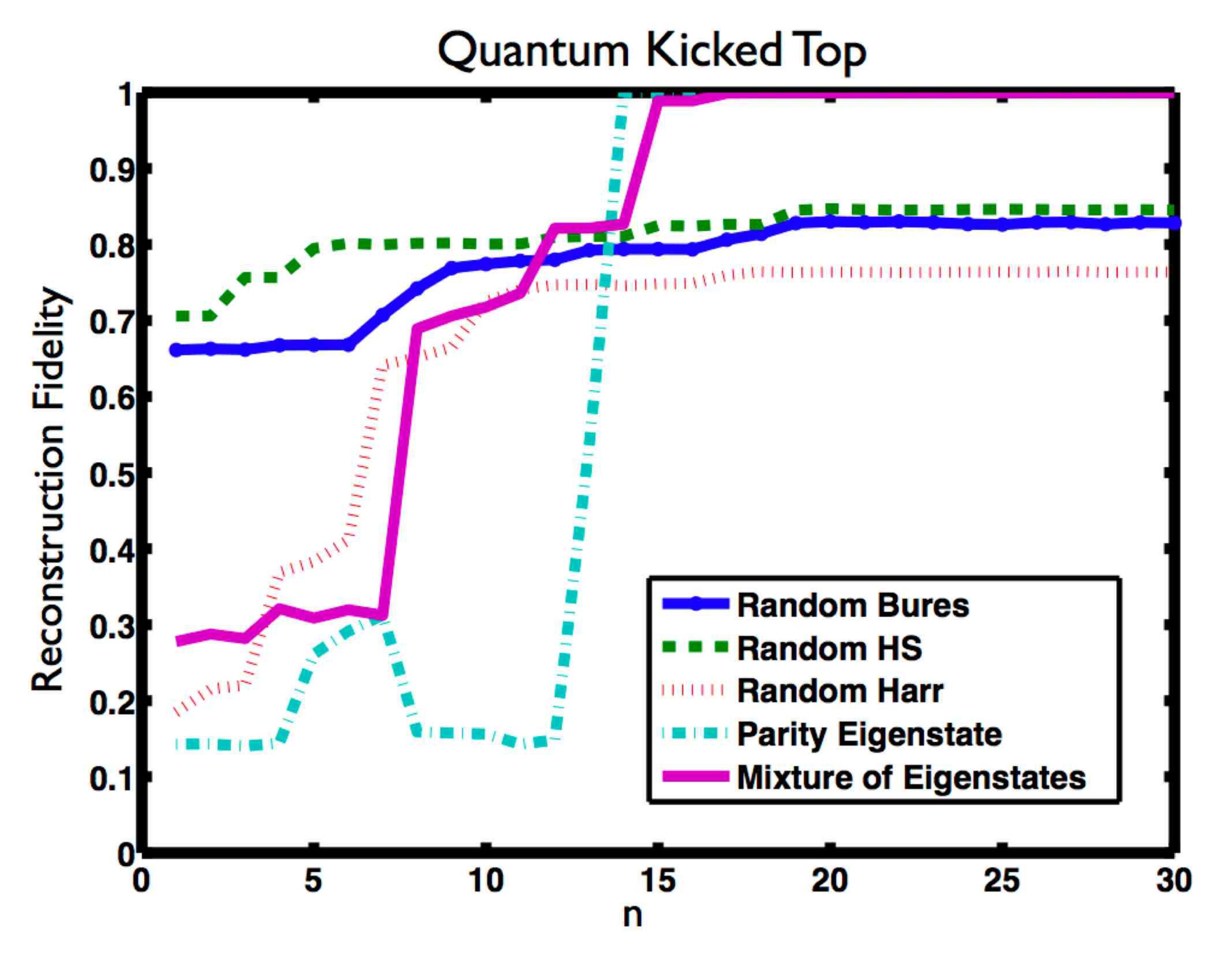}
\caption{Reconstruction fidelity for a variety of states versus $n$.  The initial measurement observable is $F_x$, and each subsequent observable whose expectation value we measure is obtained by evolving under the Floquet map of a quantum kicked top, Eq.~(\ref{eq:floquet}).  For generic random states in the whole Hilbert space (pure or mixed), the reconstruction performs poorly.  Density matrices that are invariant with respect to $\pi$-rotation around the x-axis, such as the cat state, an eigenstate of the parity operator, or an incoherent mixture of odd-parity $F_x$ eigenstates, are reconstructed with
high fidelity. }
\label{F:qkt}
\end{center}
\end{figure}

\begin{figure}[t]
\begin{center}
\includegraphics[width=8.7cm,clip]{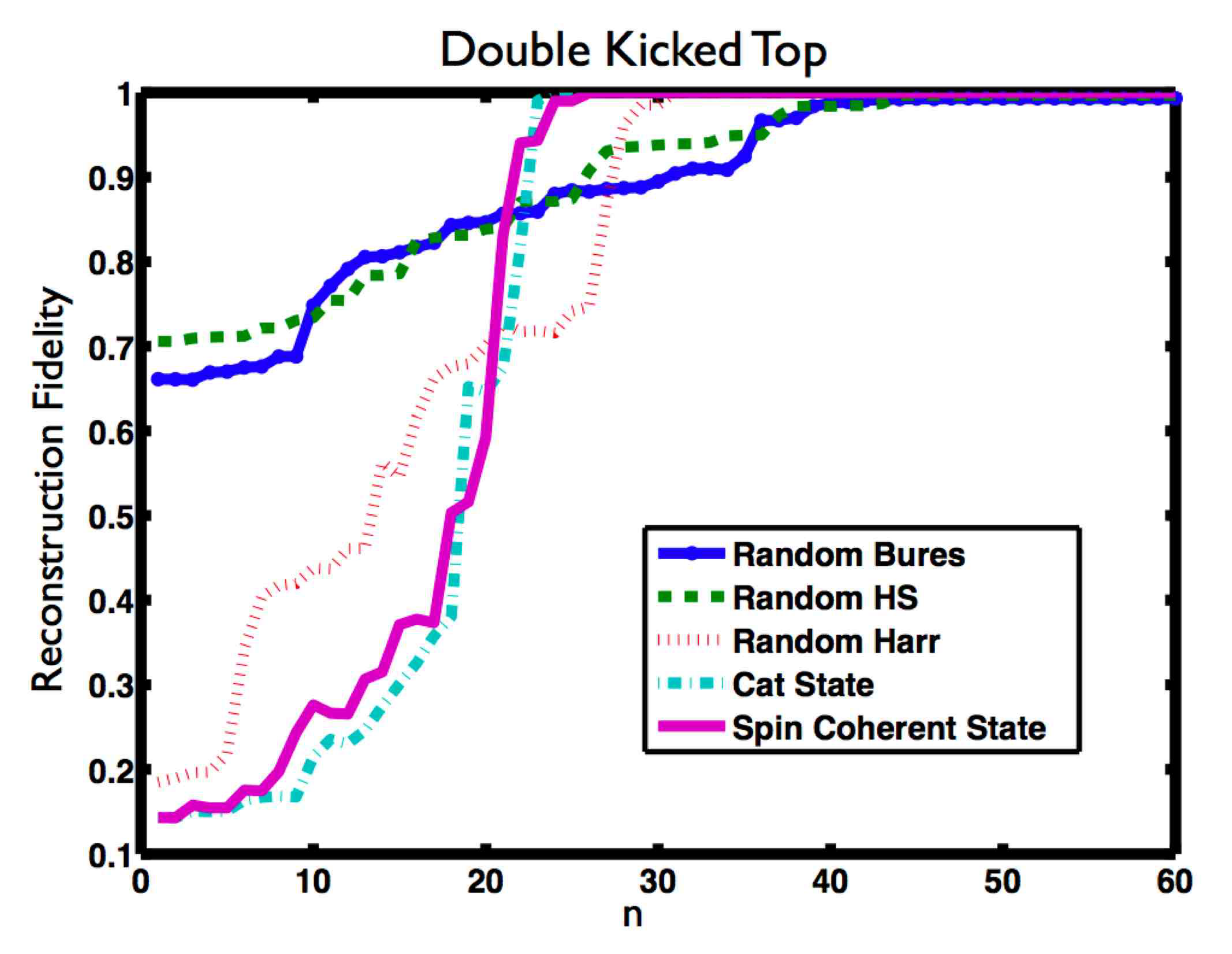}
\caption{Same as Fig.~\ref{F:qkt} when the evolution is given by the double kicked top.  All states, pure or mixed, asymptote to fidelities near unity, with the pure states reaching their maxima quicker than the mixed states.  }
\label{F:dkt}
\end{center}
\end{figure}

We present examples of this type of reconstruction in Fig.~\ref{F:qkt}.  Here we look at the QKT dynamics for a spin $F=3$ particle  (a $d=7$ dimensional Hilbert space).  We choose the parameters $\phi = 7$ and $\theta = 0.228$, values for which the classical phase space is known to be globally chaotic.  Additionally, we let the noise on the measurement approach zero. For general mixed and pure states sampled from the full Hilbert space, this reconstruction performs poorly.  However, if we restrict our attention to states that are eigenstates of parity, we reconstruct with near unit fidelity.  As a check, we find that covariance matrix $\mathcal{C}$ has rank 19.  This agrees with our predictions, since this space has a 4-fold degenerate $-1$ parity-eigenspace and a 3-fold degenerate $+1$ parity-eigenspace.  The measurement operators that preserve this symmetry must be block diagonal, lacking the $(3 \times 4) $ components from each of the two off-diagonal blocks.  For this dimension  $d^2 - d+1 - 24 = 19$.    
  
We can examine the effects of a pseudorandom unitary that satisfies Eq.~(\ref{eq:conditions}) on the whole space if we look at the ``double kicked top" where we alternate kicking about $x$ and $y$.  Here the Floquet operator has the form
\be
U_{\rm 2KT} = e^{- i \phi F_z^2 / F } e^{-i \theta_x F_x}e^{- i \phi' F_z^2 /F} e^{-i \theta_y F_y}.\label{eq:double}
\ee
If we choose $\phi = \phi'= 6$, $\theta_x = \pi /2$ and $\theta_y = 0.228$, these operators have no time-reversal symmetry, or indeed any other symmetry, and so approximate the spectrum of the CUE.  Since the double kicked top Floquet operator shares no symmetries with $F_z$ we can choose $\cO = F_z$ in order to satisfy the first condition of Eq.~(\ref{eq:conditions}).

In Fig.~\ref{F:dkt} we show the QSR performance for a time-series generated by the double kicked top in a Hilbert space of a spin-3 particle.  Here the fidelity asymptotes to unity for all of our choices of  states,  as we would expect from the simulations in Sec.~\ref{S:Reconstruct}.  The reconstruction reaches its asymptotic fidelity when we have made $d^2-d+1 = 43$ measurements.  As we saw previously, the pure states require fewer measurements to reach their asymptotic value and the QSR can be perfect in the absence of noise.

\chapter{Summary and Outlook}\label{ch:Conclusion}


In this work, we have presented a comprehensive discussion and application of a protocol to perform fast, robust, high-fidelity quantum tomography (QT) based on continuous measurement of an informationally complete set of observables.  This procedure is applicable when one has access to a large ensemble of identically prepared systems in a product state, collectively coupled to a probe field.  For weak measurement backaction, the probability distribution of parameters that defines the density matrix, conditioned on the measurement record, is Gaussian, and the problem maps onto one of classical stochastic parameter estimation.  With sufficient signal-to-noise, the density matrix can be found from a single measurement record (or an average over a few records) by solving a standard convex optimization program which produces an estimate that is Hermitian, positive semidefinite, and unit trace.

Two approaches to the estimation problem were considered. First, the maximum likelihood/least squares method relied on the fact that the noise has Gaussian statistics. It finds an estimated state by maximizing an appropriate function (the likelihood), which is related to the statistics of the noise. Second, the other method employed in this work constituted a generalization of the available compressed sensing techniques to continuous measurement QT. An estimate of a density matrix is found by minimizing its trace, which guarantees that a low rank/high purity estimate is found. These two methods were compared by subjecting them to the same numerical experiments and their range of applicability was discussed. Our results show that whenever there is prior information that the prepared state is pure or close to pure, the best choice of algorithm to estimate that state is compressed sensing. Moreover, it seems to be more robust to errors in the signal given the fact that, in its formulation, the method does not fit the data beyond some threshold value. If nothing is known about the system, however, the maximum likelihood method would be our algorithm of choice, giving the fact that the best estimate is based uniquely on the measured data, with no consideration of purity.

We have applied both QT protocols to the problem of reconstructing states encoded in hyperfine spins in cold atomic ensembles.  A key component of our procedure is to drive the system with well-chosen control fields so as to generate an informationally complete set of observables over the course of the measurement record.  We have presented here an approach based on combinations of microwave and radio-frequency-driven spin rotations that allows for controllability of the full hyperfine ground-electronic spin states of cesium atoms (16 Hilbert space dimensions). This is a more ambitious application and a generalization of a previous procedure in \cite{silberfarb05} developed by Andrew Silberfarb,  which considered reconstruction of a density matrix associated with one hyperfine manifold $F$, which was demonstrated in experiments on the 7-dimensional $F=3$ manifold of Cs atoms. 

Our protocol rests on the assumption that once the measurement record is obtained, we can invert its history to estimate the initial quantum state of the ensemble.  It is thus essential to accurately model the atomic dynamics and measurement of the observables, including known sources of imperfections.  In this context, we have given a detailed discussion of the master equation governing the atomic dynamics and a measurement model based on polarization spectroscopy. We presented a complete mathematical model of the hyperfine control problem with cesium atoms considering not only the interaction of the atoms with external magnetic and light fields, but also the decoherence effects caused by photon scattering, and the effects of inhomogeneities. We took full consideration of known imperfections in order to make the model appropriate for experimental realization, meaning that it is as close to the real experiment as possible, which is a formidable challenge. A simulation code, included in Appendix \ref{App:Code}, was developed to study the atomic system under consideration and the performance of the QT protocols by providing the necessary elements to perform the reconstructions. The code basically solves the generalized Heisenberg picture master equation, discussed in Chapter \ref{ch:Atoms}, for any particular set of control parameters and piecewise constant waveforms and returns the Heisenberg evolved measurement operators. This code is used in the laboratory to analyze the recorded data.

With these tools in hand, we proceeded to simulate the measurement record for several relevant states, with emphasis on limitations, challenges, and the steps needed to make our protocol reliable and stable. In the case of full control on the 16-dimensional spin system via RF and microwave driving, we simulated noisy measurement records and used these as inputs to our reconstruction algorithms.  We found that we can rapidly achieve average fidelities $>$0.98 at the same noise level for a measurement time of only 2 ms.  These initial results bode well for high-fidelity reconstruction of a quantum state in such a large Hilbert space. With such a tool, we can explore the implementation of qudit unitary transformations for quantum information processing \cite{merkel09}, and nontrivial dynamics, for example in quantum chaos \cite{chaudhury09}, in efficient ways. 

As a first test of a real application of these QT protocols, we studied the fidelity of preparation of an arbitrary state in an ensemble of ultra-cold cesium atoms. In the experiments, a target state was prepared by a specific set of control waveforms designed using the optimal control techniques described in \cite{merkel08}. Then, the system was driven by the control waveforms pictured in Chapter \ref{ch:Simulations} while being probed by a detuned laser field. The measured polarization data wass then analyzed using the reconstruction code shown in Appendix \ref{App:Code}, which gives an estimate of the initially prepared quantum state. Some examples of the procedure are presented in Chapter \ref{ch:Results}, intended to illustrate the performance of QT in the laboratory. The results show not only excellent agreement between theory and experiment, but also demonstrate high-fidelity QT via continuous measurement and control of a fairly large Hilbert space. Fidelities $>$0.95 are observed for particular pure states, whereas for arbitrary Haar random pure states an average of $>$0.92 was obtained. These results indicate that our QT methods can reliably be used as a diagnostic tool for other experimental applications in quantum information science and technology.

Additionally, we have studied measurement records that are derived by stroboscopically measuring the expectation values of a single observable of a system that is evolving under the repeated application of a single unitary map, as discussed in Chapter \ref{ch:RandomUnitaries}.  We have shown that this record never contains complete information about the quantum state. However, for unitary maps chosen randomly or pseudorandomly, only a vanishing fraction of the information is missing.  When combined with the constraint of positivity, this incomplete measurement record led to a protocol for quantum state reconstruction that had high-fidelity performance for typical mixed and pure quantum states.  For pure states we can achieve unit fidelity reconstruction (in the absence of noise) and for mixed states the fidelity is greater than 0.99 for $d>9$.

 A particular set of pseudo-random matrices we studied in some detail are the Floquet maps generated by the quantum delta-kicked top.  In the general case, these maps appear to be equally as effective for reconstruction as Haar-random unitary maps.  In cases where the kicked top map exhibited additional symmetry, we were able to see that our reconstruction protocol required the extra constraint that the measurement operator and states shared the symmetry as well.  A map that is chaotic on the whole phase space saturates the bound of pseudo-random unitary operators on the whole Hilbert space.

It is surprising that such a simple measurement protocol should lead to such  good average reconstruction fidelities.  The reason for this appears to be a combination of the mixing power of random evolutions and the constraints on state space associated with positivity.  We do not yet have a rigorous explanation of these results, however, because the set of positive operators is a convex cone rather than a vector subspace~\cite{zyczkowski2006}, and thus it is difficult to quantify the volume of states that have both large support in the missing subspace of $\su(d)$ and are positive.  Our conjecture is that both the increasing average fidelity and the decreasing dependence on the sampled unitary can be explained based on concentration of measure in analogy with the work in~\cite{gross10}.  Essentially, most randomly sampled states have very little support on the subspace that we do not measure.  Related  work in the field of matrix completion has shown that such ``incoherence'' between states and (incomplete) measurements can provably lead to high-fidelity state reconstructions especially when the states in question are low rank~\cite{gross11} or near to low-rank states~\cite{gross10}.  Irrespective of a rigorous proof, it is an empirical fact that our protocol works well with typical states and typical unitary maps.  We expect that as the dimension of the Hilbert space increases, almost all of the states and unitary evolutions that are sampled will be very close to a typical value, resulting in high QT fidelity. 

Another important consequence of the positivity constraint  seems to be the fact that in practice, least squares and compressed sensing perform comparably well when all experimental errors are properly modeled. From the results in \cite{gross10}, which proved that the number of measurements needed to reconstruct a high-purity state using compressed sensing is $\ll d^2$, one tends to believe that this method would achieve high-fidelity estimates in a shorter time than least squares/maximum likelihood methods. We have certainly not seen evidence of such behavior either in simulations or in the experiment. Moreover, we see that both methods give high-fidelity estimates roughly at the same time. We believe that this is a signature of the positivity constraint. In fact, positivity seems to limit the set of density matrices that are compatible with the data in such a strong way that the advantage in number of measurements needed by compressed sensing disappears. We are still trying to give a quantitative explanation of this fact.

Our approach to quantum tomography could be improved in a number of ways.  While we have found that random waveforms are sufficient for generating informationally complete measurement records, the nature of optimal waveforms (in time and/or average fidelity) remains open.  Additionally, as the protocol has analogies with classical stochastic estimation, we see potential for improving the reliability and stability of the reconstruction procedure by employing data processing tools such as Kalman filters \cite{maybeck79} and other methods of estimation theory, which we plan to explore in future studies. The question of generalizing these methods to situations in which quantum disturbance induced by the measurement process (backaction) becomes important is a possible path to continuing research in this subject. Another avenue for generalization is to develop a model and algorithm to attempt to measure the many-body state of a system through continuos measurement. Although such an approach would be intractable for moderately large numbers of particles, one can imagine approximations that allow us to estimate some particular aspect of the many-body state, e.g., correlation functions. In addition, an ultimate dream is  to implement some version of real-time adaptive QT, which, based on the current knowledge of the quantum state, would choose appropriately the next point in the control waveforms. Such a method would use more efficiently the measured data, which could make it faster and more reliable. To do so, we must have exquisite control of the system so that we do not need to fit to imperfections after the fact and a fast integrator that can respond rapidly to the real-time measurement.  Though these requirements are currently out of reach, given the rapid progress already made in theory and experiment, it is not out the question that we can push into that regime.



\appendix


	\chapter{Random Matrix Generation Methods}\label{App:RandomStates}
In this dissertation, we relied in the generation of random states and random unitary matrices to test the performance of our QT methods. We used random states generated from the Haar, Hilbert-Schmidt and Bures measures, which give us the ability to sample pure and mixed states. We employed known methods to numerically generate such states which we briefly state in this appendix \cite{Osipov09, zyczkowski01,zyczkowski11,mezzadri07,zyczkowski2006}.

Additionally, we developed two naive procedures to generate mixed states of a given rank or purity. Although those methods are not in any way optimal, we found them to be useful to test different aspects of the compressed sensing quantum tomography and illustrate its main difference with respect to the traditional least squares method. We include a brief discussion of these methods in this appendix.

\section{Random unitary matrix from the Haar measure}
For the applications discussed in Chapter \ref{ch:RandomUnitaries}, we generated unitary random matrices from the Haar measure by using a method similar to the one described in \cite{mezzadri07}. Algorithmically, we proceed as follows: 1. Generate a random matrix $A$ from the Ginibre ensemble (see for example \cite{zyczkowski11}) which is the ensemble of matrices whose entries are complex numbers drawn from a Gaussian distribution with zero mean and unit variance. 2. Normalize the first column of $A$ and apply the Gram-Schmidt orthogonalization method to the rest of the columns of $A$. An example of a Matlab code that produces such random unitaries is given below.

\lstset{basicstyle=\scriptsize,tabsize=2}
\begin{lstlisting}[language=Matlab]
function U = RandomUnitary(d)
%Returns a random unitary matrix of dimension d 
%picked from the Haar measure
A = randn(d,d)+1i*randn(d,d);
U = zeros(d);
for j=1:d
      for k=1:j-1
            A(:,j) = A(:,j)-U(:,k)*(U(:,k)'*A(:,j));
      end
      U(:,j) = A(:,j)/norm(A(:,j));
end
\end{lstlisting}

\section{Pure states from the Haar measure}
Generation of uniformly distributed random quantum pure states  is equivalent to randomly sampling points from a sphere of unit radius. The uniform measure on such a sphere is the Haar measure (for a simple and pedagogical description see \cite{mezzadri07}). It turns out that a normalized column of a matrix from the Ginibre ensemble has these properties. This is the method we use here. Algorithmically, a pure Haar-random state is found as follows; 1. The real and imaginary parts of a random-complex vector $v$ are sampled from a Gaussian distribution with zero mean and unit variance. 2. Vector $v$ is then normalized. $v$ found this way is distributed uniformly in the unit sphere. The Matlab code used in this work is shown below.

\begin{lstlisting}[language=Matlab]
function rho = HaarRandomState(d)
%Returns a random pure state of dimension d 
%picked from the Haar measure
v = randn(d,1)+1i*randn(d,1);
v = v/norm(v);
rho = v*v';
\end{lstlisting}

\section{Mixed states from the Hilbert-Schmidt measure}
The Hilbert-Schmidt inner product, defined as $\langle A,B\rangle=\Tr{(A^\dagger B)}$ for $A$, $B$ in a Hilbert space, induces a measure on the set of mixed states. Following \cite{Osipov09, zyczkowski11}, a simple algorithm is given for generating random mixed states from this measure; 1. Generate a matrix $A$ randomly from the Ginibre ensemble. 2. The matrix $\frac{AA^\dagger}{\Tr(AA^\dagger)}$ is a unit trace, Hermitian, and positive semidefinite matrix sampled from the Hilbert-Schmidt measure. The Matlab code that achieves this is given below.

\begin{lstlisting}[language=Matlab]
function rho = HSRandomState(d)
%Returns a random mixed state of dimension d 
%picked from the Hilbert-Schmidt measure
A = randn(d,d)+1i*randn(d,d);
rho = A*A';
rho = rho/trace(rho);
\end{lstlisting}

\section{Mixed states from the Bures measure}
The quantum fidelity, defined in Eq. (\ref{eq:fidelity}), induces the Bures distance, given by $D(A,B)=\sqrt{2(1-\sqrt{\mathcal{F}(A,B)})}$ for $A$, $B$ valid quantum states, which induces a measure in the set of mixed states. Following \cite{Osipov09}, a simple algorithm is given for generating random mixed states from this measure; 1. Generate a matrix $A$ randomly from the Ginibre ensemble. 2. Generate a random unitary matrix $U$ from the Haar measure. 3. The matrix $\frac{(I+U)AA^\dagger(I+U^\dagger)}{\Tr((I+U)AA^\dagger(I+U^\dagger) )}$ is a unit trace, Hermitian, and positive semidefinite matrix sampled from the Bures measures. The Matlab code that achieves this is given below.
\begin{lstlisting}[language=Matlab]
function rho = BuresRandomState(d)
%Returns a random mixed state of dimension d 
%picked from the Bures measure
A = randn(d,d)+1i*randn(d,d);
U = RandomUnitary(d);
rho = A + U*A;
rho = rho*rho';
rho = rho/trace(rho);
\end{lstlisting}

\section{Fixed-rank random mixed states}\label{App:FixedRankRandom}
In order to test the compressed sensing algorithm, we performed numerical experiments in which the initial state was randomly sampled from an ensemble of density matrices of fixed rank but arbitrary purity. We showed this results in Section \ref{sec:RankVsPurity}. Algorithmically, we proceed as follows; 1. Fix the rank, $r$, of the quantum state. 2. Choose $r$ random numbers uniformly distributed in the interval $[0,1]$; these random numbers are the unnormalized eigenvalues of the density matrix, $\tilde{\lambda}_k$, for $k=1,\ldots,r$. 3. Normalize the random numbers found in the previous step and obtain the eigenvalues of the density matrix, $\lambda_k=\tilde{\lambda}_k/\sum_{j=1}^r\tilde{\lambda}_j$, for  $k=1,\ldots,r$. 4. Generate a Haar random unitary matrix, $U$, which is assumed to be written in the $\{ \ket{F,m_F}\}$ basis. 5. Define the random state with fixed rank as $\rho_0=U^\dagger {\rm diag}(\lambda_1,\ldots,\lambda_r,0,\ldots,0)U$.  Clearly, this procedure finds some set of random initial density matrices of random purity for a given rank. The Matlab code used is shown below. 

\begin{lstlisting}[language=Matlab]
function [A]=FixedRankRandomMatrix(r)
%Returns a random density matrix A of arbitrary purity and rank r

d = 16;
TestRank = 0;
while TestRank ~= r
    AuxIndex = zeros(r,1);
    for j = 1:r
        AuxIndex(j) = randi(d);
    end
    lambda=zeros(r,1);
    for j = 1:r
        lambda(j) = rand(1);
    end
    lambda=lambda/sum(lambda);
    AuxA=zeros(d);
    for j=1:r        
        B = zeros(d);
        B(AuxIndex(j),AuxIndex(j))=1;       
        AuxA = AuxA + lambda(j)*B;
    end    
    TestRank = rank(AuxA);
end
U=zeros(d);
for k = 1:d
    vv = randn(d,1)+1i*randn(d,1); 
    for m = 1:k-1
        vv = vv-U(:,m)*(U(:,m)'*vv);
    end
    U(:,k) = vv/sqrt(vv'*vv);
end 
A=U*AuxA*U';
\end{lstlisting}

\section{Fixed-purity random mixed states}\label{App:FixedPurityRandom}
The results shown in Section \ref{sec:RankVsPurity} were generated as follows. Let $\mathcal{P}$ be the purity and $r$ be the rank of a density matrix $\rho$. Let $\{\lambda_j\}$ be the eigenvalues of $\rho$. Then, one can write the purity
\be
\mathcal{P}=\sum_{j=1}^r\lambda_j^2,
\label{app:eq:purity}
\ee
and the normalization condition
\be
1=\sum_{j=1}^r\lambda_j.
\ee
Using this normalization condition in Eq. (\ref{app:eq:purity}) to eliminate the $j=r$ eigenvalue and explicitly summing over $j=1$, one can write
\be
-2\lambda_1^2+2\left(1-\sum_{j=2}^{r-1}\lambda_j\right)\lambda_1+\mathcal{P}-\sum_{j=2}^{r-1}\lambda_j^2-\left(1-\sum_{j=2}^{r-1}\lambda_j\right)^2=0.
\ee
One needs to find a real $\lambda_1$ from this equation. Therefore, its discriminant must be positive, which leads to a condition for the $\{\lambda_j\}$ for $j=2,\ldots,r-1$
\be
-\left(1-\sum_{j=2}^{r-1}\lambda_j\right)^2+2\left(\mathcal{P}-\sum_{j=2}^{r-1}\lambda_j^2\right)\ge0,
\ee
which in turn is a quadratic equation in $\lambda_2$. We thus arrive to a recurrent set of quadratic inequalities that have the following form
\be
-(1-k)\lambda_k^2+2\left(1-\sum_{j=k+1}^{r-1}\lambda_j\right)\lambda_k+k\left(\mathcal{P}-\sum_{j=k+1}^{r-1}\lambda_j^2\right)-\left(1-\sum_{j=k+1}^{r-1}\lambda_j\right)^2\ge0,
\label{app:eq:recurrence}
\ee
for $k=r-1,r-2,\ldots,1$.

By calculating the positive roots, $\lambda_k^{+}$, of inequalities (\ref{app:eq:recurrence})  and randomly sampling the eigenvalues $\lambda_k\le\lambda^{+}_k$ one finds a set of $\{\lambda_j\}$ that fulfills Eq. (\ref{app:eq:purity}). However, there is no guaranty that the eigenvalues are positive. Enforcing that condition seems to be a very complicated problem which we only partially solved. For simplicity, we include the condition that the sum of two consecutive eigenvalues $\lambda_k+\lambda_{k-1}<1$. This condition gives a further restriction on the $k^{th}$ eigenvalue
\be
-2\lambda_k+2\left(1-\sum_{j=k+1}^{r-1}\lambda_j\right)\lambda_k+\left(\mathcal{P}-\sum_{j=k+1}^{r-1}\lambda_j^2\right)-\left(1-\sum_{j=k+1}^{r-1}\lambda_j\right)^2<0.
\ee
Finding the positive root of this inequality, $\lambda_k^{'+}$, we can sample randomly $\lambda_k$ from the interval $[\lambda_k^{'+}, \lambda_k^{+}]$. Once all $\lambda_j$ are sampled in this manner and tested to be positive, a Haar-random unitary matrix is calculated and $\rho=U^\dagger\boldsymbol{\lambda}U$. 

Clearly, this method is not the optimal way to generate a density matrix of fixed purity. Moreover, since the positivity constraint of the eigenvalues is only partial, we sometimes get negative $\lambda_j$ and thus have to throw away those eigenvalues. We generated 75000 matrices, 1000 for each rank $r=2,\ldots, 16$ for 5 different purities $\mathcal{P}=0.5,~0.6,~0.7,~0.8,~\text{and }0.9$. Out of the 75000 matrices this method generates, there is a small fraction of matrices that do not represent physical density matrices. Then, for a given purity, we choose randomly a particular rank and a particular matrix out of those 75000 matrices and check if it is a proper density matrix. If not, we pick another one of the same rank until we get a good density matrix. The procedure is repeated for all purities until we obtain 1000 random density matrices for each purity. The Matlab code that generates a fixed-purity density matrix is shown below.

\begin{lstlisting}[language=Matlab]
function [A,lambda]=FixedPurityRandomMatrix(p,r)
%This function returns most of the time a random density matrix of fixed purity p.

d=16;
if r>2   
    %For r>2
    lambda = zeros(r,1);
    for k=r-1:-1:1
        SumLambda = 0;
        SumLambdaSqr = 0;
        for j=k+1:r-1
            SumLambda = SumLambda + lambda(j);
            SumLambdaSqr = SumLambdaSqr + lambda(j)^2;
        end
        a = -(k+1);
        b = 2*(1 - SumLambda);
        c = k*(p - SumLambdaSqr) - (1 - SumLambda)^2;
        a2 = -2;
        b2 = 2*(1 - SumLambda);
        c2 = (p - SumLambdaSqr) - (1 - SumLambda)^2;
        rootH = (-b-sqrt(b^2-4*a*c))/(2*a);
        rootL = (-b2-sqrt(b2^2-4*a2*c2))/(2*a2);
        if k==1;
            lambda(k)=rootH;
        else
            if isreal(rootL);
                lambda(k)=rootL+(rootH-rootL)*rand(1);
            else
                if 1-sum(lambda) > rootH
                    lambda(k)=rootH*rand(1);
                elseif 1-sum(lambda) < rootH
                    lambda(k)= (1-sum(lambda))*rand(1);
                end
            end
        end
    end
    lambda(r)=1-sum(lambda);
    AuxIndex = zeros(r,1);
    for j = 1:r
        AuxIndex(j) = j;
    end
    AuxA=zeros(d);
    for j=1:r
        B = zeros(d);
        B(AuxIndex(j),AuxIndex(j))=1;
        AuxA = AuxA + lambda(j)*B;
    end
    if isreal(lambda)
        if lambda>=0
            U=zeros(d);
            for k = 1:d
                vv = randn(d,1)+1i*randn(d,1);
                for m = 1:k-1
                    vv = vv-U(:,m)*(U(:,m)'*vv);
                end
                U(:,k) = vv/sqrt(vv'*vv);
            end
            A=U*AuxA*U';
        else
            A=zeros(16);
        end
    else
        A=zeros(16);
    end
elseif r==2
    lambda(1)=(1+sqrt(1-2*(1-p)))/2;
    lambda(2)=1-lambda(1);AuxA=zeros(d);
    AuxIndex = zeros(r,1);
    for j = 1:r
        AuxIndex(j) = j;
    end
    for j=1:2
        B = zeros(d);
        B(AuxIndex(j),AuxIndex(j))=1;
        AuxA = AuxA + lambda(j)*B;
    end
    U=zeros(d);
        for k = 1:d
        vv = randn(d,1)+1i*randn(d,1);
        for m = 1:k-1
            vv = vv-U(:,m)*(U(:,m)'*vv);
        end
        U(:,k) = vv/sqrt(vv'*vv);
    end
    A=U*AuxA*U';
end
\end{lstlisting}

\chapter{Quantum Tomography and Simulation Code}\label{App:Code}
An extensive part of the work done in this dissertation was to develop a simulation code to study the dynamics of the hyperfine-ground manifolds of cesium atoms under the RF and microwave control scheme discussed in Chapter \ref{ch:Atoms} for realistic situations. Moreover, this code had to have the ability to produce the Heisenberg-picture evolved measurement operators $\cO(t)$ and use them to reconstruct arbitrary states. In this appendix, we show the software that achieves these objectives. The code was written in Matlab and whenever possible, parallel computing techniques were used to speed up the complicated and lengthy calculations. All simulations and data analysis shown in this dissertation were done in an 12-core 2.66 GHz MacPro running Matlab 7.11.0 (R2010b). 

We begin by showing the part of the program that allows us to estimate the unknown control parameters as described in Section \ref{sec:ParameterEstimation}. Then, we present the code that simulates the dynamics of the atomic system in either the Schr\"{o}dinger or the Heisenberg picture. The first running mode is used to simulate the \emph{forward problem}, i.e., given an initial state $\rho_0$ and a particular set of control waveforms, light intensity, and detuning, it finds the time-evolved state of the system $\rho(t)$ and simulates what the Faraday rotation signal looks like if such a measurement were to be performed on the system. The second running mode of the code is used to solve the \emph{inverse problem}, i.e., estimating the unknown density matrix at the initial time from the knowledge of the measurement record, which is the main part of this work. It produces a set of operators $\cO(t)$, independent of the initial state of the system, which are used in the reconstruction segment of the code. The tomographic reconstruction part of the code is presented in Section \ref{sec:AppQTcode}. It includes our least squares and compressed sensing methods to solve the inverse problem, i.e., it finds an estimate of the initial state of the system using the dynamics and the measured data.

\section{Parameter estimation code}
\subsection{Preparing data}
Before QT is possible, we need to do parameter estimation to determine the unknown control parameters as discussed in Section \ref{sec:ParameterEstimation}. To do so, we first correct the signals that come from the laboratory corresponding to the Larmor precession and Rabi flopping experiments, i.e., we correct for any DC offset in the signal due to an unbalanced detector, find the actual initial time of the signal which can be undetermined, and filter the signal to improve signal-to-noise. The Matlab script we wrote for that purpose is \texttt{OffsetCorrection.m}, which we transcribe below.
\clearpage
\subsubsection{DC offset correction and filtering}
\begin{lstlisting}[language=Matlab]
%This script corrects for the DC offset the experimental signal has
%due to an unbalanced detector

%INPUT DATA: RF Larmor experiments with magnetic fields in the x and y
%directions as well as the Rabi flop experiment
%Flip the signs of the experimental data to match model
SignalRFxRaw=-dataRFx(7:5000,2);
SignalRFyRaw=-dataRFy(7:5000,2);
SignaluWRaw=-datauWRabi(7:5000,2);

FM = fft(SignalRFxRaw)/length(SignalRFxRaw);
Offsetx = FM(1);
SignalRFx=SignalRFxRaw-FM(1);
SignalRFxF=filter(bzRF,azRF,SignalRFx);

FM = fft(SignalRFyRaw)/length(SignalRFyRaw);
Offsety = FM(1);
SignalRFy=SignalRFyRaw-FM(1);
SignalRFyF=filter(bzRF,azRF,SignalRFy);

FM = fft(SignaluWRaw)/length(SignaluWRaw);
SignaluW=SignaluWRaw33-(Offsetx+Offsety)/2;
SignaluWF=filter(bzuW,azuW,SignaluW);
\end{lstlisting}

Before starting the fitting process, in order to save time in the computation, we calculate all the time-independent operators needed to simulate and fit the Larmor precession and Rabi flopping experimental signals. To do so, we run the following scripts \texttt{RFLarmorPreparation.m} and  \texttt{uWRabiPreparation.m}.

\subsubsection{Time independent and decoherence terms for RF Larmor experiment}
\begin{lstlisting}[language=Matlab]
%Larmor Experiment with bias field, decoherence and RF in the x or y direction.
%This script calculates all necessary time independent operators for the RF Larmor
%calibration experiments

%AuxVar1=F, restricts evolution to manifold F=3 or 4
AuxVar1=3;
%Define OmegaRF to be resonant with F=3 manifold.
omegaRF = 1003.2; %kHz
Omega = 1;  %DO NOT CHANGE   
%Define \Delta_{4'3}=detuning of laser field
detuning = -730;%MHz %Delta 4'3 
%Define the polarization of the light vector (eL) and the three different
%polarizations for scattered light (eq) in the spherical basis for
%q=-1,0,+1, in the columns of (eq), for example, eq(:,2)=ez=[0 0 1].
eL = [1;0;0];
eq = [  1/sqrt(2) 0 -1/sqrt(2);
      -1i/sqrt(2) 0 -1i/sqrt(2);
            0     1   0       ];
%Define the total angular momentum manifold dimensions for the hyperfine
%ground states F=4 and F=3.      
F4 = 4;
F3 = 3;
dimF3 = 2*F3+1;
dimF4 = 2*F4+1;
dim = dimF3+dimF4;
%Compute the Angular momentum operators (Fx, Fy, Fz) for the F=3 and F=4 manifolds
%by using the function AngMomentum.m. Note that this operators are written in
%the basis in which Fz is diagonal and the the basis is ordered from
%|F,-F> to |F,+F>.
[Fx4,Fy4,Fz4] = AngMomentum(F4);
[Fx3,Fy3,Fz3] = AngMomentum(F3);
grel = -1.0032; %g_3/g_4 ratio of g-factors 
%Define the Linewidth (GammaLW) for the D1 transition as well as the
%the ground state splitting (GrSplit) (which in the paper is called E0). 
%In addition define j' (pj) as the electronic angular momentum (electron + 
%orbital in l=0 case). For the D1 line there are 2 excited manifolds P1/2, 
%F'=3 and F'=4, separated by the excited state splitting (ExSplit), whose
%total dimesion is 16. Finally (ExBasis) and (GrBasis) represent the basis
%vectors for the excited and ground manifolds ordered from -mf to mf. In
%particular, (GrBasis) is picked in this order:
%|4,-4>,...,|4,+4>,|3,-3>,...,|3,+3>.
GammaLW = 4.561; %MHz
ExSplit = 1167.6;%MHz
GrSplit = 9.19e3;%MHz
pJ = 1/2; 
Is = 7/2;
ExDim = 16;
GrDim = dim;
ExBasis = eye(ExDim);
GrBasis = eye(GrDim);     
%Define the detuning matrix \Delta_{F'F}. The value of the variable
%(detuning) is assigned to the detuning from F=3 to F'=4.
%%DELTA F'F
DeltapFF = zeros(2,2);
DeltapFF(1,2) = detuning; %Delta 4'3
DeltapFF(2,2) = (DeltapFF(1,2)+ExSplit); %Delta 3'3
DeltapFF(1,1) = DeltapFF(1,2)+GrSplit; %Delta 4'4
DeltapFF(2,1) = DeltapFF(2,2)+GrSplit;%Delta 3'4
%To be used later in the definition of the measurement basis, here we
%calculate the irreducible tensor coefficients C_{j'F'F}^{(k)} as they
%appear in \cite{deutsch09}. The function Wigner6jcoeff.m by Amita B Deb,
%Clarendon Lab. 2007, is used to calculate the Wigner 6J symbols that
%appear in the definition of the coefficients.
%%TENSOR COEFFICIENTS
C1 = zeros(2);
C2 = zeros(2);
C0 = zeros(2);
for pF = 4:-1:3
    for F = 4:-1:3
        C0(5-pF,5-F) = (-1)^(3*F-pF+1)*sqrt(1/3)*(2*pF+1)/(sqrt(2*F+1))...
        *Wigner6jcoeff(F,1,pF,1,F,0)*(2*pJ+1)*...
        (2*F+1)*abs(Wigner6jcoeff(pF,7/2,pJ,1/2,1,F))^2;
        C1(5-pF,5-F) = (-1)^(3*F-pF)*sqrt(3/2)*(2*pF+1)/(sqrt(F*(F+1)*(2*F+1)))*...
        Wigner6jcoeff(F,1,pF,1,F,1)*(2*pJ+1)*(2*F+1)*...
        abs(Wigner6jcoeff(pF,7/2,pJ,1/2,1,F))^2;
        C2(5-pF,5-F) = (-1)^(3*F-pF)*sqrt(30)*(2*pF+1)/...
        (sqrt(F*(F+1)*(2*F+1)*(2*F-1)*(2*F+3)))*...
        Wigner6jcoeff(F,1,pF,1,F,2)*(2*pJ+1)*(2*F+1)*...
        abs(Wigner6jcoeff(pF,7/2,pJ,1/2,1,F))^2;
    end
end
%Now we start calculating the operators that appear in the master equation.
%We do so by following the "paper" and using a similar notation.

%%MASTER EQUATION OPERATORS%%
%The first thing we calculate is the dimensionless dipole rising operator
%\vecD_{F'F}^{\dagger} (DDq). The last 3 indices are: F', F, amd q. Inside
%the for loops, there are 2 rules, (j) and (k), to asign the right value to the
%right |F',m'><F,m| component. Additionally, we calculate (Aux_eL_DDq)
%which represents \vec{\epsilon}_L\dot\vecD_{F'F}^{\dagger}. The function
%squeeze is used to elimininate any 1D dimension in the array.
DDq = zeros(ExDim,GrDim,2,2,3);
for q = -1:1:1
    for pF = 4:-1:3
        for F = 4:-1:3
            for MF = -F:1:F
                if abs(MF+q) <= pF
                   j = (pF+1)+MF+q+(4-pF)*(2*4+1);
                   k = (F+1)+MF+(F4-F)*(2*F4+1);
                   DDq(:,:,-pF+5,-F+5,q+2) = DDq(:,:,-pF+5,-F+5,q+2)+...
                   (-1)^(pF+1+pJ+Is)*sqrt((2*pJ+1)*(2*F+1))*...
                   Wigner6jcoeff(pF,7/2,pJ,1/2,1,F)*...
                   ClebschGordan(F,1,pF,MF,q,MF+q)*ExBasis(:,j)*GrBasis(k,:);
                end
            end
        end
     end
end   
Aux_eL_DDq = squeeze(dot(eL,conj(eq(:,1)))*DDq(:,:,:,:,1)+...
dot(eL,conj(eq(:,2)))*DDq(:,:,:,:,2)+dot(eL,conj(eq(:,3)))*DDq(:,:,:,:,3));
%Now that the main operators are defined, we proceed to calculate the
%effective light-shift Hamiltonian H_{eff}^{LS}, (HlsEFF). Since I defined
%the time everywhere else to be in mileseconds, we have to rescale the
%effective Hamiltonian to be in kHz instead of MHz as it was before.
%%EFFECTIVE LIGHT SHIFT HAMILTONIAN
HlsEFF = zeros(GrDim,GrDim);
for F = 4:-1:3
    for pF = 4:-1:3
        AuxHls = squeeze(Aux_eL_DDq(:,:,-pF+5,-F+5));
        HlsEFF = HlsEFF +  1/(DeltapFF(-pF+5,-F+5)+1i*GammaLW/2) * AuxHls' * AuxHls;
    end
end
HlsEFF = (Omega/2)^2*HlsEFF;       
HlsEFF = 10^3*HlsEFF; %needs to be in kHz

%GOING TO ROTATING FRAME and linear interpolation + integration
%The rotating frame is defined by the unitary
%U=exp(-i\omega_RFt(F_z^{(4)}-F_z^{(3)})). In the program
%Fzm=F_z^{(4)}-F_z^{(3)}.
Fzm = zeros(dim);
Fzm(1:9,1:9) = Fz4;
Fzm(10:16,10:16) = -Fz3; 
%After going to the rotaing frame, we must apply the RWA. The way we are
%doing so in the code is by explicitly calculating the average of
%U^{\dagger}HU. For this purpose, we use linear interpolation to calculate
%the first order apporximation to the integral \int_0^Tf(t)dt, where f(t)
%is U^{\dagger}(t)H_{eff}^{LS}U(t). Note that the linear interpolation I
%use here gives
%\int_0^Tf(t)dt\approx\frac{1}{T}\sum_j\frac{dt}{2}(f(t_{j-1})+f(t_{j})).
%Also note that \frac{1}{T}=\omega_RF in this case. Finally, we transform
%the averaged effective light-shift Hamiltonian to the superoperator
%notation (SOHlsEFF). For this, we make use of the transformation A\rhoB\rightarrow
%(A\otimes B^T)\vec{rho}, which takes the operators A and B in the standard
%notation to the superoperator picture when they appear, as they do in the
%master equation, in products with the density matrix.

%%JUMP OPERATORS
%As we did with the effective light-shift Hamiltonian, we calculate the
%jump operators in a similar way. The array (W_q) is W_q^{F_bF_a} where the
%three last indices of (Wq) are Fa, Fb and q. In addition, we define the
%projectors onto the F=3 and F=4 manifolds (PP), and the superoperator jump
%operators (SOWq). We do the latter at the same time as going to the
%rotating frame and doing the RWA. The averaging method used here is the
%same linear interpolation method used above in the light-shift
%Hamiltonian. The reason for doing this part in this way is that there is
%no other way make the RWA on the jump operators individually since they
%always appear in the master equation as various products like, for
%example, W_q\rho W_q^{\dagger}, and we must average those products to
%corectly diregard the rapidly oscillating terms.
Wq = zeros(GrDim,GrDim,2,2,3);   
Aux_eL_DDq = squeeze(dot(eL,conj(eq(:,1)))*DDq(:,:,:,:,1)+...
dot(eL,conj(eq(:,2)))*DDq(:,:,:,:,2)+dot(eL,conj(eq(:,3)))*DDq(:,:,:,:,3));
for q = -1:1:1
    for Fb = 4:-1:3
        for Fa = 4:-1:3
            for pF = 4:-1:3
                AuxW = squeeze(DDq(:,:,-pF+5,-Fb+5,q+2));
                Wq(:,:,-Fa+5,-Fb+5,q+2) = Wq(:,:,-Fa+5,-Fb+5,q+2)+(Omega/2)/...
                (DeltapFF(-pF+5,-Fa+5)+1i*GammaLW/2)* AuxW'...
                *squeeze(Aux_eL_DDq(:,:,-pF+5,-Fa+5)); %
            end
        end
    end
end
if AuxVar1==4
    %project to F=4
    HlsEFF4 =HlsEFF(1:9,1:9);
    HlsEFF = HlsEFF4;
    jmax = 2000;%number of terms in the time average
    dt = 1/(omegaRF*(jmax)); %time step in time average
    AVGHlsEFF = zeros(9);
    for j = 1:jmax
        AVGHlsEFF = AVGHlsEFF+expm(+1i*2*pi*omegaRF*(j-1)*dt*Fz4)*HlsEFF*...
        expm(-1i*2*pi*omegaRF*(j-1)*dt*Fz4)+expm(+1i*2*pi*omegaRF*(j)*dt*Fz4)*...
        HlsEFF*expm(-1i*2*pi*omegaRF*(j)*dt*Fz4);
    end
    AVGHlsEFF = dt/2*AVGHlsEFF;
    AVGHlsEFF = omegaRF*AVGHlsEFF;
    SOHlsEFF = kron(AVGHlsEFF,eye(9))-kron(eye(9),conj(AVGHlsEFF));
    %project to F=4
    Wq4 =squeeze(Wq(1:9,1:9,1,1,:));
    Wq = Wq4;
    % Wq Superoperators
    SOWq=zeros(81,81);
    for j = 1:jmax
        Uj = expm(-1i*2*pi*omegaRF*(j-1)*dt*Fz4);
        Ujp = expm(-1i*2*pi*omegaRF*(j)*dt*Fz4);
        for q = -1:1:1
            func=kron(Uj'*squeeze(Wq(:,:,q+2))*Uj,Uj.'*conj(squeeze(Wq(:,:,q+2)))...
            *conj(Uj))+kron(Ujp'*squeeze(Wq(:,:,q+2))*Ujp,Ujp.'*...
            conj(squeeze(Wq(:,:,q+2)))*conj(Ujp));
            SOWq = SOWq+func;
        end
    end
    SOWq=GammaLW*10^3*SOWq;
    SOWq = dt/2*SOWq;
    SOWq = omegaRF*SOWq;
elseif AuxVar1==3
    %project to F=3
    HlsEFF3 =HlsEFF(10:16,10:16);
    HlsEFF = HlsEFF3;
    jmax = 2000;%number of terms in the time average
    dt = 1/(omegaRF*(jmax)); %time step in time average
    AVGHlsEFF = zeros(7);
    for j = 1:jmax
        AVGHlsEFF = AVGHlsEFF+expm(-1i*2*pi*omegaRF*(j-1)*dt*Fz3)*HlsEFF*...
        expm(+1i*2*pi*omegaRF*(j-1)*dt*Fz3)+expm(-1i*2*pi*omegaRF*(j)*dt*Fz3)*...
        HlsEFF*expm(+1i*2*pi*omegaRF*(j)*dt*Fz3);
    end
    AVGHlsEFF = dt/2*AVGHlsEFF;
    AVGHlsEFF = omegaRF*AVGHlsEFF;
    SOHlsEFF = kron(AVGHlsEFF,eye(7))-kron(eye(7),conj(AVGHlsEFF));
    %project to F=3
    Wq3 =squeeze(Wq(10:16,10:16,2,2,:));
    Wq = Wq3;
    % Wq Superoperators
    SOWq=zeros(49,49);
    for j = 1:jmax
        Uj = expm(+1i*2*pi*omegaRF*(j-1)*dt*Fz3);
        Ujp = expm(+1i*2*pi*omegaRF*(j)*dt*Fz3);
        for q = -1:1:1
            func=kron(Uj'*squeeze(Wq(:,:,q+2))*Uj,Uj.'*conj(squeeze(Wq(:,:,q+2)))...
            *conj(Uj))+kron( Ujp'*squeeze(Wq(:,:,q+2))*Ujp,Ujp.'*...
            conj(squeeze(Wq(:,:,q+2)))*conj(Ujp) );
            SOWq = SOWq+func;
        end
    end
    SOWq=GammaLW*10^3*SOWq;
    SOWq = dt/2*SOWq;
    SOWq = omegaRF*SOWq;
else 
    jmax = 2000;%number of terms in the time average
    dt = 1/(omegaRF*(jmax)); %time step in time average
    AVGHlsEFF = zeros(16);
    for j = 1:jmax
        AVGHlsEFF = AVGHlsEFF+expm(+1i*2*pi*omegaRF*(j-1)*dt*Fzm)*HlsEFF*...
        expm(-1i*2*pi*omegaRF*(j-1)*dt*Fzm)+expm(+1i*2*pi*omegaRF*(j)*dt*Fzm)*...
        HlsEFF*expm(-1i*2*pi*omegaRF*(j)*dt*Fzm);
    end
    AVGHlsEFF = dt/2*AVGHlsEFF;
    AVGHlsEFF = omegaRF*AVGHlsEFF;
    SOHlsEFF = kron(AVGHlsEFF,eye(16))-kron(eye(16),conj(AVGHlsEFF));
    %projectors
    PP = zeros(dim,2);
    PP(1:9,1:9,1) = eye(dimF4); %Projector in F=4 manifold
    PP(10:16,10:16,2) = eye(dimF3);%Projector in F=3 manifold
    %%Going to the rotating frame
    %Wq Superoperators
    SOWq = zeros(dim^2,dim^2);
    for j = 1:jmax
        Uj = expm(-1i*2*pi*omegaRF*(j-1)*dt*Fzm);
        Ujp = expm(-1i*2*pi*omegaRF*(j)*dt*Fzm);
        for q = -1:1:1
            for F = 4:-1:3
                for F1 = 4:-1:3
                    func = kron(Uj'*squeeze(Wq(:,:,-F1+5,-F+5,q+2))*Uj*...
                    squeeze(PP(:,:,-F1+5)),Uj.'*conj(squeeze(Wq(:,:,-F1+5,-F+5,...
                    q+2)))*conj(Uj)*squeeze(PP(:,:,-F1+5)))+kron(Ujp'*...
                    squeeze(Wq(:,:,-F1+5,-F+5,q+2))*Ujp*squeeze(PP(:,:,-F1+5)),...
                    Ujp.'*conj(squeeze(Wq(:,:,-F1+5,-F+5,q+2)))*conj(Ujp)*...
                    squeeze(PP(:,:,-F1+5)));
                    SOWq = SOWq+func;
                end
            end
        end
    end
    for j = 1:jmax
        Uj = expm(-1i*2*pi*omegaRF*(j-1)*dt*Fzm);
        Ujp = expm(-1i*2*pi*omegaRF*(j)*dt*Fzm);
        for q = -1:1:1
            for F1 = 4:-1:3
                for F2 = 4:-1:3
                   if F1~=F2
                     func = kron(Uj'*squeeze(Wq(:,:,-F2+5,-F2+5,q+2))*Uj*...
                     squeeze(PP(:,:,-F2+5)),Uj.'*conj(squeeze(Wq(:,:,-F1+5,-F1+5,...
                     q+2)))*conj(Uj)*squeeze(PP(:,:,-F1+5)))+kron(Ujp'*...
                     squeeze(Wq(:,:,-F2+5,-F2+5,q+2))*Ujp*squeeze(PP(:,:,-F2+5)),...
                     Ujp.'*conj(squeeze(Wq(:,:,-F1+5,-F1+5,q+2)))*conj(Ujp)*...
                     squeeze(PP(:,:,-F1+5)));
                     SOWq = SOWq+func;
                   end
                end
            end
        end
    end
    SOWq = GammaLW*10^3*SOWq;
    SOWq = dt/2*SOWq;
    SOWq = omegaRF*SOWq;
end
%We created all the necessary operator that are going to be used later and
%save them in parametersRFuW.mat file.
save parametersRFLarmor SOHlsEFF SOWq F4 F3 dimF3 dimF4 dim Fx3 Fy3 Fz3 Fx4 Fy4...
Fz4 grel C1 C2 C0 DeltapFF AuxVar1 GrSplit omegaRF
clear
\end{lstlisting}

\subsubsection{Time independent and decoherence terms for Rabi flopping experiment }
\begin{lstlisting}[language=Matlab]
%Rabi Flopping Experiment with bias and uW fields and decoherence
%This script calculates all necessary time independent operators
%for the uW Rabi flop calibration experiments
AuxVar1=2;%16;
omegaRF = 1000; %kHz
Omega = 1;%DO NOT CHANGE
detuning = -730;%MHz %Delta 4'3 
%Define the polarization of the light vector (eL) and the three different
%polarizations for scattered light (eq) in the spherical basis for
%q=-1,0,+1, in the columns of (eq), for example, eq(:,2)=ez=[0 0 1].
eL = [1;0;0];
eq = [  1/sqrt(2) 0 -1/sqrt(2);
      -1i/sqrt(2) 0 -1i/sqrt(2);
            0     1   0       ];
%Define the total angular momentum manifold dimensions for the hyperfine
%ground states F=4 and F=3.      
F4 = 4;
F3 = 3;
dimF3 = 2*F3+1;
dimF4 = 2*F4+1;
dim = dimF3+dimF4;
%Compute the Angular momentum operators (Fx, Fy, Fz) for the F=3 and F=4 manifolds
%by using the function AngMomentum.m. Note that this operators are written in
%the basis in which Fz is diagonal and the the basis is ordered from
%|F,-F> to |F,+F>.
[Fx4,Fy4,Fz4] = AngMomentum(F4);
[Fx3,Fy3,Fz3] = AngMomentum(F3);
grel = -1.0032; %g_3/g_4 ratio of g-factors 
%Define the Linewidth (GammaLW) for the D1 transition as well as the
%the ground state splitting (GrSplit) (which in the paper is called E0). 
%In addition define j' (pj) as the electronic angular momentum (electron + 
%orbital in l=0 case). For the D1 line there are 2 excited manifolds P1/2, 
%F'=3 and F'=4, separated by the excited state splitting (ExSplit), whose
%total dimesion is 16. Finally (ExBasis) and (GrBasis) represent the basis
%vectors for the excited and ground manifolds ordered from -mf to mf. In
%particular, (GrBasis) is picked in this order:
%|4,-4>,...,|4,+4>,|3,-3>,...,|3,+3>.
GammaLW = 4.561; %MHz
ExSplit = 1167.6;%MHz
GrSplit = 9.19e3;%MHz
pJ = 1/2; 
Is = 7/2;
ExDim = 16;
GrDim = dim;
ExBasis = eye(ExDim);
GrBasis = eye(GrDim);     
%Define the detuning matrix \Delta_{F'F}. The value of the variable
%(detuning) is asigned to the detuning from F=3 to F'=4.
%DELTA F'F
DeltapFF = zeros(2,2);
DeltapFF(1,2) = detuning; %Delta 4'3
DeltapFF(2,2) = (DeltapFF(1,2)+ExSplit); %Delta 3'3
DeltapFF(1,1) = DeltapFF(1,2)+GrSplit; %Delta 4'4
DeltapFF(2,1) = DeltapFF(2,2)+GrSplit;%Delta 3'4
%To be used later in the definition of the measurement basis, here we
%calculate the irreducible tensor coefficients C_{j'F'F}^{(k)} as they
%appear in \cite{deutsch09}. The function Wigner6jcoeff.m by Amita B Deb,
%Clarendon Lab. 2007, is used to calculate the Wigner 6J symbols that
%appear in the definition of the coefficients.
%%TENSOR COEFICIENTS
C1 = zeros(2);
C2 = zeros(2);
C0 = zeros(2);
for pF = 4:-1:3
    for F = 4:-1:3
        C0(5-pF,5-F) = (-1)^(3*F-pF+1)*sqrt(1/3)*(2*pF+1)/(sqrt(2*F+1))*...
        Wigner6jcoeff(F,1,pF,1,F,0)*(2*pJ+1)*(2*F+1)*...
        abs(Wigner6jcoeff(pF,7/2,pJ,1/2,1,F))^2;
        C1(5-pF,5-F) = (-1)^(3*F-pF)*sqrt(3/2)*(2*pF+1)/(sqrt(F*(F+1)*(2*F+1)))*...
        Wigner6jcoeff(F,1,pF,1,F,1)*(2*pJ+1)*(2*F+1)*...
        abs(Wigner6jcoeff(pF,7/2,pJ,1/2,1,F))^2;
        C2(5-pF,5-F) = (-1)^(3*F-pF)*sqrt(30)*(2*pF+1)/...
        (sqrt(F*(F+1)*(2*F+1)*(2*F-1)*(2*F+3)))*Wigner6jcoeff(F,1,pF,1,F,2)*...
        (2*pJ+1)*(2*F+1)*abs(Wigner6jcoeff(pF,7/2,pJ,1/2,1,F))^2;
    end
end
%Now we start calculating the operators that appear in the master equation.
%We do so by following the "paper" and using a similar notation.

%%MASTER EQUATION OPERATORS%%
%The first thing we calculate is the dimensionless dipole rising operator
%\vecD_{F'F}^{\dagger} (DDq). The last 3 indices are: F', F, amd q. Inside
%the for loops, there are 2 rules, (j) and (k), to asign the right value to the
%right |F',m'><F,m| component. Additionally, we calculate (Aux_eL_DDq)
%which represents \vec{\epsilon}_L\dot\vecD_{F'F}^{\dagger}. The function
%squeeze is used to elimininate any 1D dimension in the array.
DDq = zeros(ExDim,GrDim,2,2,3);
for q = -1:1:1
    for pF = 4:-1:3
        for F = 4:-1:3
            for MF = -F:1:F
                if abs(MF+q) <= pF
                   j = (pF+1)+MF+q+(4-pF)*(2*4+1);
                   k = (F+1)+MF+(F4-F)*(2*F4+1);%
                   DDq(:,:,-pF+5,-F+5,q+2) = DDq(:,:,-pF+5,-F+5,q+2)+...
                   (-1)^(pF+1+pJ+Is) *sqrt((2*pJ+1)*(2*F+1))*...
                   Wigner6jcoeff(pF,7/2,pJ,1/2,1,F)*...
                   ClebschGordan(F,1,pF,MF,q,MF+q)*ExBasis(:,j)*GrBasis(k,:);
                end
            end
        end
     end
end
Aux_eL_DDq = squeeze(dot(eL,conj(eq(:,1)))*DDq(:,:,:,:,1)+...
dot(eL,conj(eq(:,2)))*DDq(:,:,:,:,2)+dot(eL,conj(eq(:,3)))*DDq(:,:,:,:,3));
%Now that the main operators are defined, we proceed to calculate the
%effective light-shift Hamiltonian H_{eff}^{LS}, (HlsEFF). Since I defined
%the time everywhere else to be in mileseconds, we have to rescale the
%effective Hamiltonian to be in kHz instead of MHz as it was before.

%%EFFECTIVE LIGHT SHIFT HAMILTONIAN
HlsEFF = zeros(GrDim,GrDim);
for F = 4:-1:3%%%%%%%%%%%%%%%%%%
    for pF = 4:-1:3
        AuxHls = squeeze(Aux_eL_DDq(:,:,-pF+5,-F+5));
        HlsEFF = HlsEFF +  1/(DeltapFF(-pF+5,-F+5)+1i*GammaLW/2) * AuxHls' * AuxHls;
    end
end
HlsEFF = (Omega/2)^2*HlsEFF;       
HlsEFF = 10^3*HlsEFF; %needs to be in kHz

%GOING TO ROTATING FRAME and linear interpolation + integration
%The rotating frame is defined by the unitary
%U=exp(-i\omega_RFt(F_z^{(4)}-F_z^{(3)})). In the program
%Fzm=F_z^{(4)}-F_z^{(3)}.
Fzm = zeros(dim);
Fzm(1:9,1:9) = Fz4;
Fzm(10:16,10:16) = -Fz3; 
%After going to the rotaing frame, we must apply the RWA. The way we are
%doing so in the code is by explicitly calculating the average of
%U^{\dagger}HU. For this purpose, we use linear interpolation to calculate
%the first order apporximation to the integral \int_0^Tf(t)dt, where f(t)
%is U^{\dagger}(t)H_{eff}^{LS}U(t). Note that the linear interpolation I
%use here gives
%\int_0^Tf(t)dt\approx\frac{1}{T}\sum_j\frac{dt}{2}(f(t_{j-1})+f(t_{j})).
%Also note that \frac{1}{T}=\omega_RF in this case. Finally, we transform
%the averaged effective light-shift Hamiltonian to the superoperator
%notation (SOHlsEFF). For this, we make use of the transformation A\rhoB\rightarrow
%(A\otimes B^T)\vec{rho}, which takes the operators A and B in the standard
%notation to the superoperator picture when they appear, as they do in the
%master equation, in products with the density matrix.

%%JUMP OPERATORS
%As we did with the effective light-shift Hamiltonian, we calculate the
%jump operators in a similar way. The array (W_q) is W_q^{F_bF_a} where the
%three last indices of (Wq) are Fa, Fb and q. In addition, we define the
%projectors onto the F=3 and F=4 manifolds (PP), and the superoperator jump
%operators (SOWq). We do the latter at the same time as going to the
%rotating frame and doing the RWA. The averaging method used here is the
%same linear interpolation method used above in the light-shift
%Hamiltonian. The reason for doing this part in this way is that there is
%no other way make the RWA on the jump operators individually since they
%always appear in the master equation as various products like, for
%example, W_q\rho W_q^{\dagger}, and we must average those products to
%correctly disregard the rapidly oscillating terms.
Wq = zeros(GrDim,GrDim,2,2,3); 
Aux_eL_DDq = squeeze(dot(eL,conj(eq(:,1)))*DDq(:,:,:,:,1)+...
dot(eL,conj(eq(:,2)))*DDq(:,:,:,:,2)+dot(eL,conj(eq(:,3)))*DDq(:,:,:,:,3));   
for q = -1:1:1
    for Fb = 4:-1:3
        for Fa = 4:-1:3
            for pF = 4:-1:3
                AuxW = squeeze(DDq(:,:,-pF+5,-Fb+5,q+2));
                Wq(:,:,-Fa+5,-Fb+5,q+2) = Wq(:,:,-Fa+5,-Fb+5,q+2)+(Omega/2)/...
                (DeltapFF(-pF+5,-Fa+5)+1i*GammaLW/2)* AuxW'*...
                squeeze(Aux_eL_DDq(:,:,-pF+5,-Fa+5)); 
            end
        end
    end
end
if AuxVar1==4 
    %project to F=4
    HlsEFF4 =HlsEFF(1:9,1:9);
    HlsEFF = HlsEFF4;    
    jmax = 2000;%number of terms in the time average
    dt = 1/(omegaRF*(jmax)); %time step in time average
    AVGHlsEFF = zeros(9);
    for j = 1:jmax
        AVGHlsEFF = AVGHlsEFF+expm(+1i*2*pi*omegaRF*(j-1)*dt*Fz4)*HlsEFF*...
        expm(-1i*2*pi*omegaRF*(j-1)*dt*Fz4)+expm(+1i*2*pi*omegaRF*(j)*dt*Fz4)*...
        HlsEFF*expm(-1i*2*pi*omegaRF*(j)*dt*Fz4);
    end
    AVGHlsEFF = dt/2*AVGHlsEFF;
    AVGHlsEFF = omegaRF*AVGHlsEFF;
    SOHlsEFF = kron(AVGHlsEFF,eye(9))-kron(eye(9),conj(AVGHlsEFF));
    %project to F=4
    Wq4 =squeeze(Wq(1:9,1:9,1,1,:));
    Wq = Wq4;    
    % Wq Superoperators
    SOWq=zeros(81,81);   
    for j = 1:jmax
        Uj = expm(-1i*2*pi*omegaRF*(j-1)*dt*Fz4);
        Ujp = expm(-1i*2*pi*omegaRF*(j)*dt*Fz4);
        for q = -1:1:1
           func=kron( Uj'*squeeze(Wq(:,:,q+2))*Uj,Uj.'*conj(squeeze(Wq(:,:,q+2)))...
           *conj(Uj) )+kron( Ujp'*squeeze(Wq(:,:,q+2))*Ujp,Ujp.'*...
           conj(squeeze(Wq(:,:,q+2)))*conj(Ujp) );
           SOWq = SOWq+func;
        end
    end
    SOWq=GammaLW*10^3*SOWq;
    SOWq = dt/2*SOWq;
    SOWq = omegaRF*SOWq;
elseif AuxVar1==3   
    %project to F=3
    HlsEFF3 =HlsEFF(10:16,10:16);
    HlsEFF = HlsEFF3;    
    jmax = 2000;%number of terms in the time average
    dt = 1/(omegaRF*(jmax)); %time step in time average
    AVGHlsEFF = zeros(7);
    for j = 1:jmax
        AVGHlsEFF = AVGHlsEFF+expm(-1i*2*pi*omegaRF*(j-1)*dt*Fz3)*HlsEFF*...
        expm(+1i*2*pi*omegaRF*(j-1)*dt*Fz3)+expm(-1i*2*pi*omegaRF*(j)*dt*Fz3)*...
        HlsEFF*expm(+1i*2*pi*omegaRF*(j)*dt*Fz3);
    end
    AVGHlsEFF = dt/2*AVGHlsEFF;
    AVGHlsEFF = omegaRF*AVGHlsEFF;
    SOHlsEFF = kron(AVGHlsEFF,eye(7))-kron(eye(7),conj(AVGHlsEFF));    
    %project to F=3
    Wq3 =squeeze(Wq(10:16,10:16,2,2,:));
    Wq = Wq3;    
    % Wq Superoperators    
    SOWq=zeros(49,49);
    for j = 1:jmax
        Uj = expm(+1i*2*pi*omegaRF*(j-1)*dt*Fz3);
        Ujp = expm(+1i*2*pi*omegaRF*(j)*dt*Fz3);
        for q = -1:1:1
           func=kron( Uj'*squeeze(Wq(:,:,q+2))*Uj,Uj.'*conj(squeeze(Wq(:,:,q+2)))...
           *conj(Uj) )+kron( Ujp'*squeeze(Wq(:,:,q+2))*Ujp,Ujp.'*...
           conj(squeeze(Wq(:,:,q+2)))*conj(Ujp) );
           SOWq = SOWq+func;
        end
    end
    SOWq=GammaLW*10^3*SOWq;
    SOWq = dt/2*SOWq;
    SOWq = omegaRF*SOWq;
elseif AuxVar1==2 
    %project to 2-dimensional subspace
    HlsEFF2 = [HlsEFF(9,9) HlsEFF(9,16); HlsEFF(16,9) HlsEFF(16,16)];
    HlsEFF = HlsEFF2;
    jmax = 2000;%number of terms in the time average
    dt = 1/(omegaRF*(jmax)); %time step in time average
    AVGHlsEFF = zeros(2);
    Fzm2 = [Fzm(9,9) Fzm(9,16); Fzm(16,9) Fzm(16,16)];
    for j = 1:jmax
        AVGHlsEFF = AVGHlsEFF+expm(+1i*2*pi*omegaRF*(j-1)*dt*Fzm2)*HlsEFF*...
        expm(-1i*2*pi*omegaRF*(j-1)*dt*Fzm2)+expm(+1i*2*pi*omegaRF*(j)*dt*Fzm2)*...
        HlsEFF*expm(-1i*2*pi*omegaRF*(j)*dt*Fzm2);
    end
    AVGHlsEFF = dt/2*AVGHlsEFF;
    AVGHlsEFF = omegaRF*AVGHlsEFF;
    SOHlsEFF = kron(AVGHlsEFF,eye(2))-kron(eye(2),conj(AVGHlsEFF));
    %project to 2-dimensional subspace
    Wq2 = zeros(2,2,2,2,3);
    for Fa = 4:-1:3
        for Fb = 4:-1:3
            for q = -1:1:1
                Wq2(:,:,-Fa+5,-Fb+5,q+2) =[Wq(9,9,-Fa+5,-Fb+5,q+2)...
                Wq(9,16,-Fa+5,-Fb+5,q+2); Wq(16,9,-Fa+5,-Fb+5,q+2)...
                Wq(16,16,-Fa+5,-Fb+5,q+2)];
            end
        end
    end
    Wq = Wq2;   
    %%Going to the rotating frame
    %Wq Superoperators
    SOWq = zeros(4);
    for j = 1:jmax
        Uj = expm(-1i*2*pi*omegaRF*(j-1)*dt*Fzm2);
        Ujp = expm(-1i*2*pi*omegaRF*(j)*dt*Fzm2);
        for q = -1:1:1
            for F = 4:-1:3
                for F1 = 4:-1:3
                    func = kron(Uj'*squeeze(Wq(:,:,-F1+5,-F+5,q+2))*Uj,Uj.'...
                    *conj(squeeze(Wq(:,:,-F1+5,-F+5,q+2)))*conj(Uj))+...
                    kron(Ujp'*squeeze(Wq(:,:,-F1+5,-F+5,q+2))*Ujp,Ujp.'...
                    *conj(squeeze(Wq(:,:,-F1+5,-F+5,q+2)))*conj(Ujp));
                    SOWq = SOWq+func;
                end
            end
        end
    end
    for j = 1:jmax
        Uj = expm(-1i*2*pi*omegaRF*(j-1)*dt*Fzm2);
        Ujp = expm(-1i*2*pi*omegaRF*(j)*dt*Fzm2);
        for q = -1:1:1
            for F1 = 4:-1:3
                for F2 = 4:-1:3
                    if F1~=F2
                        func=kron(Uj'*squeeze(Wq(:,:,-F2+5,-F2+5,q+2))*Uj,Uj.'...
                        *conj(squeeze(Wq(:,:,-F1+5,-F1+5,q+2)))*conj(Uj))+...
                        kron(Ujp'*squeeze(Wq(:,:,-F2+5,-F2+5,q+2))*Ujp,Ujp.'*...
                        conj(squeeze(Wq(:,:,-F1+5,-F1+5,q+2)))*conj(Ujp));
                        SOWq = SOWq+func;
                    end
                end
            end
        end
    end
    SOWq = GammaLW*10^3*SOWq;
    SOWq = dt/2*SOWq;
    SOWq = omegaRF*SOWq;
else 
    jmax = 2000;%number of terms in the time average
    dt = 1/(omegaRF*(jmax)); %time step in time average
    AVGHlsEFF = zeros(16);
    for j = 1:jmax
        AVGHlsEFF = AVGHlsEFF+expm(+1i*2*pi*omegaRF*(j-1)*dt*Fzm)*HlsEFF*...
        expm(-1i*2*pi*omegaRF*(j-1)*dt*Fzm)+expm(+1i*2*pi*omegaRF*(j)*dt*Fzm)*...
        HlsEFF*expm(-1i*2*pi*omegaRF*(j)*dt*Fzm);
    end
    AVGHlsEFF = dt/2*AVGHlsEFF;
    AVGHlsEFF = omegaRF*AVGHlsEFF;
    SOHlsEFF = kron(AVGHlsEFF,eye(16))-kron(eye(16),conj(AVGHlsEFF));
    %projectors
    PP = zeros(dim,2);
    PP(1:9,1:9,1) = eye(dimF4); %Projector in F=4 manifold
    PP(10:16,10:16,2) = eye(dimF3);%Projector in F=3 manifold
    %%Going to the rotating frame
    %Wq Superoperators
    SOWq = zeros(dim^2,dim^2);
    for j = 1:jmax
        Uj = expm(-1i*2*pi*omegaRF*(j-1)*dt*Fzm);
        Ujp = expm(-1i*2*pi*omegaRF*(j)*dt*Fzm);
        for q = -1:1:1
            for F = 4:-1:3
               for F1 = 4:-1:3
                 func = kron(Uj'*squeeze(Wq(:,:,-F1+5,-F+5,q+2))*Uj*...
                 squeeze(PP(:,:,-F1+5)),Uj.'*conj(squeeze(Wq(:,:,-F1+5,-F+5,...
                 q+2)))*conj(Uj)*squeeze(PP(:,:,-F1+5)))+kron(Ujp'*...
                 squeeze(Wq(:,:,-F1+5,-F+5,q+2))*Ujp*squeeze(PP(:,:,-F1+5)),Ujp.'...
                 *conj(squeeze(Wq(:,:,-F1+5,-F+5,q+2)))*...
                 conj(Ujp)*squeeze(PP(:,:,-F1+5)));
                 SOWq = SOWq+func;
               end
            end
        end
    end
    for j = 1:jmax
        Uj = expm(-1i*2*pi*omegaRF*(j-1)*dt*Fzm);
        Ujp = expm(-1i*2*pi*omegaRF*(j)*dt*Fzm);
        for q = -1:1:1
            for F1 = 4:-1:3
                for F2 = 4:-1:3
                  if F1~=F2
                    func=kron(Uj'*squeeze(Wq(:,:,-F2+5,-F2+5,q+2))*Uj*...
                    squeeze(PP(:,:,-F2+5)),Uj.'*conj(squeeze(Wq(:,:,-F1+5,-F1+5,...
                    q+2)))*conj(Uj)*squeeze(PP(:,:,-F1+5)))+kron(Ujp'*...
                    squeeze(Wq(:,:,-F2+5,-F2+5,q+2))*Ujp*squeeze(PP(:,:,-F2+5)),...
                    Ujp.'*conj(squeeze(Wq(:,:,-F1+5,-F1+5,q+2)))...
                    *conj(Ujp)*squeeze(PP(:,:,-F1+5)));
                    SOWq = SOWq+func;
                  end
                end
            end
        end
    end
    SOWq = GammaLW*10^3*SOWq;
    SOWq = dt/2*SOWq;
    SOWq = omegaRF*SOWq;
end
%We created all the necessary operator that are going to be used later and
%save them in parametersuWRabi.mat file.
save parametersuWRabi SOHlsEFF SOWq F4 F3 dimF3 dimF4 dim Fx3 Fy3 Fz3 Fx4 Fy4...
Fz4 grel C1 C2 C0 DeltapFF AuxVar1 GrSplit omegaRF
clear
\end{lstlisting}

\subsection{Main fitting script}
After the preparation stage of the data is finished, we proceed to fit it to our model. The script that achieves this is \texttt{Fitting.m} and it is transcribed below.
\begin{lstlisting}[language=Matlab]
%This script fits the data of the calibration runs of the 
%experiment. It fits the RF Larmor precession and the 
%microwave Rabi flopping experiments. 

%RF Larmor Precession Fitting
xxin = [8.933 0.88 0.19 2.13];%Initial seed
%xxin(1)=Omegax
%xxin(1)=factorI
%xxin(3)=sigmaDI
%xxin(4)=units
Omega0=1000; %kHz
%uW Rabi Flop Fitting
xxin2=[27.42 1000 0.002 0.000110 3.58 4.75];%Initial seed
%xxin(1)=Omegauw
%xxin(2)=Omega0
%xxin(3)=sigmauw
%xxin(4)=sigmaBias
%xxin(5)=Omega
%xxin(5)=units
NominalIntensity = 1.4; %mW/cm^2
Isat = 0.83270; %mW/cm^2
GammaLW = 4.5610; %MHz
NominalOmega = GammaLW*sqrt(0.5*NominalIntensity/Isat);
NintervalsRF=12;%Nintervals+1 points in the Gaussian distribution of intensity
NintervalsuW=12; 
TypeOfFit=1;
%TypeOfFit=1, Larmor X
%TypeOfFit=2, Larmor Y
%TypeOfFit=3, Rabi
fit=1;
%fit=0, No fitting is performed. A plot of Simulated and Experimental Signals is
%displayed for the values in xxin
%fit=1, A full fitting rutine is run using the values in xxin as the initial guess.
%No plots are displayed at this time.
options = optimset('Display','iter');
if TypeOfFit==1
    if fit==0
        LSInhomogeneityPAR(xxin,Omega0,SignalRFxF,rho0RF,bzRF,azRF,NintervalsRF,...
        NominalOmega,10*NominalIntensity,TypeOfFit,1)
    elseif fit==1
        tic
        FittedParamsX=fminsearch(@(xx)LSInhomogeneityPAR(xx,Omega0,SignalRFxF,...
        rho0RF,bzRF,azRF,NintervalsRF,NominalOmega,...
        10*NominalIntensity,TypeOfFit,0),xxin,options);
        toc
    end
elseif TypeOfFit==2
    if fit==0
        LSInhomogeneityPAR(xxin,Omega0,SignalRFyF,rho0RF,...
        bzRF,azRF,NintervalsRF,NominalOmega,...
        10*NominalIntensity,TypeOfFit,1)
    elseif fit==1
        tic
        FittedParamsY=fminsearch(@(xx)LSInhomogeneityPAR(xx,Omega0,...
        SignalRFyF,rho0RF,bzRF,azRF,NintervalsRF,...
        NominalOmega,10*NominalIntensity,TypeOfFit,0),...
        xxin,options);
        toc
    end
elseif TypeOfFit==3
    if fit==0
        TimeDependent=0;
        uWPower2InhomogeneityPAR(xxin2,SignaluWF,SignaluW,...
        rho0uW,bzuW,azuW,NintervalsuW,Phiuwin,...
        TimeDependent,1)
    elseif fit==1
        TimeDependent=0;
        tic
        FittedParamsuW=fminsearch(@(xx)uWPower2InhomogeneityPAR(xx,...
        SignaluWF,SignaluW,rho0uW,bzuW,azuW,NintervalsuW,...
        Phiuwin,TimeDependent,0),xxin2,options);
        toc
    end
end
\end{lstlisting}

\subsection{Auxiliary fitting functions}
The functions \texttt{LSInhomogeneityPAR.m} and \texttt{uWPower2InhomogeneityPAR.m} are used in the previous script to introduce the parallel computing tools. We show them below for completeness.
\subsubsection{RF Larmor experiment}
\begin{lstlisting}[language=Matlab]
function c=LSInhomogeneityPAR(xx,Omega0,SignalRawF,rho0,bz,az,Nintervals,Omega,...
NominalIntensity,TypeOfFit,ShowFig)

tf = 4;
SignalRawF = SignalRawF(1:tf*1000+1);
%xx(1)=Omega0=1000;
%xx(2)=Omegax=9;
%xx(3)=factorI=1;
%xx(4)=sigmaDI=0.2;
%xx(5)=units
initialpoint = 1-3*xx(3);
finalpoint = 1+3*xx(3);
inhomospace = (finalpoint-initialpoint)/(Nintervals);
y = initialpoint:inhomospace:finalpoint;
M = zeros(tf*1000+1,Nintervals+1);
if TypeOfFit==1%RFx
    parfor j=1:Nintervals+1
        M(:,j) = RFLarmorEvolutionX(Omega0,xx(1),Omega,xx(2)*y(j),rho0,tf-1e-3);
    end
elseif TypeOfFit==2%RFy
    parfor j=1:Nintervals+1
        M(:,j) = RFLarmorEvolutionY(Omega0,xx(1),Omega,xx(2)*y(j),rho0,tf-1e-3);
    end 
end
%Averaging and Filtering
DI = normpdf(y,1,xx(3));
dy=y(2)-y(1);
sizeM=size(M);
AuxVar=zeros(sizeM);
AuxVar(:,1)=M(:,1);
AuxVar(:,sizeM(2))=M(:,sizeM(2));
for j=2:sizeM(2)-1
    AuxVar(:,j)=2*M(:,j);
end
AuxVar=AuxVar*dy/2;
Mavg=AuxVar*DI';
MavgF=filter(bz,az,Mavg);
if ShowFig==1
    if TypeOfFit==1
        Intensity=NominalIntensity*xx(2);%
        Title = strcat('RFx Larmor. \Omega_0=',sprintf('%0.3f',Omega0),...
        'kHz, \Omega_x=',sprintf('%0.3f',xx(1)),'kHz, I=',...
        sprintf('%0.3f',Intensity),'\mu W/mm^2');
        figure, plot(0:1e-3:tf,SignalRawF*xx(4),0:1e-3:tf,MavgF),...
        xlabel('time, [ms]'),ylabel('Signal, [a.u.]'),title(Title),...
        legend('Experimental','Fitted')
    elseif TypeOfFit==2
        Intensity=NominalIntensity*xx(2);%
        Title = strcat('RFy Larmor. \Omega_0=',sprintf('%0.3f',Omega0),...
        'kHz, \Omega_y=',sprintf('%0.3f',xx(1)),'kHz, I=',...
        sprintf('%0.3f',Intensity),'\mu W/mm^2');
        figure, plot(0:1e-3:tf,SignalRawF*xx(4),0:1e-3:tf,MavgF),...
        xlabel('time, [ms]'),ylabel('Signal, [a.u.]'),title(Title),...
        legend('Experimental','Fitted')
    end
end
c=norm(xx(4)*SignalRawF-MavgF);
\end{lstlisting}

\subsubsection{Rabi flopping experiment}
\begin{lstlisting}[language=Matlab]
function [c]=uWPower2InhomogeneityPAR(xx,SignalRawF,SignalRaw,rho0,bz,az,...
Nintervals,Phiuwin,TimeDependent,ShowFig)

tf=4;
%TimeDependent=1, uw power is time-dependent
%TimeDependent=0, uw power is not time-dependent
SignalRawF=SignalRawF(1:tf*1000+1);
SignalRaw=SignalRaw(1:tf*1000+1);
DuW = 7010.364130459969;
Omegauw=xx(1);
Omega0=xx(2);
Omega=xx(5);
%xx(1)=Omegauw
%xx(2)=Omega0
%xx(3)=sigmaDI1
%xx(4)=sigmaDI1
%xx(5)=factorI
%xx(6)=units
initialpoint1 = 1-3.5*xx(3);
finalpoint1 = 1+3.5*xx(3);
inhomospace1 = (finalpoint1-initialpoint1)/(Nintervals);
y1=initialpoint1:inhomospace1:finalpoint1;
initialpoint2 = 1-3.5*xx(4);
finalpoint2 = 1+3.5*xx(4);
inhomospace2 = (finalpoint2-initialpoint2)/(Nintervals);
y2=initialpoint2:inhomospace2:finalpoint2;
%2 inhomogeneus parameters
Mc=zeros(tf*1000+1,Nintervals+1,Nintervals+1);
if TimeDependent==0
    for k=1:Nintervals+1
        parfor j=1:Nintervals+1
            Mc(:,j,k)=uWRabiEvolution(Omega0*y2(k),Omegauw*y1(j),...
            Omega,DuW,rho0,tf-1e-3);
        end
    end
elseif TimeDependent==1
    for k=1:Nintervals+1
        parfor j=1:Nintervals+1
            Mc(:,j,k)=uWRabiEvolution2T(Omega0*y2(k),Omegauw*y1(j),...
            Omega,DuW,rho0,tf-1e-3,-Phiuwin);
        end
    end
end
DI1=normpdf(y1,1,xx(3));
DI2=normpdf(y2,1,xx(4));
dy1=y1(2)-y1(1);
dy2=y2(2)-y2(1);
sizeDI1=length(DI1);
sizeDI2=length(DI2);
AuxVar1=zeros(sizeDI1,1);
AuxVar2=zeros(sizeDI2,1);
AuxVar1(1)=DI1(1);
AuxVar2(1)=DI2(1);
AuxVar1(sizeDI1)=DI1(sizeDI1);
AuxVar2(sizeDI2)=DI2(sizeDI2);
for j=2:sizeDI1-1
    AuxVar1(j)=2*DI1(j);
end
AuxVar1=AuxVar1*dy1/2;
for j=2:sizeDI2-1
    AuxVar2(j)=2*DI2(j);
end
AuxVar2=AuxVar2*dy2/2;
Mavg=zeros(length(Mc),1);
parfor j=1:length(Mc)
    Mavg(j)=AuxVar1'*squeeze(Mc(j,:,:))*AuxVar2;
end
MavgF=filter(bz,az,Mavg);
ge = 2.0023193043622;
gi = -0.00039885395;
grel =-1.0032;
GrSplit =9190;
if ShowFig==1
    Title = strcat('uW Rabi Signal. \Omega_0=',sprintf('%0.3f',Omega0),...
    'kHz, \Omega_{\mu W}=',sprintf('%0.3f',Omegauw),'kHz, \Delta_{\mu W}=',...
    sprintf('%0.3f',DuW-(4+3*abs(grel))*Omega0-7*(Omega0)^2/...
    (GrSplit*1000)*(ge-gi)^2/(7*gi+ge)^2),'kHz');
    figure, plot(0:1e-3:tf,SignalRaw*xx(6),0:1e-3:tf,Mavg),xlabel('time, [ms]'),...
    ylabel('Signal, [a.u.]'),title(Title),legend('Experimental','Fitted')
    figure, plot(0:1e-3:tf,SignalRawF*xx(6),'b-',0:1e-3:tf,MavgF,'r-'),...
    xlabel('time, [ms]'),ylabel('Signal, [a.u.]'),...
    legend('Experimental','Fitted'),...
    title('Microwave Rabi Flopping Experiment')
end
%c=norm(units*SignalRaw-Mavg)
c=norm(xx(6)*SignalRawF-MavgF);
\end{lstlisting}

\subsection{RF Larmor Simulation}
The core of the fitting process is the simulation code that calculates the Faraday rotation signal for the RF Larmor precession experiment. Here, we show the functions that do the simulations when the RF field is in the $x$ direction, \texttt{RFLarmorEvolutionX}, or the $y$ direction, \texttt{RFLarmorEvolutionY}.  
\subsubsection{RF Larmor in the $x$ direction}
\begin{lstlisting}[language=Matlab]
function [M]=RFLarmorEvolutionX(Omega0,Omegax,Omega,factor,rho0,tmax)
%This function calculates the Larmor precession evolution 
%of the system for either F=3 or F=4. The RF Larmor frequency
% is given by the field in the X direction.
load parametersRFLarmor
%%H0 Hamiltonian
ge = 2.0023193043622;
gi = -0.00039885395;
Fz3F = zeros(2*(F4+F3)+2,2*(F4+F3)+2);
Fz3F(2*F4+2:2*(F4+F3)+2,2*F4+2:2*(F4+F3)+2) = Fz3;
Fz4F = zeros(2*(F4+F3)+2,2*(F4+F3)+2);
Fz4F(1:9,1:9) = Fz4;
PpMPm = diag([1 1 1 1 1 1 1 1 1 -1 -1 -1 -1 -1 -1 -1]);
Fz4sqr = diag([diag(Fz4^2)' 0 0 0 0 0 0 0]);
Fz3sqr = diag([0 0 0 0 0 0 0 0 0 diag(Fz3^2)']);
beta = -(Omega0)^2/(GrSplit*1000)*(ge-gi)^2/(7*gi+ge)^2;
DeltaRF=omegaRF-Omega0;
xz=8*((ge-gi)/(7*gi+ge))*(Omega0/(GrSplit*1000));
DetTERMS=-DeltaRF*(Fz4F-Fz3F);
H0 = Omega0*(1-abs(grel))*Fz3F +beta*Fz4sqr-beta*Fz3sqr+...
(GrSplit*1000)/2*(1+xz^2/2)*PpMPm;
H0 = H0+DetTERMS;
if AuxVar1==4
    %project to F=4
    H04 = H0(1:9,1:9);
    H0  = H04;
    Q0  = 2*pi*(kron(H0,eye(9))-kron(eye(9),H0.'));
elseif AuxVar1==3
    %project to F=4
    H03 = H0(10:16,10:16);
    H0  = H03;
    Q0  = 2*pi*(kron(H0,eye(7))-kron(eye(7),H0.'));    
else
    Q0 = 2*pi*(kron(H0,eye(16))-kron(eye(16),H0.'));
end
deltat = 1e-3;%length of time slice
noperators = tmax*floor(1/deltat);
%%MEASUREMENT OPERATORS
O0=zeros(2*(F4+F3)+2,2*(F4+F3)+2);
a=C1(1,1)/DeltapFF(1,1)+C1(2,1)/DeltapFF(2,1);
b=C1(1,2)/DeltapFF(1,2)+C1(2,2)/DeltapFF(2,2);
O0(1:2*F4+1,1:2*F4+1)=a/b*Fz4;
O0(2*F4+2:2*(F4+F3)+2,2*F4+2:2*(F4+F3)+2)=Fz3;%defines the initial operator
if AuxVar1==4
    rho04=rho0(1:9,1:9);
    rho0=rho04;
    O0=O0(1:9,1:9);
    sorho0=reshape(rho0.',81,1);
    O0=reshape(O0.',81,1); 
elseif AuxVar1==3
    rho03=rho0(10:16,10:16);
    rho0=rho03;
    O0=O0(10:16,10:16);
    sorho0=reshape(rho0.',49,1);
    O0=reshape(O0.',49,1);
else
    sorho0=reshape(rho0.',256,1);
    O0=reshape(O0.',256,1);
end
%%RWA correction terms need this
Fx4F = zeros(2*(F4+F3)+2,2*(F4+F3)+2);
Fx4F(1:9,1:9) = Fx4;
Fx3F = zeros(2*(F4+F3)+2,2*(F4+F3)+2);
Fx3F(10:16,10:16) = Fx3;
Fy4F = zeros(2*(F4+F3)+2,2*(F4+F3)+2);
Fy4F(1:9,1:9) = Fy4;
Fy3F = zeros(2*(F4+F3)+2,2*(F4+F3)+2);
Fy3F(10:16,10:16) = Fy3;
Fz4F = zeros(2*(F4+F3)+2,2*(F4+F3)+2);
Fz4F(1:9,1:9) = Fz4;
Fz3F = zeros(2*(F4+F3)+2,2*(F4+F3)+2);
Fz3F(10:16,10:16) = Fz3;
   
M=zeros(noperators+2,1);
M(1)=sorho0'*O0;   
if AuxVar1==4
    S=eye(81);
elseif AuxVar1==3
    S=eye(49);
else
    S=eye(256);
end
Hint=0.5*Omegax*((Fx4F - abs(grel)*(1-(Omega0*(1-abs(grel)))/(2*omegaRF))*Fx3F))...
+0.25*(Omegax*DeltaRF/omegaRF)*(-abs(grel)*Fx3F-Fy4F)+(1/(16*omegaRF))*...
(-Omegax^2)*Fz4F-(grel^2/(16*omegaRF))*(-Omegax^2)*Fz3F;
if AuxVar1==4
    %project to F=4
    Hint4 = Hint(1:9,1:9);
    Hint  = Hint4;
    Qint  = 2*pi*(kron(Hint,eye(9))-kron(eye(9),Hint.'));    
elseif AuxVar1==3
    %project to F=3
    Hint3 = Hint(10:16,10:16);
    Hint  = Hint3;
    Qint  = 2*pi*(kron(Hint,eye(7))-kron(eye(7),Hint.'));
else
    Qint=sparse(2*pi*(kron(Hint,eye(16))-kron(eye(16),Hint.')));
end
Q=-1i*(Q0+Qint)+2*pi*Omega^2*(-1i*SOHlsEFF+SOWq)*factor;
R=expm(Q*deltat);
R=R';
for r=1:noperators+1
    S=S*R;
    OP=S*O0;
    M(r+1)=sorho0'*OP;
end
\end{lstlisting}

\subsubsection{RF Larmor in the $y$ direction}

\begin{lstlisting}[language=Matlab]
function [M]=RFLarmorEvolutionY(Omega0,Omegay,Omega,factor,rho0,tmax)
%This function calculates the Larmor precession evolution of the system for
%either F=3 or F=4. The RF Larmor frequency is given by the field in the Y
%direction.
load parametersRFLarmor
%%H0 hamiltonian
ge = 2.0023193043622;
gi = -0.00039885395;
Fz3F = zeros(2*(F4+F3)+2,2*(F4+F3)+2);
Fz3F(2*F4+2:2*(F4+F3)+2,2*F4+2:2*(F4+F3)+2) = Fz3;
Fz4F = zeros(2*(F4+F3)+2,2*(F4+F3)+2);
Fz4F(1:9,1:9) = Fz4;
PpMPm = diag([1 1 1 1 1 1 1 1 1 -1 -1 -1 -1 -1 -1 -1]);
Fz4sqr = diag([diag(Fz4^2)' 0 0 0 0 0 0 0]);
Fz3sqr = diag([0 0 0 0 0 0 0 0 0 diag(Fz3^2)']);
beta = -(Omega0)^2/(GrSplit*1000)*(ge-gi)^2/(7*gi+ge)^2;
DeltaRF=omegaRF-Omega0;
xz=8*((ge-gi)/(7*gi+ge))*(Omega0/(GrSplit*1000));
DetTERMS=-DeltaRF*(Fz4F-Fz3F);
H0 = Omega0*(1-abs(grel))*Fz3F +beta*Fz4sqr-beta*Fz3sqr+...
(GrSplit*1000)/2*(1+xz^2/2)*PpMPm;
H0 = H0+DetTERMS;
if AuxVar1==4
    %project to F=4
    H04 = H0(1:9,1:9);
    H0  = H04;
    Q0  = 2*pi*(kron(H0,eye(9))-kron(eye(9),H0.'));
elseif AuxVar1==3
    %project to F=4
    H03 = H0(10:16,10:16);
    H0  = H03;
    Q0  = 2*pi*(kron(H0,eye(7))-kron(eye(7),H0.'));   
else
    Q0 = 2*pi*(kron(H0,eye(16))-kron(eye(16),H0.'));
end
deltat = 1e-3;%length of time slice
noperators = tmax*floor(1/deltat);
%%MEASUREMENT OPERATORS
O0=zeros(2*(F4+F3)+2,2*(F4+F3)+2);
a=C1(1,1)/DeltapFF(1,1)+C1(2,1)/DeltapFF(2,1);
b=C1(1,2)/DeltapFF(1,2)+C1(2,2)/DeltapFF(2,2);
O0(1:2*F4+1,1:2*F4+1)=a/b*Fz4;
O0(2*F4+2:2*(F4+F3)+2,2*F4+2:2*(F4+F3)+2)=Fz3;%defines initial operator
if AuxVar1==4
    rho04=rho0(1:9,1:9);
    rho0=rho04;
    O0=O0(1:9,1:9);
    sorho0=reshape(rho0.',81,1);
    O0=reshape(O0.',81,1); 
elseif AuxVar1==3
    rho03=rho0(10:16,10:16);
    rho0=rho03;
    O0=O0(10:16,10:16);
    sorho0=reshape(rho0.',49,1);
    O0=reshape(O0.',49,1);
else
    sorho0=reshape(rho0.',256,1);
    O0=reshape(O0.',256,1);
end
%%RWA correction terms need this
Fx4F = zeros(2*(F4+F3)+2,2*(F4+F3)+2);
Fx4F(1:9,1:9) = Fx4;
Fx3F = zeros(2*(F4+F3)+2,2*(F4+F3)+2);
Fx3F(10:16,10:16) = Fx3;
Fy4F = zeros(2*(F4+F3)+2,2*(F4+F3)+2);
Fy4F(1:9,1:9) = Fy4;
Fy3F = zeros(2*(F4+F3)+2,2*(F4+F3)+2);
Fy3F(10:16,10:16) = Fy3;
Fz4F = zeros(2*(F4+F3)+2,2*(F4+F3)+2);
Fz4F(1:9,1:9) = Fz4;
Fz3F = zeros(2*(F4+F3)+2,2*(F4+F3)+2);
Fz3F(10:16,10:16) = Fz3;
M=zeros(noperators+2,1);
M(1)=sorho0'*O0;   
if AuxVar1==4
    S=eye(81);
elseif AuxVar1==3
    S=eye(49);
else
    S=eye(256);
end
Hint=0.5*Omegay* ((Fy4F-abs(grel)*(1-(Omega0*(1-abs(grel)))/(2*omegaRF) )*Fy3F))...
+0.25*(Omegay*DeltaRF/omegaRF)*(Fx4F+abs(grel)*Fy3F)+(1/(16*omegaRF))*...
(-Omegay^2)*Fz4F-(grel^2/(16*omegaRF))* (-Omegay^2)*Fz3F;
if AuxVar1==4
    %project to F=4
    Hint4 = Hint(1:9,1:9);
    Hint  = Hint4;
    Qint  = 2*pi*(kron(Hint,eye(9))-kron(eye(9),Hint.'));
elseif AuxVar1==3
    %project to F=4
    Hint3 = Hint(10:16,10:16);
    Hint  = Hint3;
    Qint  = 2*pi*(kron(Hint,eye(7))-kron(eye(7),Hint.'));
else
    Qint=sparse(2*pi*(kron(Hint,eye(16))-kron(eye(16),Hint.')));
end      
Q=-1i*(Q0+Qint)+2*pi*Omega^2*(-1i*SOHlsEFF+SOWq)*factor;
R=expm(Q*deltat);
R=R';
for r=1:noperators+1
    S=S*R;
    OP=S*O0;
    M(r+1)=sorho0'*OP;
end
\end{lstlisting}

\subsection{Rabi flop simulation}
In order to fit the Rabi flopping signal, we simulate the Faraday rotation measurement record using the function \texttt{uWRabiEvolution}.
\begin{lstlisting}[language=Matlab]
function [M]=uWRabiEvolution(Omega0,Omegauw,Omega,DuW,rho0,tmax)
%This function calculates the evolution of the system when only the microwave 
%field is on and time independent

%AuxVar2 = 1 for Intensity inhomogeneity
%AuxVar2 = 2 for uW power inhomogeneity
%AuxVar2 = 3 for Bias field inhomogeneity
load parametersuWRabi
%%H0 hamiltonian
ge = 2.0023193043622;
gi = -0.00039885395;
Fz3F = zeros(2*(F4+F3)+2,2*(F4+F3)+2);
Fz3F(2*F4+2:2*(F4+F3)+2,2*F4+2:2*(F4+F3)+2) = Fz3;
Fz4F = zeros(2*(F4+F3)+2,2*(F4+F3)+2);
Fz4F(1:9,1:9) = Fz4;
PpMPm = diag([1 1 1 1 1 1 1 1 1 -1 -1 -1 -1 -1 -1 -1]);
Fz4sqr = diag([diag(Fz4^2)' 0 0 0 0 0 0 0]);
Fz3sqr = diag([0 0 0 0 0 0 0 0 0 diag(Fz3^2)']);
beta = -(Omega0)^2/(GrSplit*1000)*(ge-gi)^2/(7*gi+ge)^2;
DeltaRF=omegaRF-Omega0;
%%AC Zeeman Shift
ACZ = zeros(16);
for m = 2:-1:-3
    V3j = zeros(16,1);
    V4k = zeros(16,1);
    jjj = m+13;
    kkk = m+6;
    V3j(jjj) = 1;
    V4k(kkk) = 1;
    ACZ = ACZ+(V3j*V3j'-V4k*V4k')*(abs(ClebschGordan(3,1,4,m,1,m+1)))^2/(m-3);
end
ACZ = (Omegauw)^2/(8*Omega0)*ACZ;%
DeltauW=DuW-(4+3*abs(grel))*Omega0+7*beta;
DetTERMS=(3.5*DeltaRF-0.5*DeltauW)*PpMPm-DeltaRF*(Fz4F-Fz3F);
H0 = Omega0*(1-abs(grel))*Fz3F+(1.5*Omega0*(1-abs(grel))+12.5*Omega0^2/...
(GrSplit*1000)*((ge-gi)/(7*gi+ge))^2)*PpMPm+beta*Fz4sqr-beta*Fz3sqr;
H0 = H0+DetTERMS;
H0 = H0+ACZ;
if AuxVar1==2
    %project 2-dimensional subspace
    H02 = [H0(9,9) H0(9,16); H0(16,9) H0(16,16)];
    H0  = H02;
    Q0  = 2*pi*(kron(H0,eye(2))-kron(eye(2),H0.'));
else
    Q0 = 2*pi*(kron(H0,eye(16))-kron(eye(16),H0.'));
end
%%time
deltat = 1e-3;
noperators = tmax*floor(1/deltat);
%%MEASUREMENT OPERATORS
O0=zeros(2*(F4+F3)+2,2*(F4+F3)+2);
a=C1(1,1)/DeltapFF(1,1)+C1(2,1)/DeltapFF(2,1);
b=C1(1,2)/DeltapFF(1,2)+C1(2,2)/DeltapFF(2,2);
O0(1:2*F4+1,1:2*F4+1)=a/b*Fz4;
O0(2*F4+2:2*(F4+F3)+2,2*F4+2:2*(F4+F3)+2)=Fz3;%defines initial operator
if AuxVar1==2
    rho02=[rho0(9,9) rho0(9,16); rho0(16,9) rho0(16,16)];
    rho0=rho02;
    O0=[O0(9,9) O0(9,16); O0(16,9) O0(16,16)];
    sorho0=reshape(rho0.',4,1);
    O0=reshape(O0.',4,1);
else
    sorho0=reshape(rho0.',256,1);
    O0=reshape(O0.',256,1);
end
%%stretch state
V44 = zeros(2*(F4+F3)+2,1);
V33 = zeros(2*(F4+F3)+2,1);
V44(2*F4+1) = 1;
V33(2*(F4+F3)+2) = 1;
Sigmax = V44*V33'+V33*V44';
%Sigmay = 1i*V44*V33'-1i*V33*V44';
%%END  
M=zeros(noperators+2,1);
M(1)=sorho0'*O0;
if AuxVar1==2
    S=eye(4);
else
    S=eye(256);
end       
Hint=0.5*Omegauw*Sigmax;
if AuxVar1==2
    %project to 2-dimensional subspace
    Hint2 = [Hint(9,9) Hint(9,16); Hint(16,9) Hint(16,16)];
    Hint  = Hint2;
    Qint  = 2*pi*(kron(Hint,eye(2))-kron(eye(2),Hint.'));
else
    Qint=sparse(2*pi*(kron(Hint,eye(16))-kron(eye(16),Hint.')));
end   
Q=-1i*(Q0+Qint)+2*pi*Omega^2*(-1i*SOHlsEFF+SOWq);       
R=expm(Q*deltat);
R=R';
for r=1:noperators+1
    S=S*R;
    OP=S*O0;
    M(r+1)=sorho0'*OP;
end
\end{lstlisting}

\section{Simulation code}
Once the control parameters have been fitted, we can proceed to simulate the forward problem and calculate the measurement record $M(t)$ and the time-evolved density matrix of the system $\rho(t)$. Moreover, we can also calculate the Heisenberg picture evolution of the initial measurement operator $\cO_0$. In this section, we transcribe the Matlab codes we wrote to achieve this.
\subsection{Main script}
The script \texttt{SimulationEvolution.m} evolves the system either in the Schr\"{o}dinger or Heisenberg picture.
\begin{lstlisting}[language=Matlab]
%This script calculates the Heisenberg or Shrodinger evolution of the hyperfine 
%ground state (16-dimensional Hilbert space) of a single cesium atom controlled
%by RF and microwave magnetic fields
tic
%Important equations
% Omega=GammaLW*sqrt(0.5*Intensity/Isat)
% Isat=0.8327 mW/cm^2
% GammaLW=4.5610 MHz
tf = 2;%ms %Final time
Omegax =  FittedParamsX(1);%kHz
Omegay = FittedParamsY(1);%kHz
Omegauw = FittedParamsuW;%kHz
Omega0 = 1000;%kHz
Omega = NominalOmega*sqrt((FittedParamsX(2)+FittedParamsY(2))/2);%MHz
DuW = 7010.364130459969;
unitsT= (FittedParamsX(4)+FittedParamsY(4))/2;%units
OffsetT= (Offsetx+Offsety)/2;% Offset correction;
%DATA INPUT: Enter the name of the data file you want to reconstruct and the target
%state rho0 for comparison
dataRecon=dataReconCat3i(7:2100,:);
rho0=rho0Cat3i;
%FILTER PARAMETERS
AZ=az;
BZ=bz;
%Filter data
SReconFilter=filter(BZ,AZ,unitsT*(-dataRecon(:,2)-OffsetT));
%Phase correction in case it is needed
xPhase=0;
yPhase=0;
uWPhase=0;
%Inhomogeneities coarse grain 
Nintervals = 12;%0 or even number
%Type of Inhomogeneity input in the model
AuxVar2 = 4;
%AuxVar2 = 1 for Intensity inhomogeneity
%AuxVar2 = 2 for uW power inhomogeneity
%AuxVar2 = 3 for Bias field inhomogeneity
%AuxVar2 = 4 for Light Intensity and Bias inhomogeneity
%AuxVar2 = 5 for uW power and Bias inhomogeneity
%AuxVar2 = 6 for Light Intensity and uW power inhomogeneity
PhaseModulation=1;
%PhaseModulation=0, No Phase Modulation, i.e., time independent evolution
%PhaseModulation=1, With Phase Modulation,  i.e., time dependent evolution
if Nintervals==0
    if PhaseModulation==1
        [OPc,Mc,rmat]=RFuWEvolution7(Omegax,Omegay,Omegauw,Omega0,Omega,DuW,...
        Phixin+xPhase,Phiyin+yPhase,-Phiuwin+uWPhase,uWrate,RFrate,rho0,tf-1e-3);
    elseif PhaseModulation==0
        [OPc,Mc,rmat]=RFuWEvolution7(Omegax,Omegay,Omegauw,Omega0,Omega,DuW,...
        zeros(501,1)+xPhase,zeros(501,1)+yPhase,zeros(501,1)+uWPhase,...
        uWrate,RFrate,rho0,tf-1e-3);
    end
    McF=filter(BZ,AZ,Mc);%Simulated measurement record for initial state rho0
    %For reference, plot filtered and unfiltered signal
    Title = strcat('Non Filtered Signal. \Omega_0=',sprintf('%0.3f',Omega0),...
    'kHz, \Omega_x=',sprintf('%0.3f',Omegax),'kHz, \Omega_y=',...
    sprintf('%0.3f',Omegay),'kHz,\Omega_{u w}=',sprintf('%0.3f',Omegauw),'kHz');
    figure, plot(1000*dataRecon(:,1),unitsT*(-dataRecon(:,2)-OffsetT),...
    0:1e-3:tf-1e-3,Mc(1:tf*1000)),legend('Experimental','Simulated'),title(Title)
    Title = strcat('Filtered Signal. \Omega_0=',sprintf('%0.3f',Omega0)...
    ,'kHz, \Omega_x=',sprintf('%0.3f',Omegax),'kHz, \Omega_y=',...
    sprintf('%0.3f',Omegay),'kHz,\Omega_{u w}=',sprintf('%0.3f',Omegauw),'kHz');
    figure, plot(1000*dataRecon(:,1),SReconFilter,0:1e-3:tf-1e-3,McF(1:tf*1000)),...
    legend('Experimental','Simulated'), title(Title)
    %Calculate least squares residue for filtered and unfiltered signals
    norm(unitsT*(-dataRecon(1:tf*1000,2)-OffsetT)-Mc(1:tf*1000))
    norm(SReconFilter(1:tf*1000)-McF(1:tf*1000))
else
    if AuxVar2==1
        spread = (FittedParamsX(3)+FittedParamsY(3))/2;%Intensity inhomogeneity
        initialpoint = 1-3.5*spread;
        finalpoint = 1+3.5*spread;
        inhomospace = (finalpoint-initialpoint)/(Nintervals);
        y=initialpoint:inhomospace:finalpoint;
        Mc=zeros(tf*1000+1,Nintervals+1);
        OPc=zeros(256,tf*1000+1,Nintervals+1);
        if PhaseModulation==1
            parfor j=1:Nintervals+1
                [OPc(:,:,j),Mc(:,j),rmat]=RFuWEvolution(Omegax,Omegay,Omegauw,...
                Omega0,Omega*sqrt(y(j)),DuW,Phixin+xPhase,Phiyin+yPhase,...
                -Phiuwin+uWPhase,uWrate,RFrate,rho0,tf-1e-3);
            end
        elseif PhaseModulation==0
            parfor j=1:Nintervals+1
                [OPc(:,:,j),Mc(:,j),rmat]=RFuWEvolution(Omegax,Omegay,Omegauw,...
                Omega0,Omega*sqrt(y(j)),DuW,zeros(501,1)+xPhase,zeros(501,1)+...
                yPhase,zeros(501,1)+uWPhase,uWrate,RFrate,rho0,tf-1e-3);
            end
        end  
        %Inhomogeneity distribution
        DI=normpdf(y,1,spread);
        dy=y(2)-y(1);
        sizeDI=length(DI);
        AuxVar=zeros(sizeDI,1);
        AuxVar(1)=DI(1);
        AuxVar(sizeDI)=DI(sizeDI);
        for j=2:sizeDI-1
            AuxVar(j)=2*DI(j);
        end
        AuxVar=AuxVar*dy/2;
        Mavg=Mc*AuxVar;
        InhomogeneityDistribution=AuxVar;   
    elseif AuxVar2==2
        spread = 0.0035;%
        initialpoint = 1-3.5*spread;
        finalpoint = 1+3.5*spread;
        inhomospace = (finalpoint-initialpoint)/(Nintervals);
        y=initialpoint:inhomospace:finalpoint;
        Mc=zeros(tf*1000+1,Nintervals+1);
        OPc=zeros(256,tf*1000+1,Nintervals+1);
        if PhaseModulation==1
            parfor j=1:Nintervals+1
                [OPc(:,:,j),Mc(:,j),rmat]=RFuWEvolution(Omegax,Omegay,Omegauw*...
                y(j),Omega0,Omega,DuW,Phixin+xPhase,Phiyin+yPhase,-Phiuwin+...
                uWPhase,uWrate,RFrate,rho0,tf-1e-3);
            end
        elseif PhaseModulation==0
            parfor j=1:Nintervals+1
                [OPc(:,:,j),Mc(:,j),rmat]=RFuWEvolution(Omegax,Omegay,Omegauw*...
                y(j),Omega0,Omega,DuW,zeros(501,1)+xPhase,zeros(501,1)+yPhase,...
                zeros(501,1)+uWPhase,uWrate,RFrate,rho0,tf-1e-3);
            end
        end 
        DI=normpdf(y,1,spread);
        dy=y(2)-y(1);
        sizeDI=length(DI);
        AuxVar=zeros(sizeDI,1);
        AuxVar(1)=DI(1);
        AuxVar(sizeDI)=DI(sizeDI);
        for j=2:sizeDI-1
            AuxVar(j)=2*DI(j);
        end
        AuxVar=AuxVar*dy/2;
        Mavg=Mc*AuxVar;
        InhomogeneityDistribution=AuxVar;
    elseif AuxVar2==3
        spread = 0.00008;%
        initialpoint = 1-3.5*spread;
        finalpoint = 1+3.5*spread; 
        inhomospace = (finalpoint-initialpoint)/(Nintervals);
        y=initialpoint:inhomospace:finalpoint;
        Mc=zeros(tf*1000+1,Nintervals+1);
        OPc=zeros(256,tf*1000+1,Nintervals+1);
        rmat=zeros(16,16,tf*1000+1,Nintervals+1); 
        if PhaseModulation==1
            parfor j=1:Nintervals+1
                [OPc(:,:,j),Mc(:,j),rmat(:,:,:,j)]=RFuWEvolution(Omegax,Omegay,...
                Omegauw,Omega0*y(j),Omega,DuW,Phixin+xPhase,Phiyin+yPhase,...
                -Phiuwin+uWPhase,uWrate,RFrate,rho0,tf-1e-3);
            end
        elseif PhaseModulation==0
            parfor j=1:Nintervals+1
                [OPc(:,:,j),Mc(:,j),rmat(:,:,j)]=RFuWEvolution(Omegax,Omegay,...
                Omegauw,Omega0*y(j),Omega,DuW,zeros(501,1)+xPhase,zeros(501,1)+...
                yPhase,zeros(501,1)+uWPhase,uWrate,RFrate,rho0,tf-1e-3);
            end
        end
        DI=normpdf(y,1,spread);
        dy=y(2)-y(1);
        sizeDI=length(DI);
        AuxVar=zeros(sizeDI,1);
        AuxVar(1)=DI(1);
        AuxVar(sizeDI)=DI(sizeDI);
        for j=2:sizeDI-1
            AuxVar(j)=2*DI(j);
        end
        AuxVar=AuxVar*dy/2;
        Mavg=Mc*AuxVar;
        InhomogeneityDistribution=AuxVar;
    elseif AuxVar2==4
        spread1 = (FittedParamsX(3)+FittedParamsY(3))/2;%Intensity
        initialpoint1 = 1-3*spread1;
        finalpoint1 = 1+3*spread1;
        inhomospace1 = (finalpoint1-initialpoint1)/(Nintervals);
        y1=initialpoint1:inhomospace1:finalpoint1;
        spread2 = 0.00008;%Bias
        initialpoint2 = 1-3.5*spread2;
        finalpoint2 = 1+3.5*spread2; 
        inhomospace2 = (finalpoint2-initialpoint2)/(Nintervals);
        y2=initialpoint2:inhomospace2:finalpoint2;
        Mc=zeros(tf*1000+1,Nintervals+1,Nintervals+1);
        OPc=zeros(256,tf*1000+1,Nintervals+1,Nintervals+1);
        if PhaseModulation==1
            for k=1:Nintervals+1
                parfor j=1:Nintervals+1
                    [OPc(:,:,j,k),Mc(:,j,k),rmat]=RFuWEvolution(Omegax,Omegay,...
                    Omegauw,Omega0*y2(k),Omega*y1(j),DuW,Phixin+xPhase,Phiyin+...
                    yPhase,-Phiuwin+uWPhase,uWrate,RFrate,rho0,tf-1e-3);
                end
            end
        elseif PhaseModulation==0
            for k=1:Nintervals+1
                parfor j=1:Nintervals+1 
                    [OPc(:,:,j,k),Mc(:,j,k),rmat]=RFuWEvolution(Omegax,Omegay,...
                    Omegauw,Omega0*y2(k),Omega*y1(j),DuW,zeros(501,1)+xPhase,...
                    zeros(501,1)+yPhase,zeros(501,1)+uWPhase,...
                    uWrate,RFrate,rho0,tf-1e-3);
                end
            end     
        end
        DI1=normpdf(y1,1,spread1);
        DI2=normpdf(y2,1,spread2);
        dy1=y1(2)-y1(1);
        dy2=y2(2)-y2(1);
        sizeDI1=length(DI1);
        sizeDI2=length(DI2);
        AuxVar1=zeros(sizeDI1,1);
        AuxVar2=zeros(sizeDI2,1);
        AuxVar1(1)=DI1(1);
        AuxVar2(1)=DI2(1);
        AuxVar1(sizeDI1)=DI1(sizeDI1);
        AuxVar2(sizeDI2)=DI2(sizeDI2);
        for j=2:sizeDI1-1
            AuxVar1(j)=2*DI1(j);
        end
        AuxVar1=AuxVar1*dy1/2;
        InhomogeneityDistribution1=AuxVar1;
        for j=2:sizeDI2-1
            AuxVar2(j)=2*DI2(j);
        end
        AuxVar2=AuxVar2*dy2/2;
        InhomogeneityDistribution2=AuxVar2;
        Mavg=zeros(length(Mc),1);
        parfor j=1:length(Mc)
            Mavg(j)=AuxVar1'*squeeze(Mc(j,:,:))*AuxVar2;
        end
    elseif AuxVar2==5
        spread1 = 0.003;%uW
        initialpoint1 = 1-3.5*spread1;%
        finalpoint1 = 1+3.5*spread1;%
        inhomospace1 = (finalpoint1-initialpoint1)/(Nintervals);
        y1=initialpoint1:inhomospace1:finalpoint1;
        spread2 = 0.00006;%Bias
        initialpoint2 = 1-3.5*spread2;%
        finalpoint2 = 1+3.5*spread2; %
        inhomospace2 = (finalpoint2-initialpoint2)/(Nintervals);
        y2=initialpoint2:inhomospace2:finalpoint2;
        Mc=zeros(tf*1000+1,Nintervals+1,Nintervals+1);
        OPc=zeros(256,tf*1000+1,Nintervals+1,Nintervals+1);
        if PhaseModulation==1
            for k=1:Nintervals+1
                parfor j=1:Nintervals+1
                    [OPc(:,:,j,k),Mc(:,j,k),rmat]=RFuWEvolution(Omegax,Omegay,...
                    Omegauw*y1(j),Omega0*y2(k),Omega,DuW,Phixin+xPhase,Phiyin+...
                    yPhase,-Phiuwin+uWPhase,uWrate,RFrate,rho0,tf-1e-3);
                end
            end
        elseif PhaseModulation==0
            for k=1:Nintervals+1
                parfor j=1:Nintervals+1
                    [OPc(:,:,j,k),Mc(:,j,k),rmat]=RFuWEvolution(Omegax,Omegay,...
                    Omegauw*y1(j),Omega0*y2(k),Omega,DuW,zeros(501,1)+xPhase,...
                    zeros(501,1)+yPhase,zeros(501,1)+uWPhase,...
                    uWrate,RFrate,rho0,tf-1e-3);
                end
            end
        end
        DI1=normpdf(y1,1,spread1);
        DI2=normpdf(y2,1,spread2);
        dy1=y1(2)-y1(1);
        dy2=y2(2)-y2(1);
        sizeDI1=length(DI1);
        sizeDI2=length(DI2);
        AuxVar1=zeros(sizeDI1,1);
        AuxVar2=zeros(sizeDI2,1);
        AuxVar1(1)=DI1(1);
        AuxVar2(1)=DI2(1);
        AuxVar1(sizeDI1)=DI1(sizeDI1);
        AuxVar2(sizeDI2)=DI2(sizeDI2);
        for j=2:sizeDI1-1
            AuxVar1(j)=2*DI1(j);
        end
        AuxVar1=AuxVar1*dy1/2;
        InhomogeneityDistribution1=AuxVar1;
        for j=2:sizeDI2-1
            AuxVar2(j)=2*DI2(j);
        end
        AuxVar2=AuxVar2*dy2/2;
        InhomogeneityDistribution2=AuxVar2;
        Mavg=zeros(length(Mc),1);
        parfor j=1:length(Mc)
            Mavg(j)=AuxVar1'*squeeze(Mc(j,:,:))*AuxVar2;
        end
    elseif AuxVar2==6
        spread1 = (FittedParamsX(3)+FittedParamsY(3))/2;%intensity
        initialpoint1 = 1-3.5*spread1;%
        finalpoint1 = 1+3.5*spread1;%
        inhomospace1 = (finalpoint1-initialpoint1)/(Nintervals);
        y1=initialpoint1:inhomospace1:finalpoint1;
        spread2 = 0.0035;%uW
        initialpoint2 = 1-3.5*spread2;%
        finalpoint2 = 1+3.5*spread2; %
        inhomospace2 = (finalpoint2-initialpoint2)/(Nintervals);
        y2=initialpoint2:inhomospace2:finalpoint2;
        Mc=zeros(tf*1000+1,Nintervals+1,Nintervals+1);
        OPc=zeros(256,tf*1000+1,Nintervals+1,Nintervals+1);
        if PhaseModulation==1
            for k=1:Nintervals+1
                parfor j=1:Nintervals+1
                    [OPc(:,:,j,k),Mc(:,j,k),rmat]=RFuWEvolution(Omegax,Omegay,...
                    Omegauw*y2(k),Omega0,Omega*y1(j),DuW,Phixin+xPhase,Phiyin+...
                    yPhase,-Phiuwin+uWPhase,uWrate,RFrate,rho0,tf-1e-3);
                end
            end
        elseif PhaseModulation==0  
            for k=1:Nintervals+1
                parfor j=1:Nintervals+1
                    [OPc(:,:,j,k),Mc(:,j,k),rmat]=RFuWEvolution(Omegax,Omegay,...
                    Omegauw*y2(k),Omega0,Omega*y1(j),DuW,zeros(501,1)+xPhase,...
                    zeros(501,1)+yPhase,zeros(501,1)+uWPhase,...
                    uWrate,RFrate,rho0,tf-1e-3);
                end
            end
        end
        DI1=normpdf(y1,1,spread1);
        DI2=normpdf(y2,1,spread2);
        dy1=y1(2)-y1(1);
        dy2=y2(2)-y2(1);
        sizeDI1=length(DI1);
        sizeDI2=length(DI2);
        AuxVar1=zeros(sizeDI1,1);
        AuxVar2=zeros(sizeDI2,1);
        AuxVar1(1)=DI1(1);
        AuxVar2(1)=DI2(1);
        AuxVar1(sizeDI1)=DI1(sizeDI1);
        AuxVar2(sizeDI2)=DI2(sizeDI2);
        for j=2:sizeDI1-1
            AuxVar1(j)=2*DI1(j);
        end
        AuxVar1=AuxVar1*dy1/2;
        InhomogeneityDistribution1=AuxVar1;
        for j=2:sizeDI2-1
            AuxVar2(j)=2*DI2(j);
        end
        AuxVar2=AuxVar2*dy2/2;
        InhomogeneityDistribution2=AuxVar2;
        Mavg=zeros(length(Mc),1);
        parfor j=1:length(Mc)
            Mavg(j)=AuxVar1'*squeeze(Mc(j,:,:))*AuxVar2;
        end
    end
    MavgF=filter(BZ,AZ,Mavg);
    figure, plot(1000*dataRecon(:,1),unitsT*(-dataRecon(:,2)-OffsetT),..
    0:1e-3:tf-1e-3,Mavg(1:tf*1000)),legend('Experiemental','Simulated'),...
    title('Not Filtered')
    figure, plot(1000*dataRecon(:,1),SReconFilter,...
    0:1e-3:tf-1e-3,MavgF(1:tf*1000)),legend('Experiemental','Simulated'),... 
    title('Filtered')
    norm(unitsT*(-dataRecon(1:tf*1000,2)-OffsetT)-Mavg(1:tf*1000))
    norm(SReconFilter(1:tf*1000)-MavgF(1:tf*1000))
end
toc
\end{lstlisting}

\subsection{Decoherence and time independent terms}
Before running the simulation script, we must calculate all time independent operators and do the RWA. We do this in a separate script in order to use time more efficiently. The script that does it is \texttt{RFLarmorPreparation.m}.
\begin{lstlisting}[language=Matlab]
%This script calculates all time-independent parts of the dynamics. The
%free Hamiltonian H0, the effective light-shift Hamiltonian Hls, and the
%Jump operators W as they appear in (\cite{paper}). In addition, It
%calculates the superoperator representation (i.e., representing a d by d
%matrix as d^2 vector whose entries are taken from the rows of the original
%matrix in order) of H0, HlsEFF and the W operators. Moreover, it
%calculates the superoperator operators in the rotating frame given by the
%bias magnetic field. This script produces the file parametersRFuW.mat
%which contains all the necessary variables for the function
%RFuWEvolution.m to work.

%Define the RF frequency OmegaRF and the detuning of the laser
%from F=3 in the ground state to F'=4 in the excited state for the D1
%transition. Note that the units this parameters should be are written
%besides them as comments.
omegaRF = 1000;%kHz 
Omega =  1; %DO NOT CHANGE
detuning = -730;%MHz %Delta 4'3 
%Define the polarization of the light vector (eL) and the three different
%polarizations for scattered light (eq) in the spherical basis for
%q=-1,0,+1, in the columns of (eq), for example, eq(:,2)=ez=[0 0 1].
eL = [1;0;0];
eq = [  1/sqrt(2) 0 -1/sqrt(2);
      -1i/sqrt(2) 0 -1i/sqrt(2);
            0     1   0       ];
%Define the total angular momentum manifold dimensions for the hyperfine
%ground states F=4 and F=3.      
F4 = 4;
F3 = 3;
dimF3 = 2*F3+1;
dimF4 = 2*F4+1;
dim = dimF3+dimF4;
%Compute the Angular momentum operators (Fx, Fy, Fz) for the F=3 and F=4 manifolds
%by using the function AngMomentum.m. Note that this operators are written in
%the basis in which Fz is diagonal and the the basis is ordered from
%|F,-F> to |F,+F>.
[Fx4,Fy4,Fz4] = AngMomentum(F4);
[Fx3,Fy3,Fz3] = AngMomentum(F3);
grel = -1.0032; %g_3/g_4 ratio of g-factors 

%%%%%%% D1 TRANSITION %%%%%%
%Define the Linewidth (GammaLW) for the D1 transition as well as the
%the ground state splitting (GrSplit) (which in the paper is called \omega_{HF}). 
%In addition define j' (pj) as the electronic angular momentum (electron + 
%orbital in l=0 case). For the D1 line there are 2 excited manifolds P1/2, 
%F'=3 and F'=4, separated by the excited state splitting (ExSplit), whose
%total dimesion is 16. Finally (ExBasis) and (GrBasis) represent the basis
%vectors for the excited and ground manifolds ordered from -mf to mf. In
%particular, (GrBasis) is picked in this order:
%|4,-4>,...,|4,+4>,|3,-3>,...,|3,+3>.
GammaLW = 4.561; %MHz
ExSplit = 1167.6;%MHz
GrSplit = 9.19e3;%MHz
pJ = 1/2; 
Is = 7/2;
ExDim = 16;
GrDim = dim;
ExBasis = eye(ExDim);
GrBasis = eye(GrDim);     
%Define the detuning matrix \Delta_{F'F}. The value of the variable
%(detuning) is asigned to the detuning from F=3 to F'=4.
%DELTA F'F
DeltapFF = zeros(2,2);
DeltapFF(1,2) = detuning; %Delta 4'3
DeltapFF(2,2) = (DeltapFF(1,2)+ExSplit); %Delta 3'3
DeltapFF(1,1) = DeltapFF(1,2)+GrSplit; %Delta 4'4
DeltapFF(2,1) = DeltapFF(2,2)+GrSplit;%Delta 3'4
%To be used later in the definition of the measurement basis, here we
%calculate the irreducible tensor coefficients C_{j'F'F}^{(k)} as they
%appear in \cite{deutsch09}, Eqs (A13). The function Wigner6jcoeff.m by Amita B Deb,
%Clarendon Lab. 2007, is used to calculate the Wigner 6J symbols that
%appear in the definition of the coefficients.
%%TENSOR COEFICIENTS
C1 = zeros(2);
C2 = zeros(2);
C0 = zeros(2);
for pF = 4:-1:3
    for F = 4:-1:3
        C0(5-pF,5-F) = (-1)^(3*F-pF+1)*sqrt(1/3)*(2*pF+1)/(sqrt(2*F+1))*...
        Wigner6jcoeff(F,1,pF,1,F,0)*(2*pJ+1)*(2*F+1)*...
        abs(Wigner6jcoeff(pF,7/2,pJ,1/2,1,F))^2;
        C1(5-pF,5-F) = (-1)^(3*F-pF)*sqrt(3/2)*(2*pF+1)/(sqrt(F*(F+1)*(2*F+1)))*...
        Wigner6jcoeff(F,1,pF,1,F,1)*(2*pJ+1)*(2*F+1)*...
        abs(Wigner6jcoeff(pF,7/2,pJ,1/2,1,F))^2;
        C2(5-pF,5-F) = (-1)^(3*F-pF)*sqrt(30)*(2*pF+1)/...
        (sqrt(F*(F+1)*(2*F+1)*(2*F-1)*(2*F+3)))*Wigner6jcoeff(F,1,pF,1,F,2)*...
        (2*pJ+1)*(2*F+1)*abs(Wigner6jcoeff(pF,7/2,pJ,1/2,1,F))^2;
    end
end
%Now we start calculating the operators that appear in the master equation.
%We do so by following the paper [Riofrio, et al. 2011] and
%using a similar notation.

%MASTER EQUATION OPERATORS
%The first thing we calculate is the dimensionless dipole rising operator
%\vecD_{F'F}^{\dagger} (DDq) as it appears in Eq (24). The last 3 indices are:
% F', F, and q. Inside the for loops, there are 2 rules, (j) and (k), to assign
%the right value to the right |F',m'><F,m| component. Additionally, we calculate
%(Aux_eL_DDq) which represents \vec{\epsilon}_L\dot\vecD_{F'F}^{\dagger}
%appearing in Eq (23). The function squeeze is used to elimininate any
%1D dimension in the array.
DDq = zeros(ExDim,GrDim,2,2,3);
for q = -1:1:1
    for pF = 4:-1:3
        for F = 4:-1:3
            for MF = -F:1:F
                if abs(MF+q) <= pF
                   j = (pF+1)+MF+q+(4-pF)*(2*4+1);
                   k = (F+1)+MF+(F4-F)*(2*F4+1);
                   DDq(:,:,-pF+5,-F+5,q+2) = DDq(:,:,-pF+5,-F+5,q+2)+...
                   (-1)^(pF+1+pJ+Is) * sqrt((2*pJ+1)*(2*F+1))*...
                   Wigner6jcoeff(pF,7/2,pJ,1/2,1,F)*...
                   ClebschGordan(F,1,pF,MF,q,MF+q) * ExBasis(:,j)*GrBasis(k,:);
                end
            end
        end
     end
end
Aux_eL_DDq = squeeze(dot(eL,conj(eq(:,1)))*DDq(:,:,:,:,1)+...
dot(eL,conj(eq(:,2)))*DDq(:,:,:,:,2)+dot(eL,conj(eq(:,3)))*DDq(:,:,:,:,3));
%Now that the main operators are defined, we proceed to calculate the
%effective light-shift Hamiltonian H_{eff}^{LS}, (HlsEFF) as it appears in Eq (23). Since I defined
%the time everywhere else to be in mileseconds, we have to rescale the
%effective Hamiltonian to be in kHz instead of MHz as it was before.

%%EFFECTIVE LIGHT SHIFT HAMILTONIAN
HlsEFF = zeros(GrDim,GrDim);
for F = 4:-1:3
    for pF = 4:-1:3
        AuxHls = squeeze(Aux_eL_DDq(:,:,-pF+5,-F+5));
        HlsEFF = HlsEFF+1/(DeltapFF(-pF+5,-F+5)+1i*GammaLW/2)*(AuxHls' * AuxHls);
    end
end
HlsEFF = (Omega/2)^2*HlsEFF;       
HlsEFF = 10^3*HlsEFF; %needs to be in kHz

%GOING TO ROTATING FRAME and linear interpolation + integration
%The rotating frame is defined by the unitary
%U=exp(-i\omega_RFt(F_z^{(4)}-F_z^{(3)})), Eq (A1a). In the program
%Fzm=F_z^{(4)}-F_z^{(3)}.
Fzm = zeros(dim);
Fzm(1:9,1:9) = Fz4;
Fzm(10:16,10:16) = -Fz3; 
%After going to the rotaing frame, we must apply the RWA. The way we are
%doing so in the code is by explicitly calculating the average of
%U^{\dagger}HU. For this purpose, we use linear interpolation to calculate
%the first order apporximation to the integral \int_0^Tf(t)dt, where f(t)
%is U^{\dagger}(t)H_{eff}^{LS}U(t). Note that the linear interpolation I
%use here gives
%\int_0^Tf(t)dt\approx\frac{1}{T}\sum_j\frac{dt}{2}(f(t_{j-1})+f(t_{j})).
%Also note that \frac{1}{T}=\omega_RF in this case. Finally, we transform
%the averaged effective light-shift Hamiltonian to the superoperator
%notation (SOHlsEFF). For this, we make use of the transformation A\rhoB\rightarrow
%(A\otimes B^T)\vec{rho}, which takes the operators A and B in the standard
%notation to the superoperator picture when they appear, as they do in the
%master equation, in products with the density matrix.
jmax = 2000;%number of terms in the time average
dt = 1/(omegaRF*(jmax)); %time step in time average
AVGHlsEFF = zeros(16);
for j = 1:jmax
    AVGHlsEFF = AVGHlsEFF+expm(+1i*2*pi*omegaRF*(j-1)*dt*Fzm)*HlsEFF*...
    expm(-1i*2*pi*omegaRF*(j-1)*dt*Fzm)+expm(+1i*2*pi*omegaRF*(j)*dt*Fzm)*...
    HlsEFF*expm(-1i*2*pi*omegaRF*(j)*dt*Fzm);
end
AVGHlsEFF = dt/2*AVGHlsEFF;
AVGHlsEFF = omegaRF*AVGHlsEFF;
SOHlsEFF = kron(AVGHlsEFF,eye(16))-kron(eye(16),conj(AVGHlsEFF));

%%JUMP OPERATORS
%As we did with the effective light-shift Hamiltonian, we calculate the
%jump operators in a similar way. The array (W_q) is W_q^{F_bF_a}, Eq (26) where the
%three last indices of (Wq) are Fa, Fb and q. In addition, we define the
%projectors onto the F=3 and F=4 manifolds (PP), and the superoperator jump
%operators (SOWq). We do the latter at the same time as going to the
%rotating frame and doing the RWA. The averaging method used here is the
%same linear interpolation method used above in the light-shift
%Hamiltonian. The reason for doing this part in this way is that there is
%no other way make the RWA on the jump operators individually since they
%always appear in the master equation in products, for
%example, W_q\rho W_q^{\dagger}, and we must average those products to
%corectly diregard the rapidly oscillating terms.
Wq = zeros(GrDim,GrDim,2,2,3);   
Aux_eL_DDq = squeeze(dot(eL,conj(eq(:,1)))*DDq(:,:,:,:,1)+...
dot(eL,conj(eq(:,2)))*DDq(:,:,:,:,2)+dot(eL,conj(eq(:,3)))*DDq(:,:,:,:,3));  
for q = -1:1:1
    for Fb = 4:-1:3
        for Fa = 4:-1:3
            for pF = 4:-1:3
                AuxW = squeeze(DDq(:,:,-pF+5,-Fb+5,q+2));
                Wq(:,:,-Fa+5,-Fb+5,q+2) = Wq(:,:,-Fa+5,-Fb+5,q+2)+(Omega/2)/...
                (DeltapFF(-pF+5,-Fa+5)+1i*GammaLW/2)* AuxW'*...
                squeeze(Aux_eL_DDq(:,:,-pF+5,-Fa+5)); 
            end
        end
    end
end
%projectors
PP = zeros(dim,2);
PP(1:9,1:9,1) = eye(dimF4); %Projector in F=4 manifold
PP(10:16,10:16,2) = eye(dimF3);%Projector in F=3 manifold
%%Going to the rotating frame 
%Wq Superoperators
SOWq = zeros(dim^2,dim^2);
for j = 1:jmax
    Uj = expm(-1i*2*pi*omegaRF*(j-1)*dt*Fzm);
    Ujp = expm(-1i*2*pi*omegaRF*(j)*dt*Fzm);
    for q = -1:1:1
        for F = 4:-1:3
            for F1 = 4:-1:3
                func = kron(Uj'*squeeze(Wq(:,:,-F1+5,-F+5,q+2))*Uj*...
                squeeze(PP(:,:,-F1+5)),Uj.'*conj(squeeze(Wq(:,:,-F1+5,-F+5,q+2)))...
                *conj(Uj)*squeeze(PP(:,:,-F1+5)))+kron(Ujp'*squeeze(Wq(:,:,-F1+5,...
                -F+5,q+2))*Ujp*squeeze(PP(:,:,-F1+5)),Ujp.'*conj(squeeze...
                (Wq(:,:,-F1+5,-F+5,q+2)))*conj(Ujp)*squeeze(PP(:,:,-F1+5)));                    
                 SOWq = SOWq+func;
            end
        end
    end
end   
for j = 1:jmax
    Uj = expm(-1i*2*pi*omegaRF*(j-1)*dt*Fzm);
    Ujp = expm(-1i*2*pi*omegaRF*(j)*dt*Fzm);
    for q = -1:1:1
        for F1 = 4:-1:3
            for F2 = 4:-1:3
                if F1~=F2
                   func = kron(Uj'*squeeze(Wq(:,:,-F2+5,-F2+5,q+2))*Uj*...
                   squeeze(PP(:,:,-F2+5)),Uj.'*conj(squeeze(Wq(:,:,-F1+5,-F1+5,...
                   q+2)))*conj(Uj)*squeeze(PP(:,:,-F1+5)))+kron(Ujp'*squeeze...
                   (Wq(:,:,-F2+5,-F2+5,q+2))*Ujp*squeeze(PP(:,:,-F2+5)),Ujp.'*...
                   conj(squeeze(Wq(:,:,-F1+5,-F1+5,q+2)))*conj(Ujp)*...
                   squeeze(PP(:,:,-F1+5)));
                   SOWq = SOWq+func;
                end
            end
        end
     end
end
SOWq = GammaLW*10^3*SOWq;
SOWq = dt/2*SOWq;
SOWq = omegaRF*SOWq;
%We created all the necessary operators that are going to be used later and
%save them in parametersRFuW.mat file.
save parametersRFuW SOHlsEFF SOWq F4 F3 dimF3 dimF4 dim Fx3 Fy3 Fz3 Fx4 Fy4...
Fz4 grel C1 C2 C0 DeltapFF omegaRF GrSplit
clear
\end{lstlisting}

\subsection{Full master equation solver}
The following script, \texttt{RFuWEvolution.m}, solves the full master equation, Eq. (\ref{eq:mastereq}), for Schr\"{o}dinger picture evolution and Eq. (\ref{eq:FullHeisenbergEvol}) for Heisenberg picture evolution for the atomic system discussed in Chapter \ref{ch:Atoms}. The fitted control parameters are input to this function and either $\cO(t)$ or $\rho(t)$ and $M(t)$ are produced as outputs.

\begin{lstlisting}[language=Matlab]
function [OP,M,rmat]=RFuWEvolution(Omegax,Omegay,Omegauw,Omega0,Omega,DuW,...
Phixin,Phiyin,Phiuwin,uWrate,RFrate,rho0,tmax)
%This function solves the full master equation and calculates the...
%Heisenberg evolution for the Faraday observables

%First, we load all the time-independent operators of the master equation
%that were calculated by the the script 'RFuWPreparation.m'. This operators
%are stored in the binary file 'parametersRFuW.mat'. Then, we rescale
%appropriately the field amplitudes that are inhomogeneous.
load parametersRFuW
%deltat is the time step for the integration in miliseconds. noperators is
%the number of time points we consider in the simulation, i.e., the number
%of discretized Heisenberg observables we measure.
deltat = 1e-3;
noperators = tmax*floor(1/deltat);
%Some basic definitions are made for some operators that will be needded
%later.
%RWA correction terms need this
Fx4F = zeros(2*(F4+F3)+2,2*(F4+F3)+2);
Fx4F(1:9,1:9) = Fx4;
Fx3F = zeros(2*(F4+F3)+2,2*(F4+F3)+2);
Fx3F(10:16,10:16) = Fx3;
Fy4F = zeros(2*(F4+F3)+2,2*(F4+F3)+2);
Fy4F(1:9,1:9) = Fy4;
Fy3F = zeros(2*(F4+F3)+2,2*(F4+F3)+2);
Fy3F(10:16,10:16) = Fy3;
Fz4F = zeros(2*(F4+F3)+2,2*(F4+F3)+2);
Fz4F(1:9,1:9) = Fz4;
Fz3F = zeros(2*(F4+F3)+2,2*(F4+F3)+2);
Fz3F(10:16,10:16) = Fz3;
%MEASUREMENT OPERATOR FOR FARADAY ROTATION As it appears in Eq (20c)
OP=zeros((2*(F4+F3)+2)^2,noperators+1); %superoperator
O0=zeros(2*(F4+F3)+2,2*(F4+F3)+2);
a=C1(1,1)/DeltapFF(1,1)+C1(2,1)/DeltapFF(2,1);
b=C1(1,2)/DeltapFF(1,2)+C1(2,2)/DeltapFF(2,2);
O0(1:2*F4+1,1:2*F4+1)=a/b*Fz4;
O0(2*F4+2:2*(F4+F3)+2,2*F4+2:2*(F4+F3)+2)=Fz3;%%%%%defines initial operator
OP(:,1)=reshape(O0.',(2*(F4+F3)+2)^2,1); %initial superoperator
%%uWTRANSITION
%The resonant microwave transition is tuned to the streched states |4,4>,
%(V44) and |3,3>, (V33), which define a two-level system with Pauli
%matrices \sigma_x=|3,3><4,4|+|4,4><3,3|, \sigma_y=-i|3,3><4,4|+i|4,4><3,3|,
%and \sigma_z=|3,3><4,4|-|4,4><3,3|.
V44 = zeros(2*(F4+F3)+2,1);
V33 = zeros(2*(F4+F3)+2,1);
V44(2*F4+1) = 1;
V33(2*(F4+F3)+2) = 1;
Sigmax = V44*V33'+V33*V44';
Sigmay = 1i*V44*V33'-1i*V33*V44';
%Here, we calculate the free Hamiltonian H_0 (H0). Here we include the
%second order Zeeman correction as it appears on the paper, Eq (A2). 
%Some definitions: ge=electron g-factor, gi=nuclear g-factor,
%PpMPm=P_4-P_3.
ge = 2.0023193043622;
gi = -0.00039885395;
Fz3F = zeros(2*(F4+F3)+2,2*(F4+F3)+2);
Fz3F(2*F4+2:2*(F4+F3)+2,2*F4+2:2*(F4+F3)+2) = Fz3;
Fz4F = zeros(2*(F4+F3)+2,2*(F4+F3)+2);
Fz4F(1:9,1:9) = Fz4;
PpMPm = diag([1 1 1 1 1 1 1 1 1 -1 -1 -1 -1 -1 -1 -1]);
Fz4sqr = diag([diag(Fz4^2)' 0 0 0 0 0 0 0]);
Fz3sqr = diag([0 0 0 0 0 0 0 0 0 diag(Fz3^2)']);
%if there is only RF on and the microwave fields are off, we go to the
%rotating frame only with respect to the RF frequency. Otherwise we use the
%expression for H0 that appears in Eq (A2)
if (Omegauw==0) && (Omegax~=0 ||Omegay~=0)%%Use RWA only in RF
    %beta here is -\alpha as defined below Eq (32). xz is the x defined
    %below the same equation. Q0 is the superoperator version of the
    %commutator 2\pi[H0,\rho] without the \rho, as it has been removed from
    %the description due to the fact that we only care about the evolution
    %of the superoperator map.
    beta = -(Omega0)^2/(GrSplit*1000)*(ge-gi)^2/(7*gi+ge)^2;
    DeltaRF = omegaRF-Omega0;%RF detuning
    xz = 8*((ge-gi)/(7*gi+ge))*(Omega0/(GrSplit*1000));
    DetTERMS = -DeltaRF*(Fz4F-Fz3F);
    H0 = Omega0*(1-abs(grel))*Fz3F +beta*Fz4sqr-beta*Fz3sqr+...
    (GrSplit*1000)/2*(1+xz^2/2)*PpMPm;
    H0 = H0+DetTERMS;
    Q0 = 2*pi*(kron(H0,eye(16))-kron(eye(16),H0.'));
else
    %beta here is -\alpha as defined below Eq (32).
    beta = -(Omega0)^2/(GrSplit*1000)*(ge-gi)^2/(7*gi+ge)^2;
    %To account for the off-resonat effect that the microwave transition has on
    %the non-resonant levels, we add to the Hamiltonian the AC Zeeman shift (ACZ)
    %correction as it appears in the second term of Eq (36). Here (V3j) is
    %|3,j> and (V4k) is |4,k>.
    %%AC Zeeman Shift
    ACZ = zeros(16);
    for m = 2:-1:-3
        V3j = zeros(16,1);
        V4k = zeros(16,1);
        jjj = m+13;
        kkk = m+6;
        V3j(jjj) = 1;
        V4k(kkk) = 1;
        ACZ = ACZ+(V3j*V3j'-V4k*V4k')*(abs(ClebschGordan(3,1,4,m,1,m+1)))^2/(m-3);
    end
    ACZ = (Omegauw)^2/(8*Omega0)*ACZ;%
    %Q0 is the superoperator version of the
    %commutator 2\pi[H0+ACZ,\rho] without the \rho, as it has been removed from
    %the description due to the fact that we only care about the evolution
    %of the superoperator map.
    DeltaRF=omegaRF-Omega0;%RF detuning
    DeltauW=DuW-(4+3*abs(grel))*Omega0+7*beta;%uW detuning
    DetTERMS=(3.5*DeltaRF-0.5*DeltauW)*PpMPm-DeltaRF*(Fz4F-Fz3F);
    H0 = Omega0*(1-abs(grel))*Fz3F+(1.5*Omega0*(1-abs(grel))+12.5*...
    Omega0^2/(GrSplit*1000)*((ge-gi)/(7*gi+ge))^2)*PpMPm+beta*Fz4sqr-beta*Fz3sqr;
    H0 = H0+DetTERMS;
    H0 = H0+ACZ;
    Q0 = 2*pi*(kron(H0,eye(16))-kron(eye(16),H0.'));
end
%I added here the capability for the code to evolve an initial density
%operator (rho0) as a function of time (rmat) in addition to evolving just
%the Heisenberg picture observables that are required for tomography. The
%function 'reshape' is used to make the superoperator version of rho0 and
%rmat is initialized at rho0.
sorho0=reshape(rho0.',256,1);
rmat=zeros(16,16,noperators+2);
rmat(:,:,1)=rho0;
%The following section calculates the correct form of the control waveforms
%that the code uses. From the input waveforms, Phixin, Phiyin and Phiuwin,
%and RFrate and uWrate, this section computes an efficient arrangement of
%the waveforms so the minimum number of matrix exponentiations are made
%during integration of the master equation.
AuxVar1=sort([RFrate uWrate]);
AvarRF=[RFrate:RFrate:floor(tmax/deltat)];
AvaruW=[uWrate:uWrate:floor(tmax/deltat)];
IndVec=union(AvarRF,AvaruW);
IndVec(length(IndVec)+1)=tmax/deltat+1;%
Phix(1)=Phixin(1);
Phiy(1)=Phiyin(1);
Phiuw(1)=Phiuwin(1);
j=1;
l=1;
for k=1:length(IndVec)-1  
    if mod(IndVec(k),AuxVar1(2))==0
        Phix(k+1)=Phixin(j+1);
        Phiy(k+1)=Phiyin(j+1);
        j=j+1;
    else
        Phix(k+1)=Phixin(j);
        Phiy(k+1)=Phiyin(j);
    end
    if mod(IndVec(k),AuxVar1(1))==0
        Phiuw(k+1)=Phiuwin(l+1);
        l=l+1;
    else
        Phiuw(k+1)=Phiuwin(l);
    end
end
%In case the microwve power changes in time, use the commented portion
%below.
T=IndVec/1000;
OmegauW = Omegauw*ones(length(T),1);
%%Time varying microwave power
%OmegauW = Omegauw + (88.6+271.3*log10(T+0.5))/1000;

%SOLVER
%Here we integrate the master equation and find the evolution of the
%Heisenberg picture observables and the initial state rho0.
S=eye(256);
for r=1:length(IndVec)
    %Hint = H_{RF}+H_{\mu w}. H_{RF} is given by Eq (A23) and includes the
    %RWA up to second order correction. H_{\mu w} is given by the first
    %term of Eq (36)
    Hint =0.5*Omegax*(cos(Phix(r))*(Fx4F-abs(grel)*(1-(Omega0*(1-abs(grel)))/...
    (2*omegaRF))*Fx3F)-sin(Phix(r))*(Fy4F+abs(grel)*( 1+(Omega0*(1-abs(grel)))/...
    (2*omegaRF))*Fy3F))+0.5*Omegay*(cos(Phiy(r))*(Fy4F-abs(grel)*...
    (1-(Omega0*(1-abs(grel)))/(2*omegaRF))*Fy3F)+sin(Phiy(r))*(Fx4F+abs(grel)*...
    (1+(Omega0*(1-abs(grel)))/(2*omegaRF))*Fx3F))+0.25*(Omegax*DeltaRF/omegaRF)*...
    (sin(Phix(r))*Fx4F-abs(grel)*cos(Phix(r))*Fx3F-cos(Phix(r))*Fy4F...
    -abs(grel)*sin(Phix(r))*Fy3F)+0.25*(Omegay*DeltaRF/omegaRF)*...
    (cos(Phiy(r))*Fx4F + abs(grel)*sin(Phiy(r))*Fx3F+sin(Phiy(r))*Fy4F+...
    abs(grel)*cos(Phiy(r))*Fy3F)+(1/(16*omegaRF))*( Omegax^2*(1-2*cos(2*Phix(r)))...
    +Omegay^2*(1-2*cos(2*Phiy(r))) + 2*Omegax*Omegay*sin(Phix(r)-Phiy(r)) )*Fz4F...
    -(grel^2/(16*omegaRF))*(Omegax^2*(1-2*cos(2*Phix(r)))+Omegay^2*...
    (1-2*cos(2*Phiy(r)))-2*Omegax*Omegay*sin(Phix(r)-Phiy(r)))*Fz3F...
    +0.5*OmegauW(r)*cos(Phiuw(r))*Sigmax + 0.5*OmegauW(r)*sin(Phiuw(r))*Sigmay;
    %Qint is the superoperator version of 2\pi[Hint,\rho].
    %In other words Qint*sorho is the same as 2\pi[Hint,\rho]
    Qint=sparse(2*pi*(kron(Hint,eye(16))-kron(eye(16),Hint.')));
    %Putting all together in superoperator notation: Q*sorho is the same as
    %the left hand side of the master equation, Eq (22) at a particular
    %time roughly given by the for-loop counter 'r'.
    Q=-1i*(Q0+Qint)+2*pi*Omega^2*(-1i*SOHlsEFF+SOWq);
    %The master equation is integrated by exponentiating Q once in the time
    %interval that the Hamiltoninan is constant (piece-wise constant), R,  and
    %multiplying R times itself as many times as time slices we consider in
    %that particular interval. S is the superoperator map integrated. OP is
    %a matrix whose colunms are the superoperator representations of the
    %Heisenberg picture observables as a function of time. Additionally,
    %\rho(t) = rmat is stored for a particular input initial state \rho_0
    R=expm(Q*deltat);
    R=R';
    if r==1
        for j=1:IndVec(r)
            S=S*R;
            OP(:,j+1)=S*squeeze(OP(:,1));
            rvecT=sorho0'*S;
            rmat(:,:,j+1)=reshape(rvecT,16,16);
        end
    else
        for j=IndVec(r-1)+1:IndVec(r)
            S=S*R;
            OP(:,j+1)=S*squeeze(OP(:,1));
            rvecT=sorho0'*S;
            rmat(:,:,j+1)=reshape(rvecT,16,16);
        end
    end
end
%After OP is calculated we calculate the measurement record as a function
%of time, M(t), Eq (2) minus the noise term, by using the fact that in
%superoperator notation the Tr(\rho_0 O_i) is the scalar product between
%the columns of OP and sorho0
M=zeros(noperators+2,1);
for j=1:noperators+2
    M(j)=sorho0'*squeeze(OP(:,j));
end
\end{lstlisting}

\section{Quantum tomography code}\label{sec:AppQTcode}
Here we show the full reconstruction code as used in this dissertation. The script is called \texttt{Reconstruction.m} and includes the two methods of QT discussed in this work.
\begin{lstlisting}[language=Matlab]
%This script calculates an estimate of an unknown quantum state given the measurement record and the Heisenberg picture operators. It uses least squares and compressed sensing methods.
var1=3;
%var1=1, standard, 2-step reconstruction
%var1=2, 1-step reconstruction
%var1=3, compressed sensing

%Dimension of Hilbert space is d
d=16;
AuxInhomoVar=length(size(OPc));
%AuxInhomoVar=2, No inhomogeneity
%AuxInhomoVar=3, 1 type of inhomogeneity
%AuxInhomoVar=4, 2 types of inhomogeneity
if AuxInhomoVar==2
    %OPF is the filtered version of OPc. Since there is no inhomogeneities
    %in this case, no averaging of Heisenberg observables is needed. The
    %function filter is used with parameters bz and az of the appropriate
    %digital filter. I restrict the number of Heisenberg operators that are
    %going to be used for the reconstruction depending on the number of
    %points in the measurement record, M.
    OPF=zeros(d^2,2001);
    for j=1:d^2
        OPF(j,:)=filter(bz,az,OPc(j,:));
    end
    OP=OPF(:,1:length(M));
elseif AuxInhomoVar==3
    %When there is 1 inhomogeneous parameter in the problem, we need to
    %average the Heisenberg operators over the distribution of that
    %inhomogeneous parameter, (InhomogeneityDistribution). The way I do it
    %is by using a first order integration rule.
    OPavg=zeros(d^2,2001);
    for j=1:2001
        OPavg(:,j)=squeeze(OPc(:,j,:))*InhomogeneityDistribution;
    end
    OPF=zeros(d^2,2001);
    for j=1:d^2
        OPF(j,:)=filter(bz,az,OPavg(j,:));
    end
    OP=OPF(:,1:length(M));
    
 elseif AuxInhomoVar==4
    %When there is 2 inhomogeneous parameters in the problem, we need to
    %average the Heisenberg operators over the distribution of those
    %inhomogeneous parameters. The way I do it
    %is by using a first order integration rule.
    OPavg=zeros(d^2,2001);
    for k=1:d^2
        parfor j=1:2001
            OPavg(k,j)=InhomogeneityDistribution1'*squeeze(OPc(k,j,:,:))*...
            InhomogeneityDistribution2;
        end
    end
    OPF=zeros(d^2,2001);
    for j=1:d^2
        OPF(j,:)=filter(bz,az,OPavg(j,:));
    end
    OP=OPF(:,1:length(M)); 
     
end

if var1==1 %Standard 2-step reconstruction
    %hb is a matrix array whose columns are the super-operator versions of
    %E_\alpha (E in the code), which are traceless, orthogonal and Hermitian basis elements as defined
    %above Eq (3). NewOP represents the projection of OP in the subspace of
    %operators spaned by {E_\alpha} which does not include the Identity matrix. If the Hilbert space
    %has dimension d, then hb is an array of d^2 by d^2-1 elements (since the
    %identity matrix is not included in the basis).
    hb = hermitian_basis_S(d);
    NewOP=real(hb'*OP);
    %Here we calculate explicitly, in matrix form, the E_\alpha as they will be
    %used later during the convex program.
    E=zeros(d,d,d^2-1);
    for j=1:255
        E(:,:,j)=reshape(hb(:,j)',d,d);
    end
    SolverType=1;
    %SolverType=1, Traditional Solver
    %SolverType=2, "low rank" solution, i.e., at most rank(NewOP') non zero velues in solution vector rhoest
    %SolverType=3, minimal norm solution, i.e., norm(rhoest) is minimun
    if SolverType==1
        %This part calculates the unconstrained least squares (maximum likelihood) estimate of the
        %density matrix as shown in Eq (7), where rhoest is \vec{r}_{ML}.
        NewR=NewOP*NewOP';
        rhoest=pinv(NewR)*NewOP*(M);
    elseif SolverType==2
        rhoest=NewOP'\M;
    elseif SolverType==3
        rhoest=pinv(NewOP')*M;
    end
    %This section solves the convex program defined in Eqs (9) and (10). It
    %uses the solver CVX that must be installed in the machine where this code
    %is intended to run. For information about CVX go to http://cvxr.com/cvx/
    %The solution of the optimization process is the variable x which is
    %\vec{\bar{\r}} in Eqs (8), (9), and (10).
    cvx_begin
    % cvx_precision high
    variable x(d^2-1);
    s = zeros(d,d);
    
    for j=1:d^2-1
        s = s+x(j)*squeeze(E(:,:,j));
    end
    
    minimize( quad_form(x-rhoest,NewR) );
    
    subject to
    eye(d)/d+s==hermitian_semidefinite(d);
    cvx_end
    %After the convex program is solved, we use x to calculate the actual
    %matrix form of the estimated density matrix, rho in the code, \bar{\rho} in the paper.
    s=zeros(d,d);
    for j=1:d^2-1
        s=s+x(j)*squeeze(E(:,:,j));
    end
    rho=eye(d)/d+s;
    %Finally, we evaluate the fidelity of the reconstruction as defined in Eq
    %(18)
    fidelity = real((trace(sqrtm(sqrtm(rho)*rho0*sqrtm(rho)))).^2)    
elseif var1==2 %1-step recontruction
    hb = hermitian_basis_S(d);
    NewOP=real(hb'*OP);
    E=zeros(d,d,d^2-1);
    for j=1:d^2-1
        E(:,:,j)=reshape(hb(:,j)',d,d);
    end
    %tic
    cvx_begin
        % cvx_precision high
        variable x(d^2-1);
        s=zeros(d,d);
        for j=1:d^2-1
            s=s+x(j)*squeeze(E(:,:,j));
        end
        minimize( norm(M-NewOP'*x,2) );
        subject to
        eye(d)/d+s==hermitian_semidefinite(d);
    cvx_end
    %toc
    s=zeros(d,d);
    for j=1:d^2-1
        s=s+x(j)*squeeze(E(:,:,j));
    end
    rho=eye(d)/d+s;
    fidelity=real((trace(sqrtm(sqrtm(rho)*rho0*sqrtm(rho)))).^2)
elseif var1==3 %Compressed sensing
    hb=zeros(d^2);
    hb(:,1)=reshape(1/sqrt(d)*eye(d),d^2,1);
    hb(:,2:d^2) = hermitian_basis_S(d);
    NewOP=real(hb'*OP);
    E=zeros(d,d,d^2);
    for j=1:d^2
        E(:,:,j)=reshape(hb(:,j)',d,d);
    end
    %tic
    cvx_begin
        % cvx_precision high
        variable x(d^2);
        s=zeros(d,d);
        for j=1:d^2
            s=s+x(j)*squeeze(E(:,:,j));
        end
        minimize( x(1) );
        subject to
                s==hermitian_semidefinite(d);
                norm(M-NewOP'*x,2)<=0.3;
    cvx_end
    %toc
    s=zeros(d,d);
    for j=1:d^2
        s=s+x(j)*squeeze(E(:,:,j));
    end
    rho=s/trace(s);
    fidelity=real((trace(sqrtm(sqrtm(rho)*rho0*sqrtm(rho)))).^2)
    HSdistance = sqrt(trace((rho-rho0)^2))
end
\end{lstlisting}
	


\end{document}